\pdfoutput=1
\documentclass[aps,rmp, twocolumn, superscriptaddress, amsmath, amssymb, floatfix]{revtex4-1}

\usepackage{natbib}

\usepackage{graphicx}
\usepackage{dcolumn}
\usepackage{bm}     
\usepackage{subfigure}
\usepackage{amsfonts}
\usepackage[usenames]{color}
\usepackage{rotating}
\usepackage{fontenc}
\usepackage{cancel}
\usepackage{cases}
\usepackage{latexsym,listings}

\usepackage{amsthm}
\usepackage{hyperref}
\usepackage[normalem]{ulem}
\usepackage{amsmath}
\usepackage{multirow}

\usepackage{chemarrow}
\usepackage{chemarr}

\definecolor{darkred}{RGB}{139,0,0}
\definecolor{chartreuse}{RGB}{127,255,0}
\definecolor{goldenrod}{RGB}{218,165,32}
\definecolor{gray}{RGB}{127,127,127}
\definecolor{Magenta}{RGB}{255, 0,255}
\definecolor{Orange}{RGB}{255,165, 0}
\definecolor{Gray}{RGB}{127,127,127}

\definecolor{red}{RGB}{0,0,0}
\definecolor{blue}{RGB}{0,0,0}

\newcommand{\be}{\begin{equation}}
\newcommand{\ee}{\end{equation}}
\newcommand{\bea}{\begin{eqnarray}}
\newcommand{\eea}{\end{eqnarray}}
\newcommand{\bw}{\begin{widetext}}
\newcommand{\ew}{\end{widetext}}
\newcommand{\mm}{\mathrm}
\newcommand{\p}{\partial}
\newcommand{\ud}{\mm{d}}

\newcommand{\nn}{\nonumber\\}
\newcommand{\bi}{\begin{itemize}}
\newcommand{\ei}{\end{itemize}}

\newcommand{\mb}{\mathbf}
\newcommand{\kout}{k_\mm{out}}
\newcommand{\kin}{k_\mm{in}}
\newcommand{\wh}{\widehat}

\newcommand{\kmean}{\langle k \rangle}

\newcommand{\mA}{{\mb A}} 
\newcommand{\mB}{{\mb B}} 
\newcommand{\mC}{{\mb C}} 
\newcommand{\mD}{{\mb D}}

\newcommand{\mj}{{\mathcal{J}}} 
 
\newcommand{\mI}{{\mb I}} 
 
\newcommand{\mL}{{\mb L}}

\newcommand{\my}{{\mb y}} 
\newcommand{\mx}{{\mb x}} 
\newcommand{\mf}{{\mb f}}
\newcommand{\mg}{{\mb g}}
\newcommand{\mz}{{\mb z}} 

\newcommand{\mF}{{\mb F}} 
\newcommand{\mG}{{\mb G}} 
 
\newcommand{\mh}{{\mb h}} 
\newcommand{\mX}{{\mb X}}

\newcommand{\mP}{{\mb P}}

\newcommand{\ms}{{\mb s}}
\newcommand{\mv}{{\mb v}}
\newcommand{\me}{{\mb e}}

\newcommand{\mxi}{{\boldsymbol \xi}}

\newcommand{\x}{{\mb x}}

\newcommand{\dx}{\bf{\delta x}}

\newcommand{\target}{\x^*}

\newcommand{\ad}{\mm{ad}}

\begin{document}
\title{Control Principles of Complex Networks}

\author{Yang-Yu Liu}  

\affiliation{Channing Division of Network Medicine, Brigham and Women's Hospital, Harvard Medical School, Boston, Massachusetts 02115, USA}
\affiliation{Center for Cancer Systems Biology, Dana-Farber Cancer Institute, Boston, Massachusetts 02115, USA}

\author{Albert-L\'{a}szl\'{o} Barab\'{a}si}
\affiliation{Center for Complex Network Research and Departments of
  Physics, Computer Science and Biology, Northeastern University,
  Boston, Massachusetts 02115, USA} 
\affiliation{Center for Cancer Systems Biology, Dana-Farber Cancer Institute, Boston, Massachusetts 02115, USA}
\affiliation{Department of Medicine, Brigham and Women's Hospital, Harvard Medical School, Boston, Massachusetts 02115, USA}
\affiliation{Center for Network Science, Central European University, Budapest 1052, Hungary}

\date{\today}

\begin{abstract}
A reflection of our ultimate understanding of a complex system is our
ability to control %
its behavior. 
Typically, control %
has multiple 
prerequisites: it
requires an accurate %
map of %
the network that governs %
the interactions between the system's 
components, %
a quantitative description of 
the dynamical laws that govern %
the temporal behavior of each component, %
and an ability to influence the state and temporal behavior of a  %
selected subset of the components.  
With deep roots in nonlinear dynamics and 
control theory, notions of
control and controllability have %
taken a new life recently in the study of complex networks, inspiring %
several fundamental questions: What are the control principles of complex
systems? How do networks %
organize themselves to balance control
with functionality? 
To address these here we %
review recent advances on the controllability and the control of
complex networks, %
exploring %
the intricate 
interplay between a system's structure, %
captured by its network
topology, and the dynamical laws that govern the interactions between
the components. 
We match %
the pertinent mathematical results %
with empirical findings and applications. 
We show that  uncovering
the control principles of complex %
systems can 
help us explore and ultimately understand the fundamental laws that
govern their behavior. %
\end{abstract} 

 \maketitle
 \tableofcontents 
 \section{Introduction}

To understand the mechanisms governing the behavior of a complex
system, we must be able to measure its state variables and to 
mathematically model the dynamics of each of the system's
components. Consequently, the traditional theory of complex systems
has predominantly focused on the measurement and the modeling
problem.  
Recently, however, questions pertaining to the control of complex
networks became an important %
research topic in %
statistical physics~\citep{Liu-Nature-11,Nepusz-NP-12,Yan-PRL-12,Cornelius-NC-2013,Sun-PRL-2013,Ruths-Science-14}.   
This %
interest is driven by 
the challenge %
to %
understand the fundamental control principles
of an arbitrary self-organized %
system. 
Indeed, there is an 
increasing realization that the design principles of many complex
systems are %
genuinely determined by the need to control their
behavior. For example, we cannot divorce the understanding of
subcellular networks %
from questions on how the %
activity or the concentrations 
of genes, proteins, and other
biomolecules are controlled. 
Similarly, the structure and the daily activity of 
an organization %
is deeply determined by governance and {\color{red}leadership} principles. %
Finally, %
to maintain the functionality of large technological systems, %
like the power grid or the Internet, and to adapt their %
functions to the shifting needs of the users, we must solve %
a host of %
control questions. These and many similar applications have led to a burst of research
activity, aiming to uncover to what degree 
the topology of a real network behind %
a complex system encodes our
ability to control it. %
The current advances %
in controlling %
complex systems were facilitated %
by %
progress in network science, offering a quantitative framework %
to understand the design
principles of complex
networks~\citep{Albert-RMP-02,Dorogovtsev-RMP-08,%
Watts-Nature-98,Barabasi-Science-99,Milo-Science-02,Newman-PNAS-06,Toroczkai-Nature-04}.  
On one end, these advances have shown that the topologies of most real
systems share numerous %
universal characteristics. Equally important was
the realization that these universal topological features
are the result of the common dynamical principles that govern their
emergence and %
growth. 
At the same time we learned that the topology fundamentally affects the
dynamical processes taking place on these networks, from
epidemic spreading~\citep{Pastor-Satorras-PRL-01,Cohen-PRL-00} to
synchronization%
~\citep{Nishikawa-PRL-03,Wang-BC-05}. Hence, 
it is fair to expect that the %
network topology of  a system %
also affects our ability to control it. %

\begin{figure}[t!]
\centering
\includegraphics[width=0.4\textwidth]{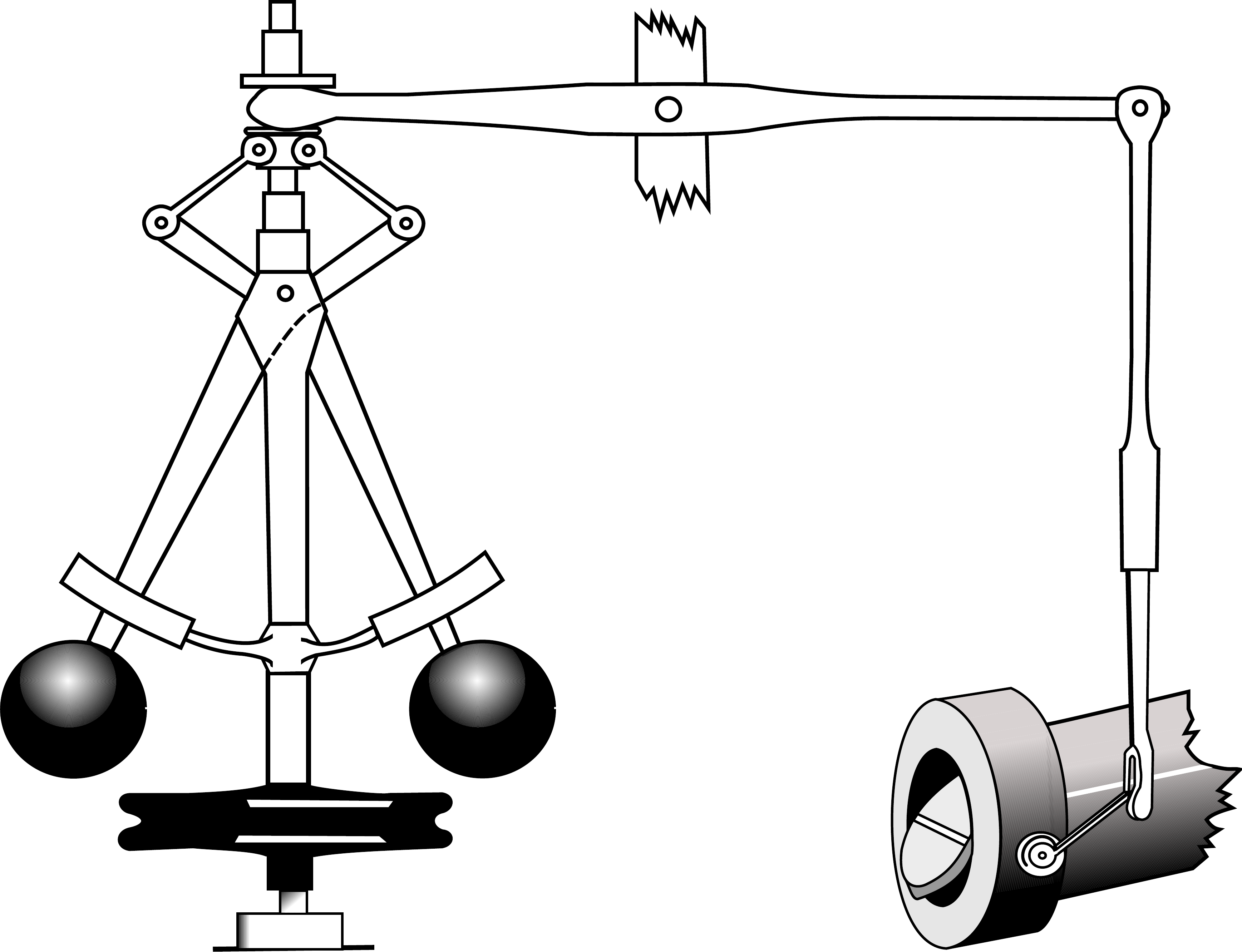}
\caption{%
Feedback control. 
A centrifugal governor represents a %
practical realization of a %
feedback process designed to control %
the speed of an engine. 
It uses velocity-dependent centrifugal force to regulate 
the release %
of fuel
(or working fluid), %
maintaining a near-constant speed of the engine. %
It %
has been frequently used in 
steam engines, %
regulating the
admission of steam into the cylinder(s). %
}\label{fig:cg}
\end{figure}

While the term ``control'' is frequently used in numerous disciplines
with rather diverse %
meanings, here %
we employ it in the strict mathematical sense of control theory,
a highly developed interdisciplinary branch of engineering and
mathematics. 
Control theory asks %
how to influence the behavior of a
dynamical system with appropriately chosen inputs so that the system's output follows a
desired trajectory or final state. %
A key notion in control theory is the %
\emph{feedback} process:  
The difference between the actual and desired output is 
applied as feedback to the system's input, forcing %
the system's
output to converge %
to the desired output. 
Feedback control has deep roots in physics and engineering. %
For example, the centrifugal governor, one of the first practical
  control devices, has been used to regulate the pressure and distance between millstones in windmills since the
  17th century and was %
used by James Watt to to maintain the steady velocity of a steam
engine.  
The feedback mechanism relies %
on a system of balls rotating
around an axis, with a velocity proportional to the engine velocity. 
When the rotational velocity increases, the centrifugal force
pushes the balls %
farther from the axis, opening valves to let %
the vapor escape. This lowers the pressure inside the boiler, slowing
the engine %
(Fig.~\ref{fig:cg}). 
The first definitive mathematical description of the centrifugal
governor used in Watt's steam engine was provided %
by James Maxwell in
1868, proposing some of the best %
known feedback control mechanisms in use today 
\citep{Maxwell-1868}.

The subsequent need to design well controlled engineered systems has resulted in
a mathematically sophisticated array of control theoretical tools,
which are today widely applied in the design of electric circuits,
manufacturing processes, communication systems, airplanes, %
spacecrafts and robots. %
Furthermore, since issues of regulation and control are central to the study of
biological and biochemical systems, the concepts and tools developed
in control theory have proven useful in the study of biological
mechanisms and disease treatment~\citep{Sontag-SB-04,Iglesias-Book-2009}. For
example, feedback control %
by transcranial electrical stimulation 
has been used %
to restore the aberrant
brain activity during epileptic seizures~\citep{Berenyi-Science-12}. 

Modern control theory %
heavily relies on the %
state space
representation (also known as the ``time-domain
approach''), where a control system is described by a set of inputs,
outputs and state variables connected by a set of %
differential (or difference) equations. 
The concept of \emph{state}, introduced into control theory by Rudolf
Kalman in 1960s, is a mathematical entity that
mediates between the inputs and the outputs of a dynamical system, 
while emphasizing 
the notions of causality and internal structure. 
Any state 
of a dynamical system can then be represented as a vector in
the state space whose axes are the state
 variables. 
The concept of the state space was inspired by the %
 \emph{phase space} concept %
used in physics, %
developed in the late
 19th century by %
Ludwig Boltzmann, Henri Poincar\'{e}, and Willard Gibbs.

For a %
nonlinear dynamical system, %
we can write the state space
model as 
\begin{subnumcases}
\mathbf{\dot{\mx}}(t) = \mathbf{f}(t, \mx(t), {\bf u}(t) ; \Theta) \label{eq:X}\\
\mathbf{y}(t) = \mathbf{h}(t, \mx(t), {\bf u}(t) ; \Theta) \label{eq:Y}
\end{subnumcases}
where the state vector $\mx(t) \in \mathbb{R}^N$ represents the 
internal state of the system at time $t$, the input vector ${\bf u}(t) \in
\mathbb{R}^M$ captures the known input signals, and the output vector
$\my(t) \in \mathbb{R}^R$ captures the set of experimentally
measured %
variables. 
The functions $\mf(\cdot)$ and $\mh(\cdot)$ are generally nonlinear, 
and $\Theta$ collects %
the system's parameters. 
Equations (\ref{eq:X}) and (\ref{eq:Y}) are called the state and output
equations, respectively, and %
describe the dynamics of a wide
range of complex systems. 
For example, in metabolic networks the state vector $\mx(t)$ represents
the concentrations of all metabolites in a cell, the inputs ${\bf
  u}(t)$ represent regulatory signals modulated through enzyme
abundance, and the outputs $\my(t)$ are experimental assays capturing
the concentrations of a particular set of secreted species or the
fluxes of a group of reactions of interest. %
In communication systems $\mx(t)$ is the amount of information
processed by a node and $\my(t)$ is the measurable traffic on
selected links or nodes.

A significant body %
of work in control theory focuses on %
linear systems~\citep{Kailath-Book-80}, %
{\color{red}
described by 
\begin{subnumcases}
\mathbf{\dot{\mx}}(t) = \mA(t) \, \mx (t) + \mB(t) {\bf u}(t) \label{eq:X_LTV}\\
\my(t) = \mC(t) \, \mx (t) + \mD(t) {\bf u}(t), \label{eq:Y_LTV}
\end{subnumcases}
where (\ref{eq:X_LTV}) and (\ref{eq:Y_LTV}) represent so-called linear
time-varying (LTV) systems. 
}
Here, 
$\mA(t) \in \mathbb{R}^{N\times N}$ is the state or system matrix,
telling us which components interact with each other and the strength
or the nature of those interactions; 
$\mB(t) \in \mathbb{R}^{N\times M}$ is the input matrix; 
$\mC(t) \in \mathbb{R}^{R\times N}$ is the output matrix; 
$\mD(t) \in \mathbb{R}^{R\times M}$ is the feedthrough or feedforward matrix. 
In case $\mA(t)$, $\mB(t)$, $\mC(t)$ and $\mD(t)$ are constant
matrices, (\ref{eq:X_LTV}) and (\ref{eq:Y_LTV}) represent a linear
time-invariant (LTI) system, which is the starting point of most
control theoretical approaches. 
Note that since we typically know %
${\bf u}(t)$ and $\mD(t)$, we can simply
define a new output vector $\tilde{\my}(t) \equiv \my(t) -\mD(t) {\bf 
  u}(t)  = \mC(t) \, \mx (t)$, allowing us to %
ignore the $\mD(t) {\bf u}(t)$ term. 

\begin{figure}[t!]
\centering
\includegraphics[width=0.45\textwidth]{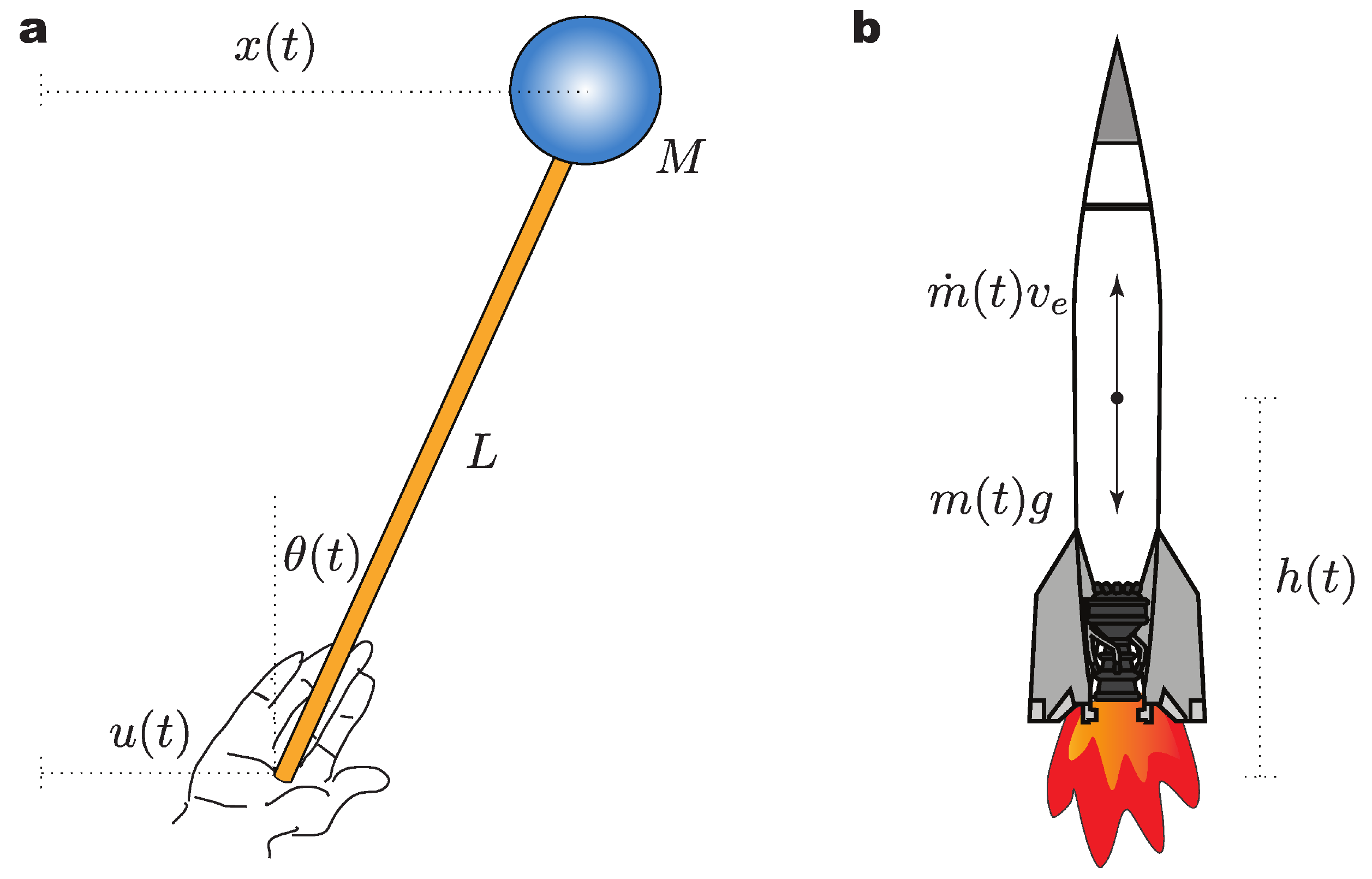}
\caption{(Color online) 
{\color{red} Two mechanical systems whose natural state space representation
with linear time-invarian (LTI) dynamics %
can be derived from Newton's laws of motion.}%
(a) %
The goal of stick balancing, a simple
{\color{red}but} much studied control problem {\color{red}(also known
  as the inverted perdulum problem)}, 
is %
to balance a stick on a palm. 
Redrawn after %
\citep{Luenberger-Book-79}. 
{\color{red}(b) 
A rocket being thrust upward. The rocket ascends from the surface of
the earth with thrust force guaranteed by %
the ejection of mass. Redrawn
after \citep{Rugh-1993}. 
}
}\label{fig:LTIexample_stick_rocket}
\end{figure}

Many nonlinear systems like (\ref{eq:X}, \ref{eq:Y}) can be linearized
around their equilibrium points, {\color{red}resulting in an LTI system}.  
For example, %
in stick balancing, a prototypical control problem~\citep{Luenberger-Book-79},  
our goal is to balance
(or control) the stick in the upright position using the horizontal position of the
hand as the control input $u(t)$. 
This mechanical system 
has a natural state
space representation %
derived from Newton's {\color{red}second} law of motion. 
Consider a stick of length $L$ whose %
mass $M$ is concentrated at the top.~\footnote{\color{red} For a more realistic
  case, treating the stick as a rigid body of uniform desnity,
  see~\citep{Stepan-2000}.} Denote the angle
between the stick and the vertical direction with $\theta(t)$. The hand
and the top of the stick have horizontal displacement $u(t)$ and
$x(t)$, respectively (Fig.~\ref{fig:LTIexample_stick_rocket}). 
The nonlinear equation of motion for this system is 
\be
 L \ddot{\theta}(t) = g \sin
\theta(t) - \ddot{u} (t) \cos \theta (t), \label{eq:stick1}
\ee 
where $g$ is the gravitational constant and 
\be 
x(t)= u(t) + L \sin \theta (t). \label{eq:stick2}
\ee
When the stick is nearly at rest in the upright vertical position 
($\theta=0$, which is an equilibrium point), $\theta$ is small, 
hence we can linearize Eqs.(\ref{eq:stick1}) and (\ref{eq:stick2}),
obtaining %
\be  \ddot{x}(t) = \frac gL [x(t) - u(t)]. 
\label{eq:stick3}
\ee
Using the state vector ${\bf x}(t) = (x(t), v(t))^\mm{T}$ with
velocity $v(t)=\dot{x}(t)$, and assuming $y(t)=x(t)$, we 
can %
rewrite the state and output equations 
in the form of an LTI system %
\begin{subnumcases}
\mathbf{\dot{\mx}}(t) = 
\begin{bmatrix}
0 & 1 \\
\frac gL & 0 
\end{bmatrix}
\mx(t) 
+ 
\begin{bmatrix}
0\\
-\frac gL 
\end{bmatrix}
u(t)  \label{eq:sticka}\\ 
y(t) = \begin{bmatrix}
1 & 0 
\end{bmatrix}
\mx(t). \label{eq:stickb}
\end{subnumcases}
%

\iffalse
\be
\begin{cases}
\mathbf{\dot{\mx}}(t) = 
\begin{bmatrix}
0 & 1 \\
\frac gL & 0 
\end{bmatrix}
\mx(t) 
+ 
\begin{bmatrix}
0\\
-\frac gL 
\end{bmatrix}
u(t)  \\ 
y(t) = \begin{bmatrix}
1 & 0 
\end{bmatrix}
\mx(t). 
\end{cases}
\ee
\fi

%
This form allows 
  us to perform
  linear controllability analysis. Indeed, as we show in
  Sec.\ref{sec:linearcontrollabilitytest}, the linearized system
  (\ref{eq:sticka}) is controllable, in line %
with our experience that we can balance a stick on our palm.

{\color{red}
Linearization of a nonlinear system around its norminal trajectory $\{ {\bf
  x}^*(t), {\bf u}^*(t)\}$ generally leads to an LTV system.   
Consider the motion of a rocket %
thrust upward, following %
  \be
  m(t) \ddot{h}(t) = \dot{m}(t) v_e - m(t) g, \label{eq:rocket1}
  \ee
 where $m(t)$ is the mass of the rocket at time $t$ and $h(t)$ is its
 altitute. The thrust force $\dot{m}(t) v_e $ follows %
 Newton's third law of motion, where $\dot{m}(t)$ denotes the mass
 flow rate and $v_e$ is the assumed-constant exit velocity of the
 exhaust (Fig.~\ref{fig:LTIexample_stick_rocket}b).  
 If we define the state vector ${\bf x}(t) = (h(t), v(t), m(t))^\mm{T}$ with
 velocity $v(t)=\dot{h}(t)$, the control input $u(t)=\dot{m}(t)$, 
 and the output $y(t)=h(t)$, 
we have the state-space representation 
\bea
 \left[\begin{array}{c}
 \dot{x}_1(t) \\
 \dot{x}_2(t)  \\
 \dot{x}_3(t) 
 \end{array}\right]
 &=&
 \left[\begin{array}{c}
 x_2(t) \\
 \frac{u(t) v_e}{x_3(t)} -g \\
 u(t) 
\end{array}\right]  \label{eq:rocket_state}\\
y(t) &=& x_1(t). 
\eea
The state equation (\ref{eq:rocket_state}) %
is clearly nonlinear. Let's consider its linearization around a
nominal trajectory that corresponds to a constant control input $u^*(t)=u_0 < 0$, i.e. a
constant mass flow rate. This nominal trajectory follows 
$x^*_1(t)=v_e [(m_0/u_0+t) \ln (1+u_0 t/m_0)] - g t^2/2$, 
$x^*_2(t)=v_e \ln(1+u_0 t/m_0) - gt$,
$x^*_3(t)=m_0 + u_0t$, 
where $m_0$ is the initial mass of the rocket. 
By evaluating the partial derivatives
$\frac{\p {\bf f}({\bf x}, u)}{\p {\bf x}}$ and $\frac{\p {\bf f}({\bf
    x}, u)}{\p u}$ at the nominal tracjectory, we obtain the
linearized state and output equations in the form an LTV system 

\iffalse
\begin{subnumcases}
\mathbf{
\dot{\mx}}_\delta(t) = 
\begin{bmatrix}
0 & 1 & 0 \\
0 & 0 & \frac{-u_0 v_e }{(m_0+u_0 t)^2} \\ 
0 & 0 & 0 
\end{bmatrix}
\mx_\delta(t) 
+ 
\begin{bmatrix}
0\\
\frac{v_e}{m_0+u_0 t} \\
1 
\end{bmatrix}
u_\delta(t)  \label{eq:sticka}\\ 
y_\delta(t) = \begin{bmatrix}
1 & 0 & 0  
\end{bmatrix}
\mx_\delta(t). \label{eq:stickb}
\end{subnumcases}
%
}
\fi

\be
\begin{cases}
\dot{\mx}_\delta(t) &=
\begin{bmatrix}
0 & 1 & 0 \\
0 & 0 & \frac{-u_0 v_e}{(m_0+u_0 t)^2} \\ 
0 & 0 & 0 
\end{bmatrix}
\mx_\delta(t) 
+ 
\begin{bmatrix}
0\\
\frac{v_e}{m_0+u_0 t} \\
1 
\end{bmatrix}
u_\delta(t)  \\ 
y_\delta(t) &= \begin{bmatrix}
1 & 0 & 0  
\end{bmatrix}
\mx_\delta(t),
\end{cases} 
\ee
where the deviation variables ${\bf x}_\delta(t) = {\bf x}(t)-{\bf x}^*(t)$, 
 $u_\delta(t)=u(t)-u^*(t)$, 
 and $y_\delta(t)=y(t)-y^*(t)={\bf x}_\delta(t)$. 
}
Notwithstanding our ability to %
design such well-controlled systems as %
a car or an airplane, we continue to lack an understanding of the control
principles that govern self-organized complex networked systems. Indeed, if
given the wiring diagram of a cell, %
we do not
understand the fundamental principles that govern its %
control, nor
do we have tools to extract them.  
Until recently the degree of penetration of control theoretical tools in the
study of complex systems %
was limited. 
The reason is that to
extract the predictive power of (\ref{eq:X}) and (\ref{eq:Y}), we need 
(i) the accurate wiring diagram of the system; 
(ii) a %
description of
the nonlinear dynamics that governs the interactions between the
components; and 
(iii) %
a precise knowledge of the system
parameters. 
For most complex systems we lack some of these
prerequisites. For example, current estimates indicate that in human
cells the available protein-protein interaction maps cover less than 20\% of all potential
protein-protein interactions~\citep{Sahni-Cell-2015}; in communication systems we may
be able to build an accurate wiring diagram, but we often lack the
analytical form of the system dynamics %
$\mf (\mx(t), {\bf u}(t);
\Theta)$;  
in biochemical reaction systems we have a good understanding of the
underlying network and dynamics, but we lack the precise 
values of the system parameters, like the reaction rate constants.  
Though progress is made on all three %
fronts, %
offering increasingly accurate data on the network structure,
dynamics, and the system parameters, 
accessing them all at once is still infeasible for most %
complex %
systems. 
Despite %
these difficulties, in the past decade we have seen
significant advances pertaining to the control of complex systems.   
These advances indicate that %
many fundamental control problems can be addressed %
without knowing all the %
details of equations (\ref{eq:X}) and
(\ref{eq:Y}). 
Hence, we do not have to wait for the description of complex systems to
be complete and accurate to address and understand the control
principles governing their behavior.

Graph theoretical methods have been successfully
applied to investigate the structural and the qualitative properties of 
dynamical systems since 1960's~\citep{Yamada-Networks-90}. 
This raises a question: Can the recent renaissance of interest in
controlling networked systems offer a better understanding of control principles
than previous graph theoretical methods? 
To answer this we must realize that 
the current interest in control in the area of complex systems 
is driven by the %
need to understand such large-scale complex networks as 
the Internet, the WWW, wireless communication networks, power grids,
global transportation systems, genome-scale metabolic networks, protein
interaction networks and gene regulatory networks, to name only a few
~\citep{Chen-IJC-14}.  
Until the emergence of network science in the 21th century we
lacked the mathematical tools to characterize the structure of these %
systems, 
not even mentioning their control principles. %
The non-trivial topology of real-world networks, uncovered and
characterized in the past two decades, brings an
intrinsic layer of complexity to most  
control problems, requiring us to rely on %
tools borrowed from many disciplines to address them.  
A typical example is the structural controllability problem of complex
networks. %
Structural control theory developed in 1970's
offered %
sufficient and necessary conditions to check if any network with LTI dynamics is structurally controllable~\citep{Lin-IEEE-74}. Yet, it
failed to offer %
an efficient algorithm to find the minimum set of driver nodes
required to %
control the network, 
nor %
an analytical framework %
to estimate the fraction of driver nodes. 
Advances on this front became possible by mapping the control problem
into well-studied network problems, like matching, and utilizing 
the notion of thermodynamic limit %
in statistical
physics and the cavity method developed in spin glass theory, tools
that were traditionally 
beyond the scope of %
control theory~\citep{Liu-Nature-11}. 
The goal of this article is to review the current advances in controlling
complex systems, %
be they of biological, social, %
or technological %
in nature. 
To achieve this we %
discuss %
a series of %
topics that are essential to understand the control principles of
networks, %
with emphasis on the impact of the network structure on control. The review
is organized
around several fundamental issues:  

(i) \emph{Controllability}. Before deciding how to control a system, 
we 
must make sure that it is 
possible to control it. %
Controllability, a key notion in modern control theory %
quantifies %
our ability to steer a dynamical system to a 
desired final state in finite time. %
We will discuss the impact of network topology on our ability to control %
complex networks, and address %
some practical issues, like %
the energy %
or effort required for control. %
(ii) \emph{Observability}. %
As a dual concept of controllability, observability describes the
possibility of inferring the initial state of a dynamical system by %
monitoring its time-dependent %
outputs. %
We will discuss different methods to identify the sensor nodes, whose
measurements over time enable us to infer the initial state of the
whole system. 
We also explore %
a closely related concept --- \emph{identifiability}, representing 
our ability to determine the system's parameters through appropriate
input/output measurements. %

(iii) \emph{Steering complex systems to desired states or
  trajectories.}  
The ultimate %
goal of control is to drive 
a complex system from its current %
state/trajectory to some %
desired final state/trajectory. This problem has %
applications %
from ecosystem management, to cell reprogramming. 
For example, 
we would like %
to design interventions that can move a cell from a disease (undesired) %
to a
healthy (desired) state. 
We discuss different ways of achieving such control: 
(a) By applying 
small perturbations to a set of physically or
experimentally feasible parameters;
(b) Via %
compensatory perturbations of state variables that exploit %
the %
basin of attraction of the desired final state; 
(c) By mapping the control problem into a %
combinatorial optimization
problem on the underlying network. %

(iv) \emph{Controlling collective behavior.} Collective behavior, a
much-studied topic in modern statistical physics, %
can result from the coordinated local activity of many interdependent
components. Examples include the emergence of 
flocking in %
mobile agents %
or %
synchronization in
coupled oscillators. %
Controlling such processes has numerous %
potential applications, from the design of flocking
robots~\citep{Olfati-Saber-IEEETAC-06}, to the treatment of
Parkinson's disease{\color{red}~\citep{Tass-PRL-98}}. %
We %
review a broad spectrum of methods to determine the conditions
for %
the emergence of collective behavior %
and discuss 
pinning control as an effective control strategy. 
Control problems are ubiquitous, %
with %
direct relevance to many
natural, social and technological phenomena. Hence the advances
reviewed here 
truly probe our fundamental understanding of the
complexity of the world surrounding us, %
potentially %
inspiring advances in numerous
disciplines. 
Consequently, our focus here is on %
conceptual advances and tools
pertaining to control, that %
apply to a wide range of
problems emerging in physical, technological, biological and social
systems. 
It is this diversity of applications that makes 
control increasingly
unavoidable in most disciplines. 

\section{Controllability of Linear Systems}\label{Sec:Controllability}

A system is %
\emph{controllable}
if we %
can drive it from any initial state to any desired final state
in finite time~\citep{Kalman-JSIAM-63}. %
Many mechanical problems can be formalized as controllability
problems (Fig.~\ref{fig:LTIexample_stick_rocket}).  
Consider, for example, the control of a rocket %
thrust upward. The rocket %
is controllable if we can find a continuous control input (thrust
force) 
that can %
move the rocket from a given initial state %
(altitute and velocity) to a desired final state. %
Another example is the balancing of a stick on our hand. We know from our experience that
this is possible, suggesting that %
the %
system must be
controllable~\citep{Luenberger-Book-79}. 
The scientific %
challenge is to decide for an arbitrary dynamical
system if it is controllable or not, given a set of inputs.

The current interest in the control of complex networked systems was %
induced by recent advances in the controllability of complex
networks~\citep{Liu-Nature-11,Posfai-SR-2013,Liu-PLOS-12,Jia-NC-2013,Gao-NC-14},
offering %
mathematical tools to identify the driver nodes, a subset
of nodes whose direct control with appropriate signals can control the state
of the full system.   
In general controllability is a prerequiste of control,
hence understanding the topological factors of the underlying network that determine a system's
controllability offers numerous %
insights into the control
principles of complex networked systems. 
As we discuss below, thanks to a convergence %
of tools from control theory, network science
and statistical physics, %
our understanding of network controllability
has advanced considerably recently. %
\subsection{Linear Time-Invariant Systems}\label{sec:controllabilitytest_LTI}

The starting point of most control theoretical approaches is
the linear time-invariant (LTI) %
control system $({\bf A},
{\bf B})$
\be 
\dot{{\bf x}}(t) = {\bf A} \, {\bf x}(t) + {\bf B} \, {\bf
  u}(t). \label{eq:LTI}
\ee 
%
%
%
\iffalse
For instance, the motion for a rocket %
thrust upward 
follows 
%
%
\be
%
 \ddot{h}(t) = f(t) - g \label{eq:rocket1}
\ee
where 
%
%
$h(t)$ is the rocket's altitude, and $f(t)$ is the thrust force per
unit mass.  
%
Using the state vector ${\bf x}(t) = (h(t), v(t))^\mm{T}$, %
the velocity $v(t)=\dot{h}(t)$, and the control input $u(t)=f(t)-g$,
we %
can write %
(\ref{eq:rocket1}) 
in the state-space representation with LTI dynamics 
\be
%
%
%
%
%
%
\dot{\mx}(t)
=
\left[\begin{array}{cc}
0 & 1 \\
0 & 0 
\end{array}\right]
\mx(t)
%
%
%
%
+ 
\left[\begin{array}{c}
0\\
1
\end{array}\right]
u(t). \label{eq:rocket2}
\ee
\fi
%

%
%
%

%
%
%
%

%
Many mechanical systems can be naturally described by LTI
dynamics, where the state vector captures the %
position and velocity of objects %
and the LTI dynamics is either directly
derived from Newton's {\color{red}Second} Law or represents some reasonable linearization
of the underlying nonlinear problem, as
illustrated by %
the stick balancing problem (\ref{eq:stick3}).

A significant fraction of the 
control theory literature deals exclusively with linear systems. There
are multiple reasons for this. 
First, %
linear systems offer an accurate model for %
some real problems, %
like %
consensus or agreement formation %
in multi-agent networks, where the state of
each agent captures %
its opinion~\citep{Tanner-IEEE-04,
  Liu-IEEE-08, Rahmani-SIAM-09, Mesbahi-Book-10}. 
Second, %
while many %
complex systems are characterized by
nonlinear interactions between the components, the first step in any
control challenge is to establish the controllability of the locally
linearized system~\citep{Slotine-Book-91}. 
Furthermore, as we show below, for systems near their equilibrium
points the linearized dynamics can 
actually characterize %
the underlying nonlinear controllability problem. 
Third, the non-trivial network topology of real-world
complex systems brings a new %
layer of complexity to controllability. 
Before we can explore the
fully nonlinear dynamical setting{\color{red}, which is mathematically
  much harder}, we must %
understand the impact of the 
topological characteristics on linear controllability, serving %
as a %
prerequisite of nonlinear controllability. %

\begin{figure}[t!]
\includegraphics[width=0.5\textwidth]{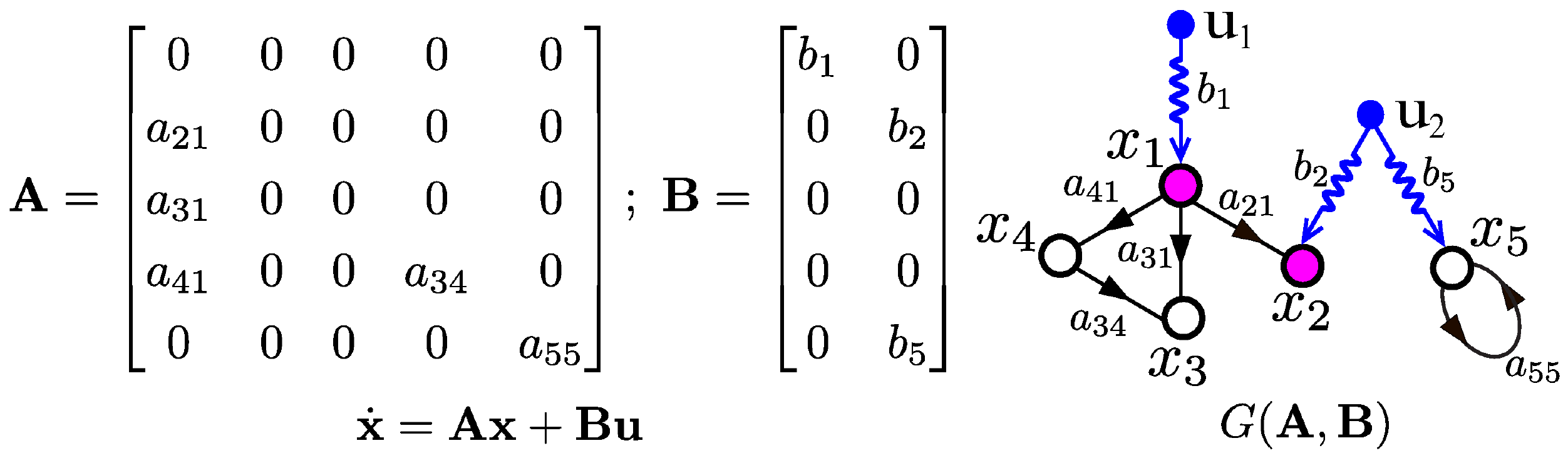}
\caption{(Color online) Graphical representation of a linear time-invariant %
  system %
(\ref{eq:LTI}). %
The state matrix ${\bf A}$ represents %
the weighted wiring diagram of the network that describes which
components interact with each other and the direction of the 
signal or information flow for each link; the input matrix
${\bf B}$ identifies the nodes (state variables) that are controlled
by an outside controller. %
The network shown in the figure is controlled by an input vector
${\bf u} = (u_1(t), u_2(t))^\mm{T}$ with two independent signals
$u_1(t)$ and $u_2(t)$.  
The three actuator nodes ($x_1, x_2$ and $x_5$) are the nodes directly
controlled by ${\bf u}(t)$. These %
actuator nodes correspond to the
three non-zero elements in ${\bf B}$.   
The two driver nodes ($x_1$ and $x_2$), representing nodes that do not share input signals,
correspond to the two columns of ${\bf B}$. 
Note that node $x_5$ is an actuator node, but not a driver node. 
}\label{fig:GraphTheoreticInterpretation}
\end{figure}

Consider the LTI dynamics (\ref{eq:LTI}) on a directed weighted
network $G({\bf A})$ {\color{red} of $N$ nodes}
(Fig.~\ref{fig:GraphTheoreticInterpretation}).  
The state variable $x_i(t)$ can denote the amount of traffic that
passes through a node $i$ on a communication network~\citep{Pastor-Satorras-Book-04}, or transcription factor
concentration in a gene regulatory
network~\citep{Lezon-PNAS-06}. %
The state matrix ${\bf A} :=(a_{ij})_{N \times N}$ 
represents the weighted %
wiring diagram of the underlying network, where $a_{ij}$
is %
the strength or weight with which %
node $j$ affects/influences node $i$:
a positive (or negative) $a_{ij}$ means the link ($j \to i$)
is excitatory (or inhibitory), and $a_{ij}=0$ if node $j$ has no
direct influence on node $i$. 
{\color{red} Consider %
$M$ independent control signals $\{u_1,
  \cdots, u_M\}$ applied to the network.}
The input matrix ${\bf B}:= (b_{im})_{N \times M}$ 
identifies the nodes that are directly
controlled, %
where $b_{im}$ represents the
strength of an external control signal $u_m(t)$ injected into node
$i$.

The input signal ${\bf u}(t) = (u_1 (t),
  \cdots, u_M (t))^\mm{T} \in \mathbb{R}^M$ can %
be imposed on all nodes or only a preselected subset of the nodes. In
general the same signal $u_m(t)$ can drive multiple nodes.  
The nodes directly
controlled by ${\bf u}(t)$ are called \emph{actuator nodes} or simply
\emph{actuators}, like %
nodes $x_1, x_2$ and $x_5$ in
Fig.~\ref{fig:GraphTheoreticInterpretation}.   
The number of actuators is given by the number of non-zero
elements in ${\bf B}$.  
The actuators that do not share
input signals, e.g. nodes $x_1$ and $x_2$ in Fig.~\ref{fig:GraphTheoreticInterpretation}, are called \emph{driver nodes} or simply \emph{drivers}. The
number of driver nodes equals the number of columns in ${\bf B}$. 

Controllability, the ability to steer a system into an
  arbitrary final state in a finite time, implies %
that we can move the %
  state variable of each node of a network %
to a predefined value, %
corresponding to the system's desired
  position in the state %
space. 
Our ability to do so is greatly determined by the
network topology. For example, if the network structure is such that a
signal cannot get from our driver nodes to a particular node, that
node, and hence the system as a whole, is uncontrollable. 
Our challenge is to %
decide when control is possible and when
is not. The answer is given by controllability tests described next.  

\subsection{Kalman's Criterion of Controllability}\label{sec:linearcontrollabilitytest}

Controllability tests allow us to check if an LTI
system is controllable from a given set of inputs. The %
best known is 
Kalman's rank condition~\citep{Kalman-JSIAM-63}, stating that 
the LTI system $({\bf A}, {\bf B})$ is controllable if and only if the
$N \times NM$ controllability matrix 
\be
{\bf \mathcal{C}} \equiv [ {\bf B}, {\bf A} \, {\bf B}, {\bf A}^2 \, {\bf B},
\ldots,  {\bf A}^{N-1} \, {\bf B}  ] \label{eq:Cmatrix0}
\ee
has full rank, i.e. 
\be
\mm{rank}\, {\bf \mathcal{C}} 
= N.  \label{eq:rankC}
\ee 

\begin{figure}[t!]
\includegraphics[width=0.45\textwidth]{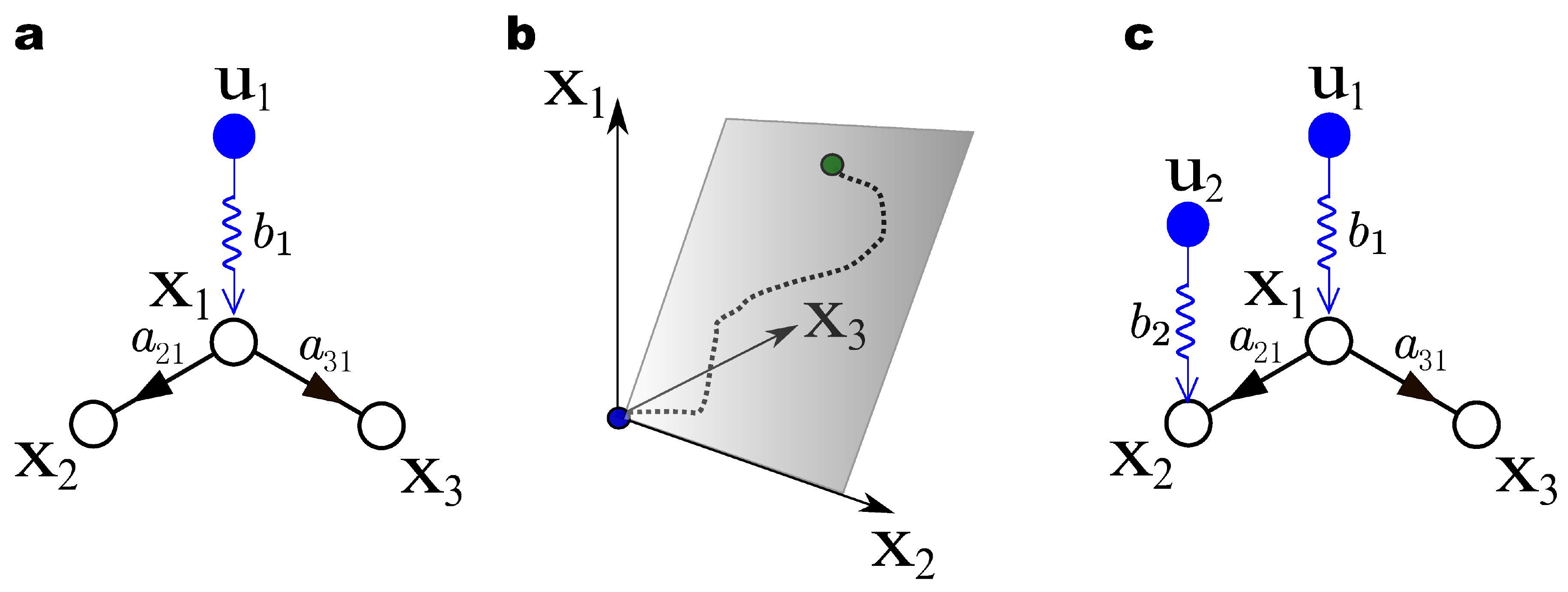}
\caption{(Color online) Controlling star networks. 
(a) Controlling the central node of a directed star does not
  assure controllability of the whole network, as shown in
  (\ref{eq:Cfig4a}). (b) Indeed, the system is stuck
in the plane %
$a_{31}x_2(t) = a_{21}x_3(t)$, hence no signal $u_1(t)$ can make the %
system leave this plane and explore the whole state space. The reason is simple: if we
change $u_1(t)$, $x_2(t)$ and $x_3(t)$ %
always evolve in a
correlated fashion, indicating that %
we are unable to control the two nodes independently
of each other. 
{\color{red}Note that while the system is not controllable in the whole
  state space, it remains %
controllable within the plane. It is natural that ensuring
controllability within a restricted subspace will require fewer driver
nodes than ensuring controllability within the whole state space~\citep{Muller-Nature-11,Liu-NatureReply-11}.
}
(c) To ensure controllability, we must inject an additional signal $u_2$ to
either $x_2$ or $x_3$, in which case, according to (\ref{eq:Cfig4c}),
the network becomes controllable.   
After \citep{Liu-NatureReply-11}.  
}\label{fig:controllablesubspace}
\end{figure}

To understand the origin of (\ref{eq:Cmatrix0}), we consider the %
formal solution of (\ref{eq:LTI}) with $\mx(0)={\bf 0}$, %
i.e. 
\be {\bf x}(t) = \int_0^{t} \exp[{\bf A} (t-\tau)] \, {\bf B} \,
{\bf u}(\tau) \,\ud \tau.\ee
If we expand $\exp[{\bf A} (t-\tau)]$ in series, we %
will realize that %
${\bf x}(t)$ is actually a linear combination of the
columns in the matrices $\{{\bf B}, {\bf A}{\bf B}, {\bf A}^2{\bf B},
\cdots\}$. 
Note that for any $N' \ge N$, we have 
$\mm{rank}\, 
[ {\bf B}, {\bf A}{\bf B}, {\bf A}^2 {\bf B},
\cdots,  {\bf A}^{N'-1} {\bf B}  ]
= \mm{rank} \, {\bf \mathcal{C}}
$. So if $\mm{rank}\, {\bf \mathcal{C}} < N$, then even the infinite series of 
$\{ {\bf B}, {\bf A}{\bf B}, {\bf A}^2{\bf B}, \cdots\}$ will not
contain a full basis to span %
the entire $N$-dimensional
state space. %
In other words, we cannot fully explore the state space, regardless of
${\bf u}(t)$, indicating that given our inputs the system is stuck in
a particular subspace, unable to %
reach an arbitrary point in the %
state space (Fig.~\ref{fig:controllablesubspace}).  
If, however, $\mm{rank}\, {\bf \mathcal{C}} = N$, then %
we can find an appropriate input vector ${\bf
  u}(t)$ to steer the system from  ${\bf x}(0)$ to an arbitrary ${\bf
  x}(t)$. Hence, the system is controllable.

One can check that in the %
stick balancing
problem (\ref{eq:sticka}), the %
controllability matrix has full rank
($\mm{rank}\, {\bf \mathcal{C}} = N=2$), 
indicating that both systems are controllable. %
In the network control problem of Fig.~\ref{fig:controllablesubspace}a 
the controllability matrix 
\be
{\bf \mathcal{C}} = 
\begin{bmatrix}
b_1 &  0           &  0  \\
0    & a_{21}b_1 &  0 \\
0    & a_{31}b_1 &  0 \\
\end{bmatrix} \label{eq:Cfig4a}
\ee
is always rank deficient, as long as the parameters $b_1$,
$a_{21}$ and $a_{31}$ are non-zero. Hence, the system is
uncontrollable.  
In contrast, for Fig.~\ref{fig:controllablesubspace}c we have 
\be
{\bf \mathcal{C}} = 
\begin{bmatrix}
b_1 & 0    &  0           &  0  &  0 &  0 \\
0    & b_2 & a_{21}b_1 &  0 &  0 &  0 \\
0    & 0    & a_{31}b_1 &  0 &  0 &  0 \\
\end{bmatrix}, \label{eq:Cfig4c}
 \ee
which %
has full rank, as long as the parameters $b_1$, $b_2$, 
$a_{21}$ and $a_{31}$ are non-zero. Hence the system is %
controllable. 

The %
example of Fig.~\ref{fig:controllablesubspace} implies that the
topology of the \emph{controlled network}{\color{red}, which consists of
  both the network itself and the control signals applied to some %
  nodes,} %
imposes some inherent limits %
on the controllability matrix: some configurations are controllable
(Fig.~\ref{fig:controllablesubspace}c), while others are not
(Fig.~\ref{fig:controllablesubspace}a). 
Thanks to the Kalman criterion, controllability 
can be easily tested %
when the dimension of the 
controllability matrix is small and its rank test can be done even
without knowing the detailed values of its non-zero matrix elements. 
For large real networks the controllability test
  (\ref{eq:rankC}) is %
difficult to perform, however. Indeed, %
there is no
  scalable algorithm to numerically determine the rank of the
  controllability matrix ${\bf \mathcal{C}}$, which has dimension $N \times NM$. 
Equally important, 
executing an accurate rank test is ill-conditioned and is
very sensitive to roundoff errors and uncertainties in the
matrix elements. %
Indeed, if we plug the numerical values of $b_i$ and $a_{ij}$ into
 (\ref{eq:Cmatrix0}), 
we {\color{red}may} obtain %
extremely
large or small matrix elements, %
{\color{red}such as} 
$a_{ij}^{N-1}$, which for large
$N$ %
are rather %
sensitive to numeric precision.
Hence, for large complex systems we need %
to determine the system's controllability without
numerically calculating the rank of the controllability matrix. As we
discuss in {\color{red}the} next section, this can be %
achieved in the context of
structural control theory.

\subsection{Structural Controllability} \label{sec:SCT}

For many complex networks %
the system parameters (e.g. the elements in
${\bf A}$) are not precisely known. 
Indeed, %
we are often unable to measure the weights of
the links, knowing only whether there is a link or not. In other
cases %
the links are time dependent, like the traffic on an internet
cable or the flux of a chemical reaction. 
Hence, 
it is hard, if not conceptually impossible, to numerically verify Kalman's rank
condition using fixed weights. %
Structural control, %
introduced by C.-T. Lin in
1970s, offers a framework to systematically %
avoid this limitation%
~\citep{Lin-IEEE-74}.%

\subsubsection{The power of structural controllability}

An LTI system $({\bf A}, {\bf B})$ is %
a \emph{structured
  system} if the elements in ${\bf
  A}$ and ${\bf B}$ are either fixed %
zeros or independent free parameters. 
The corresponding matrices $\mA$ and $\mB$ are called \emph{structured
matrices}. 
The system $({\bf A}, {\bf B})$ is %
\emph{structurally controllable} if we can %
set the nonzero elements in ${\bf
  A}$ and ${\bf B}$ such that the resulting system 
is controllable in the usual sense (i.e.,
$\mm{rank}\, {\bf \mathcal{C}} = N$). 

\begin{figure}[t!]
\includegraphics[width=0.5\textwidth]{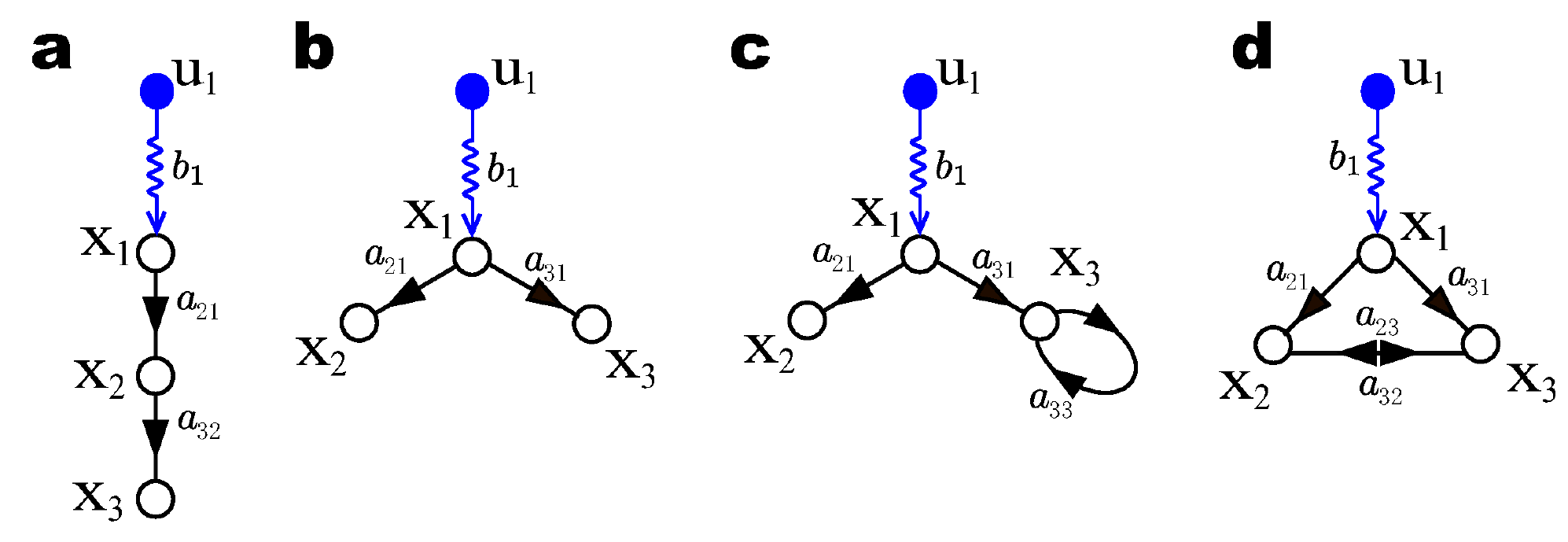}
\caption{(Color online) %
Controllability,
  structural controllability, and strong structural controllability. 
  (a) A directed path can be %
controlled by controlling the
  starting node only. The controllability is independent of the detailed (non-zero)
  values of $b_1$, $a_{21}$, and $a_{32}$, so the system is strongly
  structurally controllable. 
  (b) A directed star can never be %
controlled by controlling
  the central hub (node $x_1$) only. 
  (c) This %
network obtained by %
 adding a self-edge to the star
  shown in {\bf b}, can be %
controlled by controlling %
$x_1$ only. The controllability is independent of the detailed (non-zero) values
  of $b_1$, $a_{21}$, $a_{31}$, and $a_{33}$, so the system is strongly
  structurally controllable. 
  (d) This network is %
controllable for almost all weights
  combinations. It will be uncontrollable only in some pathological cases,
  for example when the %
weights satisfy the constraint $a_{32} a_{21}^2 = a_{23}
  a_{31}^2$ exactly. Hence, the system is structurally controllable but
  does not display strong structural controllability.}  
\label{fig:examplesSSC} 
\end{figure}

The power of structural controllability comes from the
fact that if a system is structurally controllable then it is
controllable %
for almost all possible parameter realizations~\citep{Lin-IEEE-74,Shields-IEEE-76, Glover-IEEE-76,
  Davison-Automatica-77, Hosoe-IEEE-79, Mayeda-IEEE-81,
  Linnemann-IEEE-86,Reinschke-book-88,Dion-Automatica-03}.  
To see this, denote with $\mathcal{S}$ the set of all possible LTI systems that
share the same zero-nonzero %
  connectivity pattern  
as a structurally controllable
system $({\bf A}, {\bf B})$.    
It has been shown that \emph{almost} all %
systems that belong to
the set %
$\mathcal{S}$ 
are controllable except for some pathological cases with Lebesgue measure
zero~\citep{Lin-IEEE-74,Shields-IEEE-76}.  
This is rooted in the
fact that %
if a system $({\bf A}_0,
{\bf B}_0) \in \mathcal{S}$
is uncontrollable, then for every $\epsilon > 0$ there exists a
controllable system $({\bf A}, {\bf B})$ with $||{\bf A}-{\bf A}_0|| < \epsilon$
and $||{\bf B}-{\bf B}_0|| < \epsilon$ where $||\cdot||$ denotes matrix
norm~\citep{Lee-Book-1968,Lin-IEEE-74}. 
In other words, 
an uncontrollable system in $\mathcal{S}$ %
becomes controllable if we slightly alter some of %
the link weights. %
For example, the system shown in
  Fig.~\ref{fig:examplesSSC}d is controllable for almost all
  parameter realizations, except %
when the
  edge weights satisfy the %
constraint %
$a_{32} a_{21}^2 = a_{23}
  a_{23}^2$. But these pathological cases can be easily avoided
by slightly changing one of the edge weights, hence %
this system is
structurally controllable.

Taken together, structural control tells us that we can decide a
network's controllability even if we do not know 
the precise 
weight of each edge. %
All we have to make sure is that we have an
accurate map of the system's wiring diagram, i.e., know %
which components are linked and which are not. As we demonstrate  %
in the coming
section, this framework considerably expands the practical
applicability of control tools to real systems.

\subsubsection{Graphical interpretation}
Structural control theory %
allows us to check if a controlled network is structurally controllable by simply
inspecting its topology, avoiding expensive matrix operations. %
This is possible thanks to the graphical
interpretation~\footnote{\color{red}The structural controllability
  theorem also has a pure algebraic meaning~\citep{Shields-IEEE-76},
  which plays an important role in the
characterization of strong %
structural controllability~\citep{Mayeda-SIAM-79}.} of Lin's
Structural Controllability Theorem, discussed next. %

Consider an LTI system $({\bf A}, {\bf B})$ represented
by a digraph $G({\bf A}, {\bf B})=(V, E)$ %
(Fig.~\ref{fig:GraphTheoreticInterpretation}). 
The vertex set $V= V_A \cup  V_B$ includes both the \emph{state vertices}
$V_A=\{x_1,\cdots, x_N \} \equiv \{v_1,\cdots, v_N \}$, corresponding
to %
the $N$ nodes of %
the network, and the \emph{input vertices} $V_B=\{u_1,\cdots,
u_M \} \equiv \{ v_{N+1}, \cdots, v_{N+M}\}$, corresponding to %
the $M$ input signals that are %
called %
the \emph{origins} or
\emph{roots} of the digraph $G({\bf
  A}, {\bf B})$.
The edge set $E= E_A \cup E_B$ includes both the edges among 
state vertices $E_A = \{(x_j, x_i) | a_{ij} \neq 0 \}$, corresponding
to %
the links of %
network ${\bf A}$, and the edges connecting input vertices
to state vertices $E_B = \{(u_m, x_i) | b_{im}
\neq 0 \}$. 
These definitions allow us to formulate a %
useful statement: 
The system $({\bf A}, {\bf B})$ is not
structurally controllable if and only if it has 
\emph{inaccessible nodes} or \emph{dilations}~\citep{Lin-IEEE-74}.  

\begin{figure}[t!]
\includegraphics[width=0.5\textwidth]{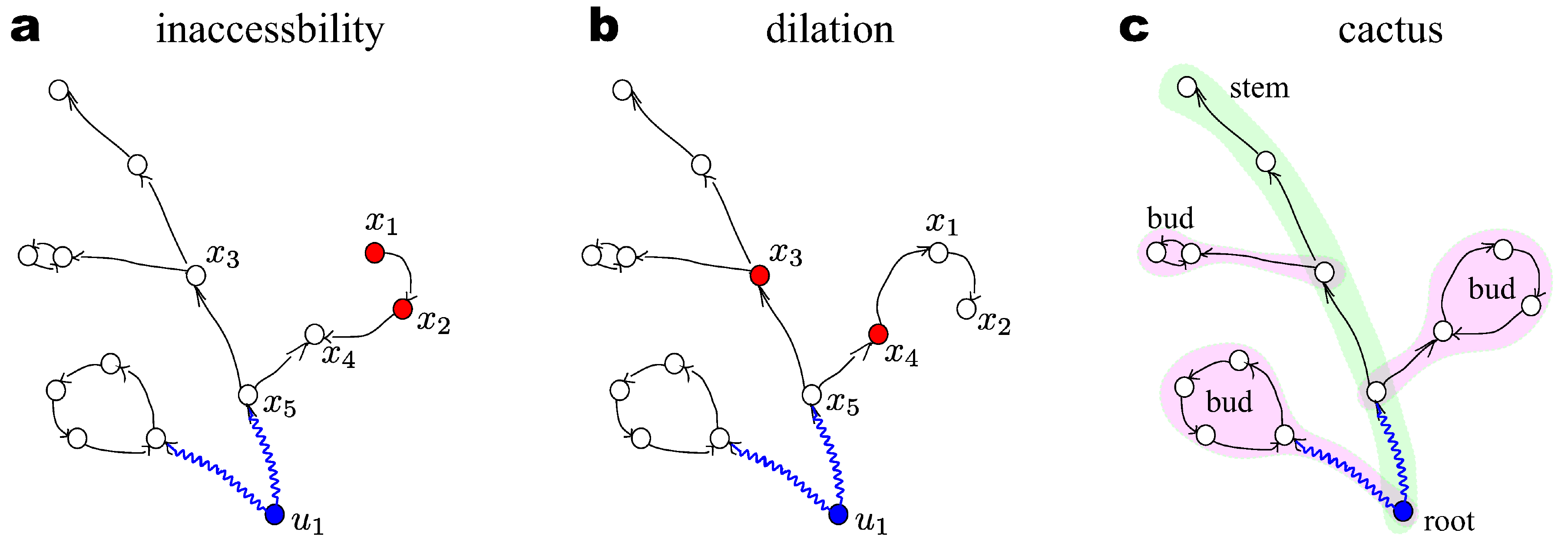}
\caption{(Color online) Inaccessibility, dilations and cacti. %
 (a) The red nodes ($x_1, x_2$) are inaccessible from the 
  input node $u_1$ (in blue), as variations in $u_1$ do not
  influence the state of $x_1$ or $x_2$. (b) The red nodes
  $S=\{x_3,x_4\}$ %
cause a dilation. %
Indeed, their %
neighborhood set $T(S)=\{x_5\}$ contains only one node, hence the size
of $T(S)$ is smaller than $S$, implying that a single node in $T(S)$
aims to control two nodes in $S$. As we showed in Eq.~(\ref{eq:Cfig4a})
and Fig.~\ref{fig:controllablesubspace}a, this is not possible. 
 (c) A cactus contains neither inaccessible nodes
  nor dilations. Note that in the cactus
  structure $T(S)=\{x_2,x_5\}$, hence there is no dilation. There is
  only one stem (shown in green) in one
  cactus. There could be multiple buds (shown in purple) in the same %
  cactus. A cactus is a minimal structure for structural
    controllability.
}\label{fig:minimal}
\end{figure}

Let us consider these two cases separately. 
A state vertex $x_i$ is %
\emph{inaccessible} if there are no directed paths reaching $x_i$ from
the input vertices (Fig.~\ref{fig:minimal}a).  
Consequently, an inaccessible node %
can not be 
influenced by %
input signals applied to the driver nodes, 
making the %
whole network %
uncontrollable.

The digraph $G({\bf A}, {\bf B})$ contains a \emph{dilation}
if there is a subset of nodes $S \subset V_A$ such that the
\emph{neighborhood set} of $S$, denoted as $T(S)$, has fewer %
nodes {\color{red}than} %
$S$ itself 
(see Fig.~\ref{fig:minimal}b). Here, 
$T(S)$ is the set of vertices $v_j$ for which there is a directed
edge from $v_j$ to some other %
vertex in $S$. Note that the input vertices are 
not allowed to belong to $S$ but may belong to $T(S)$. 
Roughly speaking, dilations are subgraphs in which a small subset of
nodes attempts to rule a larger subset of nodes. In other words,
there are more ``subordinates'' than ``superiors''. 
A controlled network containing dilations is uncontrollable. 
For example, in a
directed star configuration, where we wish to control via a central
node all the leaves, any two leaf-nodes form a dilation with the
central hub. If we control the central hub only, the system remains
uncontrollable because we cannot independently control the difference
between the two leaf nodes' states
(Fig.~\ref{fig:controllablesubspace}). In other words, we cannot
independently control two subordinates if they share the same 
superior. 

Taken together, Lin's structural controllability theorem states that
an LTI system $(\mA, \mB)$ is structurally controllable if and only if
the digraph $G(\mA, \mB)$ does not contain %
inaccessible nodes or
dilations. 
These two %
conditions %
can be
accurately checked by inspecting %
the topology of the digraph $G({\bf A}, {\bf
  B})$ without dealing with %
floating-point operations. 
Hence, this bypasses %
the %
numerical issues involved in evaluating Kalman's controllability rank
test, and also our lack of detailed knowledge on the edge weights in $G({\bf A}, {\bf
  B})$.%

An alternative graph theoretical formulation of Lin's structural
controllability %
theorem is often %
useful in practice. 
A general graph %
is \emph{covered} or \emph{spanned} by a
subgraph if the subgraph and the graph have the same vertex set. 
Typically the spanning subgraph has only a subset of links of the
original graph. 
For a digraph, a sequence of oriented edges $\{ (v_1 \to v_2), \cdots,
(v_{k-1} \to v_k)\}$, where the %
vertices $\{v_1,v_2,\cdots, v_k\}$ are
distinct, is called an \emph{elementary path}. When $v_k$ coincides
with $v_1$, the sequence of edges %
is called an \emph{elementary cycle}.  
For the digraph $G({\bf A}, {\bf B})$, 
we define the following subgraphs (Fig.~\ref{fig:minimal}c):  
(i) a \emph{stem} is an elementary path originating from an input
  vertex; 
(ii) a \emph{bud} is an elementary cycle $C$ with an additional edge
  $e$ that ends, but does not begin, in a vertex of the cycle; 
(iii) a \emph{cactus} is defined recursively: 
A stem is a cactus. Let $C$, $O$, and $e$ be, respectively, a cactus,
an elementary cycle that is disjoint with $C$, and an arc that
connects $C$ to $O$ in $G({\bf A},{\bf B})$. Then, $C \cup \{e
\} \cup O$ is also a cactus. $G({\bf A},{\bf B})$ is spanned by cacti if
there exists a set of disjoint cacti that cover all state vertices.  

Note that a cactus is a \emph{minimal} structure that contains
neither inaccessible nodes nor dilations. That is, for a given cactus,
the removal of any %
edge will result in %
either inaccessibility or
dilation, hence the controllability of the cactus is lost (Fig.~\ref{fig:minimal}). 
We can now formulate Lin's \emph{structural controllability
  theorem} as follows: %
An LTI system $({\bf A}, {\bf B})$ is structurally
  controllable if and only if %
  $G({\bf A}, {\bf B})$ is spanned by cacti~\citep{Lin-IEEE-74}. 
Later we show %
that this formulation 
helps us design an efficient algorithm to identify a minimum set of inputs
that guarantee structural controllability.

{\color{red}
\subsubsection{Strong structural controllability}

The fundamental assumption of structural control 
is that the entries of the matrices {\bf A} and {\bf B} are either %
zeros or independent free parameters. 
Therefore structural control does not require knowledge of the exact
values of parameters, any by avoiding floating-point operations, it is
not subject to numerical errors.   
However, some %
systems have interdependent parameters, making it 
uncontrollable despite the fact that it is
structurally controllable.  
For example, Fig.~\ref{fig:examplesSSC}d displays an LTI system
that is structurally controllable, but %
becomes uncontrollable when the
parameters satisfy the constraint %
$a_{32} a_{21}^2 = a_{23} a_{31}^2$. 
This leads to the notion of \emph{strong structural
  controllability} (SSC) : A system is strongly structurally
controllable if %
it remains controllable for any value (other than zero) of the
indeterminate parameters~\citep{Mayeda-SIAM-79}. %
In other words, there is no %
combination of non-zero link weights
that violates Kalman's criterion (\ref{eq:rankC}).  For example, the
LTI systems shown in Fig.~\ref{fig:examplesSSC}a and c are strongly structurally
controllable.

Both graph-theoretic~\citep{Mayeda-SIAM-79,Jarczyk-IEEE-2011} and
algebraic conditions~\citep{Reinschke-92} for SSC have been
studied. Unfortunately, those conditions do not lead to %
efficient algorithms. 
Recently, %
necessary and sufficient graph-theoretical conditions
involving constrained matchings were derived~\citep{Chapman-ACC-2013}.  
Denote a matching of size $t$ in the bipartite representation $H({\bf A})$ of the
digraph $G({\bf A})$ as $t$-matching. A $t$-matching is
\emph{constrained} if it is the only $t$-matching in $H({\bf A})$. 
A matching is called $V_\mm{s}$-less if it contains no edges
corresponding to self-loops. %
Let $\mathcal{S}$ be an
input set with cardinality $M \le N$. The corresponding structured pair $({\bf A},
{\bf B})$ is strongly structurally controllable if and only if $H({\bf
  A})$ has a constrained $(N-M)$-matching with $\mathcal{S}$ unmatched
and $H({\bf A}_{\times})$ has a constrained $V_\mm{s}$-less
$(N-M)$-matching with $\mathcal{S}$ unmatched. Here  $H({\bf
  A}_{\times})$ is formed by adding self-loops to all nodes if they
don't have one. 
The constrained matching conditions
can be applied to check if an input set is strongly structural
controllable in $\mathcal{O}(N^2)$. Though finding a minimum
cardinality input set is proven to be NP-complete, 
a greedy $\mathcal{O}(N^2)$ algorithm has been developed to provide a strongly structural
controllable input set, which is not-necessarily
minimal~\citep{Chapman-ACC-2013}.  
}

\subsection{Minimum Input Problem} \label{sec:mip}
If we wish to control a networked system, we first need to identify the set of
driver nodes that, if driven by different signals, can offer full
control over the network.
Any system is fully controllable if we control each node
individually. Yet, such full control %
is costly and typically impractical. Hence, we are particularly
interested in identifying a minimum driver node set (MDNS), %
whose control is sufficient to make the whole system controllable. 
In other words, we want
to control a system with minimal inputs. %

\subsubsection{Solution based on structural control theory}
Kalman's rank condition does not offer us the MDNS%
---  %
it only tells us if we can control a system through %
a given set of potential 
driver nodes that we must guess or select. %
Furthermore, to numerically check Kalman's rank condition, 
we have to know all the entries in ${\bf A}$ and ${\bf B}$,
which %
are often %
unknown for complex networks. 
Even if we know all the weights
(parameters) exactly, a brute-force search for the MDNS
would require us to compute the rank of almost $2^{N}$
distinct controllability matrices, a combinatorially prohibitive task for any
network %
of reasonable size. %
Yet, as we show next, we can identify the MDNS %
by mapping the control problem %
into a purely graph theoretical problem called %
maximum
matching%
~\citep{Yamada-Networks-90,Commault-02,Murota-book-09,Liu-Nature-11}. %

\begin{figure*}[t!]
\includegraphics[width=\textwidth]{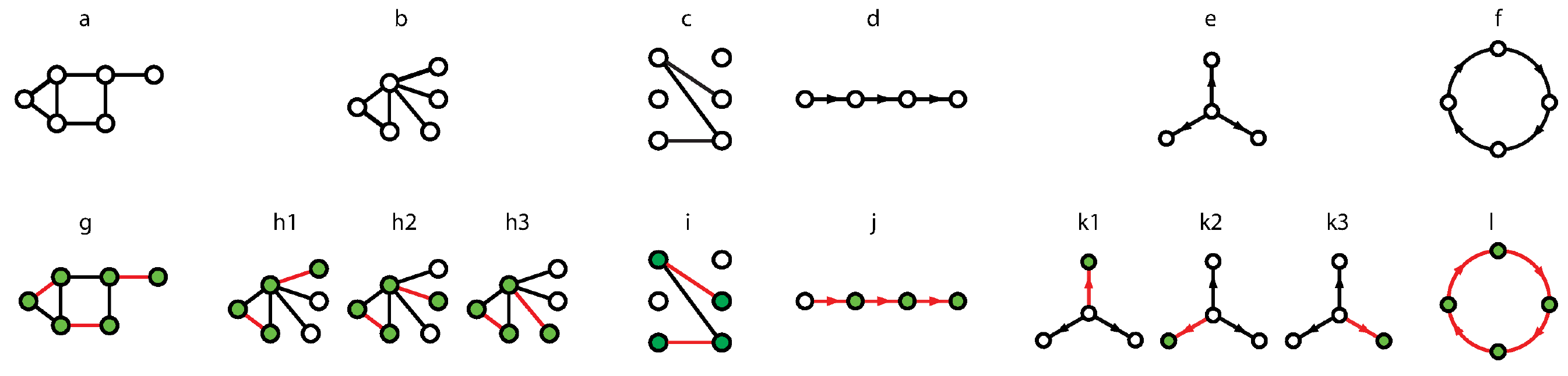}
\caption{(Color online) Matching. The figures show the 
maximum matchings of (a,b) undirected graphs, (c) a bipartite
  graph and (d,e,f) digraphs. 
For undirected or bipartite graphs, a matching represents a set of
edges without common vertices. 
For digraphs, a matching is a set of directed edges that do
not share the common start or end vertices. 
Maximum matching is a matching with the %
largest number of edges. %
On panels (g-i) edges in the matching are colored in red. Matched
  (or unmatched) nodes are shown in green (or white), respectively.
}\label{fig:matching-undirected}
\end{figure*}

Matching is a widely studied problem in graph theory, with many
practical applications~\citep{Lovasz-Book-09}. 
On undirected graphs, where it was originally defined, a matching represents %
a set of edges without
common vertices (red edges in Fig.~\ref{fig:matching-undirected}g). Maximum matching is a matching of the largest
size. For most graphs we can find %
multiple maximum matchings
(Fig.~\ref{fig:matching-undirected}h1-h3).   
The end %
vertices of a matching edge are called \emph{matched}, the remaining
vertices %
are \emph{unmatched}.  
If all vertices are matched, then the matching is \emph{perfect}
(Fig.~\ref{fig:matching-undirected}g).  
Many real world problems can be formalized as a maximum matching
problem on bipartite graphs (Fig.~\ref{fig:matching-undirected}c). 
Consider, for example, %
$M$ job applicants applying for $N$
openings. Each applicant is interested in a subset of the openings. Each
opening can only accept one applicant and an applicant can only accept
one job offer. Finding an assignment of openings to applicants such
that as many applicants as possible get a job is a classical %
maximum %
matching problem. %

\begin{figure*}[t!]
\includegraphics[width=\textwidth]{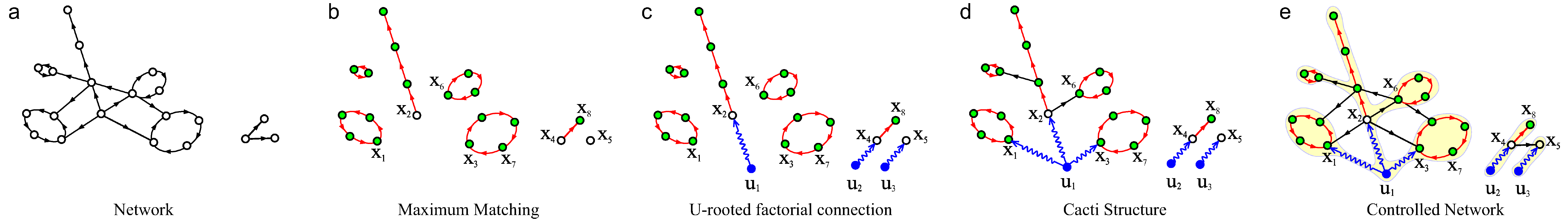}
\caption{(Color online) Graph-theoretic proof of the minimum input theorem. 
(a) A %
directed network. (b) %
The maximum
matching represents the largest set of edges without common heads or
tails. All maximum matchings can be decomposed into a set of vertex-disjoint
directed paths and directed cycles, shown in red. If a node is the head
of a matching edge, then this node is matched (shown in
green). Otherwise, it is unmatched (shown in white). 
The unmatched nodes must be directly controlled to control the whole
network, hence they are %
the driver nodes.   
(c) By injecting signals into %
driver nodes, we get a set of
directed paths whose starting points are the input nodes. The resulting 
paths are called ``stems'' and the resulting digraph is called U-rooted factorial
connection.  
(d) By ``grafting'' the directed cycles to those ``stems'', we get
``buds''. The resulting digraph is called cactus or cacti.  A cactus is a minimal structure for structural
    controllability, as removing any of its edges will cause either
    inaccessible nodes or dilations. 
(e) According to the structural controllability theorem, since there is a
cacti structure (highlighted in yellow) underlying the controlled network, the
system is structurally controllable. 
Note that (a-d) also suggests an
efficient method to identify the \emph{minimal cacti}, i.e. the cacti
structure with the minimum number of roots. %
This minimal cacti serve as the control skeleton that maintains the
structural controllability of the system. 
}\label{fig:cactus}
\end{figure*}

{\color{red}
In structural control theory, the role of matching is %
well studied and matching was originally defined in the bipartite
representation of a
digraph~\citep{Yamada-Networks-90,Commault-02,Murota-book-09}. 
The extended definition of matching on a digraph~\citep{Liu-Nature-11}
connects more naturally to the cactus structure
(Fig.~\ref{fig:cactus}), which is a fundamental notion in structural
control theory. 
}  
In a directed graph (digraph), a 
matching is defined to be a set of directed edges that do
not share common start or end vertices~\citep{Liu-Nature-11}. 
Hence, a vertex can be the starting or the end point of a red link, but
we cannot have two red links pointing to the same vertex. 
A vertex is \emph{matched} if it is the end vertex of a matching
edge. Otherwise, it is \emph{unmatched}. 
For example, in a directed path, all but the starting vertex are matched
(Fig.~\ref{fig:matching-undirected}d,j). A 
matching of maximum size %
is called a \emph{maximum
  matching}. A maximum matching is called
\emph{perfect} if all vertices are matched, like %
in a directed elementary cycle %
(Fig.~\ref{fig:matching-undirected}f,l).  
We can prove that a matching of a digraph can be
  decomposed into %
a set of directed paths and/or directed cycles
(Fig.~\ref{fig:cactus}b). 
Note that directed paths and cycles are also the basic elements of 
the cactus structure (Fig.~\ref{fig:cactus}d). 
Hence, matching in digraphs connects naturally to the cactus
structure. 

The usefulness of matching in network control comes from a %
theorem that provides %
the minimum number of driver nodes in a
network~\citep{Liu-Nature-11}.

\emph{Minimum input theorem}: \label{sec:MIT}
To fully control a directed network $G({\bf A})$, the minimum
number of inputs, or equivalently the minimum number of driver
nodes, %
is %
\be N_\mm{D} = \max \left\{ N - |M^*|,
  1\right\}, \label{eq:mit} 
\ee
where $|M^*|$ is the size of the maximum matching in $G({\bf A})$.  
In other words, 
the driver nodes correspond to %
the unmatched nodes. 
If all nodes are matched ($|M^*|=N$), we need at least one input to 
control the network, 
hence %
$N_\mm{D}=1$. 
We can choose any node as our driver node in this case.

The %
minimum input theorem maps %
an inherently dynamical problem, i.e. our ability to control a network
from a given subset of nodes, into a purely graph theoretical problem 
of finding the maximum matching of a directed network. 
Most important, it bypasses the need to search all node combinations for a
minimum driver node set, as the driver nodes are %
provided by the solution of the underlying matching problem.

\begin{figure}[t!]
\includegraphics[width=0.35\textwidth]{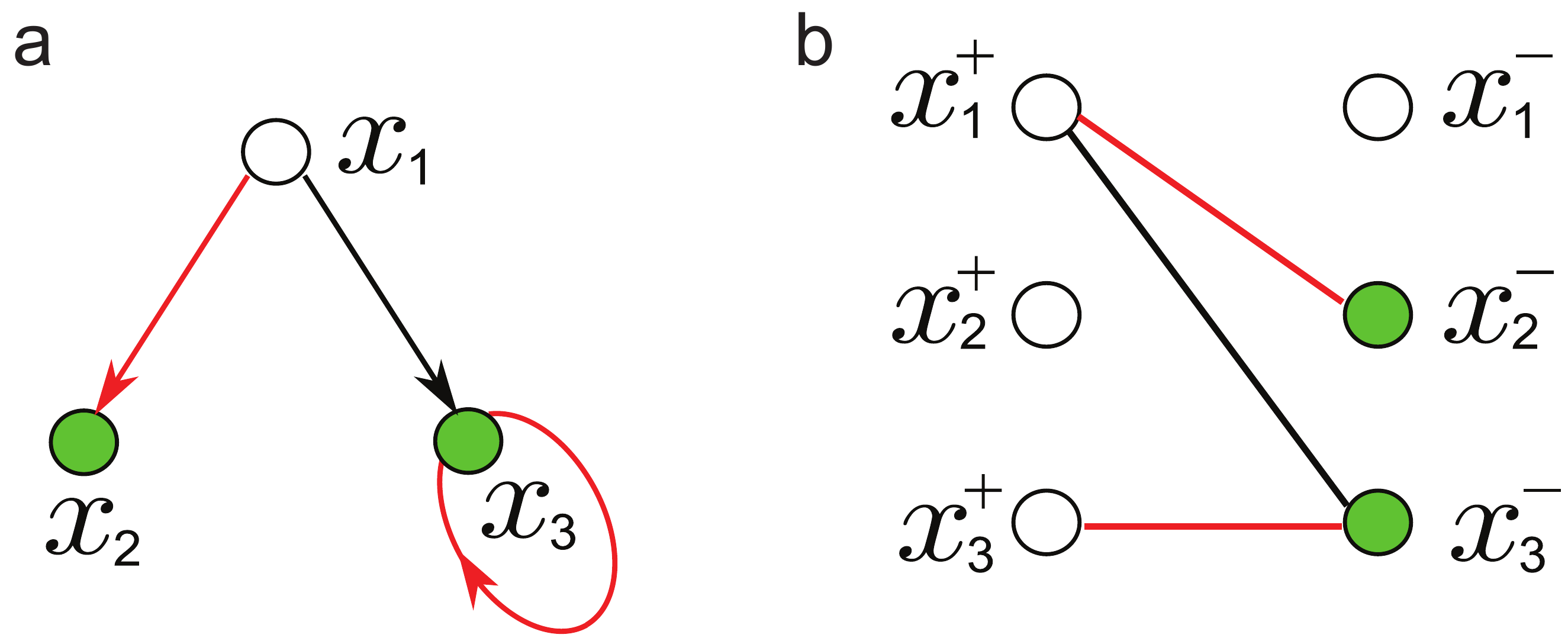}
\caption{(Color online) Maximum matching calculation. %
The maximum matching %
of the digraph (a) can be
  computed from its bipartite representation (b), which is obtained by splitting each
  node $x_i$ into two ``nodes'' ($x_i^+$
and $x_i^-$) and placing an edge $(x_j^+, x_i^-)$ in the bipartite
graph if there is a directed edge $(x_j \to  x_i)$ in the original
digraph. 
The maximum matching of %
any bipartite graph %
can be identified in polynomial time using the 
Hopcroft-Karp algorithm. Mapped back to the digraph, we obtain the
maximum matching of the original digraph and the driver nodes of the
corresponding control problem. %
}\label{fig:bipartite_matching}
\end{figure}

\emph{Maximum matching: algorithmic solution}. 
The mapping of the MDNS problem to a matching
problem via (\ref{eq:mit}) seems to map a problem of high
computational complexity %
--- an exhaustive search for the MDNS --- into another just as complicated problem,
that of finding the maximum matching for a digraph. The real value of this
mapping, however, comes from the fact that %
the maximum matching problem in a digraph in not NP-hard, but can be
solved in polynomial time. 
Indeed, the maximum matching for a %
digraph can be %
identified by mapping the digraph to its bipartite representation, as
illustrated in Fig.~\ref{fig:bipartite_matching}. 
Consider a digraph $G({\bf A})$, %
{\color{red}whose bipartite representation is} $H({\bf A}) \equiv (V_A^+
\cup V_A^-, \Gamma)$. %
{\color{red}Here, } $V_A^+ = \{
x_1^+,\cdots,x_N^+ \}$ and $V_A^- = \{ x_1^-, \cdots, x_N^- \}$ are the set of
vertices corresponding to the $N$ columns and rows of the state matrix ${\bf
  A}$, respectively. 
The edge set of this bipartite graph is $\Gamma = \{(x_j^+, x_i^-) \mid a_{ij} \neq 0
\}$. 
In other words, we split each node $x_i$ of the original digraph into two ``nodes'' $x_i^+$
and $x_i^-$. We then place %
an edge $(x_j^+, x_i^-)$ in the bipartite
graph if there is a directed edge $(x_j \to  x_i)$ in the original
digraph. 
{\color{red}Note that since we allow self-loops $(x_i \to x_i)$ in the original
  digraph, there can be edges of this type $(x_i^{+}, x_i^{-})$ in the
  bipartite graph.} 
A maximum matching of a %
bipartite graph can be found
efficiently using the %
Hopcroft-Karp algorithm, which 
runs in $O(\sqrt{V} E)$ time%
~\citep{Hopcroft-SIAM-73}. 
After running the algorithm, we can map the maximum matching in the bipartite
  representation, e.g. $(x_1^+, x_2^-), (x_3^+, x_3^-)$ in
  Fig.~\ref{fig:bipartite_matching}b, back to the maximum matching in
  the original diagraph, e.g. $(x_1, x_2), (x_3, x_3)$ in
  Fig.~\ref{fig:bipartite_matching}a, obtaining the desired maximum
  matching and hence the corresponding MDNS.

\begin{figure}[t!]
\includegraphics[width=0.4\textwidth]{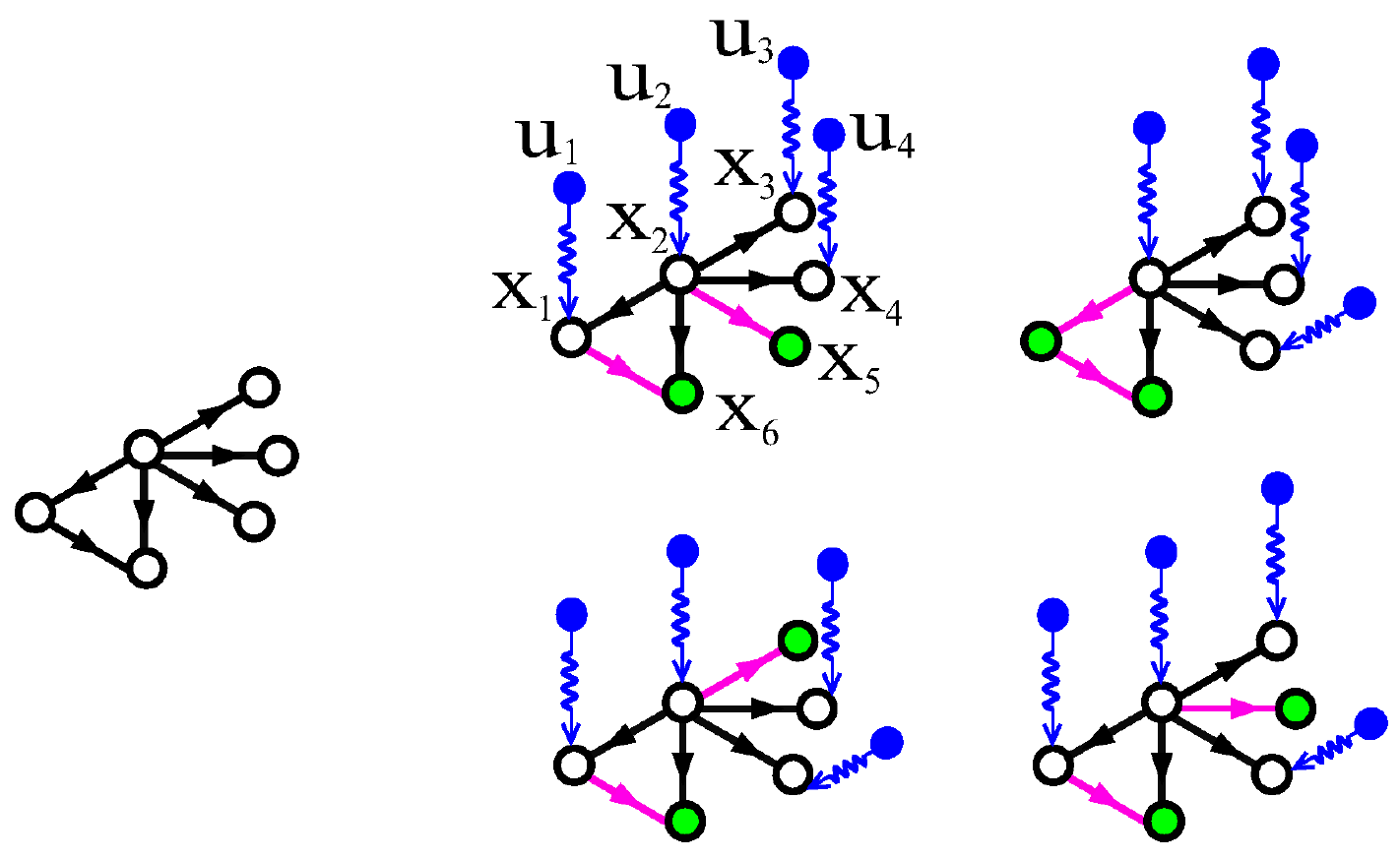}
\caption{(Color online)  Identifying the driver nodes. 
For a general directed network, like the one shown in the left panel, there could be multiple
  maximum matchings, shown in red on the right panels. Hence, we can
  identify %
multiple MDNSs (white
  nodes). To each driver node we must add a unique control signal 
necessary to ensure structural controllability. 
}\label{fig:different_matching}
\end{figure}

Taken together, the maximum matching algorithm allows the efficient
identification of the %
MDNS using the
following steps %
(Fig.~\ref{fig:different_matching}):  
(i) Start from the directed network we wish to control and %
generate its bipartite representation
(Fig.~\ref{fig:bipartite_matching}). %
Next identify
a maximum matching on the underlying bipartite graph using the Hopcroft-Karp
algorithm. 
(ii) To each unmatched node add a unique control signal, as unmatched
nodes %
represent the driver nodes. 
(iii) As there could be multiple maximum matchings for a general
digraph, multiple MDNSs exist, with the same size $N_\mm{D}$.

{\color{red}
Recently, several algorithmic approaches have been developed to
optimize the network controllability (in the sense of 
decreasing $N_\mm{D}$) via minimal structural perturbations, like adding a minimum number of edges at judiciously chosen locations
in the network~\citep{Wang-PRE-12}, rewiring redundant
edges~\citep{Hou-ISDEA-13}, and assigning the direction of
edges~\citep{Hou-CPL-12,Xiao-PRE-14}.  
}

\emph{Maximum matching: analytical solution based on the cavity method}. \label{sec:cavity}
While the maximum matching allows us to %
efficiently identify the MDNS, %
the algorithmic approach %
provides no %
physical insights about the impact of the network topology on
$N_\mm{D}$. %
For example, what network characteristics
influence $N_\mm{D}$, and how does $N_\mm{D}$ depend on them? 
Which networks are easier to control and which are harder?
To answer these %
questions we can turn to the cavity method, a
versatile tool of statistical physics~\citep{Mezard-EPJB-01,Zhou-arXiv-03,Zdeborova-JSM-06}. 
We illustrate this approach by analytically calculating %
$\overline{n}_\mm{D}$, representing the
fraction of driver nodes $n_\mm{D}$ ($\equiv
N_\mm{D}/N$) averaged over all network realizations compatible
with %
the network's degree distribution $P(k_\mm{in},
k_\mm{out})$~\citep{Liu-Nature-11}. 
We start by describing 
a matching $M$ in a digraph $G=\{V(G), E(G)\}$ by the binary
variables $s_a = s_{(i\to j)}\in \{0,1\}$ assigned to each directed
edge 
$a=(i\to j) \in E(G)$ with $s_a=1$ if $a$ belongs to the matching
$M$ 
and $s_a=0$
otherwise. According to the definition of matching in a digraph,
matching edges do not share starting or end nodes, formally resulting
in  
two constraints for %
each vertex $i \in V(G)$:  
(i) $\sum_{j \in \partial^+ i} s_{(i\to j)}  \le 1$;  
(ii) $\sum_{k \in \partial^- i} s_{(k\to i)} \le 1$  
with $\partial^- i$ and $\partial^+ i$ indicating the sets of nodes
that point to $i$ or are pointed by $i$, respectively. 
The quantity $\mathcal{E}_i(\{s\}) = 1 - \sum_{k \in \partial^- i }
s_{(k\to i)}$ tells us the state of each vertex:  
vertex $i$ is matched if $\mathcal{E}_i(\{s\}) = 0$ and unmatched 
if $\mathcal{E}_i(\{s\}) =1$.
Consequently, %
the cost (or energy) function %
gives for each matching
$M=\{s\}$  
the number of unmatched vertices 
\be
\mathcal{E}_G(\{s\}) =  \sum_{i \in V(G)} \mathcal{E}_i(\{s\}) =  N- |M|.
\ee

We define the Boltzmann probability in the space of matchings as 
\be
\mathcal{P}_G(\{s \}) = \frac{e^{-\beta \mathcal{E}_G(\{s
    \})}}{\mathcal{Z}_G(\beta)},
\ee 
where $\beta$ is the inverse temperature and $\mathcal{Z}_G(\beta)$ is
the partition function 
\be
\mathcal{Z}_G(\beta) = \sum_{\{s\}} e^{-\beta \mathcal{E}_G(\{s \})}.
\ee
%
%
%
%
%
%
\iffalse
For $\beta \to \infty$ (zero temperature limit), the internal energy
$\mathcal{E}_G(\beta)$, corresponding to %
the number of unmatched vertices, and the entropy
$\mathcal{S}_G(\beta)$, corresponding to %
the logarithm of the number of
matchings, provide %
the ground state properties, corresponding to %
the properties of the maximum matchings. 
\fi
{\color{red}
In the limit $\beta \to \infty$ (i.e. the zero temperature limit), 
the internal energy $\mathcal{E}_G(\beta)$ and the entropy
$\mathcal{S}_G(\beta)$ provide the ground state properties, i.e. the
properties of the maximum matchings. In particular,
$\mathcal{E}_G(\infty)$ represents the number of 
unmatched vertices (with respect to any maximum matching), and the
entropy $\mathcal{S}_G(\infty)$ yields the logarithm of the number of
maximum matchings. 
}

In the zero temperature limit, the average fraction of driver nodes 
is given by 
\bea
\overline{n}_\mm{D}
&=& \frac 12 \Big\{ 
\big[G(\wh{w}_2) + G(1-\wh{w}_1) - 1 \big] 
  \nn
&& + 
\big[\widehat{G}(w_2) + \widehat{G}(1-w_1) -1 \big] \nn
&& 
 + \frac{z}{2} \big[ \wh{w}_1 (1-w_2) + w_1 (1-\wh{w}_2) \big] 
\Big\}, \label{eq:nD} 
\eea
where $w_1, w_2, w_3, \wh{w}_1, \wh{w}_2, \wh{w}_3$ satisfy the
set of self-consistent equations 
\be
\begin{cases}
w_1 = H(\wh{w}_2)\\
w_2 = 1- H(1-\wh{w}_1)\\
w_3 = 1-w_2-w_1\\
\wh{w}_1 = \wh{H}(w_2)\\
\wh{w}_2 = 1- \wh{H}(1-w_1)\\
\wh{w}_3 = 1-\wh{w}_2-\wh{w}_1
\end{cases}
\ee
and 
\be
\begin{cases}
G(x)         \equiv \sum_{\kout=0}^\infty P(\kout) x^{\kout}\\ 
\widehat{G}(x) \equiv \sum_{\kin=0}^\infty \widehat{P}(\kin)
x^{\kin}\\ 
H(x)         \equiv \sum_{\kout=0}^\infty Q(\kout+1) x^{\kout}\\ 
\widehat{H}(x) \equiv \sum_{\kin=0}^\infty \widehat{Q}(\kin+1)
x^{\kin}
\end{cases}
\ee
are the generating functions, and 
$Q(\kout) \equiv \frac{\kout P(\kout)}{\langle \kout \rangle}$, 
$\widehat{Q}(\kin) \equiv \frac{\kin \widehat{P}(\kin)}{\langle \kin
  \rangle}$ 
are the out- and in- degree distributions of the node $i$ when one
selects %
uniformly at random a directed edge %
$(i\to j)$ from the digraph. 

While the cavity method does not offer a closed-form solution,
Eq.~(\ref{eq:nD}) allows %
us to systematically study the impact 
of key network characteristics, like the average 
degree $\kmean$ or the degree
exponent $\gamma$, on $\overline{n}_\mm{D}$ in the thermodynamic limit
($N\to \infty$). 
For example, for %
directed Erd\H{o}s-R\'enyi random
networks~\citep{Erdos-PMIHAS-60,Bollobas-Book-01}, 
both $P(k_\mm{in})$ and $P(k_\mm{out})$ follow a Poisson distribution,
i.e.  %
$e^{-\kmean/2} (\kmean/2)^k / k!$.
In the large %
$\kmean$ 
limit we have 
\be 
n_\mm{D} \sim %
e^{-\kmean/2}. \label{eq:nDER} %
\ee

For directed scale-free networks, we assume that $P(\kin)$ and $P(\kout)$
have the same functional form %
with power-law exponent $\gamma$ and exponential cutoff
$
P(k_\mm{in})  = C \, k_\mm{in}^{-\gamma} \, e^{-k/\kappa}$, %
$P(k_\mm{out}) = C \, k_\mm{out}^{-\gamma} \, e^{-k/\kappa}$. %
Here the normalization constant is %
$C=\left[\mm{Li}_{\gamma}(e^{-1/\kappa})\right]^{-1}$, where %
$\mm{Li}_n(x)$ is
the $n$th polylogarithm of $x$. Due to the exponential cutoff
$e^{-k/\kappa}$, the distribution is normalizable for any $\gamma$.  
One can show that %
as $\gamma \to 2$, we have $n_\mm{D} \to 1$.  
This means one has to control almost all the nodes to achieve full control over the
network. 
Therefore 
$\gamma=2$ is the critical
value for the controllability of %
scale-free networks, as only for $\gamma>2$ can we obtain %
full controllability by
controlling only a subset of the nodes. 
Note that for $\gamma \to 2$ super-hubs emerge that connect to almost all nodes in the network~\citep{Albert-RMP-02,Barabasi-Book-2015}. We know that for a star-like digraph with one central hub and $N-1$
leaves, one has to control $N_\mm{D}=N-1$ nodes (the central hub and any $N-2$
leaves). In the large $N$ limit, $N-1 \approx N$, which explains intuitively why
          we have to control almost all nodes when $\gamma \to 2$.

For scale-free networks with degree exponent
$\gamma_\mm{in}=\gamma_\mm{out}=\gamma$ generated from the static
model~\citep{Goh-PRL-01}, the parameters $\kmean$ and $\gamma$ are
independent.  
In the thermodynamic limit the degree distribution is %
$ P(k) = \frac{[m (1-\alpha)]^{1/ \alpha}}{\alpha}
\frac{\Gamma(k-1/ \alpha, m[1-\alpha])}{\Gamma(k+1)}$ where
  $\Gamma(s)$ is the
gamma function and $\Gamma(s,x)$ the upper incomplete gamma function.
In the large $k$ limit, $P(k) \sim k^{-(1+\frac{1}{\alpha})} =
k^{-\gamma}$ where  $\gamma= 1 + \frac{1}{\alpha}$.
The asymptotic behavior of $n_\mm{D}(\kmean, \gamma)$ for large $\kmean$ %
is %
\be n_\mm{D}\sim e^{-\frac 12 \left( 1  - \frac{1}{\gamma-1} \right)
  \langle k \rangle}. \label{eq:nDSF}\ee 
If %
$\gamma_\mm{in} \neq \gamma_\mm{out}$, 
the smaller of the two exponents, i.e.  %
$\min[\gamma_\mm{in}, \gamma_\mm{out}]$ %
determines the asymptotic behavior of $n_\mm{D}$. %
Equation (\ref{eq:nDSF}) indicates that as $\gamma \to 2$, %
$n_\mm{D} \to 1$, which is consistent with the result that $\gamma_\mm{c}=2$
for a purely SF network. %

\begin{figure}[t!]
\includegraphics[width=0.5\textwidth]{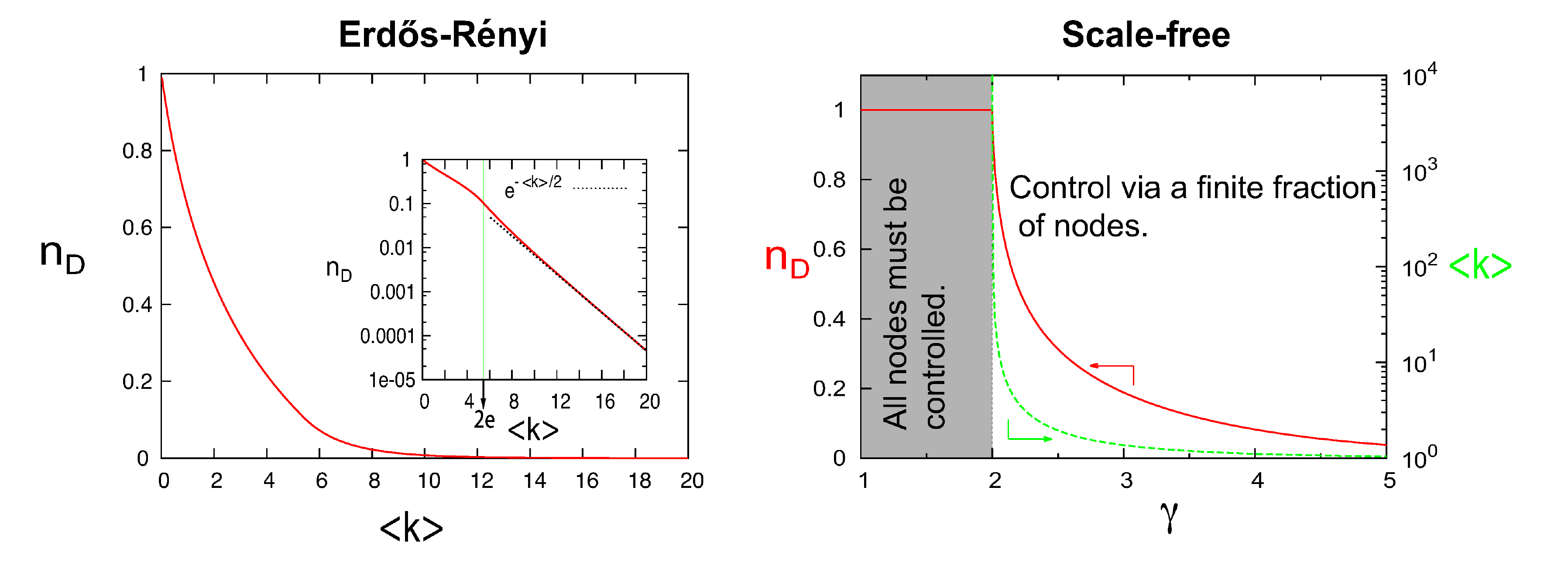}
\caption{(Color online)  Analytical results on the fraction of driver nodes ($n_\mm{D}=N_\mm{D}/N$) for
    canonical model networks. 
(a) For directed Erd\H{o}s-R\'enyi random networks, 
$n_\mm{D}$ decays exponentially for %
large $\kmean$. 
(b) For directed %
scale-free networks $n_\mm{D}$ approaches one
as the %
degree exponent $\gamma$ approaches two, indicating
that in such networks all nodes need to be controlled.   
}\label{fig:nD_analytic}%
\end{figure}

The systematic dependence of $n_\mm{D}$ on $\kmean$ and $\gamma$
prompts us to ask: How do other %
network characteristics, like degree
correlations, clustering, modularity, or the fraction of low degree
nodes, influence %
$n_\mm{D}$
~\citep{Posfai-SR-2013,Menichetti-PRL-14}. 
A combination of analytical and numerical results indicate that 
the clustering coefficient and modularity
have no discernible effect %
on $n_\mm{D}$. At the same time the symmetries of the underlying
matching problem generate %
linear, quadratic or no dependence on
degree correlation coefficients, depending on the %
nature of the underlying degree correlations~\citep{Posfai-SR-2013}.  
For uncorrelated directed networks, %
the density of nodes with $\kin,
\kout =1$ or 2 determine the size of maximum matchings ~\citep{Menichetti-PRL-14}. %
This %
suggests that random networks whose minimum
$\kin$ and $\kout$ are greater than two typically have perfect matchings and
hence can be fully controlled %
via a single control input
(i.e. $N_\mm{D}=1$), regardless of the other
properties of the degree distribution. %

\subsubsection{Solution based on PBH controllability test} \label{sec:MCP}
In structural control theory we assume that the system parameters,
like the link weights in $G({\bf A}, {\bf B})$, are %
either fixed zeroes or independent free parameters. %
This framework is ideal for many systems for which 
we only know the underlying wiring diagram (i.e. 
zero/nonzero values, indicating %
the absence/presence of
physical connections) but not the link characteristics, like their
weights. Yet, the independent free parameter assumption is 
very strong, and it is %
violated in some %
systems, like %
in undirected networks, where the state matrix ${\bf A}$ is symmetric, or
unweighted networks, where all link weights are the same. %
In such cases 
structural control theory could yield %
misleading results 
on the minimum number of driver nodes $N_\mm{D}$. 
Hence, it is important to move beyond structural control as we 
explore %
the controllability and other control related issues.
For LTI systems with exactly known system parameters
the minimum input problem can be efficiently solved using   
the Popov-Belevitch-Hautus (PBH)
controllability test. %
The PBH controllability test states %
that the system $({\bf A},
{\bf B})$ is controllable if and only if~\citep{Hautus-1969}
\be 
\mm{rank}\, [s {\bf I} - {\bf A}, {\bf B}] = N, \quad\forall s \in \mathbb{C}. 
\label{eq:PBH}
\ee  
Since the first $N\times N$ block of the $N \times (N+M)$ matrix $[s
{\bf I} - {\bf A}, {\bf B}]$ %
has full rank whenever $s$ is not an eigenvalue of $\mA$, we only need to check each
eigenvalue of $\mA$, i.e. $s\in \lambda({\bf A})$, when running the PBH test.
Note that the PBH test (\ref{eq:PBH}) and Kalman's rank condition (\ref{eq:rankC})
are equivalent. 
Yet, the advantage of the PBH test comes from the fact that it %
connects the controllability of $(\mA,\mB)$ to the eigenvalues and eigenvectors
of the state matrix $\mA$. This can be used to solve the minimum
input problem exactly. 
Indeed, the PBH controllability test suggests that $(\mA,\mB)$ is
controllable if and only if there is no left eigenvector of $\mA$
orthogonal to all the columns of $\mB$. 
In other words, 
the columns of $\mB$ must have a component in each eigendirection of
$\mA$. 
Recall that for an eigenvalue $\lambda_0 \in \lambda(\mA)$, its
\emph{algebraic multiplicity} is the {\color{red}multiplicity} of $\lambda_0$ as a
root of the characteristic polynomial $p(\lambda) = \mm{det}(\mA -
\lambda {\bf I})$. Its \emph{geometric multiplicity} is the maximal
number of linearly independent eigenvectors 
corresponding to it.
Hence, the number of control inputs must be greater than or equal to the
largest geometric multiplicity of the eigenvalues of
$\mA$~\citep{Antsaklis-Book-1997,Sontag-Book-1998,Yuan-NC-2013}. 
In other words, the minimum number
of control inputs (or equivalently the minimum number of driver nodes)
is determined by the maximum geometric multiplicity of the eigenvalues
of ${\bf A}$, i.e. 
\be
N_\mm{D} = \max_i \{ \mu(\lambda_i) \},  \label{eq:nD_exactcontrol}
\ee
where $\mu(\lambda_i)= \mm{dim} V_{\lambda_i} = N - \mm{rank}(\lambda_i
{\bf I}_N - {\bf A})$ is the geometric multiplicity of ${\bf A}$'s eigenvalue
$\lambda_i$, representing the dimension of its
eigenspace. %
Note that the
algebraic multiplicity of eigenvalue $\lambda_i$, denoted by %
$\delta({\lambda_i})$, is its multiplicity as a root of the
characteristic polynomial. In general, $\delta({\lambda_i}) \ge
\mu(\lambda_i)$. But for symmetric ${\bf A}$, which is the case for undirected
networks, we have $\delta({\lambda_i}) = \mu(\lambda_i)$.

\begin{figure}[t!]
\includegraphics[width=0.5\textwidth]{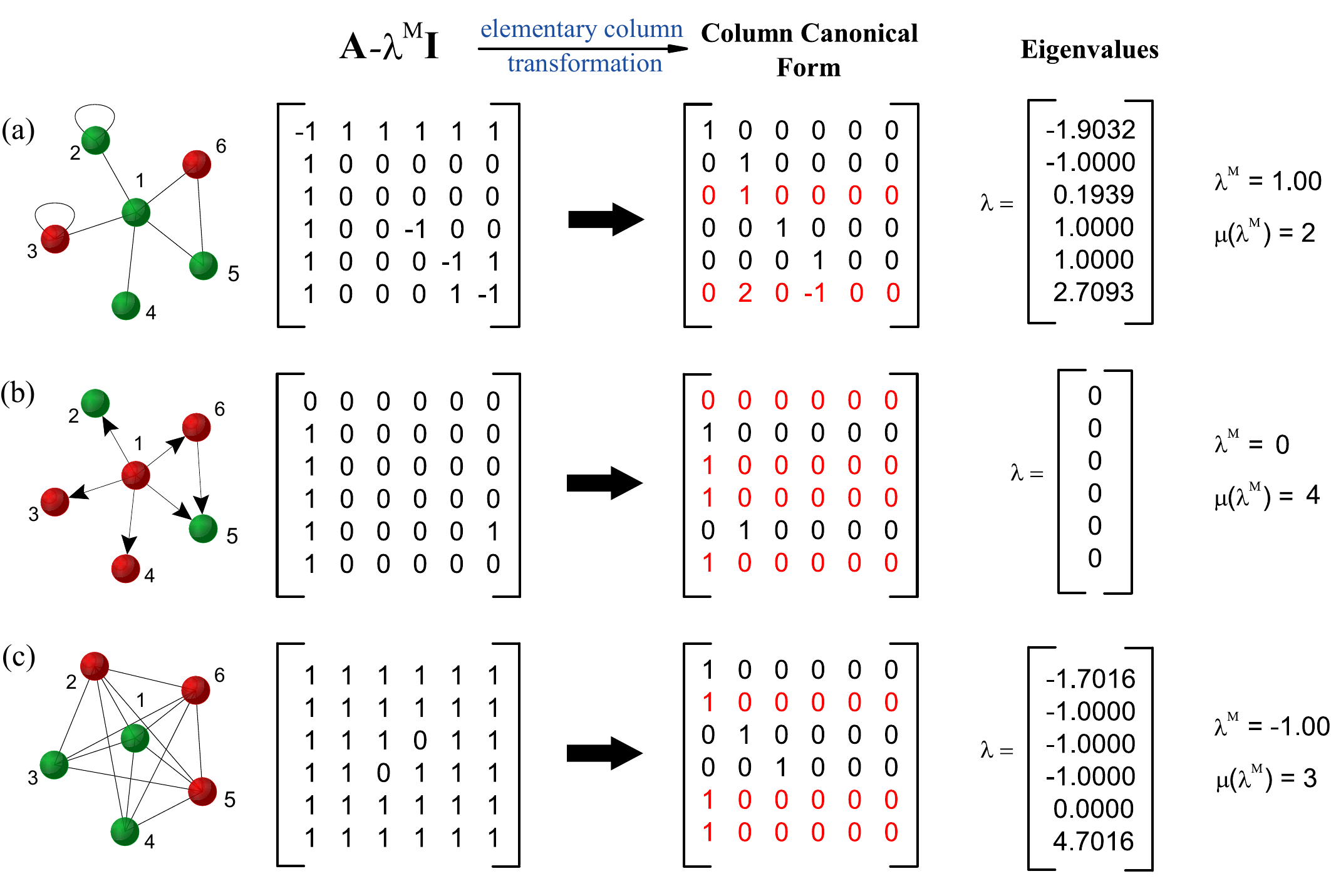}
\caption{(Color online)  Identifying a minimum set of driver nodes of small
  networks. 
For each network, we show the matrix $\mA-\lambda^M \mI$, its column canonical
form, all %
eigenvalues $\lambda$ of $\mA$, and the eigenvalue $\lambda^M$ with
the largest geometric multiplicity. %
We highlight the rows that are linearly dependent on others in the
column canonical form in red. 
The corresponding nodes are the driver nodes
(shown in red) of the corresponding %
networks. 
For undirected networks in (a) and (c), $\mu(\lambda^M)$ is equal to
the maximum algebraic multiplicity, that is, the multiplicity of
$\lambda^M$. The configuration of driver nodes is not unique as it relies
on the elementary column transformation, but the minimum number of
drivers is uniquely %
determined by the maximum geometric multiplicity
$\mu(\lambda^M)$ of matrix $\mA$. 
  After \citep{Yuan-NC-2013}.  
}\label{fig:nD-yuan}
\end{figure}

\begin{table}[h!]
\begin{center}
\caption{Eigenvalues and minimum number of driver nodes of some
    special %
graphs of $N$ nodes. For unweighted and undirected star and fully
    connected networks, the table shows the algebraic multiplicity of
    eigenvalues  %
in the parenthesis. After \citep{Yuan-NC-2013}.}
\label{table:Yuan}
\begin{tabular}{llr}
\hline
\hline
Network & Eigenvalue & $N_\text{D}$\\
\hline
Chain & $2\cos{\frac{q\pi}{N+1}}$, $q=1,\cdots,N$ & $1$ \\
Ring & $2\cos{\frac{2\pi(q-1)}{N}}$,$q=1,\cdots,N$& $2$\\
Star & $0(N-2),\pm{\sqrt{N-1}}(1)$  & $N-2$\\
Complete graph & $N-1(1),-1(N-1)$& $N-1$\\
\hline
\hline
\end{tabular}
\end{center}
\end{table}

Based on (\ref{eq:nD_exactcontrol}), we can 
develop an efficient algorithm to
identify the minimum set of driver nodes for arbitrary LTI systems
(Fig.~\ref{fig:nD-yuan}), allowing us to %
explore 
the impact of the network topology %
and link-weight distributions on
$N_\mm{D}$~\citep{Yuan-NC-2013}. 
For undirected and unweighted ER %
networks of connectivity
probability $p$, %
the results indicate that for small $p$, $n_\mm{D}$
decreases with $p$, while for sufficiently large $p$, %
$n_\mm{D}$ %
increases to %
$(N-1)/N$, which %
is exact for $p=1$ (complete graph,
see Table.~\ref{table:Yuan}). %
This approach has been recently extended to 
multiplex networks~\citep{Yuan-NJP-14}. 
\subsection{Minimal Controllability Problems} \label{sec:MCP}

{\color{blue}
Any networked system with LTI dynamics is fully controllable if we
control each node individually with an independent signal,
i.e. $M=N$. But this is costly and typically impractical for
large complex systems. Hence, we are particularly interested in
fully controlling a network with minimum number of nodes. 
Depending on the objective function and the way we ``inject'' input
signals, we can formalize different types of \emph{minimal controllability
problems} (MCPs)~\citep{Olshevsky-IEEE-14}.

\begin{figure}[t!]
\includegraphics[width=0.45\textwidth]{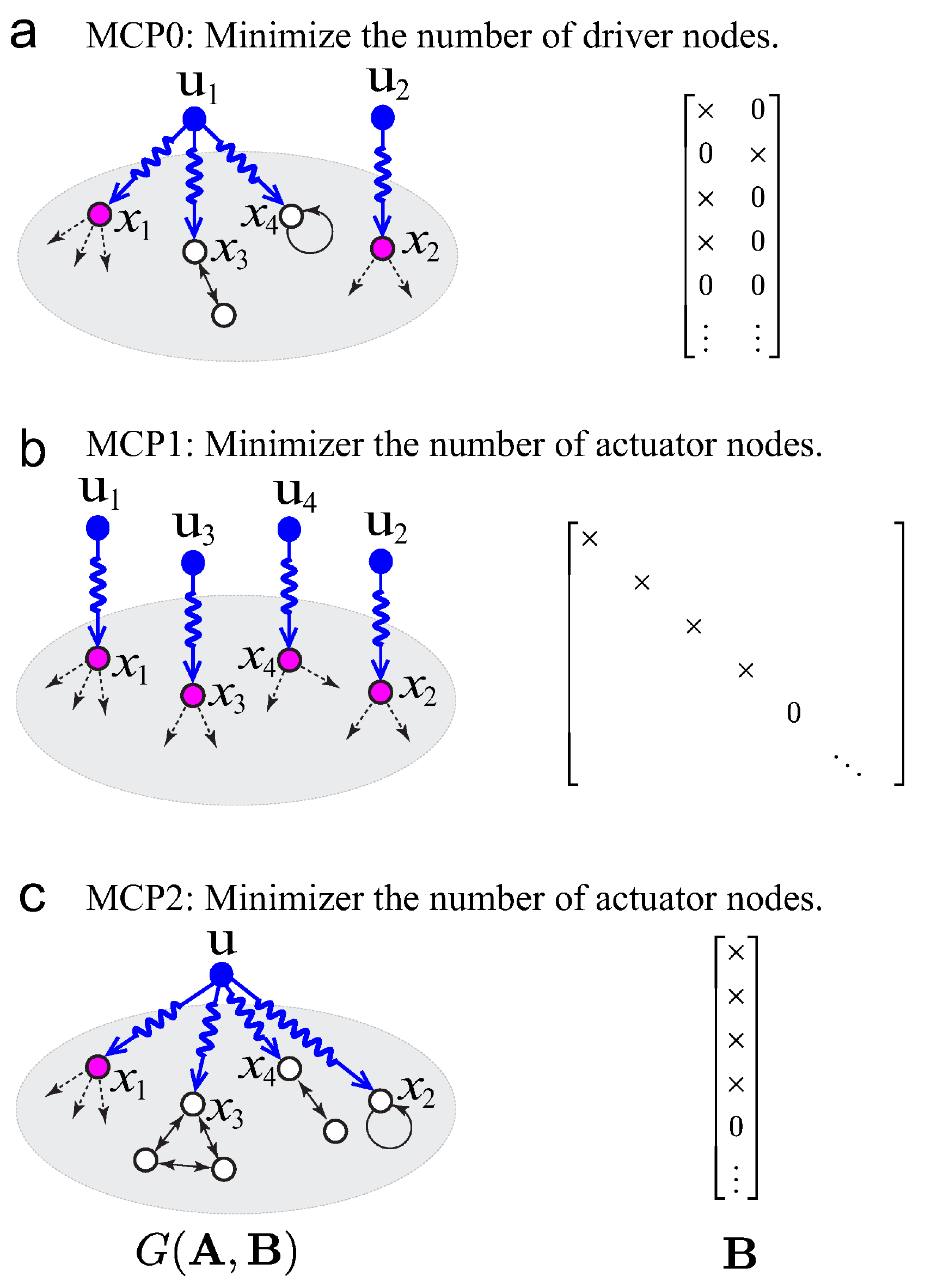}
\caption{(Color online)  \color{blue}Different minimal controllability problems
  (MCPs). For each MCP, we show the corresponding
  graph representation $G(\mA, \mB)$, and the input matrix $\mB$ (where
  $\times$'s stand for non-zero elements. 
MCP0: We aim to minimize the number of driver nodes, or equivalently,
the number of independent input signals. One signal can drive multiple nodes.
MCP1: We aim to minimizer the number of actuator nodes which
receive input signals. One signal can only drive one actuator node.
MCP2: We aim to minimizer the number of actuator nodes with only
one signal. This unique signal can drive multiple actuator nodes.
In all cases, we assume there are four actuator nodes ($x_1, x_2, x_3$
and $x_4$). We color the driver nodes in pink. 
}\label{fig:MCPs}
\end{figure}

(MCP0): One scenario is that we try to minimize the number of independent
control signals, corresponding to the number of columns in the input matrix $\mB$, or
equivalently, the number of \emph{driver nodes}~\citep{Liu-Nature-11} 
whose control is sufficient to fully control
the system's dynamics (Fig.~\ref{fig:MCPs}a). This is nothing but the minimum inputs problem
discussed in the previous subsection. 

(MCP1): We assume dedicated inputs, i.e. each control input $u_i$ can
only directly control one node (state variable).  
In the matrix form, this amounts to finding a diagonal matrix $\mB \in
\mathbb{R}^{N\times N}$ that has as few nonzero entries as possible so
that the LTI system $\dot{\mx} = \mA \mx + \mB {\bf u}$ is
controllable (Fig.~\ref{fig:MCPs}b).

(MCP2): 
We set $u_i(t)=u(t)$ and aim to find a vector ${\bf b}$ that has as
few nonzero entries as possible such that the system $\dot{\mx} = \mA
\mx + {\bf b} u$ is controllable
(Fig.~\ref{fig:MCPs}c).  %

Note that in solving MCP0, one signal can be
applied to multiple nodes. The number of \emph{actuator nodes}
(corresponding to those non-zero entries in $\mB$) is not necessarily
minimized. 
In MCP1 ${\bf u}(t)$ is a vector of control inputs, i.e. we
have multiple input signals, while in
MCP2, $u(t)$ is a scalar, i.e. there is only one input signal. In
both cases, we try to minimize the number of \emph{actuator nodes}
that are directly controlled by input signals.

Though MCP0 %
for a general LTI system is easy to
solve, MCP1 and MCP2 are NP-hard~\citep{Olshevsky-IEEE-14}. 
\iffalse
Fortunately,
he also proved that finding %
approximate solutions for both MCPs
can be done in polynomial time. %
%
This can be achieved by a %
greedy heuristic algorithm that sequentially
picks variables to maximize the rank increase of the controllability
matrix~\citep{Olshevsky-IEEE-14}.  
\fi
Yet, if we need to guarantee only %
structural controllability, MCP1 can
be easily solved~\citep{Pequito-ACC-13,Pequito-IEEE-16}. %
For a directed network $G$ with LTI dynamics the
minimum number of dedicated inputs (or actuators), %
$N_\mm{da}$, required to assure structural controllability, is %
\be
N_\mm{da} = N_\mm{D} + \beta - \alpha, 
\ee
where $N_\mm{D}$ is the minimum number of driver nodes; 
$\beta$ is the number of  \emph{root} strongly connected
components (rSCCs), which have no incoming links from other SCCs; 
and $\alpha$ is the \emph{maximum assignability index} of the bipartite
representation $\mathcal{B}(G)$ of the directed network $G$.  
An rSCC is said to be a top assignable SCC if it contains
at least one driver node with respect to a particular maximum matching
$M^*$. 
The maximum assignability index of $\mathcal{B}(G)$ is the maximum
number of top assignable SCCs that a maximum matching $M^*$ may lead
to. 
The minimum set of actuators can be found with polynomial time
complexity~\citep{Pequito-ACC-13,Pequito-IEEE-16}.%

Consider, for example, the network shown in
  Fig.~\ref{fig:GraphTheoreticInterpretation}, which has %
two possible
  maximum matchings 
$M_1 = \{ (x_1 \to x_4), (x_4 \to x_3), (x_5 \to x_5) \}$, 
$M_2 = \{ (x_1 \to x_2), (x_4 \to x_3), (x_5 \to x_5) \}$. 
Both have size $3$, hence the number of driver nodes is
$N_\mm{D}=\max\{N-|M^*|,1\}=2$, according to (\ref{eq:mit}).  
Note that the two maximum matchings will yield two minimum sets of driver
nodes, i.e. $\{x_1, x_2\}$ and $\{x_1,x_4\}$. The former is shown in
Fig.~\ref{fig:GraphTheoreticInterpretation}. 
There are two rSCCs, $\{x_1\}$ and $\{x_5\}$, each containing a single
node, hence $\beta=2$. 
The rSCC $\{x_1\}$ is a top assignable SCC, because it contains one
driver node with respect to either $M_1$ or $M_2$. 
The rSCC $\{x_5\}$ is not a top assignable SCC, because it contains no
driver nodes. 
Hence the maximum
assignability index of this system is $\alpha=1$. 
Finally, the minimum number of actuators is
$N_\mm{a}=N_\mm{D}+\beta-\alpha=3$ and there are two minimum sets of
actuators, i.e. $\{x_1, x_2, x_5\}$ and $\{x_1,x_4, x_5\}$. 

}

\subsection{Role of Individual Nodes and Links}

As we have seen %
in Sec.\ref{sec:MIT}, %
a system with $N_\mm{D}$ driver
nodes can be controlled by %
multiple driver node configurations, each
corresponding to a 
different maximum matching
(Fig.~\ref{fig:different_matching}). %
Some links may appear more often in the maximum matchings than other
links. 
This raises a fundamental question: What is the role of the individual
node (or link) in control? Are some nodes (or links) more important for
control than others? 
To answer these questions, in this section we discuss the
classification of %
nodes and links 
based on their role and importance in the control of %
a given network. %

 \subsubsection{Link classification}\label{sec:LC}

 In both natural and technological systems we need to quantify 
 how robust is our ability to control a network under unavoidable 
 link failure. To adress this question, we can use structural
 controllability to classify %
each link into one of the following three categories:  
 (1) a link is \emph{critical} if in its absence we must %
 increase the number of driver nodes to maintain full control over the
 system. In this case the link is part of \emph{all} maximum matchings of the network;  
 (2) a link is \emph{redundant} if it can be removed without
 affecting the current set of driver nodes (i.e. it does not appear in
 any maximum matching); 
 (3) a link is \emph{ordinary} if it is neither critical nor redundant
 (it appears in some but not all maximum matchings).  
Note that this classification can be efficiently done with a
polynomial-time algorithm based on Berge's
property~\citep{Regin-1994}, rather than enumerating all maximum
matchings, which is infeasible for large networks. 

We can %
compute the density of critical ($l_\mm{c}=L_\mm{c}/L$), redundant
  ($l_\mm{r}=L_\mm{r}/L$) and ordinary ($l_\mm{o}=L_\mm{o}/L$) links for
  a wide range of real-world networks. 
It turns out that most real networks have few or no critical links. 
 Most links are ordinary, meaning that they play a role in some control
 configurations, but the network can be still controlled in their absence~\citep{Liu-Nature-11}. 

\begin{figure}[t!]
 \includegraphics[width=0.45\textwidth]{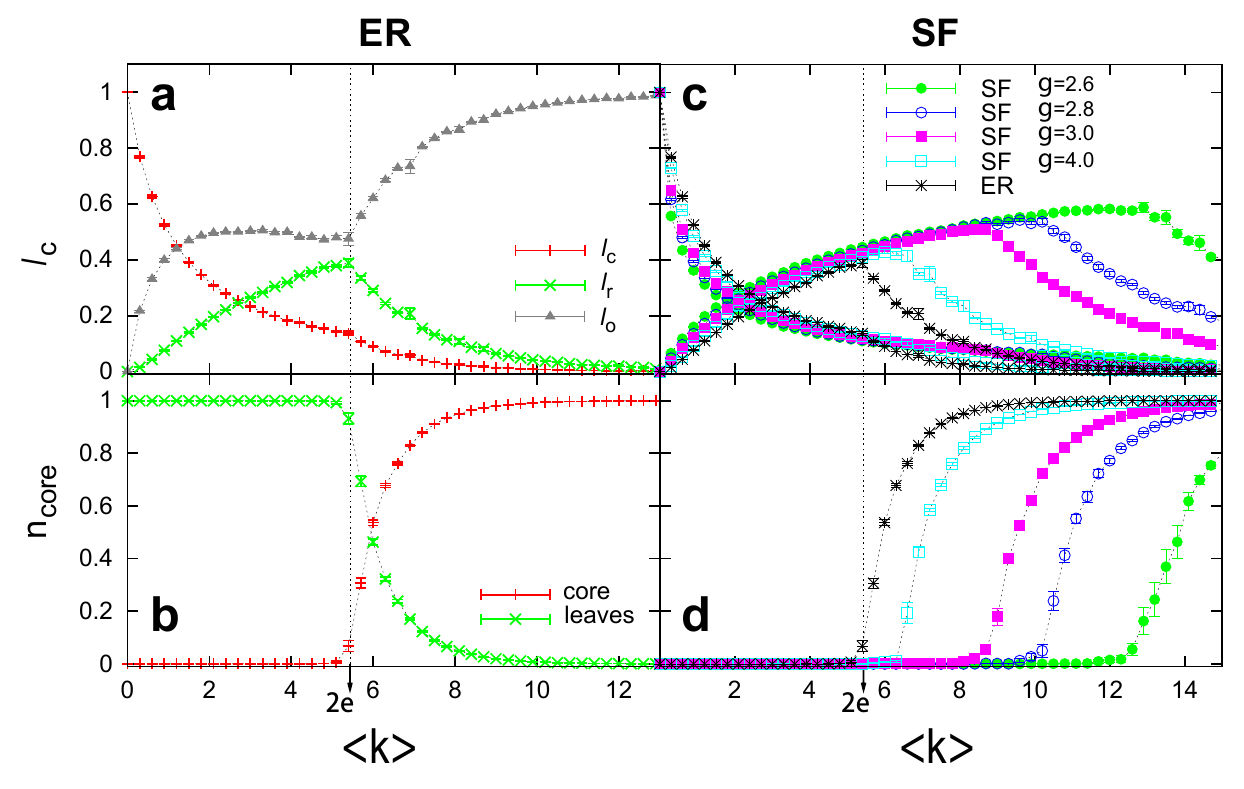}
 \caption{(Color online)  Link classification and core percolation. a, Dependence on
   $\kmean$ of the fraction of critical (red, $l_\mm{c}$), redundant
   (green, $l_\mm{r}$) and ordinary (grey, $l_\mm{o}$) links for an Erd\H{o}s-R\'enyi (ER) 
   network: $l_\mm{r}$ peaks at $\kmean=\kmean_\mm{c}=2e$ and the
   derivative of $l_\mm{c}$ is discontinuous at $\kmean=\kmean_\mm{c}$. b, Core percolation
   for the ER network occurs at $\kmean=\kmean_\mm{c}=2e$, which explains
   the $l_\mm{r}$ peak. c, d, Same as in a and b but for scale-free
   networks constructed using the static model. The ER and SF networks
   have $N=10^4$ nodes and the results are averaged over ten
   realizations with error bars defined as the standard error of the
   mean. %
 Dotted lines are only
   a guide to the eye.  %
   After \citep{Liu-Nature-11}.  
 }\label{fig:controllability-lc}
 \end{figure}

 For model networks (ER and SF), we can calculate 
$l_\mm{c}$, $l_\mm{r}$, and $l_\mm{o}$ as functions of $\kmean$ 
 (Fig.~\ref{fig:controllability-lc}). 
 The behavior of $l_\mm{c}$ is easy to understand:
 for small $\kmean$ all links are essential for
 control ($l_\mm{c}\approx 1$). As $\kmean$ increases the network's
 redundancy increases, decreasing $l_\mm{c}$. The increasing redundancy
 suggests that the density of redundant links, $l_\mm{r}$, should always
 increase with $\kmean$, but it does not: it reaches a maximum at
 $\kmean_\mm{c}$, after which it decays.   
 This non-monotonic behavior %
 results from a structural transition %
 driven by 
 core percolation~\citep{Liu-Nature-11}. %
 Here, the \emph{core} represents a compact
 cluster of nodes left in the network after applying a \emph{greedy leaf
   removal} procedure: Recursively remove in-leaf (with $\kin=1$) and
 out-leaf (with $\kout=1$) nodes' neighbors' all outgoing (or incoming)
 links. 
 The core emerges through a percolation transition (Fig.~\ref{fig:controllability-lc}b,d): for $k< \kmean_\mm{c}$, $n_\mm{core}=
 N_\mm{core}/N = 0$, so the system consists of leaves only. At
 $\kmean_\mm{c}$ a small core emerges, decreasing the number of
 leaves. For ER random networks, 
 the analytical calculations %
 predict $\kmean_\mm{c}=2\, e \approx 5.436564$, in agreement with the
 numerical result (Fig.~\ref{fig:controllability-lc}b), a value that coincides with %
 $\kmean$ %
 where $l_\mm{r}$ reaches its maximum.  Indeed, $l_\mm{r}$
 starts decaying at $\kmean_\mm{c}$ because after $\kmean_\mm{c}$ the number
 of distinct maximum matchings increases exponentially, which can be
 confirmed by calculating the ground state entropy using the cavity
 method~\citep{Liu-Nature-11}. 
 Consequently, the chance that a link does \emph{not} participate in 
   \emph{any} control configurations decreases. %
 For SF networks we observe the same behavior, with
 the caveat that $\kmean_\mm{c}$ decreases with $\gamma$ (Fig.~\ref{fig:controllability-lc}c, d). 

\subsubsection{Node Classification} \label{sec:nc}
Given the existence of multiple driver node configurations, we %
can classify nodes based on their likelihood
of being included in the minimum driver node set (MDNS): %
a node is (1) \emph{critical} if
that node must %
always be controlled to control the system, implying that %
it is part of all MDNSs; (2)
\emph{redundant} if %
it is never required for control, implying that it never %
participates in an MDNS; and (3) \emph{intermittent} if %
it is a %
driver node in some control configurations, but not in
others~\citep{Jia-NC-2013}. 

\begin{figure}[t!]
\includegraphics[width=0.45\textwidth]{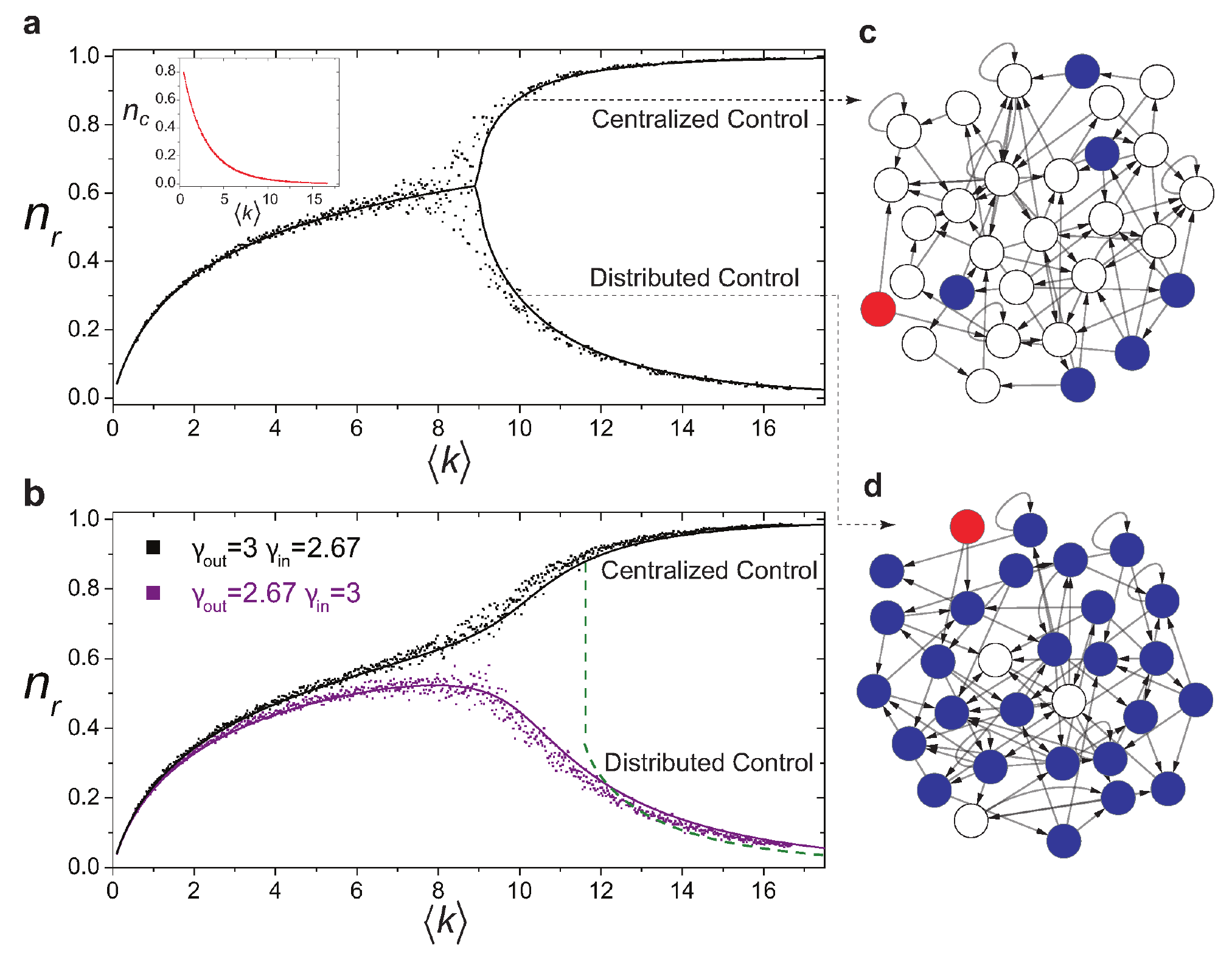}
\caption{(Color online)  Emergence of bimodality in controlling
  complex networks. %
  (a) $n_\mm{r}$ and $n_\mm{c}$ (insert) vs $\kmean$ in
  scale-free networks with degree exponents $\gamma_\mm{out} =
  \gamma_\mm{in} = 3$, displaying %
the emergence of a bimodal behavior for high $\kmean$. (b) $n_\mm{r}$ in
scale-free networks with asymmetric in- and out-degree distribution,
i.e. $\gamma_\mm{out} = 3$, $\gamma_\mm{in} = 2.67$ (upper branch) and
$\gamma_\mm{out} = 2.67$, $\gamma_\mm{in} = 3$ (lower branch). The
control mode is pre-determined by their degree asymmetry. (c,d)
Networks displaying centralized or distributed control. 
Both networks have $N_\mm{D} = 4$ and $N_\mm{c} = 1$ (red node), but they have rather
different number of redundant nodes (blue nodes), $N_\mm{r} = 23$ in (c) and
$N_\mm{r} = 3$ in (d). 
After \citep{Jia-NC-2013}.  
}\label{fig:nodeclassification}
\end{figure}

For model networks with symmetric in- and out-degree distributions, we
find that the fraction of redundant nodes ($n_\mm{r}$) 
undergoes %
a \emph{bifurcation} at a critical mean degree $\kmean_\mm{c}$: 
for low $\kmean$ the fraction of
redundant nodes ($n_\mm{r}$) is uniquely determined by $\kmean$, but beyond
$\kmean_\mm{c}$ two different solutions for $n_\mm{r}$ %
coexist, one with very high
and the other with %
very low value, leading to a bimodal behavior
(Fig.~\ref{fig:nodeclassification}a). 
Hence for large $\kmean$ (after the %
bifurcation) %
two control modes coexist%
~\citep{Jia-NC-2013}: 
(i) \emph{Centralized control}: In networks that follow the upper
branch of the bifurcation diagram most of the nodes are redundant,  as in
this case $n_\mm{r}$
is very high. This means that in these networks only a small fraction
of the nodes are involved in control %
($n_\mm{c}+  n_\mm{i}$ is very low), hence control is guaranteed by 
a few nodes in the network. A good analogy would be a %
company involved in manufacturing whose %
leadership is concentrated in the hands of a few
managers and the rest of the employees are only executors. %
(ii) \emph{Distributed control}: In networks on the lower branch $n_\mm{c}+
n_\mm{i}$ can exceed 90\%. %
Hence, most nodes participate
as 
driver nodes in some MDNSs, implying that one can engage most nodes in
control. A good analogy would be an innovation-based horizontal
organization, where any employee can take a leadership role, as the
shifting tasks require. %

For ER random networks this bifurcation occurs at $\kmean_\mm{c}=2e$, corresponding to %
the core percolation threshold~\citep{Liu-PRL-2012}.

Another way to assess a node's importance for control is to quantify 
the impact of its removal on 
controllability. 
Consider a network with minimum number of driver nodes
$N_\mm{D}$. After a node is removed (deleted), denote the minimum number of 
driver nodes with $N'_\mm{D}$. %
Once again, each node can belong to %
one of 
three categories: (1) A node is \emph{deletion critical} if in its absence
we have to control more driver nodes, i.e. $N'_\mm{D} > N_\mm{D}$. For
example, removing a node in the middle of a directed path will
increase %
$N_\mm{D}$. %
(2) A node is \emph{deletion redundant} if in its absence we have
$N'_\mm{D} < N_\mm{D}$. For example, removing a leaf node in a star will
decrease $N_\mm{D}$ by 1. (3) A node is \emph{deletion ordinary} if in its
absence 
$N'_\mm{D} = N_\mm{D}$. For example, removing the central hub in a star
will not change $N_\mm{D}$. %
The above node classification has been applied to
directed human protein-protein interaction networks, whose %
directions indicate signal flow~\citep{Vinayagam-PNAS-2016}. %
In this context %
critical nodes tend to correspond to 
disease genes, viral tagets, through which a virus 
takes control over its host, and %
targets of %
FDA approved drugs, indicating that control-based classification can
select biologically relevant proteins. %

\subsubsection{Driver node classification} 
To understand %
why a node is a driver node, 
we %
decompose the driver nodes ($N_\mm{D}$) into three groups~\citep{Ruths-Science-14}:
(1) \emph{source nodes} ($N_\mm{s}$) that have no incoming links,
hence they must %
be directly controlled, being always driver nodes;  
(2) \emph{external dilations} ($N_\mm{e}$) arise due to a surplus of sink
nodes ($N_\mm{t}$) that have no outgoing links.  Since each source
node can %
control one sink node, the number of external dilation is
$N_\mm{e}=\max(0,N_\mm{t}-N_\mm{s})$; 
(3) \emph{internal dilations} ($N_\mm{i}$) occur 
when a path must branch into two or more paths in order to
reach all nodes (or equivalently a subgraph has more outgoing links
than incoming links). 
This classification leads to %
the control profile of a network defined as 
$(\eta_\mm{s}, \eta_\mm{e}, \eta_\mm{i})=(N_\mm{s}/N,N_\mm{e}/N,N_\mm{i}/N)$, 
which quantifies the different proportions of
control-inducing structures present in a network. 
The measurements indicate that 
random network models do not reproduce the %
control profiles %
of %
real-world networks and that the control profiles of
real networks group into three well-defined clusters, dominated by %
external-dilations, sources, or %
internal-dilations~\citep{Ruths-Science-14}.

These results offer insight into the high-level organization and
function of complex networks. %
For example, %
neural and social network are source dominated, which allow
relatively uncorrelated behavior across their agents and are thus
suitable to distributed processing. 
Food webs and airport interconnectivity networks are
internal-dilation dominated. They are mostly closed systems 
and obey some type of conservation laws. 
In contrast, trust hierarchies and transcriptional systems are
external-dilation dominated. With their surplus sink
nodes, these systems display correlated behavior across their agents
that are downstream neighbors of %
a common source. 

\begin{figure}[t!]
\includegraphics[width=0.45\textwidth]{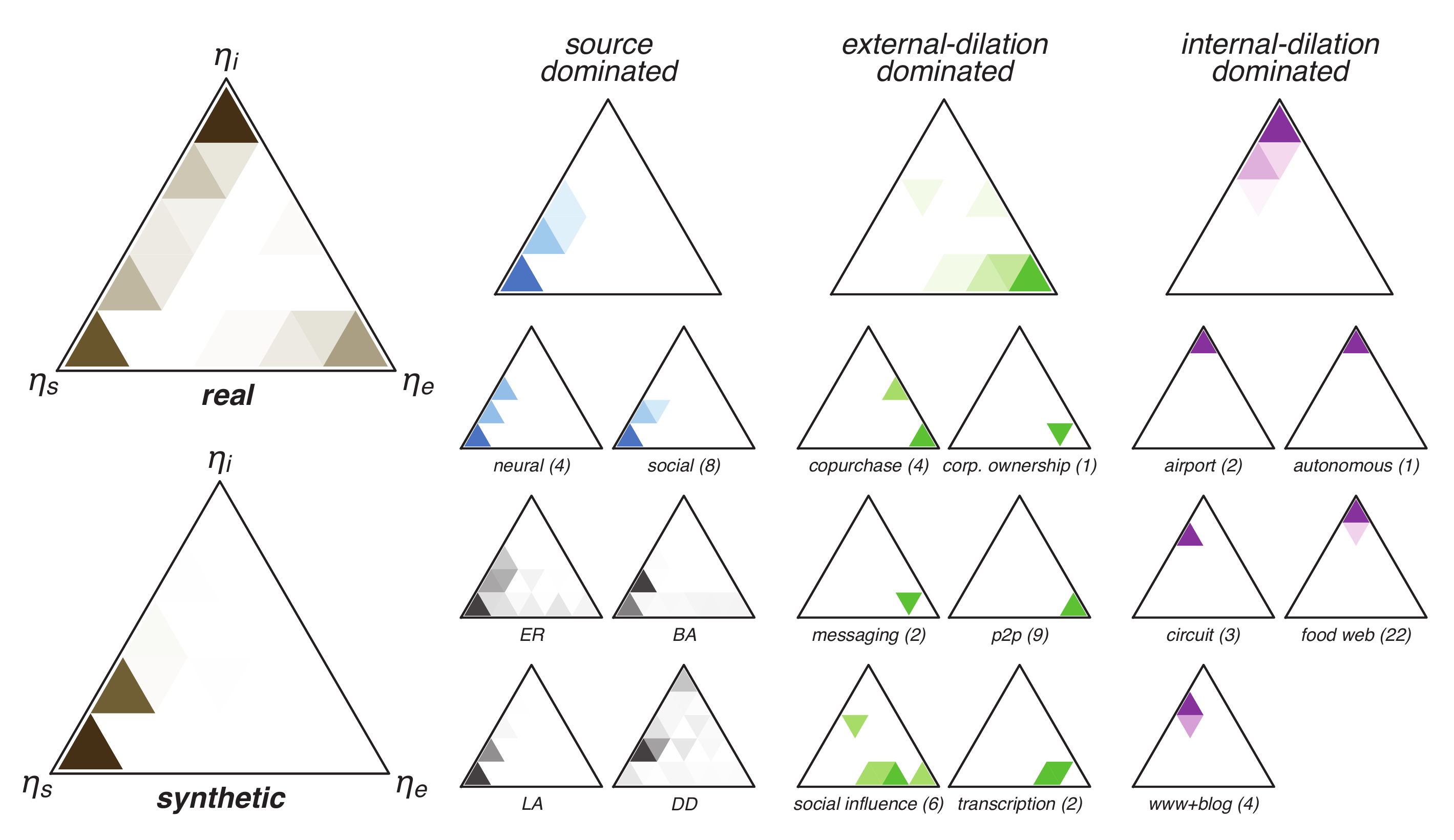}
\caption{(Color online)  Control profiles of real and model networks. %
The control profiles of real networks %
show %
a tendency to cluster around the three components $(\eta_\mm{s},
\eta_\mm{e}, \eta_\mm{i})$ of the control profile, implying that %
real networks broadly fall into three distinct classes:
external-dilation dominated, source dominated, and internal-dilation dominated. 
The coloring of each %
small heatmap %
indicates the clustering observed in a wide range %
of real networks, with numbers in parentheses indicating the number of networks
  present in each heatmap.
  Deeper shades of the heatmap represent %
a greater density of networks with control profiles located in that region. %
  After \citep{Ruths-Science-14}.  
}\label{fig:controlprofile}
\end{figure}

\subsection{Controllable Subspace, Control Centrality, {\color{red}and Structure Permeability}}\label{sec:CST}
Lin's structural controllability theorem can tell us whether an LTI
system $({\bf A}, {\bf B})$ is 
structurally controllable or not. %
If, however, the system is not structurally
controllable, the theorem does not provide further information
about %
controllability. %
Even if we are unable to %
make the system %
reach %
any point in the %
state space, we would like %
to understand which region %
of the state space is %
accessible to it, i.e., what region of the state space 
can we control it. %
For example, %
in the network of 
Fig.~\ref{fig:controllablesubspace}a %
the control input $u_1$ %
is applied to the central
hub $x_1$ of the directed star with $N = 3$ nodes. The system is
therefore %
stuck
in the plane described by 
$a_{31}x_2(t) = a_{21}x_3(t)$, %
shaded in Fig.~\ref{fig:controllablesubspace}b. 
Consequently, the network %
is not controllable in
the whole state space, but it is controllable within the %
subspace defined by the plane.

When we control a single node $i$, %
the input matrix $\mB$ reduces to a
vector ${\bf b}(i)$ with a
single non-zero entry, and the controllability matrix $\mC \in
\mathbb{R}^{N \times N}$ becomes 
$\mC(i)$. We can use %
$\mm{rank}(\mC(i))$ as a natural measure of node $i$'s ability to control the
system. If $\mm{rank}(\mC(i)) = N$, then node $i$ alone can control the whole
system. %
Any %
$\mm{rank}(\mC(i))$ %
less than $N$ yields %
the dimension of the subspace $i$ can control. 
For example,  
if $\mm{rank}(\mC(i))=1$, then node $i$ can only control itself.

In reality the system parameters (i.e. the entries of $\mA$ and $\mB$) are
often not known precisely, except the zeros that mark the absence of
connections, %
rendering the calculation of $\mm{rank}(\mC(i))$ difficult. 
This difficulty can be again avoided using structural control
theory. %
Assuming %
$\mA$ and $\mB$ are
structured matrices, i.e., their elements are
 either fixed zeros or independent free parameters, then %
 $\mm{rank}(\mC(i))$ varies as a function of the free parameters of $\mA$ and
 $\mB$. 
However, it achieves its maximum for almost all sets of values of the
free parameters except for some pathological cases with Lebesgue
measure zero.  
This maximal value is called the \emph{generic rank}{\color{red}~\citep{Johnston-IJC-84}} of the controllability
 matrix $\mC(i)$, denoted as $\mm{rank}_\mm{g}(\mC(i))$, which also represents the \emph{generic
 dimension} of the controllable subspace. 

We %
define the control capacity %
of a single node $i$, or \emph{control centrality}, as 
the \emph{generic 
dimension} of the controllable subspace
\be d_\mm{c}({\bf A},{\bf b}) = \mm{rank}_\mm{g} \, {\bf C}.\ee
Here $\mm{rank}_\mm{g} \, {\bf C}$ is the \emph{generic rank} of the controllability matrix
${\bf C}$ associated with the structured system $({\bf A}, {\bf b})$,
where we control node $i$ only~\citep{Liu-PLOS-12}. %
This definition can also be extended to the case when we control via
a group of nodes. 
The above definition corresponds directly to our intuition of how powerful a
single node is (or a group of nodes are) in controlling the whole
network. For example, if the control capacity of a single node is $N$, 
then we can %
control the whole system through it.

\begin{figure}[t!]
  \includegraphics[width=0.48\textwidth]{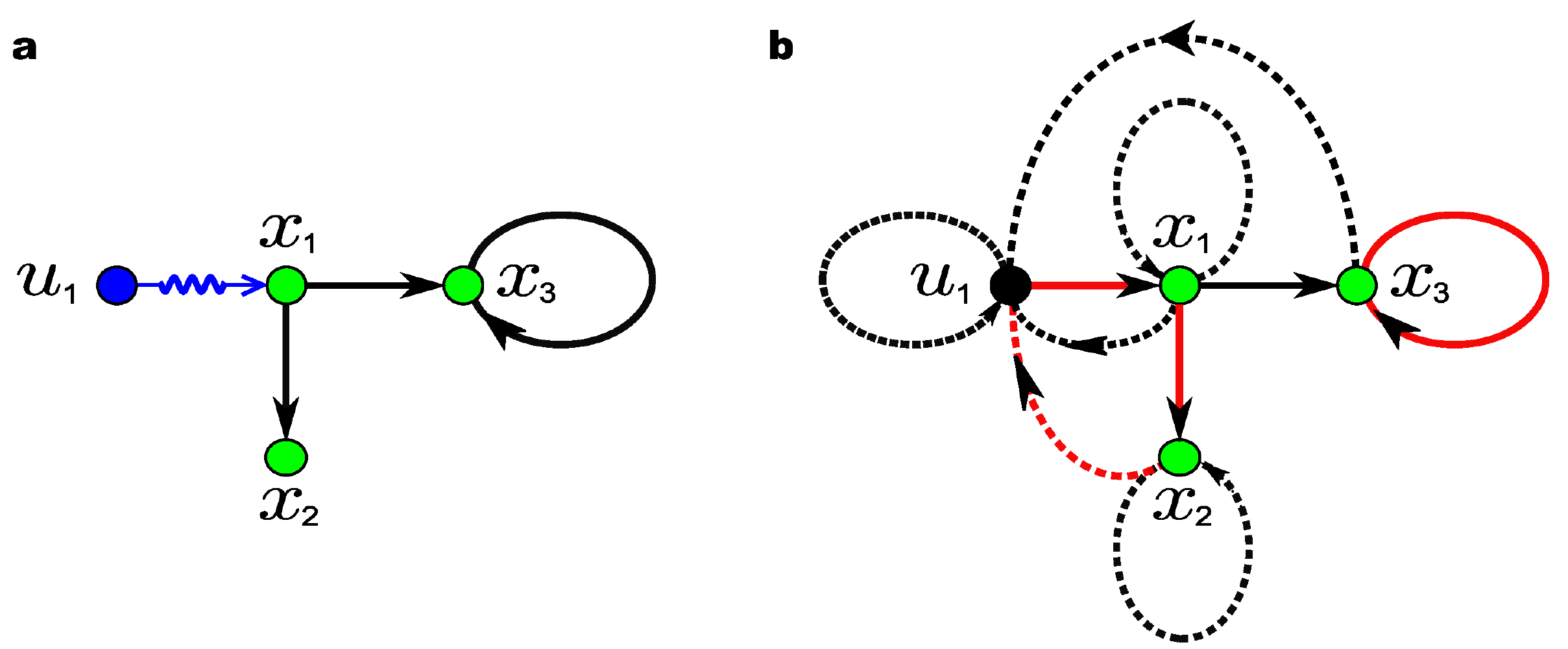}
  \caption{(Color online) Calculation of control centrality. (a) The original controlled system is
    represented by a digraph $G({\bf A}, {\bf
      B})$. (b) The modified digraph $G'({\bf A}, {\bf B})$ used in
    solving the linear programming. Dotted and solid lines are assigned with weight
    $w_{ij}=0$ and 1, respectively. The maximum-weight cycle partition is
    shown in red, which has weight 3, corresponding to the generic dimension
    of controllable subspace by controlling node $x_1$ or equivalently
    the control centrality of node $x_1$. }\label{fig:Hosoe_LP}       
\end{figure}

The calculation of $d_\mm{c}({\bf A},{\bf B})$ has a %
graph-theoretic interpretation
~\citep{Hosoe-IEEE-80}. 
Consider %
a structured system $({\bf A}, {\bf B})$, %
in which %
all state vertices are accessible, and let us denote %
with ${\cal G}$ the set of subgraphs of $G({\bf A}, {\bf B})$ which can be spanned by a
  collection of vertex-disjoint cycles and stems. In this case,
the generic dimension of the controllable subspace is 
\be d_\mm{c}({\bf A}, {\bf B}) = \max_{G \in {\cal G}}
{|E(G)|}\ee where $|E(G)|$ is the number of edges in the subgraph
$G$. This is called Hosoe's \emph{controllable subspace
  theorem}. 
Essentially, Hosoe's theorem tells us that to calculate the generic
dimension of the controllable subspace we need to find the cactus %
that contains as many edges as possible. %
Note that Hosoe's theorem applies only to a structured system $({\bf A}, {\bf B})$
that has no inaccessible state vertices. In calculating $d_\mm{c}({\bf
  A},{\bf B})$ for a general system $({\bf A}, {\bf B})$, we should only
consider the accessible part of the %
network. %
For %
a digraph %
with no directed cycles %
Hosoe's theorem further simplifies: %
the controllability of any node
equals its layer index:  
$
C_\mm{s}(i) = l_i. 
$
Here the layer index of a node is %
calculated from the unique
hierarchical structure of the digraph %
following a recursive labeling
procedure~\citep{Liu-PLOS-12}.
For general networks, we can use linear programming to %
calculate $d_\mm{c}({\bf A},{\bf B})$~\citep{Poljak-IEEE-90}. 
 We first get a new graph $G'({\bf A}, {\bf B})$ from  
 $G({\bf A}, {\bf B})$  by adding to $G({\bf A}, {\bf B})$ the edges $(v_i,
 v_{N+j})$ for $i=1,\cdots, N$, $j=1,\cdots, M$; and the loops $(v_i, v_i)$ for
 $i=1,\cdots, N+M$, if they do not exist in $G({\bf A}, {\bf B})$ (see
 Fig.~\ref{fig:Hosoe_LP}). We associate
 the weight $w_e=1$ with every original edge $e$ of $G({\bf A}, {\bf B})$ and
 the weight $w_e=0$ with every new edge. A collection of node-disjoint cycle in
 $G'({\bf A}, {\bf B})$ covering all nodes will be called a \emph{cycle
 partition}. It is easy to check that to calculate $\max_{G \in {\cal G}}
 {|E(G)|}$ is equivalent to calculating the maximum-weight cycle partition in
 $G'({\bf A}, {\bf B})$, which can then be solved by the following linear 
 programming: 
 $\max \sum_{ e \in G'({\bf A}, {\bf B})}  w_e x_e$ subject to: 
 (1)  $\sum (x_e : e \mbox{ leaves node } v_i ) = 1$ for every node $v_i
 \in G'({\bf A}, {\bf B})$;
 (2) $ \sum (x_e : e \mbox{ enters node } v_i ) = 1$ for every node $v_i
 \in G'({\bf A}, {\bf B})$;
 (3) $x_e \in \{0, 1 \}$ for every edge $e \in G'({\bf A}, {\bf B})$.

{\color{red}
Hosoe's theorem also allows us to address a problem complementary to
the notion of control centrality: identify an optimal set of driver nodes of
fixed cardinality $M$, denoted as $\Omega_\mm{D}(M)$, for a network of
size $N$ such that the dimension of the controllable subspace, denoted
as $|\mathcal{C}(M)|$, is maximized~\citep{Iudice-NC-15}. If we solve
this problem for each %
$M \in [1,N]$, we obtain a sequence of $|\mathcal{C}(M)|$. To quantify the readiness or
propensity of a network to be controllable, we can calculate the
so-called \emph{network permeability} measure~\citep{Iudice-NC-15}
\be
\mu = \frac{\int_0^N (|\mathcal{C}(M)| - M) \ud M}{\int_0^N (N-M) \ud M}.
\ee
Note that $\mu \in [0,1]$: 0 for $N$ disconnected nodes, and 1 for
networks that are completely controllable by one driver node. 
Generally, for a network with a high permeability, a large
controllable subspace can be obtained with a reasonable small set of
driver nodes.}

\subsection{Controlling Edges}%
So far we focused on nodal dynamics, where we monitored and controlled
the state of nodes. 
The sole purpose of the edges %
was to pass information
or influence between the nodes. %
In social or communication networks nodes %
constantly process the information received from their upstream neighbors and
make decisions that are communicated to their downstream neighbors. 
Most importantly, in these systems nodes can communicate different information along %
different edges. %
Hence 
the information received and passed on by a node can 
be best %
represented by 
state variables defined on the %
incoming and outgoing edges,
respectively. 
In this section we ask how to control systems characterized by %
such edge dynamics. 

To model such systems we place %
the state variables on the edges%
~\citep{Nepusz-NP-12}. 
Let ${\bf y}_i^-(t)$ and ${\bf y}_i^+(t)$ represent %
vectors consisting of the state
variables %
associated with the incoming and outgoing edges of
node $i$, respectively. Let ${\bf M}_i$ denote the %
$\kout(i) \times
\kin(i)$ matrix. The equations governing the edge dynamics 
can %
be written as 
\be
\dot{\bf y}_i^+(t) = {\bf M}_i {\bf y}_i^-(t) - \boldsymbol\tau_i \otimes
{\bf y}_i^+(t) + \sigma_i {\bf u}_i(t) \label{eq:edgedynamics}
\ee
where $\boldsymbol\tau_i$ is a vector of damping terms associated with 
the outgoing edges, {\color{red} $\otimes$ denotes the entry-wise product of two vectors of
the same size}, and $\sigma_i=1$ if node $i$ is a driver node and
0 otherwise. 
Note that even though the state variables and the
  control inputs are defined on the edges, we can still designate a
  node to be a driver node if its outgoing edges are directly
  controlled by the control inputs.
Equation (\ref{eq:edgedynamics}) states that the state variables of the outgoing
edges of node $i$ are determined by the state variables of the
incoming edges, modulated by  
a decay term. 
For a driver node, the state variables
  of its outgoing edges
  will also be influenced by the control
  signals ${\bf u}_i$. %
{\color{red}Since each node $i$ acts as a small switchboard-like
  device mapping the signals of the incoming edges to the outgoing
  edges using a linear operator ${\bf M}_i$,
  Eq. (\ref{eq:edgedynamics}) is often called 
the \emph{switchboard dynamics}.} 

There is a mathematical duality between edge dynamics on a network $G$  
and nodal dynamics on its \emph{line graph} $\mathcal{L}(G)$, which
represents the adjacencies between edges of $G$.  
Each node of $\mathcal{L}(G)$ corresponds to an edge in $G$, and each
edge in $\mathcal{L}(G)$ corresponds to a length-two directed path in
$G$.
By applying the minimum input theorem
directly to this line graph, we obtain the minimum number of %
edges we must drive to control %
the original network. However, this procedure does
not minimize %
the number of driver nodes in the original network. 
This edge control problem can be mapped to a graph theoretical problem as
follows~\citep{Nepusz-NP-12}. %
Define node $i$ to be 
(i) \emph{divergent}, if $\kout(i) > \kin(i)$;
(ii) \emph{convergent}, if $\kout(i) < \kin(i)$;
(iii) \emph{balanced}, if $\kout(i) = \kin(i)$.
A connected component in a directed network is called a \emph{balanced
component} if it contains at least one edge and all the nodes are
balanced. 
We can prove %
that the minimum set of driver nodes
 required to maintain structural controllability of the switchboard
 dynamics on a directed network $G$ can be determined by selecting the
 divergent nodes of $G$ and an arbitrary node from each balanced
 component.

The controllability properties of this edge dynamics %
significantly differ from simple nodal dynamics. For example, driver
nodes prefer hubs with large out-degree and heterogeneous networks
are more controllable, %
i.e., require fewer driver nodes, than
homogeneous networks~\citep{Nepusz-NP-12}.  
Moreover, positive correlations between the %
in- and out-degree of a node 
enhances the controllability of edge dynamics, without affecting 
the controllability of nodal dynamics~\citep{Posfai-SR-2013}. Conversely,
adding self-loops to individual nodes enhances the controllability of
nodal dynamics~\citep{Liu-Nature-11,Posfai-SR-2013}, but leaves the controllability of edge dynamics
unchanged.

\subsection{Self-Dynamics and its Impact on Controllability} %

The nodes %
of networked systems are often %
characterized
by some %
self-%
dynamics, %
e.g. a term of the form $\dot{x}_i=a_{ii}
x_i$, which captures the node's behavior in the absence of interactions
with other nodes.    
If we naively apply %
structural control theory to systems where each node has a %
self-dynamic term we obtain %
a surprising %
result %
--- %
a single control input can make an arbitrarily large linear
system controllable~\citep{Liu-Nature-11,Cowan-PL-12}.  
This result represents %
a special case of the minimum input theorem: %
The %
self-dynamics 
 contributes a self-loop to each node, %
hence each node can be
 matched by itself. Consequently, $G(\mA)$ has %
a perfect matching, independent of the network topology, 
and one input signal is sufficient to control the whole 
system~\citep{Liu-Nature-11}. 

\begin{figure}[t!]
  \begin{center}
 \includegraphics[width=.48\textwidth ]{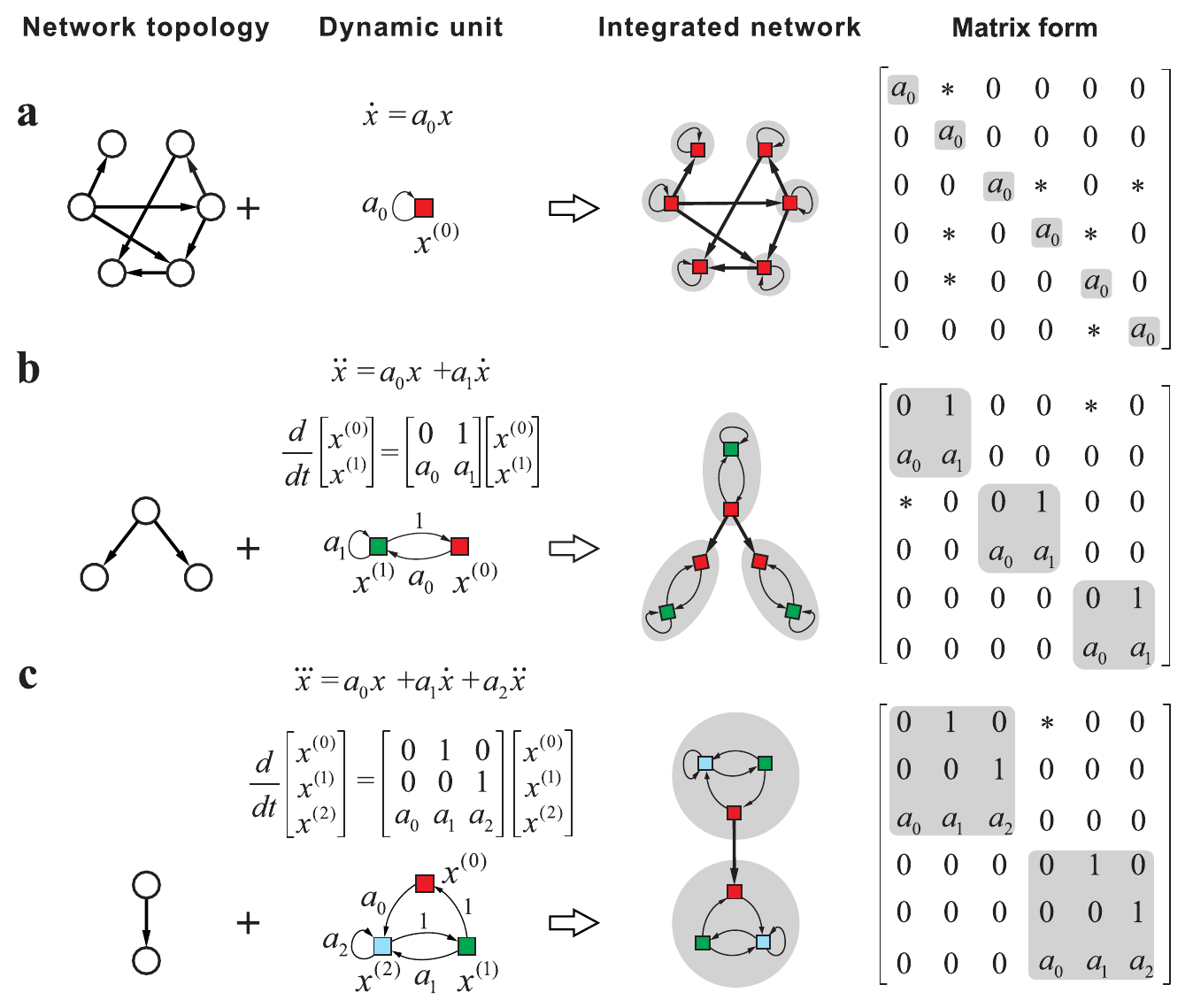}
        \vspace{-0.8cm}
  \end{center}
\caption{(Color online)  
Integrating the %
network topology with %
nodal self-dynamics.
({\bf a}) 1st-order self-dynamics: $\dot{x}=a_{0}
x$. ({\bf b}) 2nd-order self-dynamics: $\ddot{x}=a_0 x + a_1 \dot{x}$
.  ({\bf c}) 3rd-order self-dynamics: $\dddot{x}=a_0 x + a_1 \dot{x} + a_2 \ddot{x}$. To
develop a graphical representation for the %
$d$th-order individual dynamics
$x^{(d)}=a_0x^{(0)}+a_1x^{(1)}+\cdots + a_{d-1}x^{(d-1)}$, we denote
each order by a colored square. %
The couplings among orders
are characterized by links or self-loops. This graphical
representation allows the individual dynamics to be integrated with the 
network topology, 
giving rise to a unified matrix that reflects the dynamics of the whole system. In particular, each dynamic unit in the unified matrix
corresponds to a diagonal block and the nonzero elements (denoted by
$*$) outside these %
blocks stand for the couplings among
different dynamic units. Therefore, the original network %
of $N$ nodes with order $d$ self-dynamics is represented by a $dN \times dN$
matrix. 
After \citep{Zhao-SR-15}. 
} \label{fig:selfloop}
\end{figure}

To %
understand the true 
impact of self-%
dynamics on network controllability, we must %
revisit the validity of the %
assumption that the %
system parameters are independent of each other. 
As we show next, 
relaxing this assumption %
offers a more realistic characterization of %
real %
systems, for which %
not all system parameters are 
independent.
Assuming prototypical linear form of self-dynamics, e.g.,
first-order $\dot{x}=a_0 x$, second-order $\ddot{x}=a_0 x + a_1
\dot{x}$, etc, we can incorporate the linear self-dynamics with the
LTI dynamics of the network 
in a unified matrix form, as illustrated in Fig.~\ref{fig:selfloop}.  
An immediate but counterintuitive result states that %
in the absence of self-dynamics $n_\mm{D}$ is
exactly the same as in the case when each node has a self-loop with
identical weight $w$, i.e. each node is governed by precisely the same
self-dynamics. This is a direct consequence of 
the %
identity 
\bea
\text{rank}[\mB,\mA\mB,\cdots, \mA^{N-1}\mB]  \nn
= \text{rank}[\mB,(\mA+w {\bf I})\mB,\cdots, (\mA+w{\bf
  I})^{N-1}\mB],
\eea
where on the left we have %
the rank of controllability matrix in the absence of self-loops, and
on the right the same for a network where each node has an 
identical self-loop. %
For more general cases the minimum number of driver nodes $N_\mm{D}$ can be calculated
from (\ref{eq:nD_exactcontrol}), i.e. the maximum geometric
multiplicity of $\mA$'s eigenvalues. %

\begin{figure}[t!]
  \begin{center}
    \includegraphics[width=.48\textwidth]{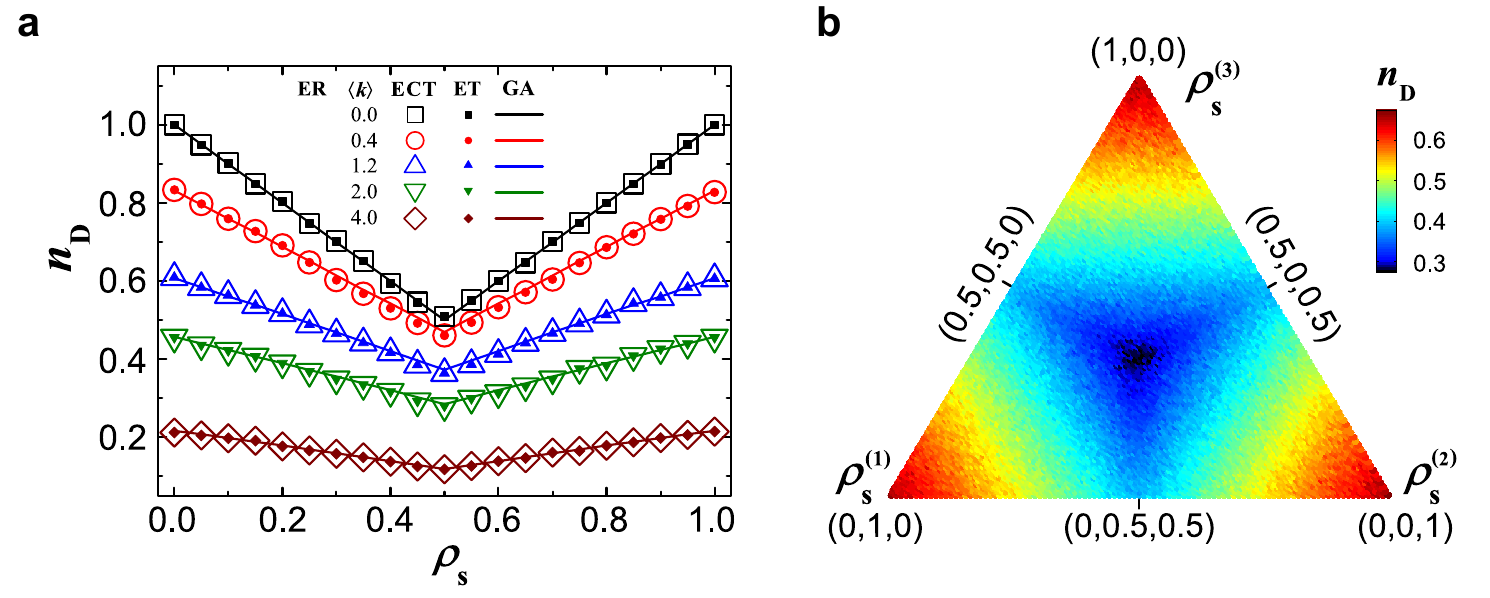}
        \vspace{-0.8cm}
  \end{center}
\caption{(Color online)  Impact of first-order individual dynamics on the
    fraction of driver nodes $n_\mm{D}$.  
The values of the off-diagonal
non-zero elements in $\mA$ are randomly chosen and hence are
independent. 
(a) $n_\text{D}$ %
in function of $\rho_\text{s}$, the density of
nodes that have the same %
type of nonzero
  self-loops. %
We observe a clear symmetry around $\rho_\mm{s}=1/2$, indicating that
$n_\text{D}$ reaches its minimum at $\rho_\mm{s}=1/2$, where the
densities of nodes with zero and non-zero self-loops are equal. %
(b) $n_\text{D}$ for an Erd\H{o}s-R\'enyi random network with three types of self-loops
$s_1$, $s_2$ and $s_3$ with densities $\rho_\text{s}^{(1)}$, $\rho_\text{s}^{(2)}$ and $\rho_\text{s}^{(3)}$,
respectively. The color bar denotes the value of $n_\text{D}$ and the coordinates in the triangle stands
for $\rho_\text{s}^{(1)}$ $\rho_\text{s}^{(2)}$ and
$\rho_\text{s}^{(3)}$. 
There is a global symmetry point where the three types of self-loops
have the same density $1/3$, and $n_\mm{D}$ reaches its minimum
value. 
After ~\citep{Zhao-SR-15}. 
}
\label{fig:1st_order}
\end{figure}

Note 
a %
remarkable symmetry in network controllability: If we exchange the
fractions of any two types of self-loops with distinct weights, %
the system's controllability, as measured by $n_\mathrm{D}$, remains
the same (Fig.~\ref{fig:1st_order}). 
For example, consider a network without 
  self-loops. Equivalently, we can assume that each node contains a
  self-loop with weight zero.
  Then we systematically add more non-zero self-loops with identical
  weights to the network. Equivalently, we are replacing the 
  zero-weight self-loops %
with non-zero self-loops. 
  $n_\mm{D}$ will first decrease as the fraction $\rho$ of
  non-zero self-loops increases, reaching 
a minimum at 
  $\rho=\frac 12$. After that, $n_\mm{D}$ increases, %
reaching its maximum %
at $\rho=1$, which coincides with %
$n_D$ observed for 
  $\rho=0$ 
  (Fig.~\ref{fig:1st_order}a).  
We can introduce more types of self-loops with different weights.  If we exchange the
fractions of any two types of self-loops, $n_\mathrm{D}$ remains
the same. 
This exchange-invariant property gives rise to a global symmetry
point, where all the different types of self-loops have equal
densities and the system displays the highest controllability (i.e., lowest number of
driver nodes). 
This symmetry-induced optimal controllability
holds for any network topology and various %
individual dynamics~\citep{Zhao-SR-15}.

\subsection{Control Energy}

Indentifying the minimum number of driver or actuator nodes sufficient 
for control %
is only the first step of the control problem. 
Once we have that, we need to ask an equally important question: How
much effort is required to %
control %
a system from a given
set of nodes? 
The meaning of the term ``control effort'' depends upon the particular
application~\citep{Kirk-book-04}. In the case of a rocket being thrust upward, the control
input $u(t)$ is the thrust of the engine, whose magnitude $|u(t)|$ is
assumed to be proportional to the rate of fuel consumption. In order
to minimize the total expenditure of fuel, the control effort can be
defined as $\int_0^T |u(t)| \ud t$, %
which is %
related to the %
energy consumed by the rocket. %
In the case of a voltage source driving a circuit containing no energy
storage elements, %
the source voltage is the control input $u(t)$ and the source current is directly proportional to $u(t)$. If
the circuit is to be controlled with minimum energy dissipation, we
can define the control effort as $\int_0^T u^2(t) \ud t$, 
which is %
proportional to the energy dissipation. 
If there are several control inputs, the general form of control
effort can be defined as 
$\int_0^{T} {\bf u}^\mm{T}(t) {\bf R}(t) {\bf u}(t)\, \ud t$, where
${\bf R}(t)$ is a real symmetric positive-definite weighting
matrix. %

Consider the LTI system (\ref{eq:LTI}) driven from an arbitrary
initial state ${\bf x}_\mm{i}$ towards a desired final state ${\bf
  x}_\mm{f}$ by the external signal ${\bf u}(t)$ in the time interval
$t \in [0, T]$. 
We define the associated control effort in the quadratic form  
\be
\mathcal{E}(T) \equiv \int_0^{T} \| {\bf u}(t) \|^2 \, \ud
t, \label{eq:costfunction}
\ee
called the %
``control energy'' in the literature~\citep{Yan-PRL-12,Yan-NP-15,Chen-arXiv-15}.
 Note that (\ref{eq:costfunction}) may not have the 
physical dimension of energy, i.e., M L$^2$ T$^{-2}$, in real control
problems. 
But for physical and electronic systems we can always assume there is
an hidden constant in the right-hand side %
of (\ref{eq:costfunction}) with
proper dimension, which %
ensures that $\mathcal{E}(T)$ has the %
dimension of energy. %
In many systems, like biological or social systems, where
(\ref{eq:costfunction}) does not correspond to energy, it captures the
effort needed to control a system.

For a fixed set of driver nodes %
the control input ${\bf u}(t)$ that can drive the system from  ${\bf x}_\mm{i}$ to ${\bf
  x}_\mm{f}$ can be chosen in many different ways, resulting in
different %
trajectories followed by the system. %
Each of these trajectories has its %
own control energy. 
Of all the possible inputs, the %
one that yields the minimum control energy is %
\be
{\bf u}(t) = {\bf B}^\mm{T} \exp({\bf A}^\mm{T} (T-t)) {\bf
  W}^{-1}(T) {\bf v}_\mm{f}, \label{eq:optimal_u}
\ee
where ${\bf W}(t)$ is %
the \emph{gramian matrix} 
\be
{\bf W}(t) \equiv \int_0^t \exp({\bf A} \tau) {\bf B} {\bf B}^\mm{T}
\exp({\bf A}^\mm{T} \tau) \, \ud \tau, \label{eq:Gramian}
\ee
which is nonsingular for any $t>0${\color{red}~\citep{Lewis-Book-2012}}. 
Note that ${\bf W}(\infty)$ is known as
the \emph{controllability Gramian}{\color{red}, often denoted with ${\bf W}_\mm{c}$~\citep{Kailath-Book-80}}. %
The energy associated with the optimal input (\ref{eq:optimal_u}) is
$\mathcal{E}(T) = {\bf
  v}_\mm{f}^\mm{T}  {\bf W}^{-1}(T)  {\bf v}_\mm{f}$, 
where ${\bf
  v}_\mm{f} \equiv {\bf x}_\mm{f} - \exp({\bf A} T) {\bf x}_\mm{i}$
represents the difference between the desired state under control and
the final state during free evolution without control. 
Without loss of generality, we %
can set the final %
state at the origin, ${\bf x}_\mm{f}  = {\bf
  0}$, and write the control energy as 
\be
\mathcal{E}(T) = {\bf x}^\mm{T}_i {\bf H}^{-1}(T) {\bf x}_\mm{i} 
\ee
where ${\bf H}(T) = \exp(-{\bf A} T) {\bf W}(T)
\exp(-{\bf A}^\mm{T} T)$ is the symmetric Gramian matrix. 
We can further define the normalized control energy as %
\be
E(T) 
\equiv \frac{\mathcal{E}(T)}{||{\bf x}_\mm{i}||^2} 
 = \frac{{\bf x}_\mm{i}^\mm{T} {\bf H}^{-1} {\bf x}_\mm{i}}{{\bf x}_\mm{i}^\mm{T} {\bf x}_\mm{i}}.
\ee
When ${\bf x}_\mm{i}$ is parallel to the direction of one of {\bf H}'s
eigenvectors, the inverse of the corresponding eigenvalue corresponds
to %
normalized energy %
associated with controlling
the system along the particular eigen-direction.

Using the Rayleigh-Ritz theorem, the normalized control energy obeys 
the %
bounds 
\be
\eta_{\max}^{-1} \equiv E_\mm{min}  \le E(T)  \le E_\mm{max} \equiv
\eta_{\min}^{-1}, 
\ee
where $\eta_{\max}$ and $\eta_{\min}$ the maximum and minimum eigenvalues
of ${\bf H}$, respectively~\citep{Yan-PRL-12}. 
Assuming linear individual dynamics characterized by
the self-loop $a_{ii} = - (a+ s_i)$ where $s_i = \sum_{j\neq i}
s_{ij}$ is the strength of node $i$ and $a$ is a %
parameter that can make the symmetric %
$\mA$ (describing an undirected network) either %
positive or %
negative definite, %
we can choose a single node %
with index $c$ %
as the driver node. 
In this case, %
the lower and
upper energy bounds follow 
\begin{equation} \label{lowerresult}
E_{\text{min}} \sim 
  \begin{cases}
     T^{-1} & \text{small $T$} \\
    \frac{1}{[(\mathbf{A}+\mathbf{A}^\text{T})^{-1}]_{cc}} & \text{large $T$, $\mathbf{A}$ is PD} \\
T^{-1} \to 0 &   \text{large $T$, $\mathbf{A}$ is semi PD}\\
\exp{\left(2\lambda_NT\right)} \to 0 &   \text{large $T$, $\mathbf{A}$ is not PD}
\end{cases},
\end{equation}

\begin{equation} \label{upperresult}
 E_{\text{max}}
\sim 
  \begin{cases}
     T^{-\theta}\;(\theta \gg 1) & \text{small $T$}\\
     = \varepsilon(\mathbf{A},c) & \text{large $T$, $\mathbf{A}$ is not ND}\\
T^{-1} \to 0 &   \text{large $T$, $\mathbf{A}$ is semi ND}\\
\exp{\left(2\lambda_1T\right)} \to 0 &   \text{large $T$, $\mathbf{A}$ is ND}
  \end{cases}.
\end{equation}
Here 
$\lambda_1 > \lambda_2 > \cdots > \lambda_N$ are the eigenvalues of
$\mA$, 
and 
$\varepsilon(\mathbf{A},c)$ is %
a positive energy %
that depends on the matrix
$\mathbf{A}$ and the choice of the controlled node $c$. 
PD (or ND) means positive-definite (or negative-definite),
respectively.
The scaling laws (\ref{lowerresult}) and (\ref{upperresult}) can be generalized to %
directed networks, in which case the decay exponents $\lambda_1$ and
$\lambda_N$ are %
replaced by $\mm{Re}\lambda_1$ and
$\mm{Re}\lambda_N$, respectively. 

Equations (\ref{lowerresult}) and (\ref{upperresult}) %
suggest that the scaling of the control energy %
is rather sensitive to the control time $T$. 
For small $T$, in which case %
we wish to get our system very fast to its
destination, both $E_\mm{min}$ and $E_\mm{max}$ decay with
increasing $T$, implying that setting a somewhat %
longer control time requires less energy. 
For large $T$, however, we reach a point 
where we cannot reduce the energy by %
waiting for longer
time. %
This occurs when the system has %
  its equilibrium point in the origin, then any attempt to steer the system
  away from the origin must overcome a certain energy barrier. %

\begin{figure}[t!]
\includegraphics[width=0.5\textwidth]{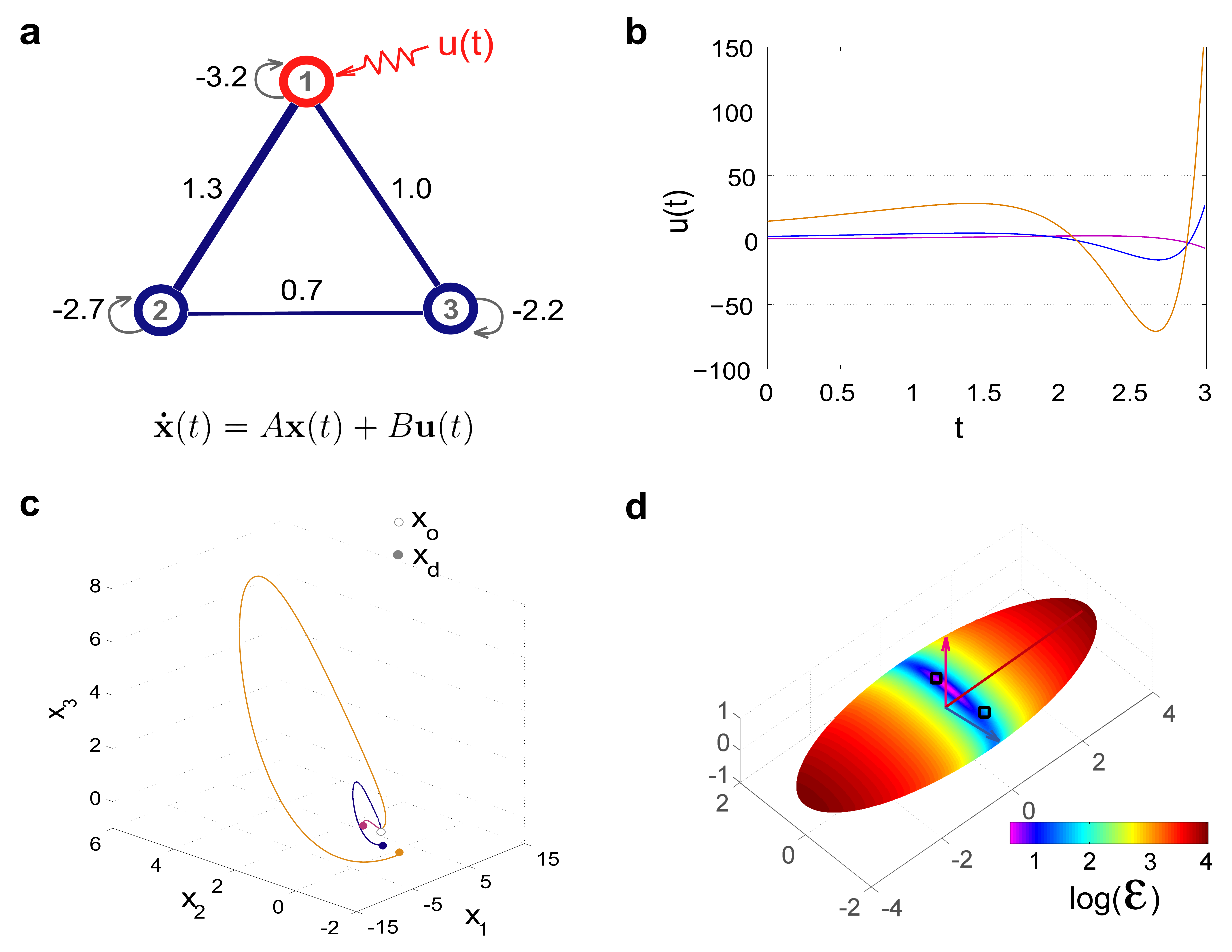}
\caption{(Color online)  Energy spectrum. 
(a) %
A three-node weighted network can be controlled via a single control input $u(t)$, injected to
the driver node shown in red. The input matrix $\mB$ is reduced to a vector $(1, 0, 
0)^\mm{T}$. Each node has a negative self-loop, which %
makes all eigenvalues of
the state matrix $\mA$ negative, hence stable. 
(b) The optimal control signals that
minimize the energies required to steer the network from the initial
state $\mx_0 = \mx(0) = (0, 0, 0)^\mm{T}$ to three different desired states
$\mx_\mm{d} = \mx(t)$ at $t=3$, with the constraint $\| \mx_\mm{d} \| = 1$. (c) The trajectories of the
network state $\mx(t)$ driven %
by the control inputs shown in
(b). (d) The energy surface for all normalized desired states, i.e.,
$\| \mx_\mm{d} \| = 1$, which is an ellipsoid spanned by the
controllability Gramian's three eigen-directions (arrows). The
ellipsoid nature of the spectrum illustrates the widely different
energies we need to move the network shown in (a) in different directions
in the state space. 
The squares correspond to
the three cases depicted in (b) and (c).
  After \citep{Yan-NP-15}.  
}\label{fig:energyspectrum-Yan}
\end{figure}

The control energy is rather sensitive to %
the direction of the
state space in which we wish to move the system%
~\citep{Yan-NP-15}. 
To see this, consider
a scale-free network with degree exponent $\gamma$. 
If we drive the system through all its %
nodes %
($N_\mm{D}=N$), the control energy spectrum, describing the
probability that moving in a randomly chosen eigen-direction will require
energy $\mathcal{E}$, follows the power
law %
$P(\mathcal{E}) \sim \mathcal{E}^{-\gamma}$. Consequently, 
the maximum energy required for
control depends sublinearly on the system size, $\mathcal{E}_\mm{max} \sim
N^{1/(\gamma-1)}$, implying that %
even in the most costly direction the
required energy grows %
slower than the system size. In other words, if we control each node, 
there are no significant energetic barriers for control.
If, however, we aim to control the system through %
a single node %
($N_\mm{D}=1$), the
control spectrum follows a power law with exponent $-1$, i.e.,
$P(\mathcal{E}) \sim \mathcal{E}^{-1}$,  which only weakly
depends on the network structure. Therefore %
the maximum
energy required for control increases as $\mathcal{E}_\mm{max} \sim
e^N$. This exponential increase means that 
steering the network in some directions is
energetically prohibitive. %
Finally, if we drive a finite fraction of nodes ($1<N_\mm{D}<N$), the control spectrum has
multiple peaks %
and the maximum energy required for control scales as $\mathcal{E}_\mm{max} \sim
e^{N/N_\mm{D}}$. Hence, as we increase the number of driver nodes, the
maximum energy decays exponentially. %

These results raise an important question: %
in case of $1<N_\mm{D}<N$, how to choose the optimal set of
$N_\mm{D}$ driver nodes such that the control energy is minimized? 
Such a combinatorial optimization problem {\color{red}(also known as the actuator
placement problem) has} not
been extensively studied in the literature.  %
Only recently has it been shown that several objective functions, i.e. 
energy-related controllability metrics  
associated with the controllability Gramian ${\bf W}_\mm{c}$ of LTI
systems {\color{red}(e.g. $\mm{Tr}({\bf W}^{-1}_\mm{c}), \log (\mm{det} {\bf
  W}_\mm{c}), \mm{rank} ({\bf W}_\mm{c})$)}, are actually 
\emph{submodular}~\citep{Summers-13,Summers-14,Cortesi-14}. 
A submodular function~\footnote{Denote $\mathcal{P}(\mathcal{S})$ as the power set (i.e. the set of all the
subsets) of a set $\mathcal{S}$. Then a submodular function is a set
function $f \,:\, \mathcal{P}(\mathcal{S})\to \mathbb{R}$ that satisfies 
$f( \mathcal{X} \cup \{x\})  - f(\mathcal{X}) \ge f( \mathcal{Y} \cup
\{x\})  - f(\mathcal{Y})$, for any $\mathcal{X} \subseteq \mathcal{Y}
\subseteq \mathcal{S}$ and $x \in \mathcal{S} \setminus \mathcal{Y}$.  
} $f$ has the
so-called %
diminishing returns property that the difference in the function value that a single
element $x$ makes
when added to an input set $\mathcal{X}$ decreases as the size of the input set
increases. 
The submodularity of %
objective functions allows for either an efficient global optimization
or a simple greedy approximation algorithm with certain performance
guarantee to solve the combinatorial optimization
problems{\color{red}~\citep{Nemhauser-MP-78}. 
In particular, the submodularity of those energy-related
controllability metrics has been explored to address the actuator
placement problem in a model of the European power
grid~\citep{Summers-13,Summers-14,Cortesi-14}.}

\subsection{Control Trajectories}
\begin{figure}[t!]
  \begin{center}
    \includegraphics[width=.48\textwidth ]{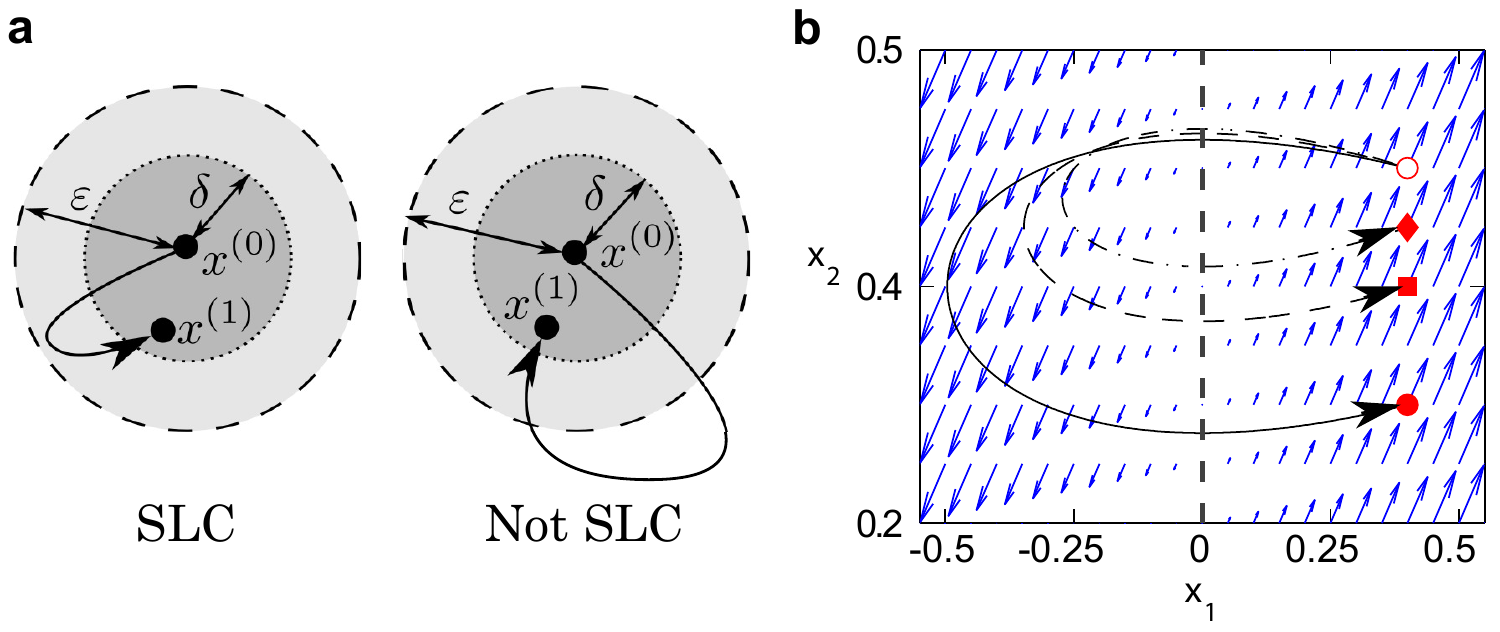}
        \vspace{-0.8cm}
  \end{center}
  \caption{(Color online)  Strictly local controllability.  
(a) Illustration of a state that is strictly locally
    controllable (left) and a state that is not  (right).
  (b) The state space of a simple LTI system with two state variables
  $x_1$ and $x_2$. 
The curves indicate control trajectories of minimal energy for given
initial state (open symbol) and final states (solid symbols). The
arrows in the background represent the vector field in the absence of
control input $u_1(t)$. 
Note that any state that is not on the line $x_1=0$ is not SLC, because the
  minimal-energy control trajectories to any neighboring final state
  with a smaller $x_2$-component will necessarily cross into the $x_1
  < 0$ half-plane. 
  After \citep{Sun-PRL-2013}. 
  }
  \label{fig:SLC}
\end{figure}

So far we have %
focused on the minimization of
  driver/actuator nodes and the energy cost of controlling LTI
  systems. The characteristics %
of the resulting control trajectories are
  also interesting and worthy of exploration~\citep{Sun-PRL-2013}. 
A state $\mx^{(0)}$ of the LTI system is called \emph{strictly locally
  controllable} (SLC) if for a ball $B(\mx^{(0)}, \varepsilon)$ centered at $\mx^{(0)}$ with
radius $\varepsilon>0$ there is a constant $\delta>0$ such that any final state $\mx^{(1)}$  inside the ball
$B(\mx^{(0)},\delta)$ can be reached from $\mx^{(0)}$ with a control
trajectory entirely inside the ball $B(\mx^{(0)},\varepsilon)$ (see
Fig.~\ref{fig:SLC}a). 
Figure~\ref{fig:SLC}b shows that in a two-dimensional LTI system
$\dot{x}_1 = x_1 + u_1(t)$, $\dot{x}_2 = x_1$, for any state in the
$x_1>0$ half-plane, the minimal-energy control trajectories to any
neighboring final state with a smaller $x_2$-component will
necessarily cross into the  $x_1<0$ half-plane.  
It has been shown that for a general LTI system whenever the
number of control inputs is smaller than the number of state variables
(i.e., $N_\mm{D} < N$), then almost all the states are not SLC~\citep{Sun-PRL-2013}. 
Therefore, the minimal-energy control trajectory is generally
nonlocal %
and remains finite even when the final state is brought
arbitrarily close to the initial state. The length $\int_{0}^{t_f} \|
\dot{\mx}(t)\| \ud t$ of such a trajectory generally increases with
the condition number of the Gramian.  
Furthermore, %
the optimal control input
(\ref{eq:optimal_u}) that minimizes the energy cost $\int_{0}^{t_f} \|
{\bf u}(t) \|^2 \ud t$ will fail in practice if the controllability
Gramian (\ref{eq:Gramian}) is ill conditioned. This can occur even
when the controllability matrix is well conditioned. 
There is a sharp transition, called the controllability transition, as a function of the number of control
inputs, below which numerical control always fails and above which it
succeeds.  
These results indicate that even for the simplest LTI dynamics, the
disparity between theory and practice poses a fundamental limit on our
ability to control large networks~\citep{Sun-PRL-2013}.

Indeed, we usually don't use the %
minimum-energy control input
(\ref{eq:optimal_u}) to steer the system to desired final states, 
simply because it is an open-loop (or non-feedback)
controller~\footnote{\color{red}An \emph{open-loop }control system
  does not use feedback. The control input to the system is determined using only the current
  state of the system and a model of the system, and is totally
  independent of the system's output. %
  In contrast, in a \emph{closed-loop} control system, the output has
  an effect on the input (through feedback) so that the input will
  adjust itself based on the output.}, which
tends to be 
very sensitive to noise. A more practical and robust strategy is to use
a simple linear feedback control %
to bring the system asymptotically towards a certain state, 
{\color{red} while minimizing %
the energy cost}.  
{\color{red} 
This is a typical objective of optimal control theory, which %
aims to design control signals that will cause
a process to satisfy some physical constraints and maximize (or
minimize) a chosen performance criterion (or cost
function)~\citep{Kirk-Book-2004,Naidu-Book-2002}.
}

 \section{Controllability of Nonlinear Systems}\label{sec:controllabilitytest_Nonlinear}

So far we focused on the controllability %
of linear systems. Yet, the dynamics of most real complex systems is %
nonlinear, prompting us to review %
the classical results on nonlinear
controllability and their 
applications to %
networked systems.

Consider a control system of the %
form 
\be  \dot{\mx} = \mf (\mx, {\bf u}), \label{eq:NonlinearODE}
\ee
where the state vector %
$\mx$ is in a smooth connected manifold $\mathcal{M}$ of dimension
$N$, and the control input ${\bf u} \in \mathcal{U}$ is a subset of $\mathbb{R}^M$. 
Note that %
(\ref{eq:NonlinearODE}) has been frequently used to model the
behavior of physical, biological and social
systems~\citep{Hermann-IEEETAC-1977}. 
Roughly speaking, %
(\ref{eq:NonlinearODE}) is
\emph{controllable} if one can steer it from any point $\mx_0 \in \mathcal{M}$
to any other point $\mx_1 \in \mathcal{M}$ by choosing ${\bf u}$ from
a set of admissible controls $\mathbb{U}$, which is a subset of
functions mapping $\mathbb{R}^+$ to $\mathcal{U}$.   

The controllability of nonlinear systems has
been extensively studied since the early
1970s~\citep{Haynes-1970,Elliot-1970,Lobry-1970,Brockett-1972,Sussmann-1972,Hermann-IEEETAC-1977,Nijmeijer-Book-1990,Isidori-Book-95,Sontag-Book-1998,Moog-Book-07,Figueiredo-1993,Rugh-1981}. 
The goal %
was to derive %
results of similar reach and generality as obtained for
linear time-invariant systems. %
However, this goal %
turned out to be too ambitious, suggesting that %
a %
general theory on nonlinear controllability may not be %
feasible. 
Fortunately, as we discuss in this section, the concerted effort on
nonlinear control has led to various weaker notions of nonlinear
controllability, %
which are %
easier to characterize and often
offer %
simple algebraic tests to explore the controllability of nonlinear
systems.  
\subsection{Accessibility and Controllability} %

As we will see in the coming sections, we
can rarely prove or test controllability of an arbitrary %
nonlinear {\color{red}system}. Instead, we %
prove and test weaker versions of controllability called 
local accessibility and local strong accessibility. %
We start by %
defining these notions.

Accessibility concerns the possibility to reach or access an open
set %
of states in the state space from a given initial state. 
If the system (\ref{eq:NonlinearODE}) is \emph{locally accessible} from
an initial state $\mx_0$ then we can reach or access the neighborhood of $\mx_0$
through trajectories that are within the neighborhood of $\mx_0$.   
Mathematically, the system (\ref{eq:NonlinearODE}) is called
\emph{locally accessible from} $\mx_0$ if for
any non-empty neighborhoods $\mathcal{V}\subset \mathcal{M}$ of $\mx_0$ and any $t_1>0$,
the \emph{reachable set} $\mathcal{R}^\mathcal{V}(\mx_0, \le t_1)$ contains a non-empty open set. 
The system is called %
\emph{locally accessible} if this %
holds for any $\mx_0$. 
Here, the reachable set $\mathcal{R}^\mathcal{V}(\mx_0, \le t_1)$
includes %
all states that can be reached from $\mx_0$ within a time 
$t_1$,
following trajectories that are within the neighborhood of $\mx_0$. 
Mathematically, the \emph{reachable set} from $\mx_0$ in
time \emph{up to} $t_1$ is defined as 
$
 \mathcal{R}^\mathcal{V}(\mx_0, \le t_1) \equiv \cup_{\tau \le t_1} 
\mathcal{R}^\mathcal{V}(\mx_0, \tau).
$
Here $\mathcal{R}^\mathcal{V}(\mx_0, \tau)$ is the \emph{reachable set} from
$\mx_0$ \emph{at time} $\tau >0$ following trajectories that remain in
$\mathcal{V}$ for $t\le \tau $.

If we look at states that can be reached \emph{exactly} at time $t_1$, then
 we have a stronger version of local accessibility. System
 (\ref{eq:NonlinearODE}) is said to be %
\emph{locally strongly
   accessible} from $\mx_0$ if at any small time $t_1>0$
   the system can reach or access the neighborhood of $\mx_0$
through trajectories that are within the neighborhood of $\mx_0$.    
Mathematically, this means that 
for any non-empty neighborhoods $\mathcal{V}$ of $\mx_0$ and any $t_1>0$
 sufficiently small, the reachable set $\mathcal{R}^\mathcal{V}(\mx_0, t_1)$ contains a non-empty open set. 
 If this holds for any $\mx_0$, then the system is called
\emph{locally
   strongly accessible}. 
Clearly, local strong accessibility from $\mx_0$ implies local
accessibility from $\mx_0$. The converse is generally not true.
Local controllability asks whether the system is
  controllable in some neighborhood of a given state. Mathematically, 
the system (\ref{eq:NonlinearODE}) is called %
\emph{locally controllable} from
$\mx_0$ if for any neighborhood $\mathcal{V}$ of $\mx_0$, the
reachable set 
$\mathcal{R}^\mathcal{V}(\mx_0, \le t_1)$ is also a neighborhood of $\mx_0$
for any $t_1$ small enough. 
The system is called %
\emph{locally controllable} if this holds for any $\mx_0$. 
Clearly, local controllability implies local
accessibility. It turns out that for a large class of systems
local controllability implies local strong accessibility. 
But the converse is not always true. 

If we do not require the trajectories of the system to remain close to
the starting point, i.e., we allow excursions, then we have the notion
of global controllability. 
System (\ref{eq:NonlinearODE}) is \emph{globally controllable} from
$\mx_0$ if the reachable set from $\mx_0$ is $\mathcal{M}$ itself,
i.e., 
$ \mathcal{R}(\mx_0) \equiv \cup_{t_1 \ge 0} 
\mathcal{R}^\mathcal{M}(\mx_0, t_1) = \mathcal{M}$. 
In other words, 
for any $\mx_1 \in \mathcal{M}$, there exists $t_1 > 0$ and ${\bf u}
: [0,t_1] \to \mathcal{U}$ such that the solution of (\ref{eq:NonlinearODE})
starting at $\mx_0$ at time 0 with control ${\bf u}(t)$ satisfies
$\mx(t_1) = \mx_1$. 
If this holds for all $\mx_0 \in \mathcal{M}$,
then the system is called \emph{globally controllable}. 
Complete algebraic characterizations of global controllability of
nonlinear systems have proved elusive. 
Weaker notions of controllability are easier to characterize than
controllability. 
For example, it can be proven that for some nonlinear systems, %
accessibility can be decided in polynomial time, while controllability
is NP-hard~\citep{Sontag-SIAM-88}. 
For complex networked systems we expect that only weaker notions of
controllability can be characterized. 
\subsection{Controllability of Linearized Control System} 
It is typically difficult to test the controllability of a nonlinear
system. %
Yet, as we discuss next, studying the controllability properties of its
linearization around an equilibrium point or along a trajectory can
often offer %
an efficient test of
local nonlinear controllability%
~\citep{Coron-Book-09}.

\subsubsection{Linearization around an equilibrium point} 

Consider an equilibrium point $(\mx^*, {\bf u}^*) \in \mathcal{M}
\times \mathcal{U}$ of the nonlinear control system
(\ref{eq:NonlinearODE}), meaning that %
$\mf(\mx^*, {\bf u}^*) = {\bf 0}$. 
Assume that $\mathcal{U}$ contains a neighborhood of ${\bf u}^*$. For
$\epsilon>0$, we define a set of control functions 
$ \mathbb{U}_\epsilon \equiv \{{\bf u}(\cdot) \in
\mathbb{U} | \| {\bf u}(t) - {\bf u}^*\| < \epsilon, t\ge 0 \}$. 
The linearized control system at
$(\mx^*, {\bf u}^*)$ is a linear control system  
$
\dot{\mx} =\mA \mx + \mB {\bf u} %
$ 
with 
\be
\mA = \frac{\p \mf}{\p \mx}(\mx^*, {\bf u}^*), \,\, \mB= \frac{\p
  \mf}{\p {\bf u}}(\mx^*, {\bf u}^*).
\ee
If the linearized control system %
is controllable (in the sense of a linear time-invariant system), then for any
$\epsilon>0$ the original
nonlinear system is locally controllable from
$\mx^*$, where the control functions ${\bf u}(\cdot)$ are taken from
the set $\mathbb{U}_\epsilon$. 

In other words, many real systems operate near some equilibrium
points and in the vicinity of such points, controllability can be
decided using the tools developed for linear systems, discussed in the
previous sections. %

\subsubsection{Linearization around a trajectory} 

We can also study the linearized control system along a
\emph{trajectory}. Consider a nonlinear control
system in the form of (\ref{eq:NonlinearODE}). 
A trajectory represents %
the path the system follows as a function of
time in the state space. It can be mathematically defined as a 
function $(\bar{\mx}, \bar{{\bf u}}) : [T_0, T_1] \to \mathcal{O}$,
where $\mathcal{O}$ is a nonempty open subset of $\mathbb{R}^N \times
\mathbb{R}^M$ and $\bar{\mx}(t_2) = \bar{\mx}(t_1) + \int_{t_1}^{t_2} \mf
(\bar{\mx}(t), \bar{{\bf u}}(t))\, \ud t$, for all $(t_1, t_2) \in [T_0,
T_1]$.
 The linearized control system of (\ref{eq:NonlinearODE}) along a
trajectory $(\bar{\mx}, \bar{{\bf u}}) : [T_0, T_1] \to \mathcal{O}$
is a linear time-varying control system 
$
\dot{\mx} =\mA(t) \mx + \mB(t) {\bf u} 
$ 
with $t\in [T_0, T_1]$, and 
\be
\mA(t) = \frac{\p \mf}{\p \mx}(\bar{\mx}(t), \bar{\bf u}(t)), \,\, \mB(t)= \frac{\p
  \mf}{\p {\bf u}}(\bar{\mx}(t), \bar{\bf u}(t)).
\ee
If the linearized control system along the trajectory $(\bar{\mx},
\bar{{\bf u}}) : [T_0, T_1] \to \mathcal{O}$ is controllable in the sense of a linear
time-varying system, then the original nonlinear system is locally
controllable along the trajectory. Once again, this means that we can
use linear control theory to explore the controllability of nonlinear
systems. 

\subsubsection{Limitations of linearization}
The linearization approaches described above may sound powerful, but
they have severe limitations. 
First, they only provide %
information about %
controllability in the immediate vicinity %
of an equilibrium point or a trajectory. Second and most important, it may be the case that
the linearized control system is not controllable, but the original
nonlinear system is actually controllable.

\begin{figure}[t!]
\includegraphics[width=0.35\textwidth]{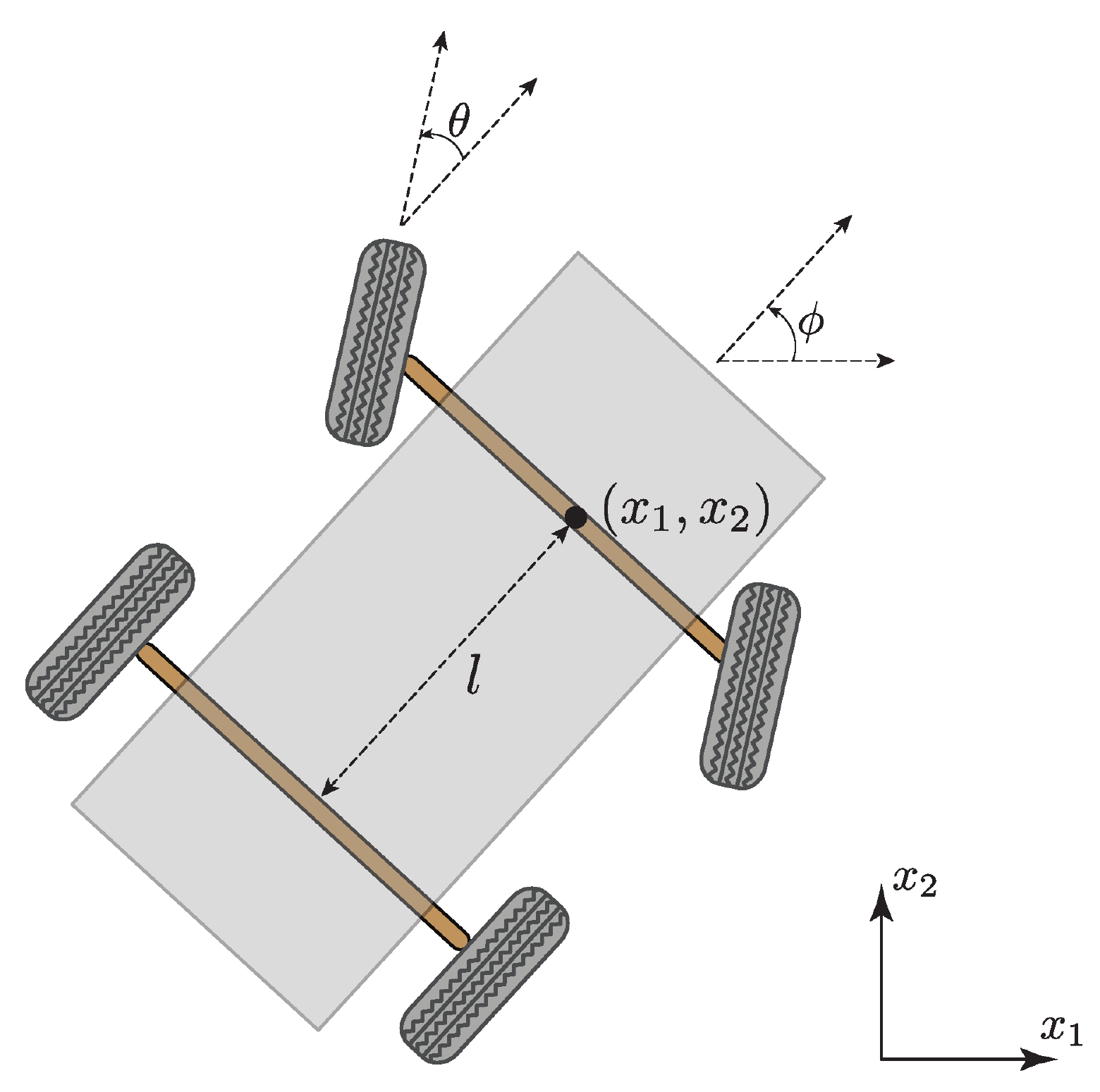}
\caption{(Color online)  Controlling a car. The figure shows a %
model of a front-wheel drive
car with 4 state variables $(x_1, x_2, \phi, \theta)$ and 2 control
inputs $(u_1, u_2)$.  
While this system is globally controllable (see Sec.~\ref{subsec:nonlineartest}), %
its linearized dynamics %
around the origin is not
controllable.
Figure redrawn from
~\citep{Sontag-Book-1998}.  
}\label{fig:car_nonlinearcontrol}
\end{figure}

Consider, for example,  a %
model of a front-wheel drive
car with four state variables: the positions ($x_1,x_2$) of the center
of the front axle, the orientation $\phi$ of the car, and the angle
$\theta$ of the front wheels relative to the car orientation 
(Fig.~\ref{fig:car_nonlinearcontrol}).
There are two control inputs $(u_1, u_2)$, where $u_1$, the %
steering velocity, represents %
the velocity with which the steering wheel is
turning, %
and $u_2$ is the %
driving velocity.  
Assuming that the front and rear wheels do not slip and
that the distance between them is $l=1$, %
the car's equations of motion have the form~\citep{Nelson-Book-67,Sontag-Book-1998}
\be
\begin{pmatrix}
\dot{x}_1 \\
\dot{x}_2 \\
\dot{\phi} \\
\dot{\theta}
\end{pmatrix}
 = u_1 
\begin{pmatrix}
0 \\
0 \\
0 \\
1 
\end{pmatrix}
+ u_2 
\begin{pmatrix}
\cos(\theta+\phi) \\
\sin(\theta+\phi) \\
\sin \theta  \\
0 
\end{pmatrix}.
\label{eq:frontwheeldrive_car}
\ee
The linearization of (\ref{eq:frontwheeldrive_car}) around the origin
is %
\be
\begin{pmatrix}
\dot{x}_1 \\
\dot{x}_2 \\
\dot{\phi} \\
\dot{\theta}
\end{pmatrix}
 = 
\begin{pmatrix}
u_2 \\
0 \\
0 \\
u_1 
\end{pmatrix},
\ee
which is %
uncontrollable, because $x_2$ and $\phi$ are time-invariant and not
controlled by any of the system's inputs. %
Yet, from our driving experience we
know that a car %
is controllable. We will prove that this system is indeed
globally controllable in Sec.~\ref{subsec:nonlineartest}.

System (\ref{eq:frontwheeldrive_car}) belongs to an
especially interesting class of nonlinear systems, %
called control-affine systems, where $\mf (\mx, {\bf u})$ is linear in the
control signal ${\bf u}$ 
\be 
\dot{\mx} = \mf (\mx ) + \sum_{i=1}^M \mg_i (\mx) u_i.  \label{eq:affine}
\ee
Here, $\mf$ is called 
the \emph{drift vector field}, or simply \emph{drift}; and  $\mg_1, \cdots, \mg_M$
are called the \emph{control vector fields}. 
The system (\ref{eq:affine}) is called \emph{driftless} if $\mf(\mx)\equiv{\bf 0}$,
which arises in kinematic models of many mechanical systems, e.g., in
(\ref{eq:frontwheeldrive_car}).   
Control-affine systems %
are natural generalization of %
linear time-invariant systems. 
Many nonlinear controllability results were obtained for
them. Hereafter, we will focus on %
control-affine systems, referring %
the reader to
\citep{Hermann-IEEETAC-1977,Sontag-Book-1998} for more general %
nonlinear systems.       
\subsection{Basic concepts in differential geometry}%
Before we discuss the nonlinear tests for accessibility and
controllability, we need %
a few concepts in differential
geometry, like Lie brackets and distributions. 

\subsubsection{Lie brackets} %
For nonlinear control systems, both controllability and accessibility 
are intimately tied
to Lie brackets. 
The reason is simple. In the nonlinear framework, the directions in
 which the state may be moved %
around an initial state $\mx_0$ are %
those belonging
 to the Lie algebra generated by %
vector fields $\mf(\mx_0,{\bf
   u})$, when ${\bf u}$ varies in the set of admissible controls
 $\mathbb{U}$~\citep{Isidori-Book-95,Sontag-Book-1998}.  
Here 
the Lie algebra $\mathcal{A}$ generated by a family
 $\mathcal{F}$ of vector fields is the set of Lie brackets $[\mf, \mg]$ 
 with $\mf,\mg \in \mathcal{F}$, 
 and all %
vector fields that can be obtained by iteratively
 computing Lie brackets. 
In turn, %
a Lie bracket is %
the derivative of a vector field with respect to another.

Consider two vector fields ${\bf f}$ and ${\bf g}$ on an
open set $D \subset \mathbb{R}^N$. The Lie bracket operation generates
a new vector field $[\mf, \mg]$, defined as  
\be
[\mf, \mg ] (\mx) \equiv \frac{\p {\bf g}}{\p \mx} {\bf f}(\mx)- \frac{\p {\bf
    f}}{\p \mx} {\bf g}(\mx) 
\ee
where $\frac{\p {\bf g}}{\p \mx}$ and  $\frac{\p {\bf f}}{\p \mx}$
are %
the Jacobian matrices of $\mg$ and $\mf$, respectively. 
Higher order Lie brackets can be recursively defined as 
\bea
\ad_\mf^0 \mg (\mx) &\equiv&  \mg (\mx), \\
\ad_\mf^k \mg (\mx) &\equiv&  [\mf, \ad_\mf^{k-1} \mg ] (\mx), \quad
\forall k \ge 1
\eea
where ``ad'' denotes ``adjoint''.

\begin{figure}[t!]
\includegraphics[width=0.35\textwidth]{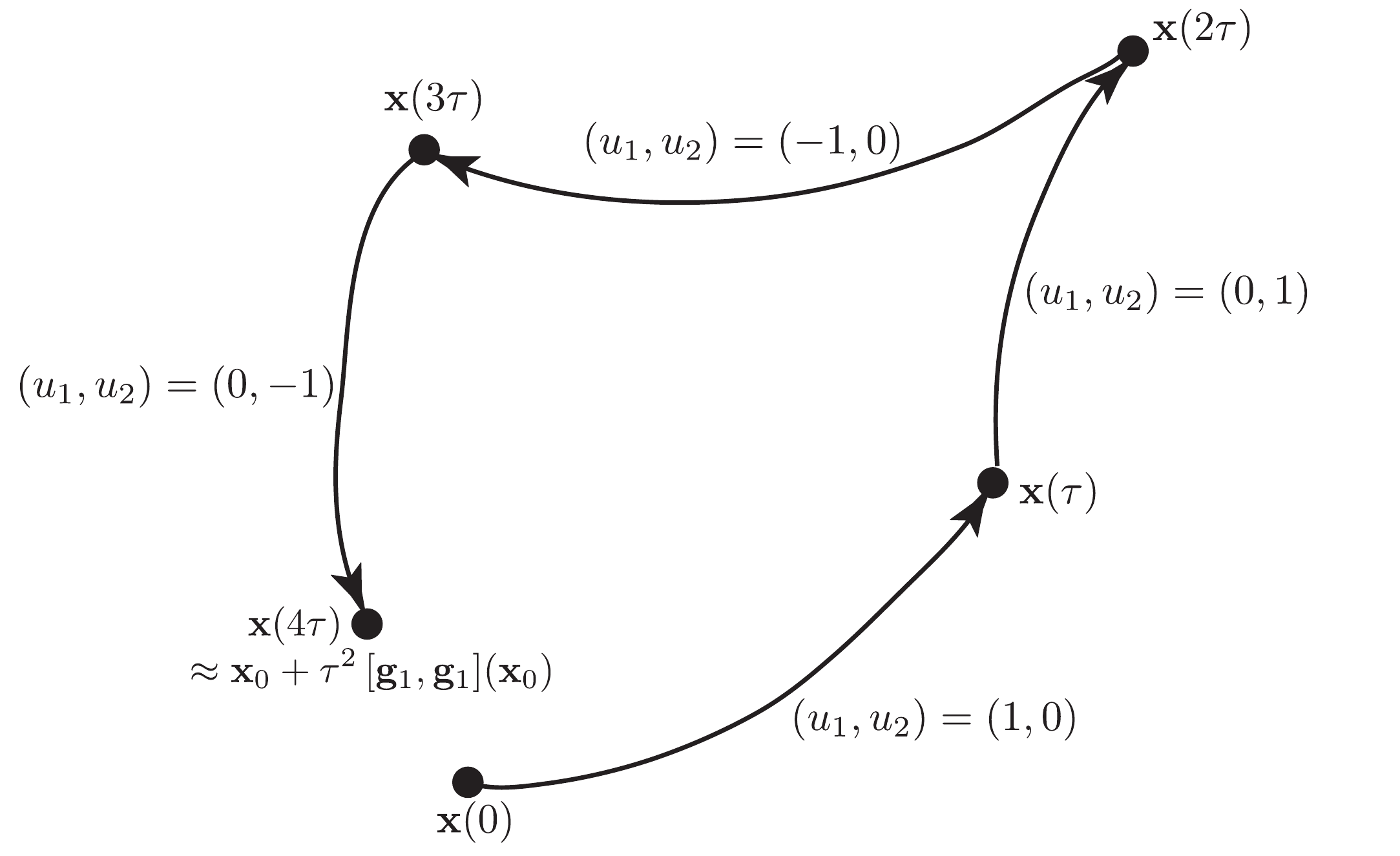}
\caption{Lie bracket. The physical meaning of a Lie bracket can be demonstrated
  by applying the piece-wise constant control inputs
  (\ref{eq:piece-wise-constant-control}) to a two-inputs driftless
  control-affine {\color{red}system} $\dot{\mx}=\mg_1(\mx) u_1 + \mg_2(\mx) u_2$. 
Up to terms of order $\tau^2$, the difference between the final state
$\mx(4\tau)$ and the initial state $\mx_0$ is given by the Lie bracket $[\mg_1,
\mg_2](\mx_0)$. 
}\label{fig:lie}
\end{figure}

To understand the physical meaning of the Lie bracket, %
consider 
the following piece-wise constant
control inputs 
\be
{\bf u}(t) = 
\begin{cases}
(1,0)^\mm{T}, & t \in [0, \tau) \\
(0,1)^\mm{T}, & t \in [\tau, 2\tau) \\
(-1,0)^\mm{T}, & t \in [2\tau, 3\tau) \\
(0,-1)^\mm{T}, & t \in [3\tau, 4\tau) 
\end{cases} \label{eq:piece-wise-constant-control}
\ee
applied onto a two-inputs control-affine system 
\be
\dot{\mx} = \mg_1(\mx) u_1 + \mg_2(\mx) u_2 \label{eq:affine2}
\ee
with initial state
$\mx(0)=\mx_0$~\citep{Sastry-Book-99}. 
The piece-wise constant control inputs
  (\ref{eq:piece-wise-constant-control}) can be considered as a 
sequence of ``actions'' applied for example to %
a car ($\mg_1$,  $\mg_2$,
  reverse-$\mg_1$, reverse-$\mg_2$). 
In the limit $\tau \to 0$ the
final state reached %
at $t=4\tau$ is %
\be
\mx(4\tau)  = \mx_0 + \tau^2 \left(
\frac{\p {\bf g}_2}{\p \mx} {\bf g}_1(\mx_0)- \frac{\p {\bf
    g}_1}{\p \mx} {\bf g}_2(\mx_0) \right) + O(\tau^3). 
\ee
We see that up to terms of order $\tau^2$, the state change is
exactly along the direction of the Lie bracket 
$[\mg_1,
\mg_2](\mx_0)$ (see Fig.~\ref{fig:lie}).  

Consider two %
examples that demonstrate the meaning of Lie
brackets. First, %
the Brockett system %
is one of the simplest driftless control-affine systems~\citep{Brockett-1982} 
\be \label{eq:Brockett}
\begin{cases}
\dot{x}_1(t) = u_1 \\
\dot{x}_2(t) = u_2 \\
\dot{x}_3(t) = u_2 x_1 - u_1 x_2  
\end{cases}
\ee
which can %
be written in the form of (\ref{eq:affine2}) using %
$\mg_1(\mx) = (1,0,-x_2)^\mm{T}$ and $\mg_2(\mx)=(0,1,x_1)^\mm{T}$, or
equivalently 
$\mg_1 = \frac{\p}{\p x_1} - x_2 \frac{\p}{\p x_3}$, 
and
$\mg_2 = \frac{\p}{\p x_2} + x_1 \frac{\p}{\p x_3}$. 
These two operators $\mg_1$ and $\mg_2$ have a nontrivial Lie bracket 
$[\mg_1, \mg_2] (\mx) = \mg_3(\mx) = 2 (0,0,1)^\mm{T}$,
or equivalently $\mg_3 = 2\frac{\p}{\p x_3}$. %
Consider the system (\ref{eq:Brockett}) initially at the origin, hence
$\mg_1({\bf 0}) = (1,0,0)^\mm{T}, \mg_2({\bf 0}) = (0,1,0)^\mm{T}$. If we
again apply the control
sequence (\ref{eq:piece-wise-constant-control}) with time interval $\tau=1$, we can
check that the final state reached %
at $t=4$ is $(0,0,2)^\mm{T}$, which is precisely
captured by %
$[\mg_1, \mg_2]({\bf 0})$.%

Note that for the Brockett system we have 
$[\mg_1, [\mg_1, \mg_2]] (\mx) = [\mg_2,
[\mg_1, \mg_2]] (\mx) = {\bf 0}$. 
A similar three-dimensional Lie algebra, called the
  Heisenberg algebra, also arises in quantum mechanics. Hence the Brockett
  system is also known as the Heisenberg system
~\citep{Bloch-Book-03}. %
Note, however, that %
the commutation relations obeyed by %
the Heisenberg algebra do %
not always apply to %
general nonlinear 
  systems. 
To see this consider again %
the model of a front-wheel drive
car (\ref{eq:frontwheeldrive_car}), representing another two-input control-affine
system, where the two control vector fields $\mg_1=(0,0,0,1)^\mm{T}$
and $\mg_2=(\cos(\theta+\phi), \sin(\theta+\phi), \sin\theta,
0)^\mm{T}$ can be interpreted %
as the actions $steer$ and $drive$, respectively. 
Some Lie brackets from $\mg_1(\mx)$ and 
$\mg_2(\mx)$ are %
\bea
\mg_3 (\mx) & \equiv & [\mg_1, \mg_2] (\mx)= \begin{pmatrix}
-\sin(\theta+\phi) \\
\cos(\theta+\phi) \\
\cos \theta  \\
0 
\end{pmatrix}, \label{eq:wriggle}\\
\mg_4 (\mx)&\equiv & [[\mg_1, \mg_2], \mg_2] (\mx)= \begin{pmatrix}
-\sin \phi \\
\cos \phi \\
0 \\
0 
\end{pmatrix} \label{eq:slide}. 
\eea
Equation (\ref{eq:wriggle}) can be interpreted as  
$[steer, drive] = wriggle$, arising from the %
sequence of actions $(steer, drive, reverse\, steer,
reverse\, drive)$, which is %
what we do in order to get a car out
of a tight parking space.   
Similarly, (\ref{eq:slide}) can be interpreted as  
$[wriggle, drive] = slide$, arising from the %
sequence of actions $(wriggle, drive, reverse\, wriggle, reverse\, drive)$, which
is what we do during %
parallel parking. 
Equations (\ref{eq:wriggle}, \ref{eq:slide}) indicate
  that starting from only two control inputs: \emph{steer} and \emph{drive}, we
can %
``generate'' other actions, e.g., \emph{wriggle} and
\emph{slide}, which allows us to %
fully control the car.

The above two examples demonstrate that by applying the right sequence
of control inputs we can steer the system along a 
direction that the system does not have direct control
over. 
In general, by choosing more elaborate sequences of control inputs
we can steer a control-affine system in directions precisely captured by higher-order Lie brackets,
e.g., $[\mg_2, [\mg_1, \mg_2]]$, $[[\mg_1,\mg_2], [\mg_2, [\mg_1,
\mg_2]]]$, etc. 
If the system of interest has %
a drift term $\mf$, we also have to
consider Lie brackets involving $\mf$. 
This is %
the reason why nonlinear controllability is closely
related to the Lie brackets.

\subsubsection{Distributions}

To discuss the nonlinear tests of accessibility and controllability, we
need %
the notion of \emph{distribution} in the sense of
differential geometry. %
A distribution can be roughly considered as the nonlinear version of the
controllability matrix of a linear system.  

Consider $m$ vector fields $\mg_1, \mg_2, \cdots,
\mg_m$ on an open set $\mathcal{D} \subset \mathbb{R}^N$. We denote
\be
\Delta(\mx) = \mm{span} \{\mg_1(\mx), \mg_2(\mx), \cdots,
\mg_m(\mx) \}
\ee
as the vector space %
spanned by the vectors 
$\mg_1(\mx), \mg_2(\mx), \cdots,
\mg_m(\mx)
$ at any fixed $\mx \in \mathcal{D}$.
Essentially, we assign a vector space $\Delta(\mx)$ to each point
$\mx$ in the set %
$\mathcal{D}$. The
collection of vector spaces $\Delta(\mx)$, $\mx \in \mathcal{D}$ is called a
\emph{distribution} and referred to by 
\be 
\Delta = \mm{span} \{\mg_1, \mg_2, \cdots,
\mg_m \}.
\ee
If the vectors 
$\mg_1(\mx), \mg_2(\mx), \cdots,
\mg_m(\mx)
$ are linearly independent for any $\mx$ in %
$\mathcal{D}$, 
then the dimension of
$\Delta(\mx)$ is constant and equals $m$. 
In this case we call $\Delta$ a
\emph{nonsingular} distribution on $\mathcal{D}$. 
For example, in the Brockett system we have 
$\mg_1(\mx)=(1,0,-x_2)^\mm{T}$,
$\mg_2(\mx)=(0,1,x_1)^\mm{T}$,
$\mg_3(\mx)=[\mg_1, \mg_2](\mx)=(0,0,2)^\mm{T}$. Since $\mg_1(\mx),
\mg_2(\mx), \mg_3(\mx)$ are linearly independent for all $\mx \in
\mathbb{R}^3$, we conclude that the distribution $\Delta= \mm{span}
\{\mg_1, \mg_2, \mg_3 \}$ is nonsingular. 
Similarly, in the front-wheel drive car system of Fig.~\ref{fig:car_nonlinearcontrol}, $\mg_1(\mx),
\mg_2(\mx), \mg_3(\mx)$ and $\mg_4(\mx)$ are linearly independent for all $\mx \in
\mathbb{R}^4$, hence the distribution $\Delta= \mm{span}
\{\mg_1, \mg_2, \mg_3, \mg_4 \}$ is nonsingular. 
Note that a nonsingular distribution is analogous to a full rank matrix.

\subsection{Nonlinear Tests for Accessibility}

\subsubsection{Accessibility}

Roughly speaking, accessibility concerns whether we can %
access all %
directions of %
the state space from
any given state. %
The accessibility of control-affine
  systems can be checked using a simple algebraic test based on Lie
  brackets. 

For control-affine systems (\ref{eq:affine}), we
denote $\mathcal{C}$ as the linear combinations of recursive Lie %
brackets of the form  
\be
[\mX_k, [\mX_{k-1}, [\cdots, [\mX_2, \mX_1]\cdots]]], k = 1, 2, \cdots,
\ee
where $\mX_i$ is a vector field in the set $\{\mf, \mg_1, \cdots, \mg_M\}$. 
As the linear space $\mathcal{C}$ is a Lie algebra, %
it is closed under the Lie bracket operation. In other words, $[\mf,\mg]\in
\mathcal{C}$ whenever $\mf$ and $\mg$ are in $\mathcal{C}$. 
Hence $\mathcal{C}$ is called as the \emph{accessibility algebra}.

The \emph{accessibility distribution} $C$ is the distribution
generated by the accessibility algebra $\mathcal{C}$:
\be
C(\mx) = \mm{span} \{ \mX(\mx)  | \mX \in \mathcal{C} \}.
\ee 

Consider a control-affine
system (\ref{eq:affine}) and a state $\mx_0 \in \mathcal{M} \subset \mathbb{R}^N$. If 
\be
\mm{dim}\, C(\mx_0) = N \label{eq:ARC}
\ee
then the system is \emph{locally accessible} from $\mx_0$. Equation (\ref{eq:ARC}) 
is often called %
the \emph{accessibility rank condition} (ARC) at
$\mx_0$.  
If it holds for any $\mx_0$, then the system is called locally
  accessible. %

Interestingly, the sufficient ARC is ``almost'' necessary for accessibility. Indeed,
if the system is accessible then ARC holds for all $\mx$ in an
open and dense subset of $\mathbb{R}^N$~\citep{Isidori-Book-95,Sontag-Book-1998}.

The computation of the accessibility distribution $C$ is nontrivial,
because it is not known a priori how many (nested) Lie brackets of the vector
fields need to be computed until the ARC holds. 
In practice, a systematic search must be performed by starting with
$\{\mf, \mg_1, \cdots, \mg_M\}$ and iteratively generating new,
independent vector fields using Lie brackets. 
This can be achieved by constructing the Philip Hall basis of the Lie
algebra, which essentially follows a breadth-first search and the
search depth is defined to be the number of nested levels of bracket
operations~\citep{Serre-Book-92,Duleba-Book-98}.

In general, accessibility does not imply controllability, which is why
accessibility is a weaker version of controllability. 
Consider %
a simple dynamical system %
\be 
\begin{cases}
\dot{x}_1 &= x_2^2 \\
\dot{x}_2 &= u
\end{cases}
\ee
which can be written in the control-affine form (\ref{eq:affine}) with
$\mf(\mx)=(x_2^2, 0)^\mm{T}$ and $\mg(\mx) = (0,1)^\mm{T}$. We can
compute some Lie brackets: $[\mf, \mg](\mx)=-(2x_2, 0)^\mm{T}$, 
$[\mf, [\mf, \mg]](\mx)=(2, 0)^\mm{T}$. Since $[\mf, [\mf,
\mg]](\mx)$ is independent from $\mg(\mx)$, we conclude that $\mm{dim} C(\mx)
= 2$, for any state in $\mathbb{R}^2$, indicating that 
the system is locally
accessible. But the system is not locally controllable: $\dot{x}_1=x_2^2
>0$ for all $x_2\neq 0$, i.e., $x_1$ always grows as 
long as the system is not at the $x_2$-axis. In other words, the drift
vector field $\mf$ always steers the system to the right unless
$x_2=0$.

If we compute the accessibility distribution $C$ for a linear system
$\dot{\mx} = \mA \mx + \mB {\bf u} = \mA \mx + \sum_{i=1}^M {\bf b}_i
u_i $ where $\mB=[{\bf b}_1, \cdots, {\bf b}_M]$, we find that
$C(\mx_0)$ is spanned by $\mA \mx_0$ together with the constant vector fields ${\bf b}_i, \mA {\bf
  b}_i, \mA^2 {\bf b}_i, \cdots$, for $i=1,\cdots, M$. More precisely, 
\be
C(\mx_0) = \mm{span} \{ \mA \mx_0 \} + \mm{Im}(\mB, \mA \mB, \mA^2 \mB,
\cdots, \mA^{N-1}\mB), \label{eq:Clinear}
\ee
where $\mm{Im}()$ stands for the image or column space of a matrix.  
Note that the term $\mm{span} \{ \mA \mx_0 \}$ does not appear in 
Kalman's controllability matrix %
(\ref{eq:Cmatrix0}). Only at $\mx_0={\bf 0}$, Eq.(\ref{eq:Clinear}) reduces to Kalman's
controllability matrix. %
This shows that %
accessibility is indeed weaker than controllability, because the
former does not imply %
the latter while the latter induces the former.  

\subsubsection{Strong accessibility}
A nonlinear test %
for strong
accessibility tells us %
whether we can reach
  states in the neighborhood of the initial state exactly at a given
  small time.
Define $\mathcal{C}_0$ as the \emph{strong accessibility algebra},
i.e., the smallest algebra which contains $\mg_1,
\mg_2, \cdots, \mg_M$ and satisfies $[\mf, {\bf w}] \in
\mathcal{C}_0$, $\forall {\bf w} \in \mathcal{C}_0$. Note that
$\mathcal{C}_0 \subset \mathcal{C}$ and $\mathcal{C}_0$ does not
contain the drift vector field $\mf$. 
Define the corresponding \emph{strong accessibility distribution} 
\be
C_0(\mx) = \mm{span} \{ \mX(\mx)  | \mX \in \mathcal{C}_0 \}.
\ee 
If 
$\mm{dim}\, C_0(\mx_0) = N$ %
then the system is locally strongly accessible from $\mx_0$. 
If this holds for any $\mx_0$, then the system is called \emph{locally
  strongly accessible}. 
If we compute the strong accessibility distribution $C$ for a linear
system $(\mA, \mB)$, we will find that 
\be
C_0(\mx_0) = \mm{Im}(\mB, \mA \mB, \mA^2 \mB,
\cdots, \mA^{N-1}\mB). \label{eq:C0linear}
\ee
Then $\mm{dim}C_0(\mx_0) = N$ is %
equivalent with %
Kalman's
rank condition (\ref{eq:rankC}). In other words, strong accessibility and
controllability are equivalent notions for linear systems.

\subsection{Nonlinear Tests for Controllability}\label{subsec:nonlineartest}

For general nonlinear systems, we lack %
conditions that are both
sufficient and necessary for controllability. %
Yet, as we discuss next, we have 
some sufficient conditions %
that are believed to be almost %
necessary as well.

Consider a special class of control-affine system (\ref{eq:affine})
with $\mf(\mx) \in \mm{span}\{\mg_1(\mx),\cdots,\mg_M(x)\}$ for all
$\mx \in \mathcal{M} \subset \mathbb{R}^N$. 
In other words, the drift vector field $\mf(\mx)$, which
  describes the intrinsic dynamics of the system, can be spanned by
  the control vector fields $\mg_1(\mx),\cdots,\mg_M(x)$. %
Then, if $\mm{dim} C(\mx_0) = N$, the system is \emph{locally controllable} from
$\mx_0$. If this holds for all $\mx \in \mathcal{M}$, then the system
is \emph{globally controllable}. %

Driftless systems ($\mf(\mx)\equiv {\bf 0}$), like %
the %
front-wheel drive car system (\ref{eq:frontwheeldrive_car}), naturally fall into this class.  
To see this, we 
recognize %
that the determinant of the matrix formed by the %
vectors $\mg_1(\mx),
\mg_2(\mx), \mg_3(\mx)=[\mg_1,\mg_2](\mx)$ and $\mg_4(\mx)=[[\mg_1,
\mg_2], \mg_2](\mx)$, i.e., 
\be
\det
\begin{pmatrix}
0 & \cos(\theta+\phi) & -\sin(\theta+\phi) & -\sin \phi \\
0 & \sin(\theta+\phi) & \cos(\theta+\phi) & \cos \phi \\
0 & \sin \theta & \cos \theta & 0 \\
1 & 0 & 0 & 0
\end{pmatrix}
\ee
is identically equal to 1, regardless of $\mx$, implying that %
$\mm{dim} C(\mx_0) = N =4$ for all $\mx_0 \in \mathbb{R}^4$. 
Hence the front-wheel drive car systems is globally controllable,
in line %
with our physical intuition and experience.

For control-affine systems that do not fall into the above two
classes, Sussmann provided a general set of sufficient
conditions~\citep{Sussmann-1987}. 
We call a Lie bracket computed from $\{\mf,\mg_1, \cdots, \mg_M\}$
  \emph{bad} if it contains an odd number of $\mf$ factors and an
  even number of each $\mg_k$ factors. Otherwise we call it
  \emph{good}. The degree of a bracket is the total number of vector
  fields from which it is compuated. %
Denote with $\sum_M$ %
the permutation group on $M$ symbols. For
$\sigma \in \sum_{M}$ and ${\bf b}$ a Lie bracket computed from $\{\mf,\mg_1,
  \cdots, \mg_M\}$, define $\bar{\sigma}({\bf b})$ as the bracket
  obtained by fixing $\mf$ and changing $\mg_k$ by $\mg_{\sigma(k)}$,
  $1 \le k \le M$. 
The control-affine system (\ref{eq:affine}) is locally controllable
from $\mx_0$ if $\mm{dim} C(\mx_0) = N$ and every bad bracket ${\bf
  b}$ has the property that $\beta({\bf b})(\mx_0) \equiv
\sum_{\sigma\in \Sigma_M} \bar{\sigma} ({\bf b}) (\mx_0)$ is a linear
combination of good brackets, evaluated at $\mx_0$, of degree lower
than ${\bf b}$.

\subsection{Controllability of Nonlinear Networked Systems}

\subsubsection{Neuronal network motifs} %
\begin{figure}[t!]
\includegraphics[width=0.4\textwidth]{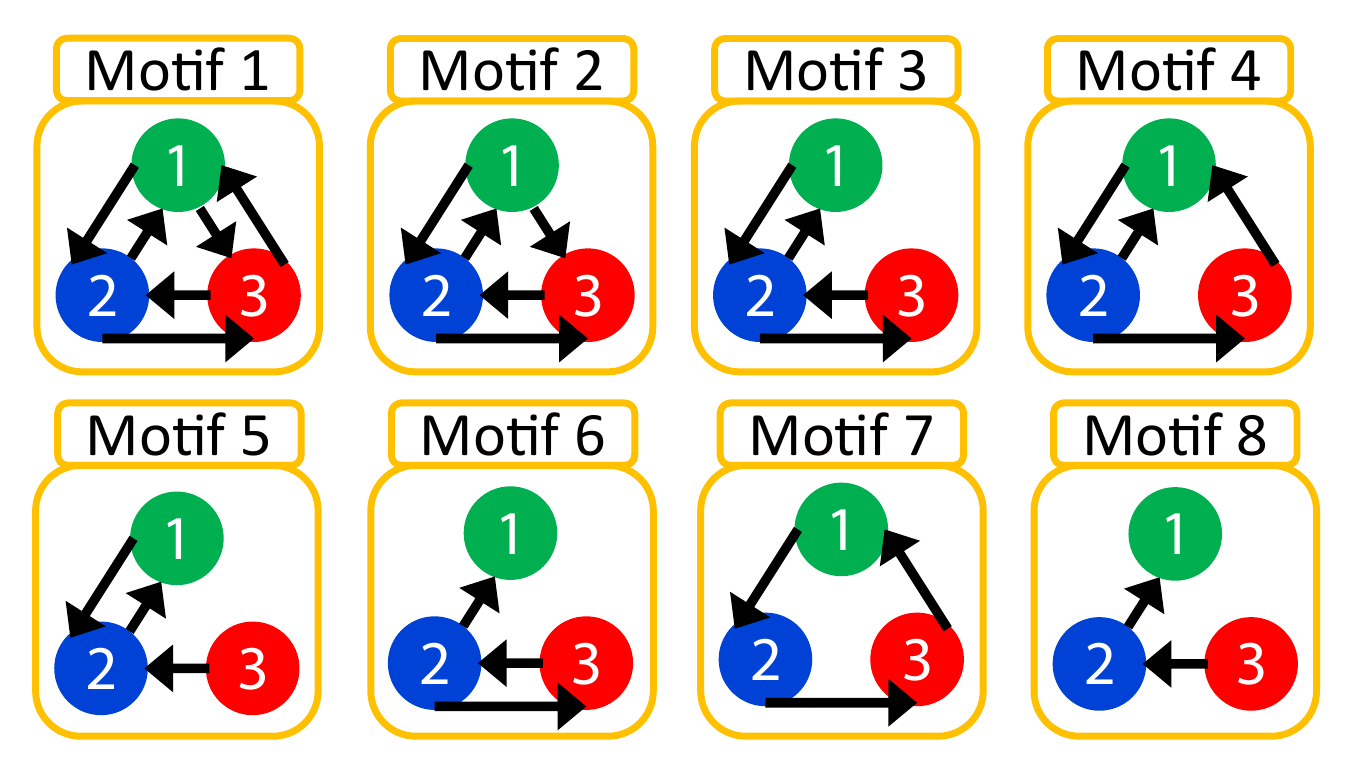}
\caption{(Color online)  Symmetries and controllability.  
The eight different three-node neuronal network motifs
  studied in \citep{Whalen-PRX-15}.
Those motifs display a variety of symmetries. For example, Motif 1 has
a full ${\bf S}_3$ symmetry, and Motif 3 has a reflection ${\bf S}_2$
symmetry across the plane through node 2. Not all symmetries have the same effect on network controllability.
}
\label{fig:neuron_nonlinearcontrol}
\end{figure}

While most complex systems are described by nonlinear {\color{blue}\emph{
  continuous-time}} dynamics defined over a network, 
there has been little attention paid so far to %
the controllability of such systems, {\color{blue} due to obvious
  mathematical challenges. 
Controllability studies of continuous-time nonlinear dynamics are
still} limited to very simple %
networks consisting of a few 
nodes, like %
neuronal network motifs governed by %
Fitzhugh-Nagumo dynamics~\citep{Whalen-PRX-15}.  
These offered an opportunity to study %
the impact of structural symmetries on nonlinear
controllability. %
The three-node neuronal motifs shown in
Fig.~\ref{fig:neuron_nonlinearcontrol} can have multiple %
symmetries. 
Yet, 
not all symmetries have the same effect on network controllability. 
For example, with identical nodal and coupling parameters, Motif 1 has
a full ${\bf S}_3$ symmetry, rendering the poorest controllability over
the entire range of coupling strengths. 
Similarly, no controllability is obtained from node 2 in Motif 3,
which has a reflection ${\bf S}_2$ symmetry across the plane through
node 2. 
Surprisingly, the rotational ${\bf C}_3$ symmetry in Motif 7 does not
cause loss of controllability at all. 
Note that %
symmetries have an impact on network controllability %
in 
linear systems as well. For example, in the case of a directed star
with LTI dynamics for which we
control the central hub (Fig.~\ref{fig:controllablesubspace}), a 
symmetry among the leaf nodes renders the system %
uncontrollable.  

Extending this analysis to larger %
networks with symmetries remains a %
challenge, however. 
Group representation theory might offer tools %
to gain insights into the impact of symmetries on the controllability of
nonlinear networked systems~\citep{Whalen-PRX-15}. 
Note, however, that %
for large real networks such symmetries are less
frequent.

{\color{blue}
\subsubsection{Boolean networks}

The controllability of Boolean networks, a class of 
  \emph{discrete-time} nonlinear systems that are often used to model
  gene regulations, has been intensively
  studied~\citep{Akutsu-JTB-07,Cheng09}. 
We can prove that finding a control strategy leading to the desired
final state is NP-hard for a general Boolean network and this problem
can be solved in polynomial time only if the network has a tree
structure or contains at most one directed cycle~\citep{Akutsu-JTB-07}. 
Interestingly, based on semi-tensor product of
matrices~\citep{Cheng-JSSC-07} and the matrix expression of Boolean
logic, the Boolean dynamics can be exactly mapped into the standard
discrete-time linear dynamics~\citep{Cheng09}.  
Necessary and sufficient conditions to assure controllability of
Boolean networks can then be proved~\citep{Cheng09}. 
Despite the formally simplicity, the price we need to pay is that the
size of the discrete-time linear dynamical system is $2^N$, where $N$
is the number of nodes in the original Boolean network. 
Hence, the controllability test will be computationally intractable
for large Boolean networks. 
}

\section{Observability}\label{sec:observability}

Before controlling a system, it is useful to know its position in the
state-space, allowing us to decide %
in which direction we should steer
it to accomplish the control objective. 
The position of a system in
the state-space can be identified %
only if we can measure the state
of all components separately, like %
the concentration of each metabolite
in a cell, or the current on each transmission line of a power
grid. Such detailed measurements are %
often infeasible and impractical. Instead, in practice we
must rely on a subset of well-selected accessible %
variables (outputs)
which can be used to observe the system, i.e. to estimate the state of 
the system. 
A system is said to be \emph{observable} if it is possible to
recover the state of the whole system from the measured variables
inputs and outputs). This is a fundamental and
primary issue in most complex systems.

In general, we can observe a system because its components form a
network, hence the state of the nodes depend on the state of their neighbors'. %
This offers the possibility %
to estimate all unmeasured
variables from the measured ones. If %
the inputs and model of the system
are known, observability can be equivalently defined as the
possibility to recover the initial state $\mx(0)$ of the system from
the output variables. 
To be specific, let us assume that we
have no knowledge of a system's initial state ${\bf x} (0)$, but we
can monitor some of its %
outputs ${\bf y}(t)$ %
in some time interval. The observability problem aims to
establish a relationship between the outputs ${\bf y} (t)$, the state vector
${\bf x} (t)$, and the inputs ${\bf u} (t)$ such that the system's initial state ${\bf x} (0)$
can be inferred. %
If no such relation exists, the system's initial state
cannot be estimated from the experimental measurements, i.e., the
system is not observable. 
In other words, if the current value of at least one state variable cannot be
determined through the outputs sensors, then it remains 
unknown to the
controller. This may disable feedback control, which %
requires reliable real-time estimates of the system's state. 
Note that observability and controllability %
are mathematically dual concepts. 
Both concepts were first
introduced by Rudolf Kalman for linear dynamical systems~\citep{Kalman-JSIAM-63}, 
and were extensively explored in nonlinear dynamical systems by many
authors~\citep{Hermann-IEEETAC-1977,Diop-ECC-91,Diop-IEEE-91,Sontag-IEEE-91,Isidori-Book-95,Gildas-Book-07}. 

In this section, we first discuss %
methods that %
test the observability of linear and nonlinear control systems. 
We also discuss the parameter identifiability problem, 
which is a special case of the observability problem. 
Finally, we introduce %
a graphical approach to identify the minimum
set of sensor nodes that assure the observability of nonlinear
  systems~\citep{%
    Khan-IEEE-08,Khan-11,
    Siddhartha-01,Letellier-PRE-2005a,Letellier-CNSNS-2006, 
    Letellier-PRE-2010,Aguirre-JPA-2005} and %
its application to 
metabolic networks~\citep{Liu-PNAS-13}.  

\subsection{Observability Tests}

\subsubsection{Linear systems} 
\begin{figure}[t!]
\includegraphics[width=0.2\textwidth]{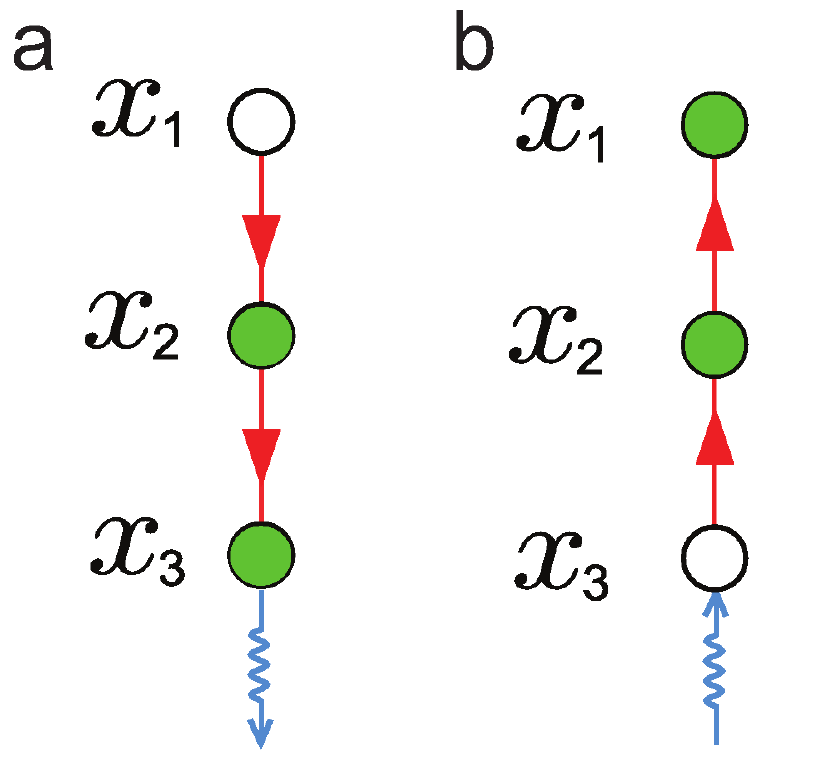}
\caption{(Color online)  Duality principle. If a system follows %
the LTI dynamics (\ref{eq:LTI_X}), the observability of the network $\mA$
  shown in (a) can be addressed by studying
  the controllability of the transposed network $\mA^\mm{T}$ shown in
  (b), obtained by reversing the direction of each link. This is a general property of all networks.  
}
\label{fig:dual} 
\end{figure}

For linear systems there is an exact duality between %
controllability and observability. 
To see this, %
consider 
an LTI control system  
\begin{subnumcases}
\mathbf{\dot{\mx}}(t) = \mA \, \mx (t) + \mB {\bf u}(t) \label{eq:LTI_X}\\
\my(t) = \mC \, \mx (t). \label{eq:LTI_Y}
\end{subnumcases}
The duality principle states that an LTI system $(\mA, \mB, \mC)$ is
observable if and only if its dual system $(\mA^\mm{T},
\mC^\mm{T},\mB^\mm{T})$ is controllable. %
Mathematically, the duality can be seen and proved from the structure of the
controllability Gramian and the observability Gramian. 
In terms of network language the duality principle has a straightforward interpretation: %
The linear observability of a network $\mA$ can be addressed by studying
  the controllability of the transposed network $\mA^\mm{T}$, which is
  obtained by flipping the direction of each link in $\mA$ (Fig.~\ref{fig:dual}). 

Thanks to the duality principle, %
many observability tests can be mapped into controllability tests. For example, according to Kalman's rank
condition, %
the system $(\mA, \mB, \mC)$ is observable
if and only if the
\emph{observability matrix} 
\be
{\bf O} = 
\begin{bmatrix} 
\mC \\ \mC\mA \\ \mC\mA^2 \\ \vdots \\ \mC\mA^{N-1} 
\end{bmatrix} \label{eq:Omatrix}
\ee 
has full rank, i.e., $\mm{rank}\, {\bf O} =
N$~\citep{Kalman-JSIAM-63,Luenberger-Book-79}. 
 This rank condition is based on the fact that if the $N$
rows of ${\bf O}$ are linearly independent, then each of the $N$ state
variables can be determined by linear combinations of the output
variables ${\bf y}(t)$.

\subsubsection{Nonlinear systems}

Consider a nonlinear control system with inputs ${\mb u}(t) \in \mathbb{R}^K$ and outputs ${\mb
  y}(t)  \in \mathbb{R}^M$: 
\be
\begin{cases}
\dot\mx(t) = \mf(t,\mx(t),{\mb u}(t)) \\
\my(t) = \mh(t,\mx(t),{\mb u}(t)) 
\end{cases}
\label{eq:ode}
\ee
where %
$\mf(\cdot)$ and
$\mh(\cdot)$ are some %
nonlinear functions.

Mathematically, we can quantify observability from either 
an algebraic viewpoint~\citep{Diop-ECC-91, Diop-IEEE-91,Moog-Book-07} or a differential geometric
viewpoint~\citep{Hermann-IEEETAC-1977}. 
Here we focus on the former. 
If a system is \emph{algebraically observable}, then there are
algebraic relations between the state variables and the successive
derivatives of the system's inputs and outputs~\citep{Diop-ECC-91,
  Diop-IEEE-91}. These algebraic relations guarantee that 
the system is observable and %
forbid symmetries.
A family of symmetries is equivalent to infinitely many trajectories of the state
variables that fit the same specified input-output behavior, in which
case the system is not observable.  
If the number of such trajectories is finite, the system is called \emph{locally
observable}. If there is a unique trajectory, the system is \emph{globally
observable}.%

Consider, for example, the dynamical system defined by the equations 
\be
\begin{cases} 
\dot{x}_1 = x_2 x_4 + u \\
\dot{x}_2 = x_2 x_3\\
\dot{x}_3 = 0 \\
\dot{x}_4 = 0 \\
y = x_1 
\end{cases}
\ee
The system has a family of symmetries $\sigma_\lambda$: $\{
x_1, x_2, x_3, x_4 \}$ $\to$ $\{
x_1, \lambda x_2, x_3, x_4/\lambda \}$, %
so that the input $u$ and the output $y$ and all their derivatives are
independent of $\lambda$~\citep{Anguelova-Thesis-04}. This means that we cannot distinguish whether the system
is in state $(x_1, x_2, x_3, x_4)^\mm{T}$ or its symmetric counterpart
$(x_1, \lambda x_2, x_3, x_4/\lambda )^\mm{T}$, because they are both 
consistent with the same input-output behavior. Hence we cannot
uncover the system's internal state by monitoring $x_1$ only.

The algebraic observability of a rational system is determined by the
dimension of the space spanned by the gradients of the Lie-derivatives 
\be
L_f \equiv    \frac{\p}{\p t} 
       + \sum_{i=1}^N f_i \frac{ \p}{ \p x_i} 
       + \sum_{j \in \mathbb{N}} \sum_{l=1}^K u_l^{(j+1)} \frac{ \p}{ \p u_l^{(j)}}
\ee
of its output functions $\mh(t,\mx(t),{\mb u}(t))$. The observability problem
can be further reduced to the so-called rank test: the system (\ref{eq:ode})
is \emph{algebraically observable} if and only if the $NM \times N$ 
Jacobian matrix 
\be
{\bf J} %
= \begin{bmatrix}
\frac{\p L_f^0 h_1}{\p x_1} & \frac{\p L_f^0 h_1}{\p x_2} & \cdots & \frac{\p L_f^0 h_1}{\p x_N}\\
\cdots & \cdots & \cdots & \cdots\\
\frac{\p L_f^0 h_M}{\p x_1} & \frac{\p L_f^0 h_M}{\p x_2} & \cdots & \frac{\p L_f^0 h_M}{\p x_N}\\
\vdots & \vdots & \vdots & \vdots\\
\frac{\p L_f^{N-1} h_1}{\p x_1} & \frac{\p L_f^{N-1} h_1}{\p x_2} & \cdots & \frac{\p L_f^{N-1} h_1}{\p x_N}\\
\cdots & \cdots & \cdots & \cdots\\
\frac{\p L_f^{N-1} h_M}{\p x_1} & \frac{\p L_f^{N-1} h_M}{\p x_2} & \cdots & \frac{\p L_f^{N-1} h_M}{\p x_N}\\
\end{bmatrix}\label{eq:J}
\ee
has full rank~\citep{Diop-ECC-91,
  Diop-IEEE-91}, i.e., 
\be
\mm{rank}\, {\bf J} = N. \label{eq:rankJ}
\ee
Note that for an LTI system (\ref{eq:LTI_X},\ref{eq:LTI_Y}), the Jacobian matrix (\ref{eq:J}) reduces to
the observability matrix (\ref{eq:Omatrix}).

For rational dynamic systems, the algebraic observability test can be
performed %
using an algorithm developed by %
Sedoglavic%
~\citep{Sedoglavic-JSC-02}.  
The algorithm offers a %
generic rank computation of the Jacobian
matrix (\ref{eq:J}) using the techniques of symbolic calculation,
allowing us to test local algebraic observability for
rational systems in polynomial time. 
This algorithm certifies that a system is locally observable, but %
its answer for a non-observable system is probabilistic with high probability of
success. A system that is found %
non-observable %
can be further analyzed to identify %
a family of symmetries,
which can %
confirm the system is truly non-observable. %

\subsection{Minimum sensor problem}\label{sec:msp}

In complex systems, the state variables are rarely independent of
each other. %
The interactions between the system's components induce intricate
interdependencies among them.  

Hence a well-selected subset of state variables can contain sufficient
information about the remaining %
variables %
to reconstruct the system's complete internal state, %
making the system observable~\citep{Liu-PNAS-13}.

We assume that we can monitor a selected subset of state variables,
i.e. $\my(t) = (\cdots, x_i(t), \cdots)^\mm{T}$, corresponding to the
states of several nodes that we call \emph{sensor nodes} or just
\emph{sensors}. 
Network observability %
can then be posed as
follows: Identify the minimum set of sensors from whose measurement
we can infer %
all other state variables. 
 For linear systems, this problem can be %
solved using %
the duality principle and solving the minimum
    input problem of the transposed network $\mA^\mm{T}$. For general
  nonlinear systems this trick does not work.
While \eqref{eq:rankJ} offers a formal answer to the observability issue and can
be applied to small engineered systems, it has notable practical
limitations for large and complex systems. First, it can only confirm
if a specific set of sensors can be used to observe a system or not,
without telling us how to identify them.
Therefore, a brute-force search for a minimum sensor set requires us to
inspect via \eqref{eq:rankJ} %
about $2^N$ sensor combinations, a
computationally prohibitive task for large %
systems. 
Second, the rank test of the Jacobian matrix via symbolic computation
is computationally limited to small systems~\citep{Sedoglavic-JSC-02}. 
Hence, the fundamental question of identifying the minimum set of
sensors through which we can observe a large complex system remains an
outstanding challenge.

\begin{figure}[t!]
\includegraphics[width=0.5\textwidth]{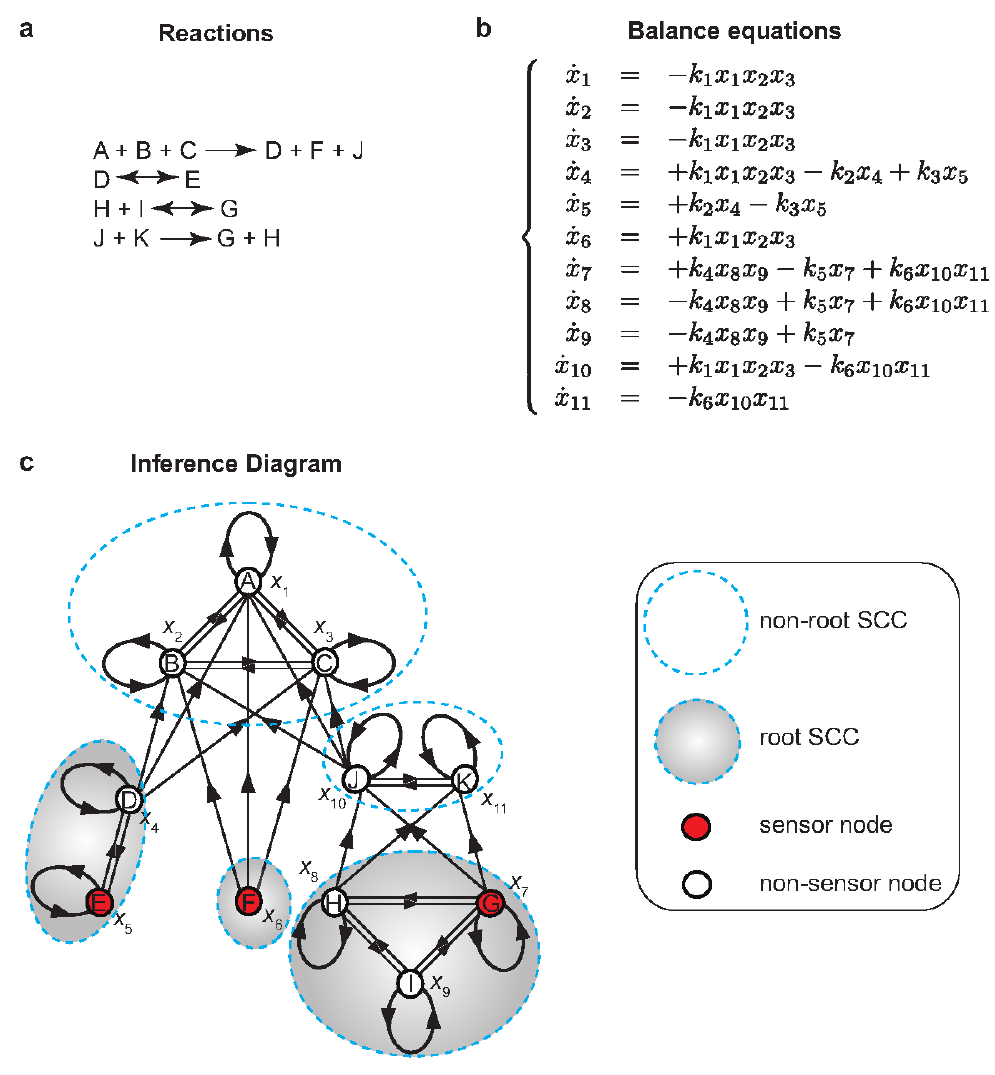}
\caption{(Color online)  The graphical approach to determine the minimum sensors of a chemical reaction system. %
(a) A chemical reaction system with eleven species
(A,B,$\cdots$, J,K) involved in four reactions. Since two reactions
are reversible, we have six elementary reactions. 
(b) The balance equations of the chemical reaction system shown
in (a). The concentrations of the eleven species are denoted by $x_1, x_2,
\cdots, x_{10}, x_{11}$, respectively. The rate constants of the six
elementary reactions are given by $k_1, k_2, \cdots, k_6$, respectively.
The balance equations are derived using the mass action kinetics.   
(c) The inference diagram is constructed by drawing a directed link
($x_i \to x_j$) as long as $x_j$ appears in the RHS of $x_i$'s balance 
equation shown in (b). Strongly connected components (SCCs) are marked with dashed
circle. Root SCCs, which have no incoming links, are shaded in
grey. A potential minimum set of sensor nodes, whose measurements
allows us to reconstruct the state of all other variables (metabolite
concentrations), are shown in red. After \citep{Liu-PNAS-13}.}
\label{fig:inferringdiagram} 
\end{figure}

To resolve these limitations, we can exploit the dynamic interdependence of
the system's components through a graphical
representation%
~\citep{Lin-IEEE-74,Reinschke-book-88,Murota-book-09,Siddhartha-01,Khan-11}. %
The procedure consists of the following steps~\citep{Liu-PNAS-13}: 

(i) \emph{Inference diagram:} 
Draw a directed link $x_i \to x_j$ if $x_j$ appears in $x_i$'s
differential equation (i.e., if $\frac{\partial f_i}{\partial x_j}$ is not identically zero),
implying that one can retrieve some %
information on $x_j$ by monitoring $x_i$
as a function of time. 
Since the constructed network captures the information flow to infer 
the state of individual variables, we call it the \emph{inference diagram}
(Fig.~\ref{fig:inferringdiagram}c). 

(ii) \emph{Strongly connected component (SCC) decomposition:} Decompose
the inference diagram into a unique set of maximal SCCs (dashed
circles in Fig.~\ref{fig:inferringdiagram}c), i.e. the largest
subgraphs chosen such that in each of them there is a directed path
from every node to every other node
~\citep{Cormen-Book-90}. 
Consequently, each node in an SCC contains some information about
all other nodes within the SCC. 

(iii) \emph{Sensor node selection:} %
Those SCCs that have no incoming edges are referred to as \emph{root
  SCCs} (shaded circles in Fig.~\ref{fig:inferringdiagram}c). 
We must choose at least one node from each root SCC to ensure the %
observability of the whole system. For example, %
the inference diagram of Fig.~\ref{fig:inferringdiagram}c contains three root SCCs; hence we
need at least three sensors to observe the system. 

The graphical approach (GA) described above can be %
used to determine
whether a variable provides full observability of small dynamic
systems~\citep{Letellier-PRE-2005a,Aguirre-Chaos-2008}. 
As these systems have only a few state variables, steps (ii) and
(iii) are often not necessary.      
For large networked systems, the GA is very powerful because it reduces the observability issue, a
dynamical problem of a nonlinear system with many unknowns, to a
property of the static graph of the inference diagram, which can be %
accurately mapped for an increasing number of complex
systems, from biochemical reactions to ecological systems. %

We can prove that monitoring the root SCCs identified by
the GA are \emph{necessary} for observing any nonlinear dynamic
system~\citep{Liu-PNAS-13}.  
In other words, the number of root SCCs yields a strict lower bound for the size of the
minimum sensor set. 
Consequently, any state observer (i.e. a dynamical device that aims to
estimate the system's internal state) will fail if it doesn't monitor
these sensors. %

If the dynamics is linear, %
the duality principle maps the
minimum sensor problem into the minimum input problem and predicts not only the necessary, but also the sufficient
sensor set for observability. %
Numerical simulations on model networks
suggest %
that for linear systems the
sufficient sensor set is noticeably larger than the
necessary sensor set predicted by GA~\citep{Liu-PNAS-13}. 
This is because that any symmetries in the state variables
leaving the inputs, outputs, and all their derivatives
invariant will make the system unobservable
~\citep{Sedoglavic-JSC-02}. %
For structured linear systems, the symmetries correspond to a particular
topological feature, i.e., dilations, which can be detected from the
inference diagram. Yet, %
for general nonlinear systems, the symmetries can not be easily
detected from the inference diagram only. 

For linear systems the minimum
sensor set predicted by the GA is generally not sufficient for full
observability.  
Yet, for large 
nonlinear dynamical systems the symmetries in state
variables are extremely rare, especially when the number of state
variables is big, hence the sensor set predicted by GA is often not
only necessary but also sufficient for observability~\citep{Liu-PNAS-13}. %

To better understand network observability, next we apply the
developed tools to %
biochemical and technological networks.

\subsubsection{Biochemical reaction systems}

Consider a biochemical reaction system of $N$ species
$\{\mathcal{S}_1, \mathcal{S}_2, \cdots, \mathcal{S}_N\}$ involved in $R$
reactions $\{\mathcal{R}_1, \mathcal{R}_2, \cdots, \mathcal{R}_R\}$ with 
\be 
\mathcal{R}_j:\,\, \sum_{i=1}^{N} \alpha_{ji} \mathcal{S}_i \rightarrow
\sum_{i=1}^N \beta_{ji} \mathcal{S}_i, \label{eq:reaction}
\ee 
where $\alpha_{ji} \ge 0$ and $\beta_{ji}\ge 0$ are the
stoichiometry coefficients. For example, (\ref{eq:reaction}) captures
the reaction 2 H$_2$ + O$_2$ = 2 H$_2$O with $\alpha_{11}=2$,
$\alpha_{12}=1$ and $\beta_{11}=2$. 
Under the continuum hypothesis and the well-mixed
assumption the system's dynamics is described
by \eqref{eq:ode}, where $x_i(t)$ is the concentration of species $\mathcal{S}_i$ at time $t$,  
the input vector ${\bf u}(t)$ represents %
regulatory signals or external nutrient
concentrations, and %
the vector $\my(t)$ captures the set of experimentally
measurable species concentrations or reaction fluxes.   
The vector $\mv(\mx) =
(v_1(\mx), v_2(\mx), \cdots, v_R(\mx))^\mm{T}$ is often called the flux
vector, which follows the %
mass-action kinetics~\citep{Heinrich-book-96,Palsson-Book-06}
\be
v_j(\mx) = k_j \prod_{i=1}^N x_i^{\alpha_{ji}} \label{eq:MAK}
\ee
with rate constants $k_j>0$.
The system's dynamics is therefore described by the balance equations
\be 
\dot{x}_i = f_i({\bf x}) = \sum^R_{j=1} \Gamma_{ij} \, v_j(\mx), \label{eq:balance}
\ee 
where $\Gamma_{ij} =
\beta_{ji} - \alpha_{ji}$ are the elements of the $N \times R$
stoichiometric matrix $\boldsymbol\Gamma$.   
The RHS of \eqref{eq:balance} represents a
sum of all fluxes $v_j$ that produce and consume the species
$\mathcal{S}_i$. 

Assuming that the outputs ${\mb y}(t)$ are just the concentrations of a
particular set of sensor species that can be experimentally measured,
then observability problem aims to identify \emph{a minimum set of sensor
species from whose measured concentrations we can determine all other
species' concentrations}. 
In this context, the advantage of GA is that it does not require the
system's kinetic constants (which are largely unknown \emph{in vivo}),
relying only on the topology of the inference diagram. For a metabolic
network or an arbitrary biochemical reaction system, the topology of
the inference diagram is uniquely determined by the full reaction
list, which is relatively accurately known for several model 
organisms~\citep{Schellenberger-10}. 
Applying GA to biochemical reaction systems offers several interesting
results, elucidating the principles behind biochemical network 
observability~\citep{Liu-PNAS-13}:    

a) Species that are not reactants in any reaction, being instead
\emph{pure
  products}, will be root SCCs of size one. Consequently, %
they are always sensors, and must be observed by the external observer (e.g., $x_6$ in Fig.~\ref{fig:inferringdiagram}c). 

b) For root SCCs of size larger than one (e.g. $\{x_4, x_5\}$ and
$\{x_7, x_8, x_9\}$ in Fig.~\ref{fig:inferringdiagram}c), \emph{any} node could be chosen as
a sensor.  %
Given that some root SCCs are %
quite large, 
and typically we only need to monitor one node for each root SCC,
the number of sensor nodes is thus considerably reduced.

c) %
A minimum set of sensors consists of all pure products and one node
from each root SCC of size larger than one %
(e.g. $\{x_5, x_6, x_7\}$ in Fig.~\ref{fig:inferringdiagram}c). 

d) Since any node in a root SCC can be selected as a sensor node,  
there are $\Omega_\mm{s} =
\prod_{i=1}^{N_\mm{root-SCC}} n_i$  equivalent sensor node combinations, representing the product of all root SCCs' sizes. 
For example, in 
Fig.~\ref{fig:inferringdiagram}c we have three root SCCs with sizes $n_i = 1,2,3$,
hence $\Omega_\mm{s}=1\times 2 \times 3= 6$. 
This multiplicity offers significant flexibility in selecting experimentally accessible sensors.

It turns out that the minimum set of sensors obtained by GA almost
always achieve full observability for the whole system, except in some
pathological cases~\citep{Liu-PNAS-13}. 
The sufficiency of the sensors predicted by GA 
is %
unexpected because substantial details
   about the system's dynamics are ignored in GA, hence offering an
   exact proof that the
   \emph{sufficiency} of the predicted sensors 
   for observability is a %
difficult, if not an impossible,
   task. %
Note, however, that the rigorous proof of sufficiency and the systematic search for
exceptional cases %
making a system unobservable remain %
open questions. 
\subsubsection{Power grid}
\begin{figure}[t!]
\includegraphics[width=0.4\textwidth]{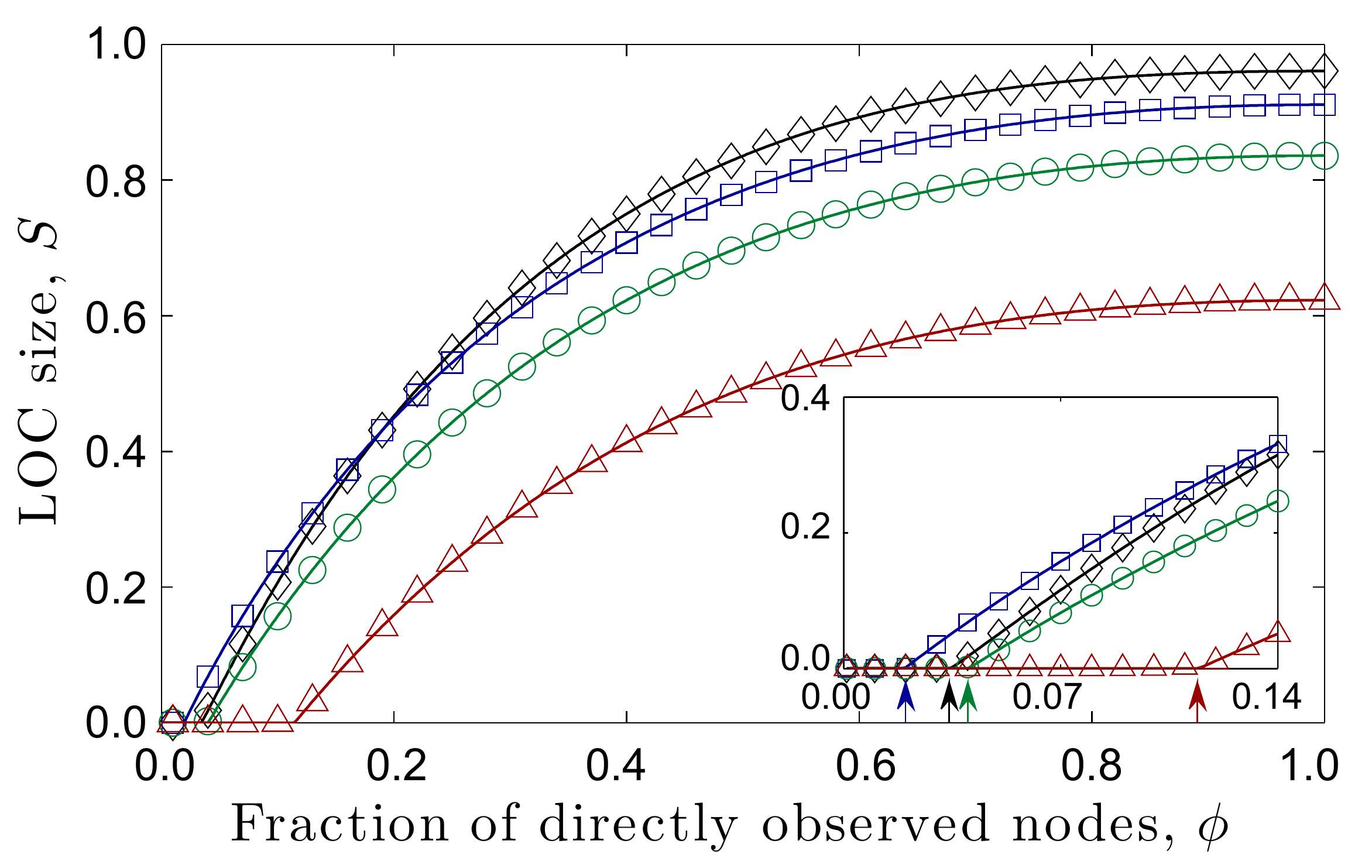}
\caption{(Color online)  Observability transitions in the power grid. (a)
  Fraction of the largest observable component %
as a function of the
  fraction of directly observed nodes ($\phi$) in networks with
  prescribed degree distributions of the power grids of Eastern North
  America (black), Germany (red), Europe (green), and Spain (blue). 
  The
  continuous lines are %
analytical predictions, and the
  symbols represent the average over ten  $\mbox{$10^6$-node}$ random
  networks for $10$ independent random PMU placements each. The inset
  shows a magnification around the transitions, with the analytically predicted
  thresholds $\phi_c$ indicated by arrows.  
After \citep{Yang-PRL-12}.
\label{fig:not}}
\end{figure}

In the power grid, %
the state variables represent %
the voltage of all nodes, which in practice can be determined by phasor
measurement units (PMUs). 
Since a PMU can measure the real time %
voltage and line
currents of the corresponding node, a PMU placed on a
node $i$ will determine the state variables of both node $i$ and all of its
first nearest neighbors. %
In this case the observability problem 
can be mapped to a purely graph theoretical
problem. %
The random placement of PMUs leads to a 
network observability
transition~\citep{Yang-PRL-12}, which is a new type of percolation transition that 
characterizes the emergence of
macroscopic observable components in the network as the number of
randomly placed PMUs increases (Fig.~\ref{fig:not}). 
Using the generating function formalism~\citep{Newman-PRE-01}, we can
{\color{red} analytically calculate the expected size of the largest
observable component for networks with any prescribed degree
distribution. This has been demonstrated for %
real power grids~\citep{Yang-PRL-12}.  
Moreover, it has been found that} the percolation threshold
decreases with the increasing %
average degree or degree
heterogeneity~\citep{Yang-PRL-12}. 

\begin{figure}[t]
  \begin{center}
\includegraphics[width=0.4\textwidth]{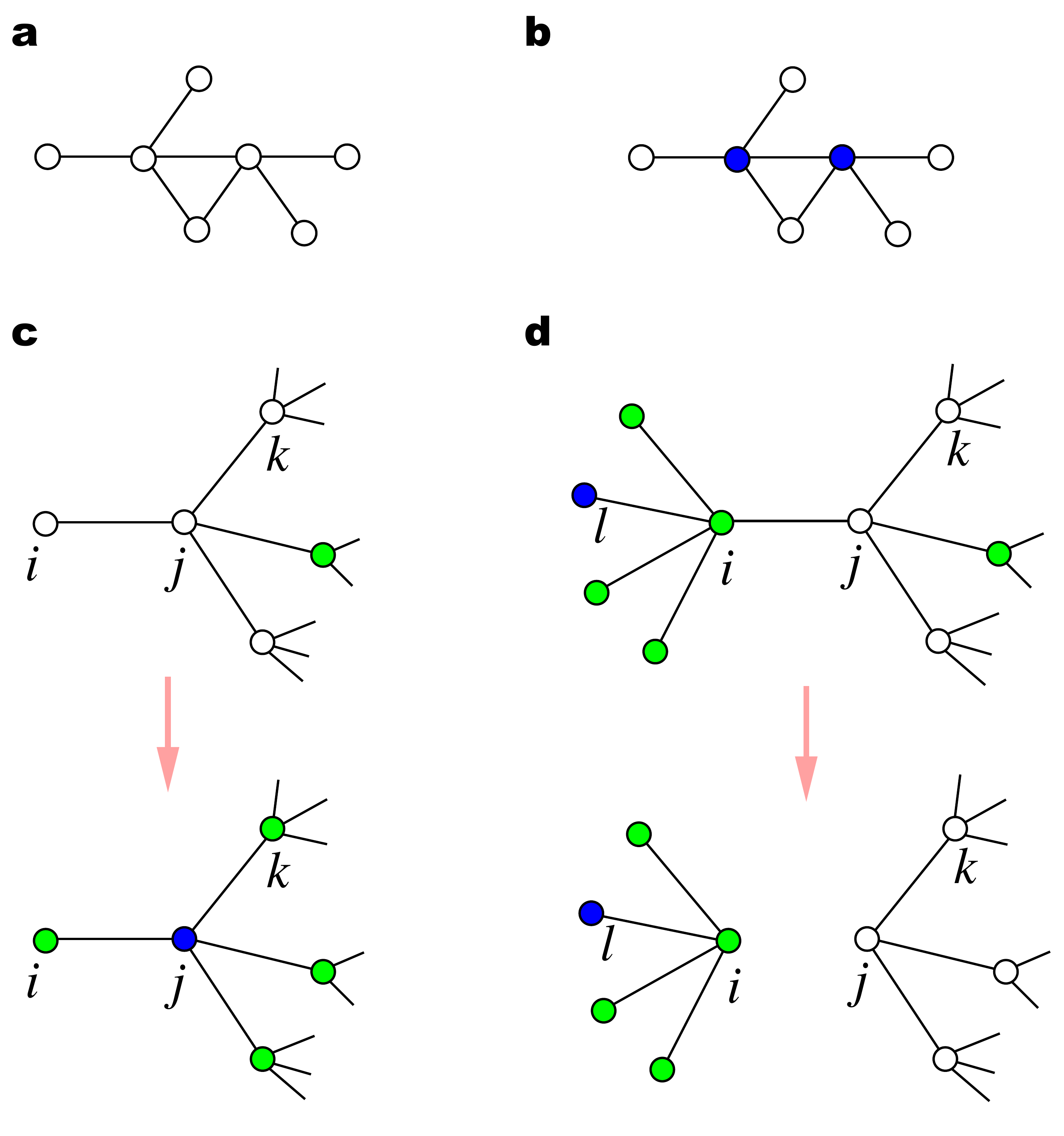}
  \end{center}
  \caption{(Color online)  
    \label{fig:GLR-MDS} 
Dominating set and generalized leaf removal process. 
(a-b) Dominating set. A dominating set of a graph
    $G=(V,E)$ is a subset $D$ of $V$ such that every vertex not in $D$
    is adjacent to at least one vertex in $D$. 
A minimum dominating set (MDS, shown in blue) is a dominating set of the smallest
size.
(c-d) 
Generalized leaf removal (GLR) process. If a
    network is sufficiently sparse, then its minimum dominating set (MDS) can be found exactly
    using GLR, consisting of two basic operations illustrated 
    in (c) and (d). 
Blue circles denote nodes occupied with sensor nodes. 
White circles denote empty (i.e. non-occupied) and unobservable
nodes. 
Green circles denote empty but observable nodes. 
(c) For an empty leaf node $i$, its only adjacent node $j$ must be
occupied, 
i.e. be chosen as a sensor node. Consequently %
all adjacent nodes of $j$ are observed. Node $j$ and its adjacent
nodes can be removed from the network to simplify the MDS problem. 
(d) If an empty observed node $i$ has only a single unobserved 
    adjacent node $j$, then it must be an optimal strategy not to
    occupy node $i$. Hence, the link between $i$ and $j$ can be
    removed from the network to simplify the MDS problem.
After \citep{Zhou-arXiv-14}.
  }
\end{figure}

The random placement of PMUs apparently will not solve the minimum
sensor problem. 
For a power grid, the problem of identifying the minimum set of sensor
nodes is reduced to the minimum dominating set (MDS)
problem: Identify a
minimum node set $D\subseteq V$ for a graph $G = (V, E)$ such that
every node not in $D$ is adjacent to at least one node in $D$ (Fig.~\ref{fig:GLR-MDS}a,b). %
 Consider a undirected network $G$. %
Node $i$ is either empty (with occupation state $c_i = 0$) or occupied by
sensors (with $c_i = 1$). In other words, if $c_i=1$ then node $i$ can
be considered a sensor node. 
Node $i$ is called observed if it is a sensor node itself or it is not a
sensor node but adjacent to one or more sensor nodes. Otherwise node
$i$ is unobserved. 
The MDS problem requires us to occupy a
minimum set $D$ of nodes so that %
all %
$N$ nodes of $G$ are %
observed.~\footnote{\color{red}Interestingly, the MDS problem can also
  be formalized as a control problem on a undirected network by assuming that
  every edge in a network is bi-directional and every node in the MDS
  can control all of its outgoing links
  separately~\citep{Nacher-NJP-12}. 
This formulation has recently been applied to analyze
  biological networks~\citep{Wuchty-PNAS-14,Nacher-Methods-15}.}  
The MDS problem for a general graph is NP-hard, and %
the best polynomial algorithms 
can only offer %
dominating sets with sizes not
exceeding $\log N$ times of the minimum size of the dominating sets~\citep{Lund-Book-94,Raz-1997}. 
If the underlying network has no core, we can %
solve exactly the MDS problem in polynomial time %
using a generalized leaf-removal (GLR) process
(Fig.~\ref{fig:GLR-MDS}c,d). 
The GLR process can be recursively applied to simplify the
network $G$. 
If eventually all the nodes are removed, then the set of nodes
occupied during this process must be an MDS and choosing them as
sensor nodes will make the whole network
observable~\citep{Zhou-arXiv-14}. 
If, however, the final simplified network is non-empty, then there
must be some nodes that are still unobserved after the GLR
process. The subnetwork induced by these unobserved nodes is referred
to as the \emph{core} of the original network $G$. %
For networks with an
extensive core, 
a belief-propagation algorithm, rooted %
in spin glass theory, can offer %
nearly optimal solutions, which %
also performs well on 
real-world networks~\citep{Zhou-arXiv-14}. %
{\color{red}
Recently, %
probabilistic methods have been developed to
approximate the size of the MDS in scale-free
networks~\citep{Molnar-SR-14}.  
}

\subsection{Target observability} 

In many applications it is overkill %
to observe the full system,
but it is sufficient to infer the state of a 
subset of %
\emph{target variables}. Such target variables could for example correspond to the
concentrations of metabolites whose activities
are altered by a disease~\citep{Barabasi-Nature-11}, 
representing potential biomarkers.   
In case those target variables cannot be directly measured, we can invoke 
\emph{target observability}, and aim to identify the optimal
sensor(s) that can infer the state of the target variables. 
These could represent 
the optimal experimentally accessible biomarkers for 
a disease. 
The graphical approach discussed above helps us select such optimal sensors: %
a) The state of a target node $x_\mm{t}$ can be observed from a
sensor node $x_\mm{s}$ only if
there is a directed path from $x_\mm{s}$ to $x_\mm{t}$ in
the inference diagram. 
For example, in Fig.~\ref{fig:inferringdiagram}c, $x_4$ can only be inferred from $x_5$ while 
$x_1$ can be inferred from any other nodes. 
b) There are important differences in the complexity of the
inference process, which depends on %
the size of the subsystem we need to infer for a given sensor choice.  
The SCC decomposition of the inference diagram indicates that 
to observe %
$x_\mm{t}$ from $x_\mm{s}$, we need to reconstruct 
$\mathcal{N}_\mm{s} = \sum_{n_i \subset \mathcal{S}_{\mm{s}}} n_i$
metabolite concentrations, where 
$\mathcal{S}_{\mm{s}}$ denotes the set of all SCCs that are reachable
from $x_\mm{s}$, and $n_i$ is the size of the $i$-th SCC.  
This formula can be %
extended to multiple targets. 
c) To identify the optimal sensor node for any target node, we can minimize  
$\sum_{n_i \subset \mathcal{S}_{\mm{s}}} n_i$, which is the %
minimum amount of information required for the inference process. 
For example, if 
$x_\mm{t}$ is inside an SCC of size larger than one
(e.g., $x_1$ in Fig.~\ref{fig:inferringdiagram}c), then the optimal sensor can be any 
other node in the same SCC (e.g., $x_2$ or $x_3$ in
Fig.~\ref{fig:inferringdiagram}c). 
If all other nodes in the same SCC is experimentally inaccessible, 
then the optimal
sensor node belongs to the smallest SCC that
points to $x_i$ (e.g., $x_6$ in Fig.~\ref{fig:inferringdiagram}c). 
Note that this minimization procedure can be implemented for
any inference diagram in polynomial time. 
Hence the graphical approach can aid the efficient selection of optimal sensors for any targeted
node, offering a potentially indispensable tool for biomarker design.

{\color{red}
 \subsection{\color{red}Observer Design}
 The observability test and the graphical approach mentioned above do
 not tell us how to reconstruct
 the state of the system from %
 measurements. To achieve
 this we must design an \emph{observer}, a dynamic device that runs a replica 
 of the real system, adjusting its state from the available outputs to
 uncover the missing variables. 

For an %
LTI system (\ref{eq:LTI_X}, \ref{eq:LTI_Y}), we can easily design the
so-called \emph{Luenberger
  observer}~\citep{Luenberger-IEEE-64,Luenberger-IEEE-66,Luenberger-IEEE-71} 
 \be
 \dot{\mz}(t) 
 = \mA \, \mz(t) + {\bf L} \left[ \my(t) - {\bf C}\, \mz(t) \right] +
 \mB \, {\bf u}(t) \label{eq:z}
 \ee
 where the $N\times K$ matrix ${\bf L}$ is to be specified later.  
 Note that %
with initial condition $\mz(0)=\mx(0)$, the Luenberger observer
 will follow $\mz(t)=\mx(t)$ exactly for all $t>0$. Because $\mx(0)$
 is typically unaccessible, we start from $\mz(0)\neq \mx(0)$ and hope
 that $\mz(t)$ will asymptotically converge to $\mx(t)$, i.e. the state of the observer tracks the state of the
 original system. This can be achieved by choosing a proper ${\bf L}$
 matrix such that the matrix $\left[ \mA - {\bf L} \, {\bf C} \right]$ is
 asymptotically stable, in which case the error vector ${\bf e}(t) = \mz(t)
 - \mx(t)$, satisfying 
 $\dot{\bf e}(t) 
 = \left[ \mA - {\bf L} \, {\bf C} \right]  {\bf e}(t), 
$
will converge to zero with rate determined by the largest eigenvalue
of $\left[ \mA - {\bf L} \, {\bf C} \right]$. 
For nonlinear systems the observer design is rather involved and still
an open challenge~\citep{Friedland-Book-1996,Besancon-Book-2007}. %
}

\subsubsection{Parameter Identification}%

Most modeling efforts assume that the system parameters, like 
the rate constants of biochemical reactions, are known. Yet, for most complex
systems, especially in biological context, the system parameters are usually unknown or are only known
approximately. Furthermore, the known parameters are typically estimated in vitro, and their
in vivo relevance is often questionable. %
This raises a natural question: Can we determine the model parameters through
appropriate input/output measurements, like monitoring the
concentrations of properly selected chemical species? This problem is
called \emph{parameter identification} (PI) in control
theory~\citep{Bellman-MB-70,Glad-PDC-90,Pohjanpalo-MB-78,Saccomani-Automatica-03,Ljung-87}.

We can formalize the parameter identifiability problem 
as the observability problem of an extended system as follows%
~\citep{Anguelova-Thesis-04}.
For this we consider the system parameters $\Theta$ as special state variables
with time-derivative zero ($\ud \Theta /\ud t = 0$). We can extend the state vector
to include a larger set of state variables, i.e., $(\mx(t), \Theta)$, allowing
us to formally determine whether/how the system parameters can be
identified from the input-output behavior by checking the
observability of the extended system. 
{\color{red}Consequently, PI can be considered as a special observer
  design problem.}

{\color{red}

\subsubsection{Network Reconstruction}

When the system parameters contain information about the network
structure, the corresponding PI problem can be generalized to a
network reconstruction (NR) problem. Consider a network 
whoe state variables are governed by a set of
ODEs 
\be
\dot{x}_i(t) = \sum_{j=1}^N a_{ij} f_{ij}(x_i(t), x_j(t)) + u_i(t), \label{eq:system}
\ee
where $i=1,\cdots, N$; the coupling functions $f_{ij} : \mathbb{R}
\times \mathbb{R} \to \mathbb{R}$ capture the interactions between
nodes: self interactions when $i=j$ or pairwire interactions when
$i\neq j$. The term $u_i(t) \in \mathbb{R}$ represents either known signals
or control inputs that can affect node $i$'s state. The interaction
matrix $\mathbf{A}=[a_{ij}] \in \mathbb{R}^{N\times N}$ captures the
directed interactions between the nodes: $a_{ji} \neq 0$ if 
node $j$ directly affects node $i$'s
dynamics. Given measured temporal data $\{ x_i(t), u_i(t)\}_{i=1}^N$,
$\forall t \in [t_0, t_1]$, NR aims to recover
some properties of the $\mathbf{A}$ matrix, e.g. 
its \emph{sign pattern} $\mathbf{S}= [s_{ij}] = [\mm{sign}(a_{ij})] \in \{-1,0,1\}^{n
  \times n}$, \emph{connectivity pattern} $\mathbf{C}= [c_{ij}] =
[|s_{ij}|] \in \{0,1\}^{n \times n}$, \emph{adjacency pattern} $\mathbf{K}=[k_{ij}] =[c_{ij} (1-\delta_{ij})] \in \{0,1\}^{n \times n}$ %
($\delta_{ij}$ is the Kronecker delta)
or \emph{in-degree sequence} ${\bf d}=[d_i]=[\sum_j c_{ij}] \in \mathbb Z^n$. 
Note that PI aims to recover
the $\mathbf{A}$ matrix itself.

There are three principally different NR approaches, which assume various levels of a priori
knowledge about the system%
~\cite{Timme:14}. 

\emph{Driving-response}. %
Here we try to measure and evaluate the
collective response of a networked system to external perturbations or
driving. As the response depends on both the external driving signal
(which unit is perturbed, when and how strong is the perturbation,
etc.), and the (unknown) structural connectivity of the network,
sufficiently many driving-response experiments should reveal the
entire network. This approach is relatively simple to implement and
the required computational effort scales well with the system size. 
It has been well established for the reconstruction of gene regulatory 
networks
~\cite{Gardner:03, Tegner:03, Yu:10, Yu:10b}. 
Yet, this approach requires us to %
measure and drive the
dynamics of all units in the system, which is often infeasible. %
The collective dynamics suitable for the driving-response experiments
also needs to be simple (i.e., to exhibit a stable fixed point or periodic
orbits, or to allow the system to be steered into such a state). 
For systems exhibiting more complex features, e.g. chaos,
bifurcations, multi-stability, this approach is not applicable.
If the system exhibits the same fixed point %
for different constant
inputs (as some biological systems that have ``perfect adaptation''),
it is impossible to reconstruct the network using driving-response
experiments %
\cite{prabakaran2014paradoxical}.

\emph{Copy-synchronization}: This approach sets up a copy %
of the original system and updates its interaction matrix 
continuously until the copy
system synchronizes its trajectories with the original
system~\cite{Yu-PRL-06}. 
We expect the final interaction 
matrix of the copy system %
to converge to that of the original system. Unfortunately, sufficient
conditions for the convergence of this approach have not been fully
understood and the %
approach is model dependent. Knowing the details of the coupling
functions $f_{ij}(x_i, x_j)$ is crucial to set up the copy
system. Furthermore, $f_{ij}(x_i, x_j)$ needs to be Lipschitz
continuous. %
These constraints significantly narrow the applicability of this
approach.

\emph{Direct approach}: This approach relies on the evaluation of
temporal derivatives from time series data~\cite{Shandilya-NJP-11}. 
Exploiting smoothness assumptions, it %
finds the unknown
interaction matrix by solving %
an optimization problem
(e.g., $\ell_1$ or $\ell_2$-norm minimization). 
The rationale %
is as follows. If the time derivatives of the state
variables are evaluated,  and if the system coupling functions are also known, then the only remaining unknown
parameters %
are the edge weights or interaction strengths $a_{ij}$'s. Repeated evaluations of (\ref{eq:system}) at  
different sufficiently closely spaced times $t_m \in \mathbb{R}$
comprise a simple and implicit restriction on the interaction matrix $A$. %
This approach serves as a simple starting strategy of NR. 
Yet, it has an fundamental drawback --- there is no reason why the
true interaction matrix should be optimal in some %
a priori metric. 
Moreover, it may suffer from the poor evaluation of time derivatives of
noisy time series data.

All three approaches suffer from %
one common issue: %
The necessary and sufficient conditions under which they succeed are unknown. %
An important exception is the \emph{Modular Response Analysis} method
\cite{kholodenko2002untangling,sontag2008network}, which is a special
driving-response approach, and %
guarantees to
recover the interaction matrix using steady-state
data collected from sufficiently many  perturbation experiments. One
drawback of this method is that it assumes the system is not \emph{retroactive}
\cite{sontag2002differential,sontag2011modularity}.
Here, retroactivity  manifests as ``load'' or ``impedance'' effects that might be hard to anticipate if we have no a-priori knowledge of the system dynamics.%

Recently, two classes of fundamental limitations of NR were characterized by deriving
necessary (and in some cases sufficient) conditions to reconstruct any desired property of
the interaction matrix~\citep{Marco-NR-15}. %
The first class of fundamental limitations is due to our uncertainty
about the coupling functions $f_{ij}(x_i, x_j)$, leading to a natural
trade-off: the more information we want to reconstruct about the
interaction matrix the more certain we need to be about the coupling
functions. 
 For example, %
it is possible to reconstruct the
 adjacency pattern $\mathbf{K}$ without knowing exactly 
 the coupling functions. But, in order to reconstruct the interaction matrix $\mathbf{A}$ itself, it is
 necessary to know these functions exactly. 
In this sense, if we are uncertain about the coupling
functions, NR is easier than PI.
The second class of fundamental limitations originates solely from
uninformative temporal data, i.e. $\{ x_i(t), u_i(t)\}_{i=1}^N$,
 $\forall t \in [t_0, t_1]$. This leads to a rather counterintuitive
 result: regardless of how much information we aim to reconstruct
 (e.g. edge weights, sign pattern or connectivity pattern), the
 measured temporal data needs to be equally informative. This happens even if
 we know the coupling functions exactly. %
Hence, in the sense of informativeness of the measured data, reconstructing any property of the interaction matrix
is as difficult as reconstructing the interaction matrix itself,
i.e. NR is as difficult as PI. 
A practical solution to circumvent this limitation without acquiring
more temporal data (i.e. performing more experiments, which are
sometime either infeasible to too expensive), prior knowledge of the
interaction matrix, e.g. the bounds of the edge weights, is extremely
useful~\citep{Marco-NR-15}. 

}

\section{Towards Desired Final States or Trajectories} %

A significant
body of work %
in %
control theory deals with the design of 
control inputs that can
move the system from a given initial state to a %
desired final state  %
in the state space
~\citep{Sontag-Book-1998}.  
For linear dynamics, %
Equation (\ref{eq:optimal_u}) 
provides the optimal input
signal to take an arbitrary linear system %
into an arbitrary final state using the minimum %
  control energy $\int_0^T \|{\bf u}(t)\|^2 \ud t$.
For nonlinear dynamics we lack %
a ready-to-use solution, and
finding one %
can be very difficult. %
Yet, solving such nonlinear control %
problems has %
important %
applications %
from robotics to ecosystem management, and from cell
reprogramming to drug discovery. %
For example, 
in robotics engineers frequently encounter the so-called 
motion- or path-planning problem,
needing to %
decompose a %
 desired
movement %
into discrete motions that satisfy specific movement constraints
and possibly optimize some aspect of the trajectory. %
The parallel parking problem is a typical example, 
requiring us 
to determine the sequence of motions %
a car must follow %
in order to parallel park into a parking space. 

In many cases, we are interested in steering the system towards a desired
trajectory or attractor, instead of %
a desired final state.  
A trajectory or an orbit of a dynamical system is a collection of points
(states) in the state space. 
For example, a \emph{periodic orbit} %
repeats itself in time with period $T$, so that $\mx(t) = \mx(t+nT)$
for any integer $n\ge 1$. %
Roughly speaking, an attractor %
is a closed subset $\mathcal{A}$ of a dynamical
system's %
state space such that for ``many'' choices of initial states the system will
evolve towards states in
$\mathcal{A}$~\citep{Milnor-Scholarpedia-06}. 
Simple attractors 
correspond to fundamental geometric objects, like %
points, lines, surfaces, spheres, toroids, manifolds, 
or their simple combinations. %
Fixed (or equilibrium) point and limit cycle are common simple
attractors. 
Fixed points are defined for mappings
  $x_{n+1}=f(x_n)$, where $x$ is a fixed point if $x=f(x)$, whereas
    equilibrium points or equilibria are defined for flows (ODEs)
    $\dot{\mx}=\mf(\mx)$, where $\mx$ is an equilibrium point if
    $\mf(\mx)=0$.  
A limit cycle is a periodic orbit of the dynamic system that is
isolated.  
{\color{red} 
 An attractor is called \emph{strange} if it has a fractal structure
 that cannot be easily described as fundamental geometric objects or
 their simple combinations. A strange attractor often emerges in
 chaotic dynamics. 
}

In this section we briefly %
review progress made in several %
directions with the common goal of controlling some dynamical systems:  
(a) Control of chaos, which requires us to transform a chaotic motion
into a periodic trajectory using open-loop control~\citep{Hubler-HPA-88}, Poincar\'{e} map linearization~\citep{Ott-PRL-90} or time-delayed
feedback~\citep{Pyragas-PLA-1992}. 
(b) Systematic design of %
compensatory perturbations of state variables that take advantage of
the full basin of attraction of the desired final state~\citep{Cornelius-NC-2013}.  
(c) %
Construction of  the attractor network~\citep{Lai-NSR-2014,Wang-IF-2014}; %
(d) Mapping the control problem into a %
combinatorial optimization problem on the underlying networks~\citep{Fiedler-JDDE-2013,Mochizuki-JTB-2013}. %
\subsection{Controlling Chaos}

A deterministic dynamical system is said to be \emph{chaotic}  if its
evolution is highly %
sensitive to its initial
conditions. This sensitivity means that %
arbitrary small measurement errors in %
the initial conditions grow
exponentially with time, %
destroying the long-term predictability of the system's future
state. %
This phenomenon, known as %
the \emph{butterfly effect}, is often considered %
troublesome%
~\citep{Lorenz-JAS-63}.  %
Chaotic behavior %
commonly emerges in natural and engineered systems, being encountered %
in %
chemistry, %
nonlinear optics, electronics, fluid dynamics, meteorology,
and biology~\citep{Strogatz-Book-94}.   
It has been realized that well-designed control laws can %
overcome the butterfly effect, 
forcing chaotic systems 
to follow some desired
behavior~\citep{Hubler-HPA-88,Ott-PRL-90,Pyragas-PLA-1992,Zoltan-94,Sass-JPA-96}. %
Next, we %
review several %
key methods devised for the control of chaotic systems from the
control theoretical
perspective~\citep{Boccaletti-PR-00,Fradkov-ARC-05,Chen-Book-1998}. 
\subsubsection{Open-loop Control}\label{sec:olc}

Since the late 1980s, a series of methods have emerged to manipulate 
chaotic systems towards %
a desired ``goal dynamics'' $\mg(t)$
~\citep{Hubler-HPA-88}. 
Consider a controlled system %
\be
\dot{\mx} = \mF(\mx) + \mB {\bf u}(t)  \label{eq:Hubler_0}
\ee
where $\mx \in \mathbb{R}^N$ is the state vector, ${\bf u}(t) \in
\mathbb{R}^M$ is the control input.  
In contrast with the network-based problems discussed earlier, here we 
assume that all state variables are controlled ($M=N$) and $\det \mB \neq 0$. 
The goal is to design ${\bf
u}(t)$ so that $\mx(t)$ converges to a desired trajectory $\mg(t)$,
i.e., $|\mx(t) - \mg(t)| \to 0$ as $t\to \infty$. 
We can use %
open-loop control for this purpose, using the control input %
called the \emph{Hubler action},       
\be
{\bf u}(t) = \mB^{-1} \left[ \dot{\mg}(t) - \mF (\mg(t)) \right], \label{eq:Hubler}
\ee
which ensures that $\mx(t) = \mg(t)$ is a solution of the controlled
system. 
In this case, the error $\me(t) = \mx(t) - \mg(t)$ satisfies 
\be
\dot{\me}(t) = \mF(\me(t) + \mg(t)) - \mF(\mg(t)),
\ee
which can be linearized as 
$
\dot{\me}(t) = \mA(t) \me(t)
$, 
where $\mA(t) = \frac{\p \mF(\mx)}{\p \mx} |_{\mx=\mg(t)}$. 
If the linearized system is \emph{uniformly asymptotically stable}, i.e., 
its equilibrium point $\me^*={\bf 0}$ is stable for all
  $t>0$, %
then the error $\me(t)$ converges to zero, and %
$\mx(t)$ converges to the desired trajectory $\mg(t)$. 
We call the regions of the state space from which %
the controlled orbits %
converge to the goal trajectory $\mg(t)$ %
\emph{entrainment regions}.

Note that the method (\ref{eq:Hubler_0})-(\ref{eq:Hubler}) is not tailored to %
chaotic systems, but %
potentially works for any %
nonlinear system. %
It has several disadvantages, though: (i) 
the open-loop control
(\ref{eq:Hubler}) requires a priori knowledge of the dynamics,
which is often not precisely known for complex systems;
(ii) the applied controls are not always small, %
requiring high control energy; (iii) the convergence of
$|\mx(t) - \mg(t)| \to 0$ for $t\to \infty$ depends on the detailed
functional form of $\mF(\mx)$ and the initial condition $\mx(0)$, 
hence this method is not guranteed to work for arbitrary systems. 
\subsubsection{Linearization of the Poincar\'{e} map: OGY method}%

\begin{figure}[t!]
\centering
\includegraphics[width=0.4\textwidth]{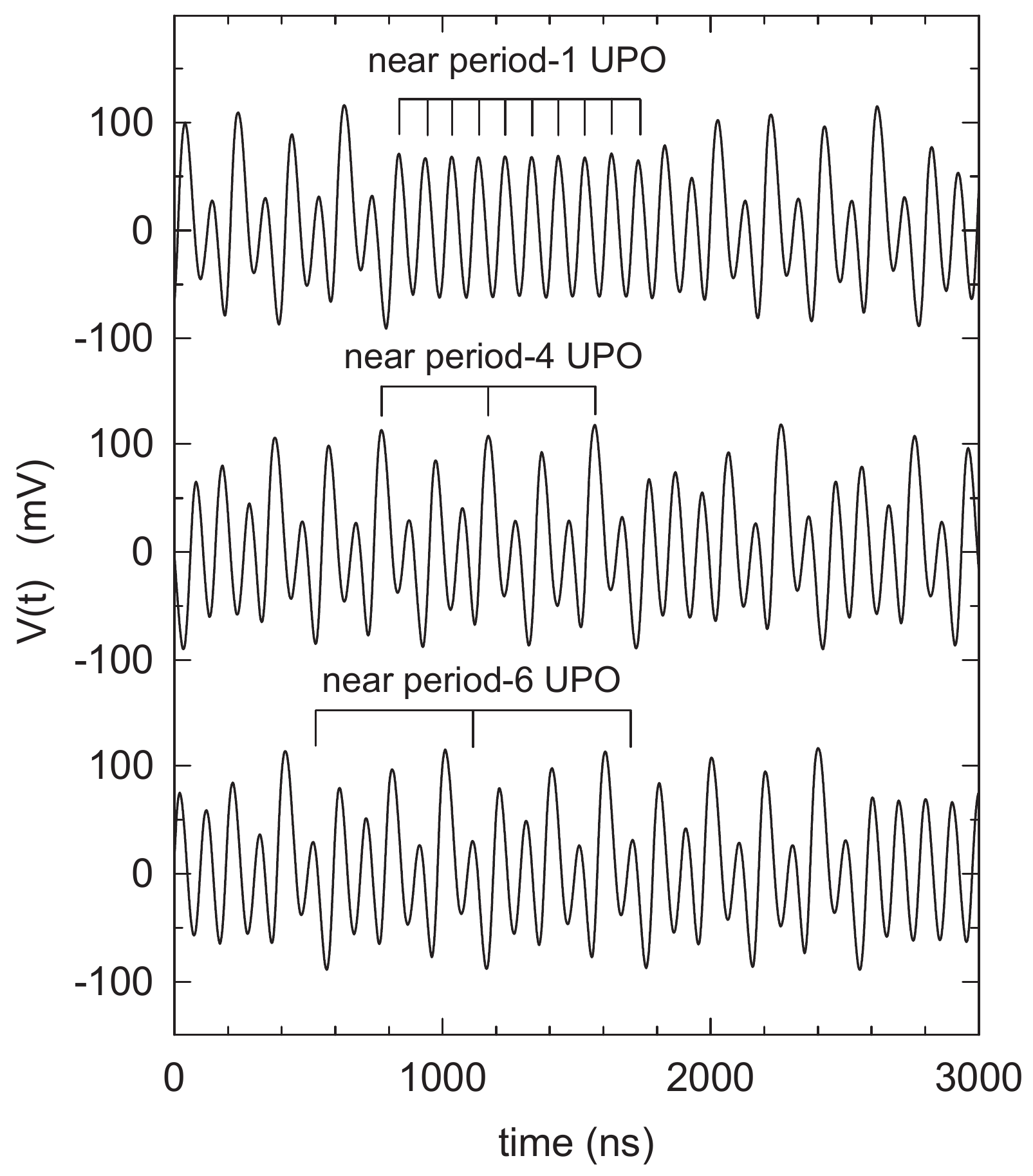}
\caption{%
Chaotic behavior in a nonlinear electronic circuit. The
vertical axis measures the voltage drop $V(t)$ across a 50$\Omega$
resistor, being %
proportional to the current in the circuit. 
The system ergodically visits the unstable periodic orbits (UPOs)
embedded in the chaotic attractor. The plot shows three such %
UPOs. 
After \citep{Sukow-Chaos-97}. 
} 
\label{fig:UPO}
\end{figure}

The OGY method proposed by Ott, Grebogi and Yorke
~\citep{Ott-PRL-90} exploits 
the %
observation that %
typically an infinite number of unstable
periodic orbits (UPOs) are embedded in a chaotic attractor
(Fig.~\ref{fig:UPO}).  
Therefore we %
can obtain %
a desired periodic motion by making only small perturbations
to an accessible system parameter.   

The OGY method can be summarized as follows: First, we determine and
examine some of the low-period UPOs embedded in the chaotic
attractor. Second, we choose a desired UPO.   
Finally, we design %
small time-dependent
parameter perturbations %
to stabilize this %
pre-existing UPO.

This method is not only %
very general and practical, but also 
suggests that in some systems the presence of chaotic behavior can be an %
advantage for control. 
Indeed, if the attractor of a system is not chaotic but has a stable periodic orbit
(SPO), then small parameter perturbations can only slightly change the existing
orbit. 
Therefore, given %
that any one of the infinite number of UPOs
can be stabilized, we can always choose the UPO that achieves the
best system performance. 
Hence, chaotic behavior offers us a diverse and rich %
landscape for the desired dynamic
behavior of the system. 

\begin{figure}[t!]
\centering
\includegraphics[width=0.3\textwidth]{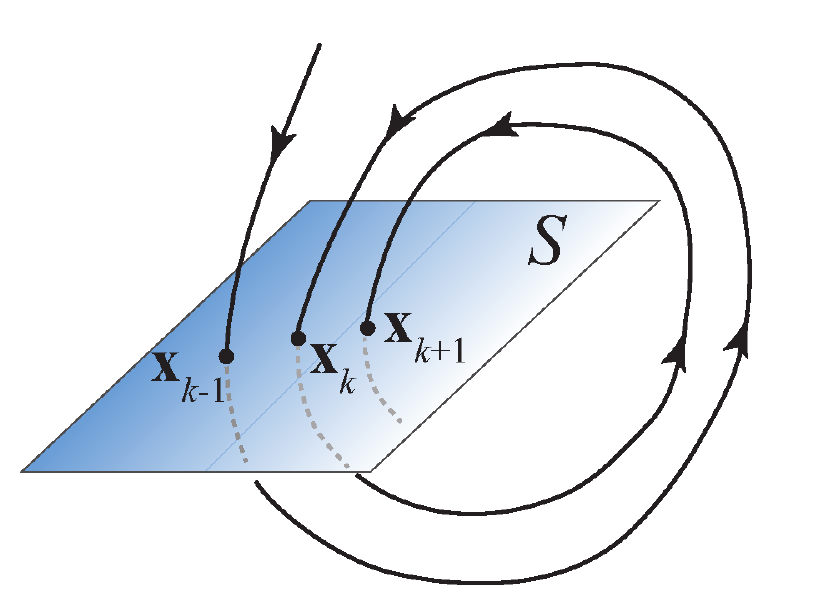}
\caption{(Color online)  Poincar\'{e} map. 
In a continuous dynamical system the Poincar\'{e} map is the intersection
of a periodic orbit in the state space with a certain
lower-dimensional subspace, called the Poincar\'{e} section $S$, 
transversal to the flow of the system. 
In the Poincar\'{e} section $S$, the Poincar\'{e} map $\mx \mapsto {\bf F}
(\mx, u)$ projects point $\mx$ onto point ${\bf F}(\mx, u)$, i.e.,
$\mx_{k} = {\bf F}(\mx_{k-1}, u_{k-1})$, $\mx_{k+1} = {\bf F}(\mx_k,
u_k), \cdots$.
} 
\label{fig:poincaremap}
\end{figure}

To demonstrate this method, let us consider a nonlinear
continuous-time dynamical system  
\be
\dot{\mx} = \mf (\mx, u) \label{eq:chaoscontrol_ODE}
\ee
where $\mx \in \mathbb{R}^N$ is the state vector and $u \in
\mathbb{R}$ represents a tunable parameter, which can be considered as 
a control input. %
Our task is to reach a desired trajectory $\mx^*(t)$ %
that satisfies (\ref{eq:chaoscontrol_ODE}) with $u=0$. 
To achieve that we first %
construct a surface $S$, called a Poincar\'{e} section, 
which 
passes through the %
point $\mx_0 = \mx^*(0)$ transversally to the trajectory $\mx^*(t)$ (see
Fig.~\ref{fig:poincaremap}). 
Consider a map $\mx \mapsto {\bf F} (\mx, u)$, where ${\bf F}(\mx, u)$
is the point of first return to the Poincar\'{e} section of the solution of
(\ref{eq:chaoscontrol_ODE}) that begins at the point $\mx$ and was
obtained for the constant input $u$. 
Since we can integrate (\ref{eq:chaoscontrol_ODE}) forward in time
from $\mx$, the map $\mx \mapsto {\bf F} (\mx, u)$, %
called the \emph{Poincar\'{e} map}, must exist.   
Note that even though we may not be able to write down the map ${\bf F}$
explicitly, the knowledge that it exists is still useful~\citep{Shinbrot-Nature-1993}. 
By considering a sequence of such maps, we get a discrete system 
\be
\mx_{k+1} = {\bf F}(\mx_k, u_k), \label{eq:chaoscontrol_ODE2} 
\ee
where $\mx_k = \mx(t_k)$, $t_k$ is the time %
of the $k$-th intersection of the Poincar\'{e} section $S$, and $u_k$ is the value 
of control $u(t)$ over the interval between $t_k$ and $t_{k+1}$.  

A key step in the OGY method is to %
linearize the discrete system (\ref{eq:chaoscontrol_ODE2}) as 
\be
\mz_{k+1} = \mA \mz_k + \mB u_k, \label{eq:chaoscontrol_ODE3} 
\ee
where $\mz_k = \mx_k - \mx_0$, $\mA = \left.\frac{\p \mF}{\p
    \mx}\right|_{\mx_0}$ is the Jacobian matrix, and $\mB =
\left.\frac{\p \mF}{\p u}\right|_{\mx_0}$ is a column vector.

To stabilize the linear system (\ref{eq:chaoscontrol_ODE3}) and hence steer
the original system to a desired periodic orbit that passes through $\mx_0$,
the OGY method employs a linear state feedback control law 
\be
u_k = 
\begin{cases}
\mC \mz_k & \mm{if} \,\, | \mz_k | \le \delta \\
0 & \mm{otherwise}
\end{cases}, \label{eq:chaoscontrol_OGY}
\ee
where $\delta > 0$ is a sufficiently small parameter. 
Note that the control is only applied in some
neighborhood of the desired trajectory, which ensures the 
smallness of the
control action. 
This piecewise-constant small action control is a key feature %
of the OGY method.   
To guarantee the efficiency of the method, the matrix $\mC$ must be %
chosen so that in the linear closed-loop 
system %
$\mz_{k+1} = (\mA + \mB \mC)\, \mz_k$, 
the %
norm  
$| (\mA + \mB \mC) \mz | \le \rho | \mz|$
decreases, where $\rho < 1$. 
Extensive numerical simulations %
have corroborated the practical utility of the OGY method. 
Furthermore, %
the OGY method was proven to be effective in experimental
systems as well, allowing %
the stabilization of unstable periodic orbits %
in a chaotically oscillating magnetoelastic ribbon, a driven diode
circuit, a multimode laser with an intracavity crystal, a thermal
convection loop, and the Belousov-Zhabotinsky reaction~\citep{Boccaletti-PR-00}. %
Slow convergence was often reported, a price we must pay 
to achieve global stabilization of a nonlinear system with small
control action~\citep{Fradkov-ARC-05}. 

The advantage of the OGY method is that it does not require %
prior knowledge of the system's dynamics. %
Instead, we just rely on %
the system's behavior %
to learn the necessary %
small perturbation %
to nudge %
it towards %
a desired trajectory.   
This is similar to the %
balancing of a stick on our palm, %
which can be achieved without knowing %
Newton's {\color{red}second} law of motion and the stick's detailed equation of motion. 
Indeed, in the OGY method, both $\mA$ and $\mB$ in
(\ref{eq:chaoscontrol_ODE3}) can be extracted purely from observations
of the trajectory on the chaotic attractor~\citep{Shinbrot-Nature-1993}.
Finally, 
the OGY method can be extended to %
arbitrarily high dimensional systems, without assuming %
knowledge of
the underlying dynamics~\citep{Auerbach-PRL-92}. 

\begin{figure}[t!]
\centering
\includegraphics[width=0.5\textwidth]{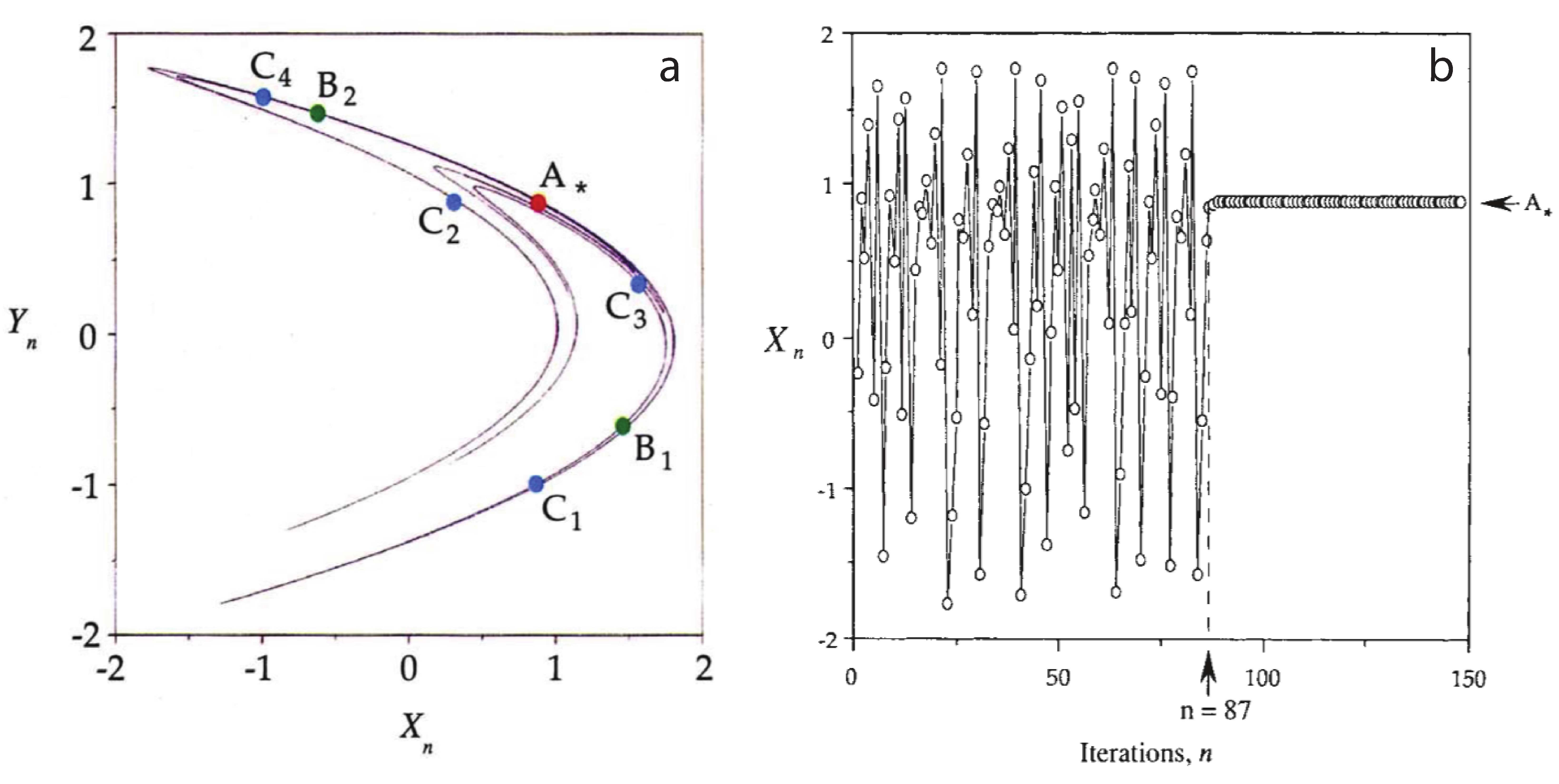}
\caption{(Color online)  Controlling chaos. The image shows the use of the
  Ott-Grebogi-Yorke (OGY) method to control the chaotic behavior in
  the H\'{e}non map $X_{n+1} = p + 0.3 Y_n - X_n^2, Y_{n+1}=X_n$ where the parameter $p$ is set to 
$p_0=1.4$. 
(a) The H\'{e}non attractor contains period-1 point $A^*$, which is
revisited in each %
map iteration, period-2
points $B_1$ and $B_2$, which are revisited every other map iteration,
i.e., $B_1 \to B_2 \to B_1 \to B_2 \to \cdots$, and period-4 points $C_1, C_2, C_3$ and
$C_4$, which are cycled through every four map iterations. 
(b) The result of stabilizing the periodic orbit $A^*$ of the
H\'{e}non attractor by tuning $p$ by less than 1\% around $p_0$. %
The arrow indicates the time step at which the small
perturbation is initiated. 
For the first 86 iterations, the trajectory moves chaotically
on the attractor, never falling within the desired small region about
$A^*$. %
On the 87th iteration, following the application of the control
perturbation, the state falls within the desired
region, and %
is held near $A^*$. 
After \citep{Shinbrot-Nature-1993}.
} 
\label{fig:Henon}
\end{figure}

\subsubsection{Time-delayed feedback: Pyragas method} %

The Pyragas method %
employs continuous
feedback %
to synchronize the current state of a system with a
time-delayed version of itself, offering an alternative approach to stabilizing a desired UPO embedded in a chaotic
attractor~\citep{Pyragas-PLA-1992}. 
Consider the nonlinear system (\ref{eq:chaoscontrol_ODE}). 
If %
it has a desired UPO $\Gamma = \{ \mx^*(t) \}$ with
period $T$ for ${\bf u} = {\bf 0}$, then we can use the %
feedback control  
\be
{\bf u}(t) = K \left[ \mx(t) - \mx(t-\tau) \right], \label{eq:Pyragas-u}
\ee
where $K$ is the feedback gain and $\tau$ is the delay time, to stabilize the desired UPO. 
If $\tau=T$ and the solution $\mx(t)$ of the
closed-loop system (\ref{eq:chaoscontrol_ODE}, \ref{eq:Pyragas-u}) begins    
on the UPO, then it remains on the UPO for all $t \ge 0$. 
Surprisingly, $\mx(t)$ can converge to the UPO even if initially is 
not on the UPO, i.e., $\mx(0) \notin
\Gamma$. 
Considering that not all the state variables are experimentally
accessible, we can rewrite (\ref{eq:Pyragas-u}) as 
\be
u(t) = K \left[ y(t) - y(t-T) \right] \label{eq:Pyragas-u2}
\ee
for a desired UPO of period $T$. Here $y(t)=\mh(x(t)) \in \mathbb{R}$
is an experimentally accessible output signal.
The %
advantage of the time-delayed feedback control law
(\ref{eq:Pyragas-u2}) is that it does not require rapid switching or
sampling, nor does it require a reference signal corresponding to the
desired UPO. 
Unfortunately, the domain of system parameters over which control can
be achieved via (\ref{eq:Pyragas-u2}) is limited. Furthermore, the method fails
for highly unstable orbits.   
Note, however, that an extended variant of the Pyragas method, %
using a %
control law whose form is %
closely related to the amplitude of light reflected from a Fabry-P\'{e}rot
interferometer 
can stabilize
highly unstable orbits~\citep{Socolar-PRE-94}. %

Despite the simple form %
(\ref{eq:Pyragas-u},\ref{eq:Pyragas-u2}) of the %
control signal, the analytical study of the closed-loop system is %
challenging. 
Indeed, while there are extensive numerical and experimental results pertaining to the
properties and application of the Pyragas method, %
the sufficient conditions that guarantee its 
applicability %
remain unknown~\citep{Fradkov-ARC-05}.  

{\color{red}
Note that similar to the Pyragas method, a geometric method of
stabilizing UPOs~\citep{Zoltan-94,Sass-JPA-96}
also uses time-delays. This method is based on some observations about
the geometry of the linearized dynamics around these orbits in the phase
space. It does not require explicit knowledge of the dynamics (which
is similar to the OGY method), but only %
experimentally accessible state information within a short period of the system's
immediate past. More specifially, it %
requires a rough location of
the UPO and a single parameter easily computed from four data
points. This geometric method %
does not have the problems of the Pyragas method in stabilizing UPOs. 
The drawback of this geometric method is that it has only been
formulated for 2D maps and 3D flows. 
}
\subsection{Compensatory Perturbations of State Variables}\label{sec:sean}

The control tools described above were mainly designed for low dimensional
dynamical systems with a simple structure. Most complex systems are
high-dimensional, however, consisting of 
a network of components %
connected by
nonlinear interactions. We need, therefore, tools 
to 
bring a %
networked system to a desired target
state. %
A recently proposed method %
can work even
when the target state is not directly accessible due to certain
constraints \citep{Cornelius-NC-2013}. 
The basic insight of the approach is 
that each desirable state has a ``basin of
attraction'', representing a region of initial conditions
whose trajectories converge to it. %
For a system that is in %
an undesired
state, we need to identify %
perturbations to the
state variables that can 
bring the system to the basin of attraction of the desired
target state. Once there, the system will evolve spontaneously to the
target state.  
Assume that a %
physically admissible perturbation fulfills
some constraints that can be represented  by vector expressions of the form 
\begin{equation} \label{constraints}
\mathbf{g}(\x_0, \x_0') \leq {\bf 0} \,  \mbox{ and } \, \mathbf{h}(\x_0, \x_0') = {\bf 0},
\end{equation}
where the equality and inequality %
apply to each component individually. %
To iteratively identify compensatory perturbations we use the
following procedure: %
Given the current initial state of the network, $\x_0'$, we integrate the
system's dynamics over a time window $t_0 \leq t \leq t_0 + T$  to
identify the time when the orbit is closest to the target, $t_\mathrm{c} \equiv
\mm{arg} \min \lvert \target - \x(t) \rvert$. 
We then integrate the variational equation up to $t_\mathrm{c}$ to
obtain the corresponding variational matrix, ${\bf M}(t_\mathrm{c})$, which maps 
a small change $\dx_0$ in the initial state of the network to a change
$\delta {\bf x}(t_\mathrm{c})$ in the resulting perturbed orbit at $t_\mathrm{c}$ 
according to $\delta {\bf x}(t_\mathrm{c}) = {\bf M}(t_\mathrm{c}) \cdot \dx_0$. This
mapping is used to select an incremental perturbation $\dx_0$ %
that minimizes the distance between the
perturbed orbit and the target at time $t_\mathrm{c}$, subject to the
constraints (\ref{constraints}) %
and additional constraints on $\dx_0$ to ensure the validity of the
variational approximation.

\begin{figure}[t!]
\includegraphics[width=0.45\textwidth]{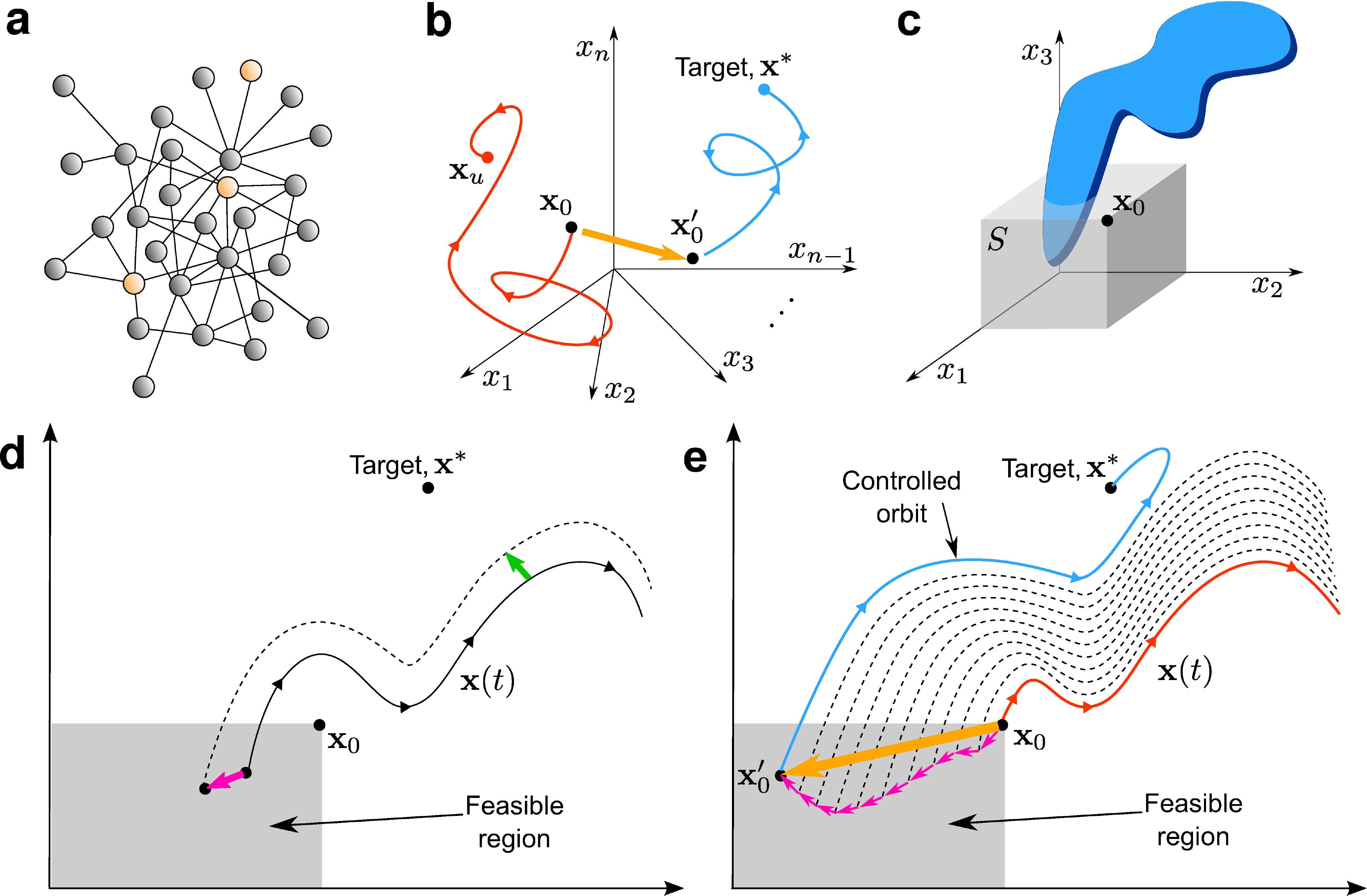}
\caption{(Color online)  Steering a network %
using compensatory perturbations of state variables. 
  ({\bf a}) %
The {\it control set} (shown in
  yellow) is a set of nodes that are accessible to compensatory
  perturbations. 
({\bf b}) In the absence of
  control, the network %
is in %
an initial state $\mx_0$ and evolves to an
  undesirable equilibrium $\mx_\mathrm{u}$ (red curve). By perturbing
  the initial state (orange arrow), the network reaches a new state
  $\mx_0'$, which %
evolves to the desired
  \emph{target} state (blue curve). 
({\bf c}) Typically, the compensatory
  perturbations must obey %
some constraints. In this example, we can only
  perturb three out of $N$ dimensions, %
  corresponding to the three-node control set (shown in yellow), and the state variable
  of each control node can only be reduced. 
These constraints form a cube (grey volume) within the
  three-dimensional subspace of the control nodes.
 The network can be steered to the target state if and only if the
 corresponding slice of the target's basin of attraction (blue volume)
 intersects this cube. 
({\bf d}) %
Along each orbit there is a point %
that is closest to the target
state. We seek to identify a
  perturbation (magenta arrow) to the initial condition that brings the
  closest point closer to the target (green arrow).
 ({\bf e})
  This process is repeated 
(dashed curves), until we identify %
a perturbed state
  $\mx_0'$ that is in the attraction basin of the target state, hence
  the system %
automatically evolves to the target state.  
 This results in a compensatory perturbation $\mx_0 \rightarrow \mx_0'$
 (orange arrow). 
After \citep{Cornelius-NC-2013}.
}
\label{Fig:compensatory_perturbation} 
\end{figure}

The selection of $\dx_0$ is performed via a nonlinear optimization 
that can be efficiently solved using %
sequential quadratic programming. %
The initial condition is then updated %
$\x_0' \rightarrow \x_0'+\delta \x_0$, and we test whether the new
initial state lies in the target's basin of attraction by integrating
the system dynamics over a long time 
$\tau$. If the system's orbit reaches a small ball of radius $\kappa$
around $\target$ within this time, we declare success  and recognize 
$\x_0 - \x_0'$ as a compensatory perturbation (for the updated
$\x_0'$). If not, we calculate the time of closest approach of the
new orbit and repeat the procedure. 
Similar to the open-loop control of chaos discussed in
Sec.~\ref{sec:olc}, the approach based on compensatory perturbation 
potentially
works for any %
nonlinear system. %
It has been successfully
applied to the mitigation of cascading failures in a power grid and
the identification of %
drug targets in a cancer signaling
network%
~\citep{Cornelius-NC-2013}. 
Yet, the %
approach requires a priori knowledge of the detailed model describing
the dynamics of the system we wish to control, %
a piece of knowledge we often lack in 
complex systems. 
With an imperfect model, a compensatory perturbation may steer
the %
system into a different basin of attraction than the desired
one. 
Studying the dependence of the success rate of this approach
on the parameter uncertainty and system noise remains an analytically
challenging issue. 
Moreover, it is unclear how to choose the optimal control set consisting of one or
more nodes accessible to compensatory perturbations so that some
control objectives, like the number of control nodes or the amount of
control energy, are minimized.  

\subsection{Small Perturbations to System Parameters}\label{sec:lai}
The control tool described %
above perturbs the state variables of a
networked system. %
In analogy with the %
OGY method~\citep{Ott-PRL-90}, we can also control a networked system 
via small perturbations to its %
parameters.   
Note that %
networked systems are typically high-dimensional, to which the %
control methodologies developed for chaotic system %
do not immediately apply. 
Yet, %
we can control complex networked systems
via perturbations to the \emph{system parameters}~\citep{Lai-NSR-2014}, %
an approach %
complementary to the %
approaches based on perturbations of the \emph{state variables}. %
The key step of this approach is to choose a set of experimentally
adjustable parameters and determine whether small perturbations to
these parameters can steer the system towards the desired attractor~\citep{Lai-PLA-96,Lai-NSR-2014}. 
Depending on the physical constraints the
control parameters %
obey, the directed control path from the undesired
attractor to the desired attractor can either be via a direct
connection 
or %
via intermediate attractors along the control %
path. 
If there are no feasible control paths reaching the desired attractor,
then we can not steer the system to that attractor, hence control is
not possible.

Considering each attractor as a node, and
the control paths as directed %
edges between them, %
we can construct an
``attractor network'', whose properties determine the controllability 
of the original dynamic network~\citep{Lai-NSR-2014}. %
For a given nonlinear system, the attractor network can be
constructed as follows. First, we identify all possible attractors 
and choose a set of system parameters that can be
experimentally perturbed. Second, we set the system into a specific 
attractor $a$, and determine the set of attractors into which the system can
evolve from the original attractor $a$ with a reasonable combination
of the adjustable parameters. This effectively draws a link from attractor
$a$ to all other %
attractors reachable %
by feasible parameter perturbations. Finally, 
we repeat this procedure for all attractors, obtaining the 
attractor network%
~\citep{Lai-NSR-2014}.

\begin{figure}[t!]
\includegraphics[width=0.5\textwidth]{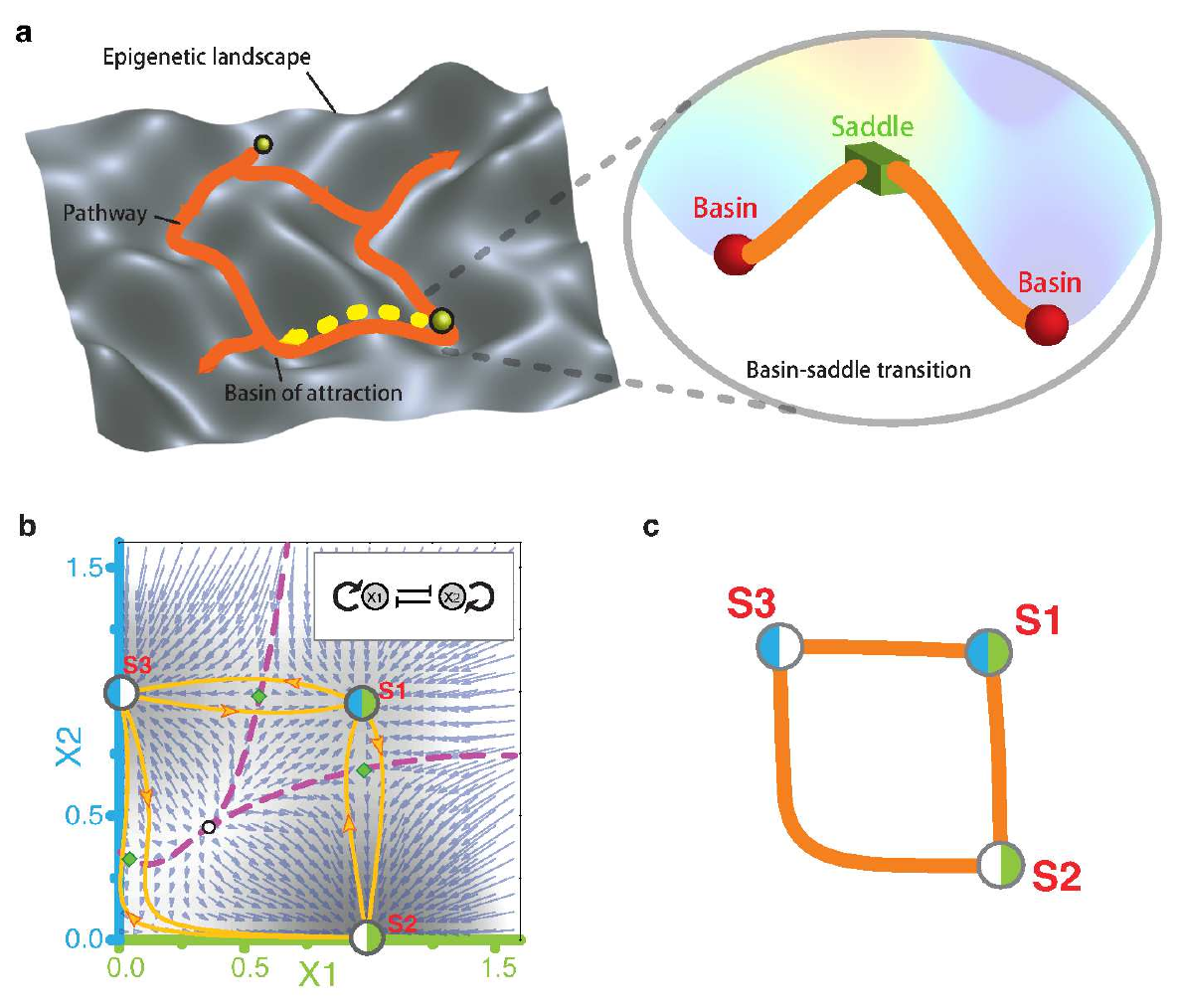}
\caption{(Color online)  Epigenetic state network (ESN). (a) On the epigenetic landscape, a
  minimal energy path connects two neighboring attractors through an unstable transition point
  (first-order saddle point). 
  The landscape can be represented by a network, where nodes are attractors or
  basins of attraction and edges are minimal energy paths connecting the neighboring
  attractors. 
(b) The vector field of a mutually
  inhibitive two-gene circuit (inset). Nodes S1, S2 and S3
  are fixed-point attractors. 
The pie diagram of each attractor
  represents the %
expression pattern of the two genes. 
The first-order saddle points (green diamond) are surrounded by forward
  and backward optimal paths (dark blue) connecting two neighboring
  attractors. 
(c) The ESN constructed from (a) by connecting neighboring
  attractors. 
After \citep{Wang-IF-2014}.
}
\label{fig:ESN} 
\end{figure}

To illustrate the construction of such an attractor network, consider
the epigenetic state network (ESN) that describes the phenotypic
transitions on the epigenetic landscape of a cell
(Fig.~\ref{fig:ESN}). 
In the epigenetic landscape, two neighboring fixed-point attractors,
corresponding to stable cell phenotypes, are connected by a minimal
energy path through an unstable transition point (first-order saddle
point)~\citep{Wang-PNAS-11,Wang-IF-2014}. 
The number of fixed points (nodes) and
 saddle points (edges) grows exponentially with the number of genes
 (dimensionality). 
We can
 rely on a conditional root-finding algorithm%
~\citep{Wang-IF-2014} to construct this epigenetic state network (ESN). 
The obtained ESN captures the global
architecture of stable cell phenotypes, helping us translate the metaphorical
Waddington epigenetic landscape
concept~\citep{Waddington-Book-57,Slack-NRG-2002} into a mathematical
framework of cell phenotypic transitions. 

\subsection{Dynamics and Control at Feedback Vertex Sets}\label{sec:fvs}
\begin{figure}[t!]
\includegraphics[width=0.35\textwidth]{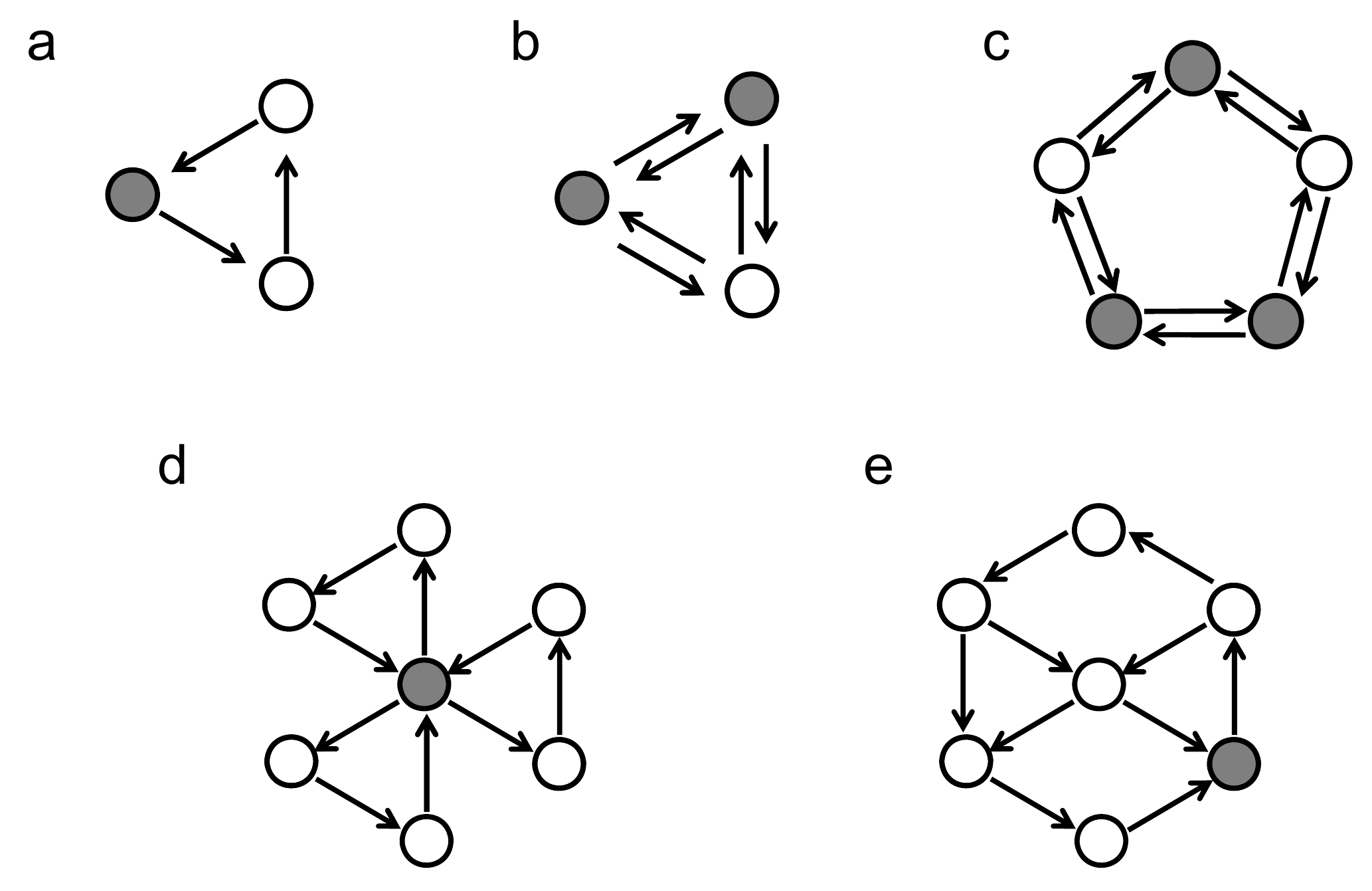}
\caption{Feedback vertex set (FVS). This figure show examples of FVSs in directed graphs, whose
  removal render the graphs acyclic. The
  gray vertices represent a %
choice of a minimal FVS in each panel 
(a)-(e). 
 Controlling the dynamics of the nodes in an FVS %
allows us to switch the dynamics of the whole system from one attractor to
 some other attractor. 
After \citep{Mochizuki-JTB-2013}.
}
\label{fig:FVS} 
\end{figure}

For %
regulatory networks described as %
a digraph of %
dependencies, it has been recently shown that open-loop control
applied to a %
feedback vertex set (FVS) will force the remaining network to stably
follow the desired trajectories~\citep{Fiedler-JDDE-2013,Mochizuki-JTB-2013}.   
An FVS is a subset of nodes in {\color{red}the absence of which} %
the digraph %
becomes acyclic, i.e., contains no directed cycles (Fig.~\ref{fig:FVS}). 
Unlike the %
approaches discussed in Sec.~\ref{sec:sean} and
Sec.~\ref{sec:lai}, this approach has rigorous analytical underpinnings.  

Consider a general non-autonomous nonlinear networked system
\be
\dot{x}_i = F_i(t, x_i, x_{\mathcal{I}_i}) \label{eq:ODE_FVS}
\ee
where $i=1, \cdots, N$, and $\mathcal{I}_i$ denotes the set of upstream
or incoming neighbors of node $i$, i.e., $j\in \mathcal{I}_i$ if there is directed
edge $(j \to i)$ %
in the network. %
The corresponding network is often called the \emph{system digraph}, which is a
transpose of the \emph{inference diagram} introduced in %
observability %
(see Sec.~\ref{sec:msp}).

\begin{figure*}[ht!]
\includegraphics[width=\textwidth]{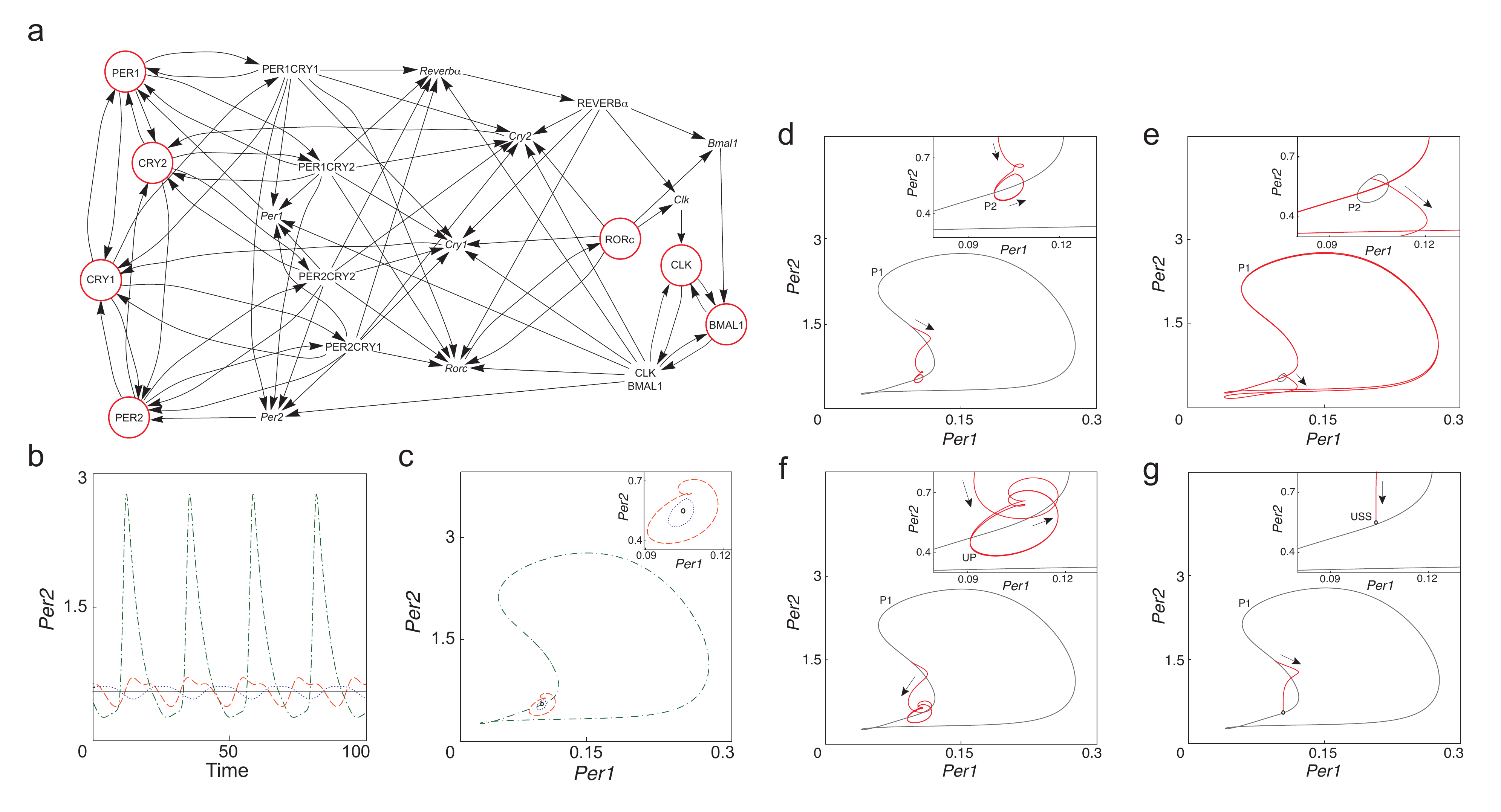}
\caption{(Color online)  Controlling a system through its feedback vertex set (FVS).  
(a) A regulatory
network with 21 variables describes the mammalian circadian rhythms in mice~\citep{Mirsky-PNAS-09}. A minimal FVS of
seven elements, denoted as $\mathcal{I}$, is
highlighted %
by red circles. (b) Trajectories of two stable periodic orbits,
period1 (P1, dotted
and broken curve) and period2 (P2, dotted curve), one unstable
periodic orbit (UP, broken curve) and
one unstable stationary state (USS, solid line), represented by time tracks of the
variable Per2. (c) Trajectories of the same solutions in the phase plane of the
two variables Per1 and Per2, which are not in the FVS. 
(d-g) Numerical trajectories of successful open loop controls of
circadian rhythms via the full feedback vertex set
$\mathcal{I}$. Zooms into P2, UP, and USS are shown as top-right
insets. The resulting trajectory of the control experiment is always
the red solid curve. (d) ``From P1 to P2''. The stable cycles P1 and P2
are shown by gray solid curves. (e) ``From P2 to P1''. Gray solid: P1
and P2. (f) ``From P1 to UP''. Gray solid: P1 and UP. (g) ``From P1 to
USS''. Gray solid: P1, open dot: USS. 
After \citep{Mochizuki-JTB-2013}.
}
\label{fig:FVS_control} 
\end{figure*}

An open-loop control applied to %
the 
nodes of an FVS will %
completely control the dynamics of
  those nodes and hence effectively remove all the incoming
 links to them. Consequently, those nodes will not %
be influenced by 
other nodes. %
They will, however, continue to influence %
other nodes and 
 drive the whole system to a desired attractor. 
Consider, for example, the gene regulatory network of circadian
rhythms in mice, consisting of 21 nodes
(Fig.~\ref{fig:FVS_control}a). 
In general there can be multiple minimal FVS's for a given digraph.
One such minimal FVS of size seven, i.e. $\mathcal{F}=\{$PER1,
PER2, CRY1, CRY2, RORc, CLK, BMAL1$\}$, is highlighted %
in red in Fig.~\ref{fig:FVS_control}a.   
The associated dynamical system can be described by a set of ODEs
involving %
21 variables and hundreds of
parameters. 
Under a particular choice of parameters, the system has %
several invariant sets: (i) two stable periodic oscillations (P1 and P2); 
(ii) one unstable periodic oscillation (UP); and 
(iii) one unstable stationary
point (USS) (Fig.~\ref{fig:FVS_control}b,c). 
Let us %
aim to steer the system from P1 to
P2. To achieve this, we first need to calculate the time tracks of the
seven FVS nodes on the desired invariant set P2, denoted as
$x_{i}^\mm{P2}, i \in \mathcal{F}$, which can be done by numerically integrating the ODEs.
Then we prescribe the time
tracks of the seven nodes in $\mathcal{F}$ to follow their desired values $x_{i}^\mm{P2}$. This way, we
effectively remove any influence from the other 14 nodes to the nodes in $\mathcal{F}$. 
The dynamics of the remaining 14 nodes $x_{i}, i \notin \mathcal{F}$, are
determined %
by the remaining 14 ODEs of the system, where the initial
state of these remaining nodes is chosen to coincide with an arbitrary point
on the P1 trajectory. 
As shown in Fig.~\ref{fig:FVS_control}d,  the trajectories of the
remaining 14 nodes deviate from the original stable
periodic orbit P1 and quickly converge to the competing orbit P2. The
whole system eventually displays periodic oscillation on the P2
orbit. 

In the above example, the identified FVS is a minimal one,
i.e., any subset of $\mathcal{F}$ is not an FVS. Yet, a \emph{minimal}
FVS is not guaranteed to be the \emph{minimum} one that contains the
least number of nodes. 
Naturally, it will be more desirable to identify and control the nodes
in the minimum FVS. Unfortunately, finding the minimum FVS of
a general digraph is an %
NP-hard problem~\citep{Karp-Book-1972}. %

{\color{blue}
This FVS-based open-loop control %
can be applied to a wide range of nonlinear
dynamical systems. %
It requires only a few conditions (e.g. continuous, dissipative and
decaying) on the nonlinear functions $F_i$ that are very mild and
satisfied by many real systems~\citep{Fiedler-JDDE-2013}.
For systems associated with a digraph $G(V,E)$, we
can rigorously prove that clumping the dynamics of a subset of
nodes $S \subseteq V$ will control the
rest of the %
network towards %
the desired attractor for \emph{all} choices of nonlinearities $F_i$
that satisfy the above-mentioned conditions \emph{if and only if }$S$
is an FVS in $G$~\citep{Fiedler-JDDE-2013}.   
Yet, there do exist specific systems (with certain nonlinearity $F_i$)
where clumping a reduced FVS (i.e. removing one or more nodes from an
FVS) is sufficient to control the system to a desired attractor. In
other words, for a specific system, clumping an FVS might be not
necessary. It would be a natural starting point, though.  
}

%
%
%
%
%
%
%
%
%

\iffalse
Note that to apply the three approaches discussed in the previous sections,
namely the compensatory perturbations of
state variables (Sec.~\ref{sec:sean}), attractor network based on small perturbations of
system parameters (Sec.~\ref{sec:lai}), and the FVS-based
open-loop control (Sec.~\ref{sec:fvs}), 
%
we %
need %
a detailed
knowledge of the system dynamics, including all system %
parameters. In
many cases, we lack such a piece of knowledge. %
\fi
{\color{blue}
Note that to apply the two approaches discussed in the previous subsections,
namely the compensatory perturbations of
state variables (Sec.~\ref{sec:sean}), and attractor network based on small perturbations of
system parameters (Sec.~\ref{sec:lai}), we need a detailed
knowledge of the system dynamics, including all system parameters. In
many cases, we lack such a piece of knowledge. 
In contrast, to apply the FVS-based open-loop control
(Sec.~\ref{sec:fvs}), we just need the trajectories of FVS nodes on
the desired attractors. 
We do not have to know full dynamics, nor the exact parameter
values. We just need to assure a few mild conditions on the nonlinear
functions $F_i$ are satisfied. 
}

\section{Controlling Collective Behavior}
Dynamical agents interacting through complex networks can display a
wide range of %
collective behavior, from 
synchronization 
to %
flocking among many interacting agents. 
In particular the study of network-mediated synchronization has a long history,
with %
applications 
from biology to neuroscience,
engineering, computer science, economy and social
sciences~\citep{Arenas-PR-2008}. 
Flocking %
has also gained significant attention in the past
two decades, capturing %
phenomena %
from the coordinated motion of birds or fish
to %
self-organized networks of mobile
agents. %
Applications range from massive distributed sensing using
mobile sensor networks to the self-assembly of connected mobile networks,
and military missions such as reconnaissance, surveillance,
and combat using cooperative unmanned aerial vehicles~\citep{Olfati-Saber-IEEE-07}.
These problems pose, however, a number of fundamental questions
pertaining to the control of self-organized networks.

If we aim to achieve a desired collective behavior, it is often infeasible to directly
control all nodes of a large network. %
This difficulty is partially alleviated by the notion of \emph{pinning
control}~\citep{Wang-PA-02,Wang-IJBC-02}, which relies heavily on %
feedback processes. 
In pinning control a feedback control input
is applied to %
a small subset of nodes 
called pinned nodes, which %
propagates to the rest of the network
through the edges. %
The design and implementation of feedback control %
must take into
account both the individual dynamics of the components and the network topology. 
Conceptually, %
pinning control is similar to the minimum controllability problem of
a linear system discussed in Sec.~\ref{Sec:Controllability}. The key difference is
that, instead of fully controlling a system, pinning control aims to
control only the system's collective behavior, like synchronization or
flocking. %
Pinning control has been extensively applied to %
the synchronization of coupled oscillators and flocking %
of interacting
agents~\citep{Wang-PA-02,Wang-IJBC-02,Li-IEEE-04,Sorrentino-PRE-07,Chen-Chaos-2008,Zou-EPL-08,Porfiri-Automatica-08,Yu-Automatica-09,Yu-SIAM-2013,Bernardo-Scholarpedia-14}. 
In this section we review some fundamental results on controlling %
the collective behavior of complex networked systems. 
We pay particular attention to %
the pinning control of synchronization and flocking. %
Synchronization of coupled oscillators is typically studied on fixed
network topology. 
We build on the master stability formalism to explore %
pinning synchronization, focusing on local
and global stability conditions %
and adaptive strategies. %
Flocking of multi-agent systems are typically associated
with switching or time-varying network topology, because 
the agents, like %
robots, vehicles or animals, are often mobile. 
To illustrate this we discuss 
the %
Vicsek model of flocking behavior, emphasizing
its control theoretical interpretation. 
Finally, we review key protocols that can induce %
flocking in %
multi-agent systems. %
\subsection{Synchronization of coupled oscillators}

Consider a static network of $N$ identical nodes (oscillators) with 
nearest-neighbor coupling:
\bea
\dot\mx_i 
&=& \mf(\mx_i) + \sigma \sum_{j=1}^N a_{ij} w_{ij} [\mh(\mx_j) - \mh(\mx_i)] \nn
&=& \mf(\mx_i) - \sigma \sum_{j=1}^N g_{ij} \mh(\mx_j),  \label{eq:xi}
\eea
where $\mx_i \in \mathbb{R}^d$ is the $d$-dimensional state vector of
the $i$th node,  %
$\mf(\mx_i): \mathbb{R}^d \to \mathbb{R}^d$
determines the individual dynamics of each node,  
$\sigma$ is the coupling strength, also called the \emph{coupling gain}, ${\bf A} = (a_{ij})$ is the ${N \times N}$ adjacency matrix of
the network, $w_{ij} \ge 0$ is the %
weight of link $(i,j)$. 
The output function $\mh(\mx): \mathbb{R}^d \to \mathbb{R}^d $ is used
to couple the oscillators and is identical for all %
oscillators. 
For example, if we use ${\bf h}({\bf x})=(x, 0,
0)^\mm{T}$ for a three-dimensional oscillator, like the %
Lorenz or R\"ossler oscillator, it means that the oscillators are coupled only
through their $x$-components. 
In general, ${\bf h}({\bf x})$ can be any linear or nonlinear
mapping  of the state vector ${\bf x}$. 
${\bf G} = (g_{ij})$  is the ${N \times N}$ coupling matrix of the
network ($g_{ij}= - a_{ij} w_{ij}$ for $i\neq j$ and 
$g_{ii}=  -\sum_{j=1,j\neq i}^N g_{ij}$). If $w_{ij}=1$ for all links,
${\bf G}$ is %
the Laplacian matrix $\mL$ of the
network. Note that ${\bf G}$ is not necessarily symmetric.  

The system (\ref{eq:xi}) %
is synchronized when %
the trajectories of all nodes 
converge to %
a common %
trajectory, i.e.  
\be \lim_{t\to \infty} \|  {\bf x}_i (t) -  {\bf x}_j (t) \|   = 0 
\ee 
for all $i,j=1,\cdots,N$. 
Such synchronization behavior %
describes %
a continuous system that has a
uniform movement, %
used to model synchronized neurons, lasers and electronic circuits~\citep{Pecora-PRL-1998}.

Due to the diffusive coupling, the completely synchronized state  
$
\mx_1(t) = \mx_2(t) = \cdots = \mx_N(t) = \ms(t)
$ 
is a natural solution of Eq.~(\ref{eq:xi}).  
This also defines a linear
invariant manifold, %
called the \emph{synchronization manifold}, where all the oscillators evolve
synchronously as %
$\dot{\ms}=\mf(\ms)$.  
Note that $\ms(t)$ may be an equilibrium point, a periodic orbit, or
even a chaotic solution.

Despite the fact that the completely synchronized state is a natural solution of
Eq.~(\ref{eq:xi}), it may not emerge spontaneously. For example, if
the coupling gain $\sigma$ is close to zero, the oscillators tend
to behave independently. %
If the coupling gain $\sigma$ is too strong,
the oscillators may not synchronize either. 
Our goal is to identify the %
conditions under which the system (\ref{eq:xi}) 
can %
synchronize. 
A broad spectrum of methods allows us to address this question 
~\citep{Wu-IJBC-94,Pecora-PRL-1998,Mauricio-PRL-02,Belykh-PD-04,Chen-PRE-07,Russo-CS-09}.  
The best-known method, discussed next, is based on the calculation of the eigenvalues of
the coupling matrix.  %

\subsubsection{Master stability formalism and beyond}
Consider the stability of 
the synchronization manifold in the presence of a small perturbation
$\mx_i(t) = \ms(t) + \delta \mx_i(t)$.
By expanding $\mf(\mx_i)$ and $\mh(\mx_i)$ to the first order of $\delta \mx_i$, we obtain a linear variational
equation for $\delta \mx_i(t)$,  
\be
\delta \dot{\mx}_i = \mj(\ms) \delta \mx_i  - \sigma 
\sum_{j=1}^N g_{ij} \mathcal{E}(\ms) \delta\mx_j ,
\label{eq:mx}
\ee
with Jacobian matrices $\mj(\ms) = \frac{\partial \mf(\mx)}{\partial
  \mx}\vert_{\mx=\ms}$ and $\mathcal{E}(\ms) = \frac{\partial
  \mh(\mx)}{\partial \mx}\vert_{\mx=\ms}$.
Let $\delta\mX \equiv [\delta\mx_1, \cdots, \delta\mx_N]^\mm{T}$. Then
formally we have 
\be
\delta\dot{\mX} =\left[ \mI \otimes \mj(\ms) - \sigma
  \mG \otimes \mathcal{E}(\ms) \right] \, \delta\mX
\label{eq:0}
\ee
where $\mI$ is the $N\times N$ identity matrix and  $\otimes$ is the %
Kronecker product (a.k.a. matrix direct product).

The key idea of the master stability
formalism is that we need to consider only variations that
are transverse to the synchronization manifold, as variations along
$\ms(t)$ leave the system in the synchronized state~\citep{Pecora-PRL-1998,Mauricio-PRL-02}. If these
transverse variations %
damp out, then the synchronization manifold is stable. To separate out
the transverse variations, we can project $\delta\mX$ into the eigenspace spanned by the
eigenvectors ${\bf e}_i$ of the coupling matrix $\mG$, i.e., $
\delta\mX = (\mP \otimes \mI_\mm{d}) \, {\mb \Xi}$ with $\mP^{-1}\,\mG\, \mP
= \widehat{\mG}=\mm{Diag}(\lambda_1, \lambda_2, \cdots,
\lambda_N)$. 
Then we have 
  \be
  \dot{\mb \Xi} = \left[ \mI \otimes \mj(\ms) - \sigma
    \widehat{\mG}\otimes \mathcal{E}(\ms) \right] \, {\mb \Xi}, \label{eq:Xi}
 \ee
which results in %
a block diagonalized variational equation with $N$ blocks,
corresponding to %
$N$ decoupled eigenmodes. Each block has the form 
\be
\dot{\mxi}_i = \left[ \mj(\ms) - \sigma \lambda_i \mathcal{E}(\ms)
\right] \, \mxi_i ,\label{eq:deltay}
\ee
where $\mxi_i$ is the eigenmode associated with the eigenvalue $\lambda_i$ of
$\mG$. 
Note that in deriving %
(\ref{eq:Xi}) we have implicitly
assumed that the coupling matrix $\mG$ is 
diagonalizable, which is %
always true for symmetric $\mG$. Thus each
eigenmode of the perturbation is decoupled from 
the others, and will damp out independently
and simultaneously. If $\mG$ is not diagonalizable, we can %
transform $\mG$
into the Jordan canonical form. In this case, some eigenmodes of the  perturbation may suffer from a long
transient~\citep{Nishikawa-PRE-06}. 

We can order the eigenvalues of $\mG$ such that 
$0=\lambda_1 \le \mm{Re} \lambda_2 \le \cdots \le \mm{Re} \lambda_N$. 
Because the row sum of $\mG$ is zero, the minimal eigenvalue
$\lambda_1$ is always zero with the corresponding eigenvector ${\bf
  e}_1=(1,1,\ldots, 1)^\mm{T}$. 
Hence the first eigenmode
$
\dot{\mxi}_1 = \mj(\ms)  \, \mxi_1
$
corresponds to the perturbation parallel to the synchronization
manifold. 
Due to the Gerschgorin Circle Theorem, all other eigenvalues must have
non-negative real parts. %
The corresponding $(N-1)$
eigenmodes are transverse to the synchronization manifold and must %
decay 
to have a stable synchronization manifold. 

The form of each block in %
(\ref{eq:deltay}) is the same
up to the scalar multiplier $\sigma \lambda_i$. 
This leads to %
the %
variational equation, 
called the \emph{master stability equation}, 
\be
\dot{\mxi} =\left[ \mj - (\alpha + i \beta) \mathcal{E} \right] \mxi.
\ee
For small $\mxi$ we have %
$||\mxi(t)||
\sim\exp[\Lambda(\alpha,\beta) t]$, which decays exponentially if the
maximum Lyapunov characteristic exponent $\Lambda(\alpha,\beta)<0$. 
Consequently, $\Lambda(\alpha, \beta)$ is called the \emph{master stability function} (MSF). 
Given a coupling strength $\sigma$, %
the sign of the MSF in the point $\sigma
\lambda_i$ in the complex plane 
reveals the stability of that eigenmode. If all %
eigenmodes are
stable (i.e. $\Lambda(\sigma \lambda_i) < 0$ for all $i$'s), then the synchronization manifold
is stable at that coupling strength.  
Note that since the master stability formalism only assesses the linear stability
  of the synchronized state, it only yields the necessary, but not the
  sufficient condition for synchronization. %

For undirected and unweighted
networks, the coupling matrix $\mG$ is symmetric and all its
eigenvalues are real, %
simplifying the stability analysis. 
In this case, depending on $\mj$ and $\mathcal{E}$, the MSF
$\Lambda(\alpha)$ can be classified as follows:

(i) Bounded: $\Lambda(\alpha) < 0$ for $\alpha_1 < \alpha < \alpha_2$. This usually happens when
$\mh(\mx) \neq \mx$. %
The linear stability of the synchronized manifold requires that 
$
\alpha_1 < \sigma \lambda_2 \le \cdots \le \sigma\lambda_N <
\alpha_2$. 
This condition can be only fulfilled for %
$\sigma$ when the eigenratio $R$ satisfies 
\be
R \equiv \frac{\lambda_N}{\lambda_2} < \frac{\alpha_2}{\alpha_1}.\label{ieq:R1}
\ee
The beauty of this inequality comes from the fact that its r.h.s. %
depends only on the dynamics while its l.h.s. %
depends only on the network
structure. 
If $R> \alpha_2/\alpha_1$, %
for any $\sigma$ the synchronization manifold is unstable, indicating
that %
it is impossible to
synchronize the network. 
If $R< \alpha_2/\alpha_1$, the synchronization manifold is
stable for  
$\sigma_\mm{min}= \alpha_1/\lambda_2 < \sigma <
\sigma_\mm{max} = \alpha_2/\lambda_N.$ 
The \emph{synchronizability} of the network can be quantified by the relative
interval $\sigma_\mm{max} / \sigma_\mm{min} =
\alpha_2 / (\alpha_1 R)$. A network is %
more synchronizable for higher $\sigma_\mm{max} / \sigma_\mm{min}$ %
(or smaller $R$).%

(ii) Unbounded: $\Lambda(\alpha) < 0$ for 
$\alpha > \alpha_1$. The stability
criteria of the synchronized manifold is %
$
\alpha_1 < \sigma \lambda_2  \le \cdots \le
\sigma\lambda_N$, 
which is true if 
\be
\sigma > \sigma_\mm{min}=
\alpha_1 / \lambda_2.  \label{ieq:R2}\ee 
The larger is $\lambda_2$ the smaller is the 
synchronization threshold $\sigma_\mm{min}$, hence the more
synchronizable is the network.

Inequalities~(\ref{ieq:R1}) and (\ref{ieq:R2}) demonstrate that %
the MSF framework %
provides an objective criteria ($R$ or
$\lambda_2$) to assess the synchronizability of
complex networks based on the spectrum of the coupling matrix $\mG$ only,
without referring to specific oscillators and output 
functions. 
The MSF framework allows us to address the impact of the network topology and edge weights on
synchronizability%
~\citep{Arenas-PR-2008}.
Consequently, there have been numerous numerical attempts to relate 
the spectral properties of network
models %
to a single structural characteristic of networks, like %
mean degree, degree heterogeneity, %
path lengths, 
clustering coefficient, degree-degree correlations, etc.%
~\citep{Arenas-PR-2008}. 
The outcome of these analyses is occasionally 
confusing,  %
because in a networked environment it is %
usually impossible %
to isolate a single structural characteristic while keeping
the others fixed. %
Overall, several network characteristics 
can influence %
synchronizability, 
but none of them is an %
exclusive factor in the observed dependencies. 

The fundamental limitation of MSF 
is that it only assesses the \emph{linear} or \emph{local}
stability of the synchronized state, which is %
a \emph{necessary},
but not a \emph{sufficient} condition for
synchronization~\citep{Arenas-PR-2008}.  
To obtain a sufficient condition, one can use global stability
analysis, like Lyapunov's direct
method~\citep{Wu-IJBC-94,Wu-CSI-95a,Wu-CSI-95b,Wu-CSI-95c,Belykh-PD-04,Belykh-PD-04b,Belykh-IJBC-05,Belykh-Chaos-06,Chen-CSII-06,Chen-PRE-07,Chen-CSI-08,Li-PRE-09}
or contraction
theory~\citep{Lohmiller-Automatica-98,Wang-BC-05,Li-PRE-07,Russo-CS-09,Zahra-IEEE-15,Pham-NN-07,Tabareau-PCB-10}. 
\subsubsection{Pinning synchronizability} %
If a network of coupled oscillators can not synchronize spontaneously,
we can design %
controllers that, applied %
to a subset of \emph{pinned} nodes  
$\mathcal{C}$, %
help %
synchronize the network. 
Hence the pinned nodes %
behave like \emph{leaders}~\citep{Wang-PA-02,Li-IEEE-04,Wang-BC-05,Wang-IEEE-06}, forcing %
the remaining \emph{follower} nodes to %
synchronize. %
This procedure, known as \emph{pinning 
synchronization}, is fundamentally different from
  \emph{spontaneous synchronization} of coupled oscillators, where we
  don't specify the synchronized trajectory $\ms(t)$, hence the system
  ``self-organizes'' %
into the
  synchronized trajectory under appropriate conditions. In pinning
  synchronization, we choose the
  desired trajectory $\ms(t)$, aiming %
to achieve
  some desired control objective, and this trajectory must be explicitly taken into
  account in the feedback controller design. %
Note that in literature pinning
  synchronizability is often called pinning controllability. Here we
  use the term synchronizability to avoid %
confusion with
the classical notion of controllability discussed in
Secs.~\ref{Sec:Controllability} and ~\ref{sec:controllabilitytest_Nonlinear}.

A controlled network is %
described by  
\be
 \dot\mx_i  = \mf(\mx_i) - \sigma \sum_{j=1}^N g_{ij} \mh(\mx_j) +
 \delta_i {\mb u}_i(t), \label{eq:ps}
\ee
where $\delta_i = 1$ for %
pinned nodes 
and 0 otherwise, and 
\be
{\mb u}_i(t) = 
 \sigma  [{\bf p}_i(\ms(t)) - {\bf p}_i(\mx_i(t))]
\label{eq:pinning-linearfeedback}
\ee
is the $d$-dimensional linear feedback controller~\citep{Wang-PA-02,Li-IEEE-04}, %
${\bf p}_i(\mx(t))$ %
is the \emph{pinning
  function}  
that controls the input of node $i$, 
and $\ms(t)$ is the desired synchronization %
trajectory satisfying
$\dot{\ms}(t) = \mf(\ms(t))$. 
Note that in the fully %
synchronized state
$\mx_1(t)=\mx_2(t)=\cdots=\mx_N(t)=\ms(t)$, we have ${\mb u}_i(t) =
{\bf 0}$ for all nodes. 
The form of the linear feedback controller
(\ref{eq:pinning-linearfeedback}) implies that the completely
synchronized state is a natural solution %
of the
controlled network (\ref{eq:ps}).

Similar to %
spontaneous synchronization, we must %
derive 
  the necessary and sufficient conditions for pinning synchronization.
These conditions are more important from the control perspective, %
  because they are the prerequisite for the %
design of any practical controller.  
If we focus on the \emph{local} (or \emph{global}) stability of the
synchronized manifold of the controlled network (\ref{eq:ps}), we
obtain the \emph{necessary} (or \emph{sufficient}) condition for pinning synchronization, describing
the \emph{local} (or \emph{global}) pinning
synchronizability.
\emph{Local pinning synchronizability}: 
Given the presence of inhomogeneous dynamics at the controlled and
 uncontrolled nodes, the MSF approach can not be directly applied to
 the controlled network (\ref{eq:ps}). 
Instead, %
we first 
introduce a virtual node whose dynamics follows %
$\dot{\ms}(t) = \mf(\ms(t))$, representing %
the desired synchronization
solution~\citep{Sorrentino-PRE-2007,Zou-EPL-08}.  
The extended system now has $N+1$ nodes: $\my_i (t)=\mx_i (t)$ for
$i=1,\cdots,N$; and $\my_{N+1}(t)=\ms (t)$. %
The virtual node is connected to each pinned node.

We choose the pinning function 
\be
{\bf p}_i(\mx) = \kappa_i
\mh(\mx)
\label{eq:pinningfunction} 
\ee with %
\emph{control gains} $\kappa_i > 0$, 
parameters %
that capture %
the relationship between the magnitude
  of $\mh(\mx)$ and %
${\bf p}_i(\mx)$.  
By defining the pinning function via (\ref{eq:pinningfunction}) 
we can then rewrite (\ref{eq:ps}) in the form of (\ref{eq:xi}), with an
effective coupling matrix satisfying the zero row-sum condition,
allowing us to apply 
the MSF approach. 
Indeed, plugging (\ref{eq:pinningfunction}) into (\ref{eq:pinning-linearfeedback}),
we have $
{\mb u}_i(t) = 
\sigma  \kappa_i [ \mh(\ms(t)) - \mh(\mx_i(t)) ]
$ and 
(\ref{eq:ps}) becomes 
\be
 \dot\my_i  = \mf(\my_i) - \sigma \sum_{j=1}^{N+1} m_{ij} \mh(\my_j) \label{eq:ps1}
\ee
where 
\be
{\bf M} = 
\begin{bmatrix}
g_{11} + \delta_1 \kappa_1 & g_{12} & \cdots & g_{1N} & - \delta_1 \kappa_1 \\ 
g_{21} & g_{22} + \delta_2 \kappa_2 & \cdots & g_{3N} & - \delta_2 \kappa_2 \\ 
\vdots &    \vdots            & \ddots &  \vdots &\vdots \\
g_{N1} & g_{N2} &  \cdots & g_{NN} + \delta_N \kappa_N & - \delta_N \kappa_N \\ 
0 & 0 & \cdots & 0 & 0 
\end{bmatrix} 
\ee
is the effective coupling
  matrix of the %
$(N+1)$-dimensional extended system. 
Apparently, ${\bf M}$ is a zero row-sum matrix, hence %
we can sort its eigenvalues as %
$0=\lambda_1 \le \mm{Re} \lambda_2 \le \cdots \le \mm{Re}
\lambda_{N+1}$.  
We can now %
apply the MSF 
approach to numerically explore the local %
stability of the synchronization manifold of
the controlled network (\ref{eq:ps1}). %

\begin{figure}[t!]
  \begin{center}
   \includegraphics[width=0.5\textwidth]{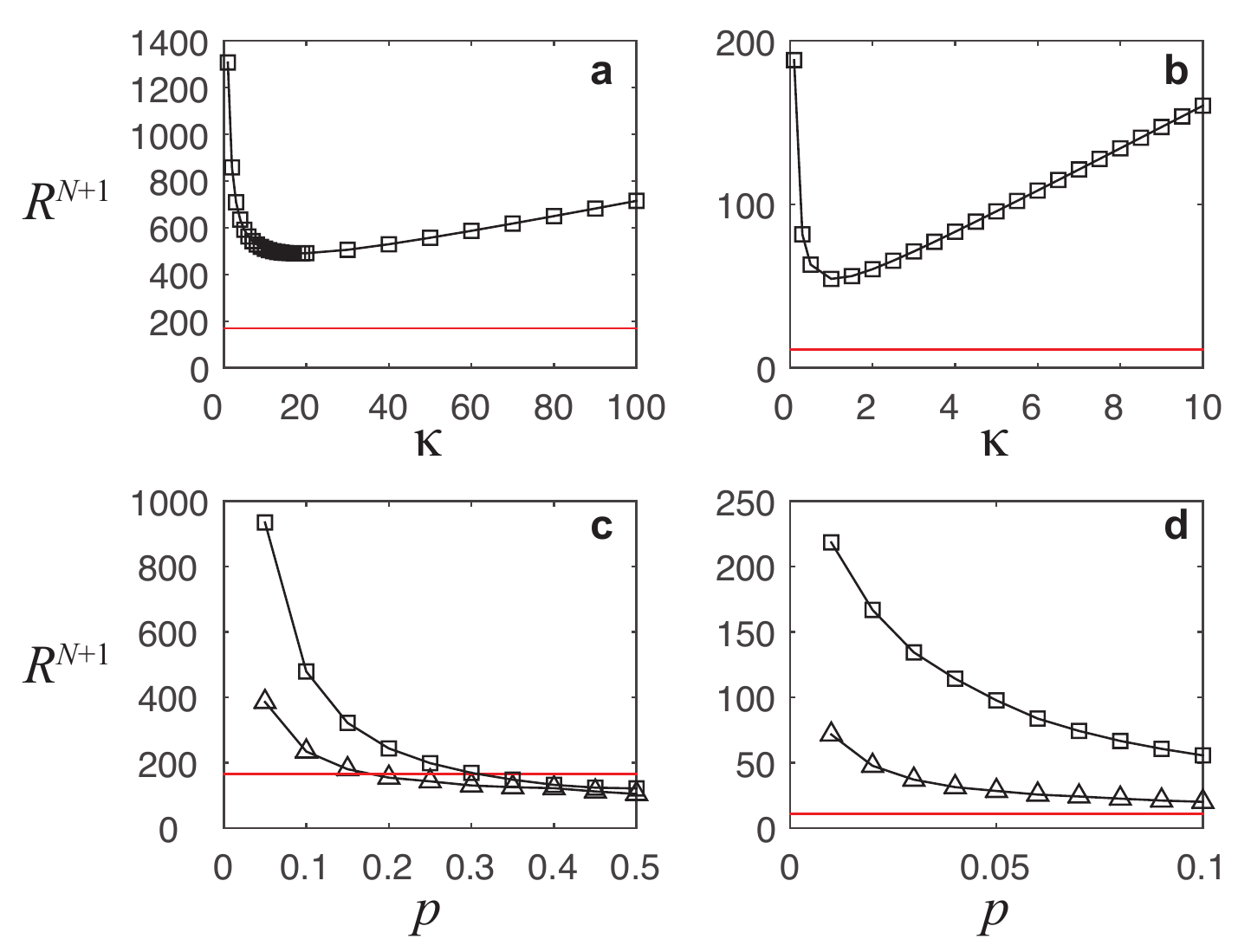}
  \end{center}
  \caption{(Color online)  %
Local pinning
    synchronizability of scale-free networks. The local pinning
    synchronizability is quantified by the eigenratio $R^{N+1}=\mm{Re}
    \lambda_{N+1}/ \mm{Re} \lambda_2$ of the extended
    system (\ref{eq:ps1}).  
    The calculation was performed %
for $N=10^3$ identical R\"{o}ssler oscillators coupled in
    $x$ and $z$ directions, with coupling gain $\sigma$ and a $p$
    fraction of pinned nodes, placed on a Barab\'{a}si-Albert (BA) scale-free
    network with mean degree $\kmean=4$.    
(a-b) We choose $p=0.1$ fraction of nodes to pin and study the impact
of control gain $\kappa$ on local pinning
    synchronizability with  coupling gain $\sigma=0.3$ (a) and $2.8$
    (b), respectively. 
We find that in both cases the eigenratio $R^{N+1}\equiv \text{Re}
\lambda_{N+1}/ \text{Re} \lambda_{2}$ of the new coupling matrix ${\bf
  M}$ is minimized and hence the local pinning synchronizability is
maximized around a specific $\sigma$-dependent value of the
control gain $\kappa$. 
(c-d): We study the impact of the fraction of pinned nodes on local pinning
    synchronizability: (c) $\sigma=0.3, \kappa=10$. (d) $\sigma=2.8, \kappa=1.5$.
The horizontal continuous lines (red) represent the eigenratio $R^{N}$
of the corresponding uncontrolled system (\ref{eq:ps}). 
We find that the number of pinned nodes, regardless if they are chosen
randomly %
or selectively %
within the network, has a monotonic impact on the pinning
synchronizability. Controlling more nodes %
always enhances the
network pinning synchronizability. 
In all plots squares represent the case of random pinning, i.e., a $p$
fraction of nodes is randomly chosen to be pinned. 
In (c) and (d), triangles represent the case of selective pinning,
where nodes have been sorted in the order of decreasing degree and the
top $p$ fraction of the nodes are chosen to be pinned.  
After~\citep{Sorrentino-PRE-2007}.
  }    \label{fig:pinning_kappa_p}
\end{figure}

The role of the control gain ($\kappa_i$), coupling gain ($\sigma$),
and %
the number and locations of the pinned nodes, on %
local pinning
synchronizability %
has been systematically studied %
~\citep{Sorrentino-PRE-2007}.  
Consider for example a Barab\'{a}si-Albert (BA) scale-free network of $N$
identical R\"{o}ssler oscillators coupled in $x$ and $z$
directions.    
By assuming $\kappa_1=\cdots=\kappa_N=\kappa$, it was 
found  
that for a wide range of coupling gain $\sigma$, the eigenratio $R^{N+1}\equiv \text{Re}
\lambda_{N+1}/ \text{Re} \lambda_{2}$ of the new coupling matrix ${\bf
  M}$ is minimized and hence the local pinning synchronizability is
maximized around a specific $\sigma$-dependent value of the
control gain $\kappa$. 
In other words, %
too large or too small control gain %
can reduce the network pinning synchronizability (Fig.~\ref{fig:pinning_kappa_p}a,b). 
In contrast, the number of pinned nodes, regardless if they are chosen
randomly %
or selectively %
within the network, has a monotonic impact on pinning
synchronizability: Controlling more nodes %
always enhances the
network pinning synchronizability, in line %
with our intuition (Fig.~\ref{fig:pinning_kappa_p}c,d). 
Furthermore, selective pinning, when %
the nodes are 
chosen in the order of decreasing degree, yields 
better synchronizability than random
  pinning.  

\emph{Global pinning synchronizability}: 
By describing the time evolution of the controlled network
(\ref{eq:ps1}) in terms of the error dynamics, we can map the global
pinning synchronizability of (\ref{eq:ps1}) to the global asymptotic
stability of the synchronized manifold, which can %
be studied via Lyapunov stability
theory.  

If the desired asymptotic trajectory is an equilibrium point
($\dot{\ms}=\mf(\ms)={\bf 0}$), 
we can derive sufficient conditions for globally stabilizing the
pinning controlled network%
~\citep{Li-IEEE-04}.  
For a more general desired trajectory, 
it has been shown that a single feedback controller can pin a %
complex network to a homogenous solution, without assuming symmetry,
irreducibility, or linearity of the couplings~\citep{Chen-CSI-07}. 
If the oscillator dynamics $\mf(\mx)$ fulfills
\be
\mf(\mz_1) - \mf(\mz_2) = \mathcal{F}_{\mz_1, \mz_2} (\mz_1 - \mz_2),
\quad \forall  \mz_1, \mz_2 \in \mathbb{R}^d, \label{eq:Fz1z2}
\ee
 where $\mathcal{F}_{\mz_1, \mz_2} \in \mathbb{R}^{d\times d}$ is
bounded, i.e., there exists a positive constant $\alpha$ such that for
any $\mz_1, \mz_2 \in \mathbb{R}^d$, $||\mathcal{F}_{\mz_1, \mz_2} ||
\le \alpha$, then we can derive tractable sufficient conditions for global pinning
synchronizability in terms of the network topology, the oscillator
dynamics, and the linear state feedback~\citep{Porfiri-Automatica-08}.  
Note that condition (\ref{eq:Fz1z2}) applies to a large
variety of chaotic oscillators%
~\citep{Jiang-CSF-03}. %
The results indicate that for a
connected network, even for a limited number of pinned nodes, global
pinning synchronizability can be achieved by properly selecting the
coupling strength and the feedback gain
~\citep{Chen-CSI-07}.

If $\mh(\mx)= {\bf \Gamma} \mx$ and 
the oscillator dynamics 
$\mf(\mx)$
satisfies  
\be
(\mx-\my)^\mm{T} [\mf(\mx, t)-\mf(\my, t)] \le (\mx-\my)^\mm{T} {\bf
  K} {\bf \Gamma} (\mx-\my) \label{eq:QUAD-0}
\ee
for a constant matrix ${\bf K}$, sufficient conditions for global pinning
synchronizability can also be
derived~\citep{Yu-Auto-09,Song-IEEE-10,Yu-SIAM-2013}.  
Note that the %
condition (\ref{eq:QUAD-0}) is 
so mild that many systems, from Lorenz system to Chen system, L\"{u}
system, recurrent neural networks, Chua's circuit 
satisfy this condition%
~\citep{Yu-SIAM-2013}. %
Counterintuitively, it was found that for undirected networks, the small-degree nodes, instead
of hubs, should be pinned first when the
coupling strength $\sigma$ is small~\citep{Yu-Auto-09}. %
For directed networks, %
nodes with very small
in-degree or large out-degree should be pinned
first~\citep{Yu-SIAM-2013}.
This result can be understood by realizing that %
low in-degree nodes %
receive less
information from other nodes and hence are less ``influenced'' by
others.  
In the extreme case, nodes with zero in-degree
will not be ``influenced'' by any other
nodes, hence they must %
be pinned first. 
On the other hand, large out-degree nodes can
influence many other nodes, hence 
it makes sense to pin them first. %
\subsubsection{Adaptive pinning control}
Implementing the linear feedback pinning controller (\ref{eq:pinning-linearfeedback}) %
requires %
detailed knowledge of the global network topology. This
  is because we have to check whether there are possible coupling and
  control gains that ensure %
pinning synchronizability. %
Yet, in practice we do not always have access to 
the global
  network topology. %
Given this limitation, recently %
adaptive control has been proposed for pinning
  synchronization, in which case a controller
  adapts to a controlled system with parameters that vary in time, or
  are initially uncertain, without requiring %
a detailed knowledge of the
  global network topology%
~\citep{Wang-IEEE-06,Zhou-Automatica-08,Wang-EPJB-08,DeLellis-ISCAS-2010,Wang-CNSNS-2010,DeLellis-CSI-2011}.   
As we discuss next, many different %
strategies have been designed to tailor the control
gains, coupling gains, %
or to rewire the network
topology to ensure pinning synchronizability.

(i) \emph{Adaptation of control gains:} 
To adapt the control gain $\kappa_i$ (\ref{eq:pinningfunction}),
representing %
the ratio between the pinning
function and output function, we choose 
the control input %
${\bf u}_i(t) = - \delta_i 
\kappa_i(t) (\mx_i(t) - \ms)$, and the control gains as~\citep{Zhou-Automatica-08,Wang-EPJB-08} %
\be
\dot{\kappa}_i(t) = q_i |{\bf e}_i(t)|. \label{eq:adaptivepinning-1}
\ee
In other words, %
the control
  gain $\kappa_i$ varies in time and adapts %
to the error vector ${\bf
    e}_i(t) \equiv {\bf s}(t ) - \mx_i (t)$, that 
describes the deviation %
of the oscillator $i$
from %
the reference signal ${\bf s}(t)$. %
If the  individual dynamics $\mf(\mx)$ %
satisfies the Lipschitz condition, then the global stability
of this adaptive strategy can be assured. %

(ii) \emph{Adaptation of coupling gains:}
The coupling gain
$\sigma_{ij}$, %
defining the mutual coupling strength between node
pair $(i,j)$, can also be adapted using~\citep{DeLellis-ISCAS-2010}%
\be
\dot{\sigma}_{ij}(t) = \eta_{ij} |{\bf e}_i(t) - {\bf e}_j(t)|^2.\label{eq:adaptivepinning-2}
\ee
This %
strategy is very effective in controlling networks 
of quadratic %
dynamical systems, where the dynamics $\mf(\mx,t)$ of each
oscillator satisfies %
$
(\mx-\my)^\mm{T} [\mf(\mx,t) - \mf(\my, t)] - (\mx-\my)^\mm{T}  \Delta
(\mx-\my) \le - \omega (\mx-\my)^\mm{T}  (\mx-\my) 
$.
Here, $\Delta$ is an $d\times d$ diagonal matrix and $\omega$ is a
real positive scalar. 

Note that the %
adaptive strategies (\ref{eq:adaptivepinning-1}) and
(\ref{eq:adaptivepinning-2}) are based on the local error vectors of
nodes or between neighboring nodes, hence they %
avoid the need for 
a prior tuning of the control or coupling gains. This is attractive in
many circumstances. However, these adaptive strategies still require 
a prior selection of the pinned nodes based on some knowledge of the
network topology. This limitation can be avoided by choosing 
pinned nodes in an adaptive fashion, as we discuss next.

(iii) \emph{Adaptive selection of pinning nodes:}
Adaptive pinning %
can be achieved by assuming %
the pinning node indicator $\delta_i$ to be %
neither fixed nor binary. A common approach %
is to introduce 
\be
\delta_i(t) = b_i^2(t) 
\ee
where $b_i(t)$ satisfies the %
dynamics 
\be
\label{eq:snapping}
\ddot{b}_i+\zeta\dot{b}_i+\frac{\ud U(b_i)}{\ud b_i}=g(| {\bf e}_i|). %
\ee
In other words, 
$b_i(t)$ follows 
the dynamics of a unitary mass in a
potential $U(b_i)$ subject to an external force $g$ that is a
function of the pinning error ${\bf e}_i$ and a linear damping term
described by $\zeta\dot{b}_i$. This is termed as the
\emph{edge-snapping mechanism}.
For convenience, $U(\cdot)$ can be chosen as a double-well
potential: $U(z)=k\, z^2(z-1)^2$, where the parameter $k$ defines the
height of the barrier between the two wells. 
Then (\ref{eq:snapping}) has only two stable equilibria, %
$0$ and $1$, describing whether %
node $i$ is %
pinned or not, respectively. 
Sufficient conditions for %
the edge snapping mechanism (\ref{eq:snapping}) to drive the network to a
steady-state pinning configuration have been derived~\citep{DeLellis-CSI-2011}. %
The key advantage of the adaptive selection of pinning
  nodes is that we don't have to choose the %
nodes we need to
  pin before we design the controller. Instead, we can select %
them as we go %
  in an adaptive fashion.

(iv) \emph{Adaptation of the network topology:}
We can ensure synchronization by adapting the network topology. %
Specially, we can set each off-diagonal element of the Laplacian matrix of the
network as 
\be
\mathcal{L}_{ij}(t)=-\sigma_{ij}(t) \, \alpha_{ij}^2 (t), 
\label{eq:snapping_netw}
\ee
where $\sigma_{ij}(t)$ is the mutual coupling strength between node pair
$(i,j)$, which is adapted as in \eqref{eq:adaptivepinning-2}. 
The weight $\alpha_{ij}(t)$ is associated to every undirected
edge of the target pinning edge and is adapted as 
\be
\dot{\alpha}_{ij}+\nu
\dot{\alpha}_{ij}+\frac{\mathrm{d}
U(\alpha_{ij})}{\mathrm{d}\alpha_{ij}}=c( |{\bf e}_{ij}|),\quad i,j=1,\ldots,N,\:i\ne j,\label{eq:snapping_netwo} 
\ee
where ${\bf e}_{ij}(t)={\bf e}_j (t)-{\bf e}_i (t)$, and $U(\cdot)$ can be again
chosen as a double-well potential so that (\ref{eq:snapping_netwo})
has only two stable equilibria, 0 and 1. 
In this case, the target network topology evolves in a decentralized
way. The local mismatch of the trajectories can be considered as 
an external forcing on the edge dynamics (\ref{eq:snapping_netwo}), inducing the activation of the
corresponding link, i.e. $\alpha_{ij}=1$.

The above adaptive strategies %
  cope better when %
pinning controllability using 
 a non-adaptive or static approach is initially not feasible. 
They are also %
successful in ensuring %
network synchronization %
in the presence of perturbations or deterioration, like link failures~\citep{Jin-IEEE-12}. 

Taken together, we have multiple strategies to %
force a networked system to 
  synchronize. %
The discussed tools have %
a wide range of
  applications for systems in which a synchronized state is
  desired. 
In some cases synchronization can be %
harmful, like in the case of %
synchronized clients or
routers that cause %
congestion in %
data traffic on the %
Internet~\citep{Li-IEEE-03}, or in schizophrenia. In this case %
the synchronized state can be %
destroyed by the addition of a single link with inhibitory
coupling %
~\citep{Slotine_contractionanalysis}.

\subsection{Flocking of multi-agent dynamic systems}

The flocking %
of birds, shoaling %
of fish, swarming %
of insects, and
herding %
of land animals are %
spectacular manifestations of coordinated collective behavior of multi-agent
systems. These phenomena have fascinated scientists from %
diverse disciplines, from ecologists to physicists, social and computer
scientists
~\citep{Olfati-Saber-IEEETAC-06,Vicsek-PR-2012}.  %
Many %
models have been proposed to reproduce the behavior of such
self-organized systems. 
The first widely-known flocking simulation was %
primarily motivated by the visual
appearance of a few dozen coherently flying objects, e.g., imaginary
birds and spaceships \citep{Reynolds-87}. 
Yet, the %
quantitative %
interpretation of the emerging behavior of
huge flocks in the presence of perturbations was possible only
following the development of a %
statistical physics-based interpretation
of flocking obtained through the Vicsek model%
~\citep{Vicsek-PRL-1995}. 
As we discussed next, 
the Vicsek model and its variants 
can be interpreted as decentralized feedback control system with
time-varying network structure, offering %
a %
better understanding of the origin of %
collective behavior~\citep{Jadbabaie-IEEE-2003,Moreau-IEEE-05,Ren-IEEE-2005,Olfati-Saber-IEEETAC-06}.

 \subsubsection{Vicsek Model and the Alignment Problem}%
\begin{figure}[t!]
  \begin{center}
   \includegraphics[width=0.25\textwidth]{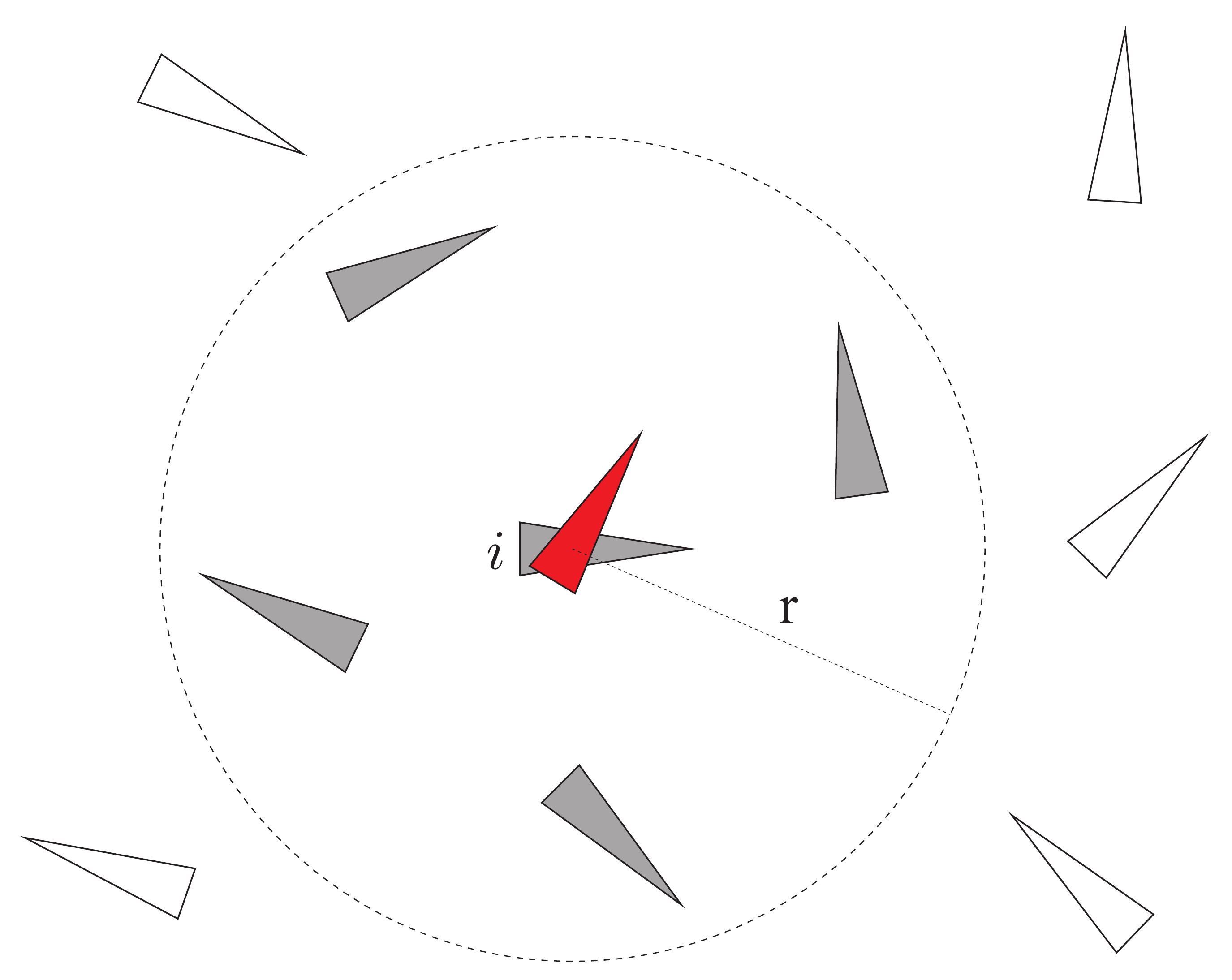}
  \end{center}
  \caption{(Color online) Vicsek model. The direction of agent $i$ at
    time $t+1$ (shown in red) is the average of its own direction and the
    directions of all other agents at a %
distance less than $r$ to
    agent $i$ at time $t$ (shown in grey). Agents outside this
    circle (shown in white), do not contribute to the direction of agent $i$ at time $t+1$.}    \label{fig:VM}
\end{figure}

The Vicsek model explains %
the origin 
of \emph{alignment}, %
a key feature of flocking behavior~\citep{Vicsek-PRL-1995}. 
It is a discrete-time stochastic model, in which autonomous agents
move in a plane with
a constant speed $v_0$, %
initially following randomly chosen %
directions. %
The position $\mx_i$ of agent $i$ changes as 
\be
  \mx_i(t+1)=\mx_i(t)+{\bf v}_i(t+1), 
  \label{eq:SVM}
\ee
where the velocity of each agent %
has the same absolute value %
$v_0$. 
The direction %
of agent $i$ is updated using a local rule that depends %
on the average of its own direction %
and the directions of its ``neighbors'', i.e. all %
agents within a %
distance $r$ from agent $i$ (Fig.\ref{fig:VM}). 
In other words, 
\begin{equation}
 \theta_i(t+1)= \langle \theta_i(t) \rangle_r +\Delta_i(t).
 \label{eq:UjThi}
 \end{equation}
Here $\langle \theta_i(t) \rangle_r \equiv \arctan \left[\langle
    \sin \theta(t) \rangle_r / \langle
     \cos \theta(t) \rangle_r  \right]$ denotes the average direction of
the agents (including agent $i$) %
within a circle of radius
$r$. The interaction radius $r$ can be set as the unit %
distance, %
$r=1$. 
The origin of the alignment rule (\ref{eq:UjThi}) can be the 
stickiness of the agents, hydrodynamics, could be pre-programmed,
or based on information processing~\citep{Vicsek-PR-2012}.   
The perturbations are contained in %
$\Delta_i(t)$, which is a
random number taken from a uniform distribution in the interval
$[-\eta/2, \eta/2]$. Therefore the final direction of agent $i$ is
obtained after rotating the average direction of the neighbors with a
random angle. 
These random perturbations can %
be rooted in %
any stochastic or deterministic
factors that affect the motion of the flocking agents.  

The Vicsek model has %
three parameters: (i) the agent density $\rho$
(number of agents in the %
area $L^2$);
(ii) the speed $v_0$ and (iii) the magnitude %
of perturbations $\eta$. 
The model's order parameter is the normalized average velocity 
\be
\phi \equiv \frac{1}{N v_0}\left| \sum_{i=1}^N {\bf v}_i
\right|. 
\ee

\begin{figure}[t!]
  \begin{center}
   \includegraphics[width=0.5\textwidth]{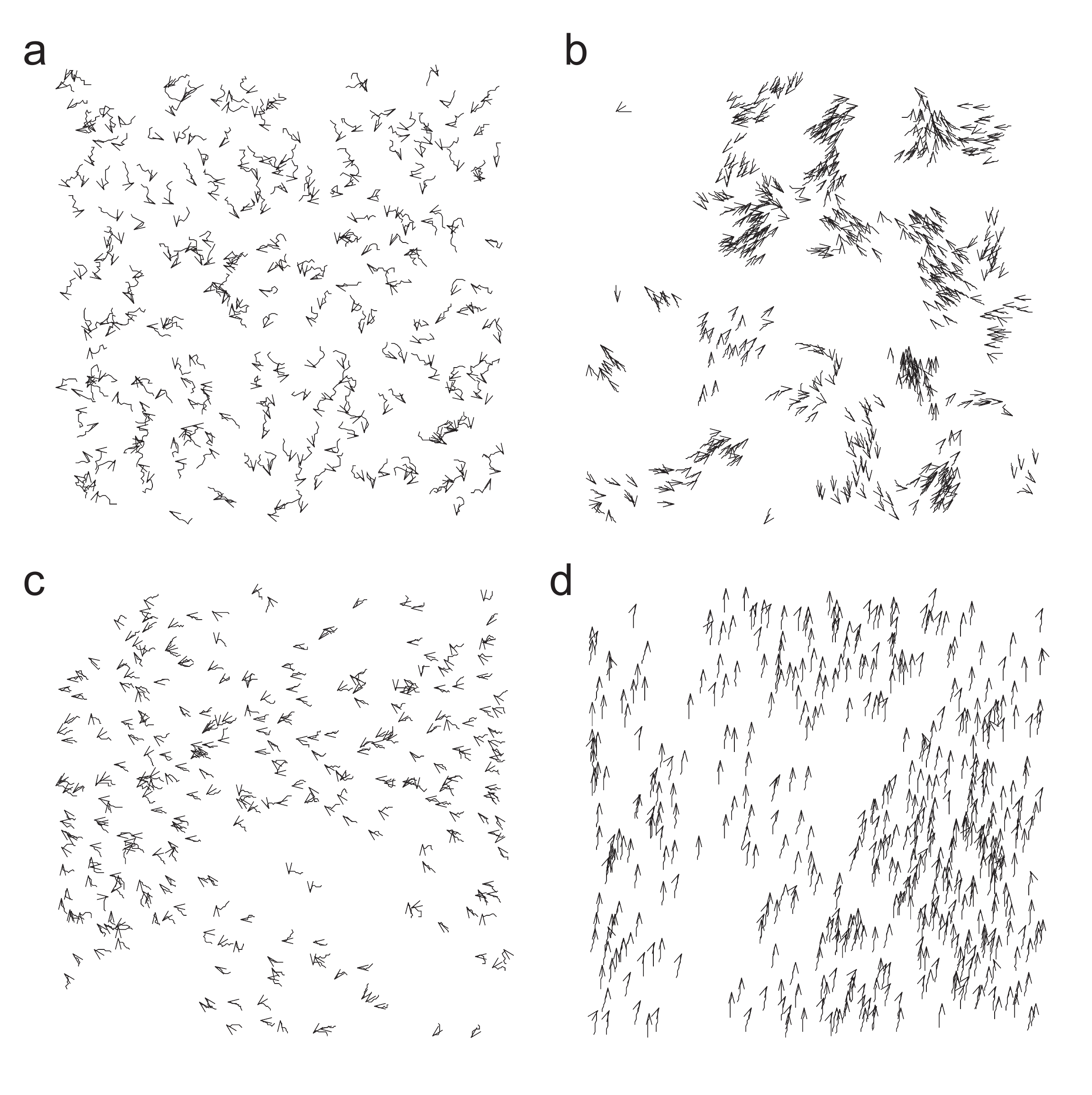}
  \end{center}
  \caption{%
Emergence of order in the Vicsek Model. %
The panels show the agent velocity %
for varying values of the density and the noise level. The
actual velocity of an agent is indicated by a small arrow while their
trajectories for the last 20 time steps are shown as short
continuous curves.The number of agents is $N=300$, and the %
absolute velocity is $v_0=0.03$. (a) At $t=0$, the positions and the direction of
velocities are randomly distributed. $L=7, \eta=2.0$ 
(b) For small densities ($L=25$) and noise ($\eta=0.1$)
level, the agents %
form groups that move together %
in random directions. 
(c) At higher densities ($L=7$) and noise ($\eta=2.0$) the agents move
randomly with some correlation. 
(d) When the density is large ($L=5$) and noise is
small ($\eta=0.1$), the motion becomes ordered on a macroscopic scale
and all 
agents tend to move in the same spontaneously selected direction. 
After \citep{Vicsek-PRL-1995}.
  }    \label{fig:SVM}
\end{figure}

For small speed $v_0$, if we decrease the magnitude of perturbations
$\eta$, %
the Vicsek model 
displays a continuous %
phase transition from a disordered phase %
(zero average velocity $\phi$, implying that %
all agents move independently of each
other, Fig.~\ref{fig:SVM}b) to an ordered phase %
when almost all agents move in the same direction, through a
spontaneous symmetry breaking of the rotational symmetry
(Fig.~\ref{fig:SVM}d). 
This much studied kinetic phase transition takes place 
despite the fact
that each agent's set of nearest neighbors change with time as the
system evolves and the absence of centralized coordination. %

Numerical results indicate %
that the phase transition is second-order %
and the normalized average velocity $\phi$ %
scales as
\be
\phi \sim  [\eta_c(\rho) - \eta]^\beta, 
\ee 
where the critical exponent $\beta \approx 0.45$ and $\eta_c(\rho)$ is
the critical noise for $L\to \infty$~\citep{Vicsek-PRL-1995}. 
Many studies have explored %
the nature of the above phase
transition (whether it is first or second order), finding that %
two factors play an important role: (i) the precise way that the noise is
introduced into the system; and (ii) the speed $v_0$ with which the
agents move
~\citep{Gregoire-PRL-04,Aldana-PRL-07,Pimentel-PRE-08,ALDANA-IJMPB-09,Baglietto2009527}.    

\begin{figure}[t!]
  \begin{center}
   \includegraphics[width=0.5\textwidth]{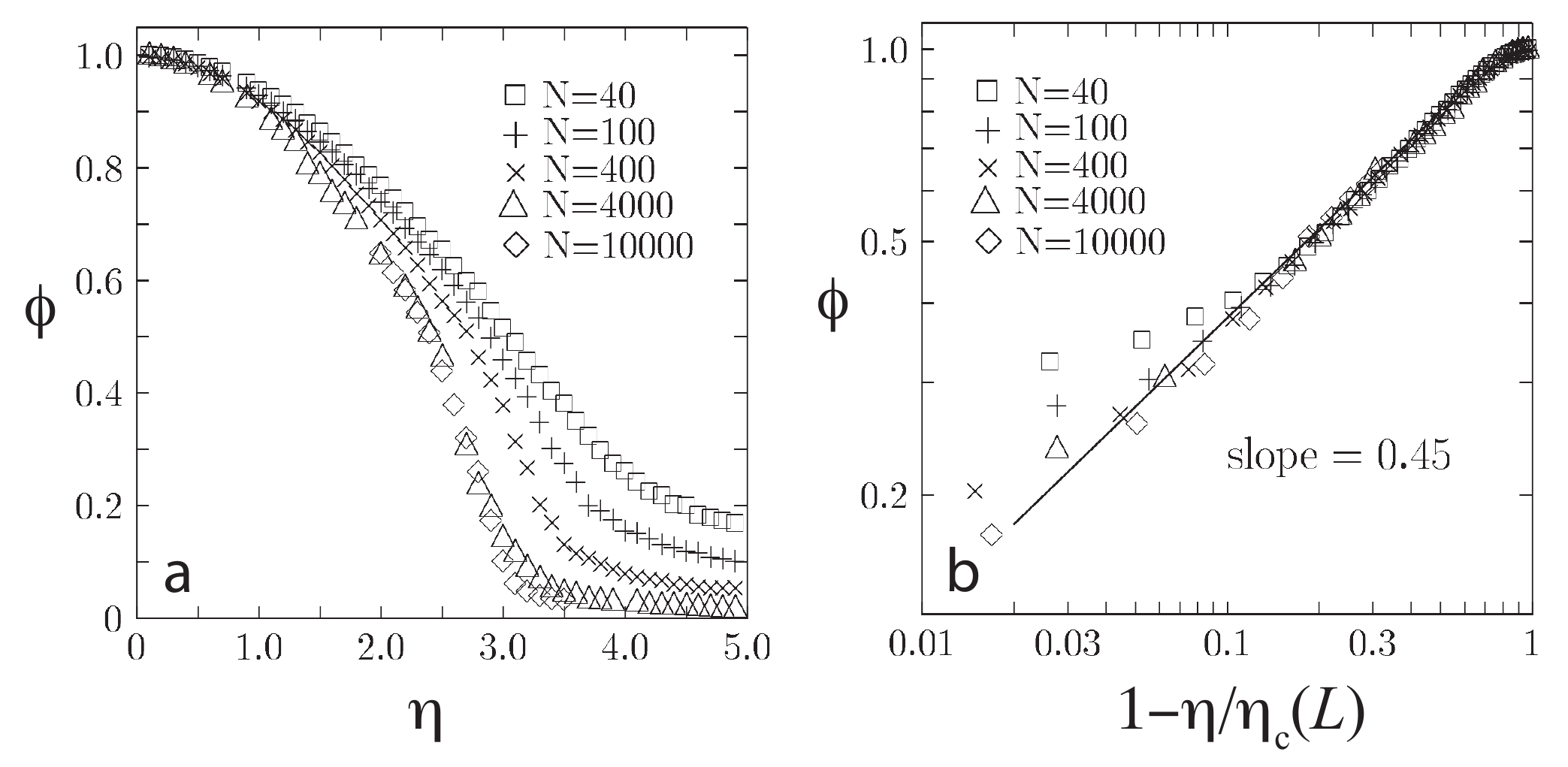}
  \end{center}
  \caption{%
Kinetic phase transition in the Vicsek model.
(a) The normalized average velocity ($\phi$) versus the magnitude of
perturbations (noise $\eta$) in cells of various sizes ($L$) with a fixed density
$\rho=N/L^2=0.4$.    
As $\eta$ decreases, $\phi$ increases, implying the emergence of order
in the Vicsek model. 
(b) Dependence of $\phi$ on $[\eta_\mm{c}(L)-\eta]/\eta_\mm{c}$ in
log-log scale. The slope of the lines is associated with the critical
exponent $\beta$ for which we get $\beta=0.45\pm
0.07$. 
The scaling behavior of $\phi$ observed in such a kinetic phase
transition is analogous to what we often observe in continuous phase
transitions in equilibrium systems. 
After \citep{Vicsek-PRL-1995}.
  }    \label{fig:SVM_scaling}
\end{figure}

The Vicsek model raises %
a fundamental control problem: Under
what conditions can the multi-agent system display a particular
collective behavior? %
Behind each flock of collectively moving agents, like biological
organisms or  
robots, there is 
a dynamically changing or temporal network, where two
agents are connected if they interact, e.g. if their distance is
under a %
certain threshold. 
Since the agents are moving, 
the network of momentarily
interacting units evolves in time in a %
complicated fashion. %

To offer a control theoretical explanation for the
emergence of the ordered phase in the Vicsek model, we %
consider 
the following updating rule %
~\citep{Jadbabaie-IEEE-2003}: 
\be
 \theta_i(t+1)= %
\frac{1}{1+k_i(t)} 
\left( \theta_i(t) + \sum_{j\in \mathcal{N}_i(t)} \theta_j(t) \right).
 \label{eq:Jadbabaie}
\ee
Though the scalar average in (\ref{eq:Jadbabaie}) is fundamentally different from the
vectorial average in (\ref{eq:UjThi}), this updating rule still
captures the essence of the Vicsek model in the absence of
perturbation.  
More importantly, (\ref{eq:Jadbabaie}) can be considered as 
a decentralized feedback
control system 
\be
 \boldsymbol\theta(t+1)= \boldsymbol\theta(t) + {\bf u}(t)
\ee
with the control input 
\be
{\bf u}(t) = - (\mD_{\sigma(t)} + {\bf I})^{-1} {\bf L}_{\sigma(t)} \boldsymbol\theta(t). \label{eq:SVM-u}
\ee
Here %
${\bf L}_p = \mD_p -\mA_p$ is the Laplacian matrix of graph $G_p$
with $p\in \mathcal{P}$. $\mA_p$ is the adjacency matrix of graph $G_p$ and $\mD_p$ is a 
diagonal matrix whose $i$th diagonal element is the degree of node $i$
in the graph $G_p$. $\sigma(t):{0,1,\cdots} \to \mathcal{P}$ is a switching signal whose
value at time $t$ is the index of the interaction graph at time $t$,
i.e., $G(t)$. 

If $r$ is %
small, some agents/nodes are always %
isolated, %
implying that $G(t)$ is never connected. 
If $r$ is %
large, then $G(t)$ is always a complete graph. %
The situation of %
interest is between the two extremes. 
The goal is to show that for any
initial set of agent directions %
$\boldsymbol\theta(0)$ %
and for a large class of switching signals 
the directions %
of all agents will converge to the same steady state $\theta_\mm{ss}$,
reaching %
alignment %
asymptotically. %
Mathematically, this means that %
the state vector $\boldsymbol\theta(t)$ converges to a
vector of the form $\theta_\mm{ss} {\bf 1}$ with $\theta_\mm{ss}$ %
the steady state direction, i.e.,  
\be
\lim_{t\to \infty}
\boldsymbol\theta(t) = \theta_\mm{ss} {\bf 1}, \label{eq:alignment}\ee 
where ${\bf 1} \equiv
(1,\cdots,1)_{N\times 1}^\mm{T}$, representing %
                                the case when %
all agents move in the same direction.

If $G(t)$ is connected for
all $t\ge 0$, then we can prove that %
alignment will be asymptotically reached~\citep{Jadbabaie-IEEE-2003}. %
But this condition %
is very stringent.  
It can be relaxed by considering that the agents %
are linked together across a time interval, i.e., the
collection or union of graphs encountered
along the interval is \emph{connected}. %
It has been proven that if the $N$ agents are linked together for each time
interval, then the alignment will be asymptotically reached~\citep{Jadbabaie-IEEE-2003}. %
This result has been %
further extended by %
proving that if the collection
of graphs is \emph{ultimately connected}, i.e., there exists an initial
time $t_0$ such that over the infinite interval $[t_0, \infty)$ the
union graph $\mathcal{G} = \cup_{t=t_0}^\infty G_t$ is connected, then the
alignment is asymptotically reached~\citep{Moreau-IEEE-05}. 

Though the control theoretical analysis~\citep{Jadbabaie-IEEE-2003,Moreau-IEEE-05,Ren-IEEE-2005} %
is deterministic, ignoring the presence of noise,  
it offers rigorous theoretical explanations, based on the
  connectedness of the underlying graph, for some fundamental aspects of
  the Vicsek model. For example, by applying the nearest 
neighbor rule, all agents tend to align %
the same direction
despite the absence of centralized coordination and despite the fact
that each agent's set of nearest neighbors changes in time. %
These control theoretical results suggest that to understand the
effect of additive noise, we should focus on how noise inputs effect
connectivity of the associated neighbor graphs. For example, %
the numerical finding %
that, for a fixed noise beyond %
a critical agent density %
all agents eventually become aligned, can be adequately explained by %
percolation theory of random graphs~\citep{Jadbabaie-IEEE-2003}. 
\subsubsection{Alignment via pinning} 
While the virtue of the Vicsek model is its ability to spontaneously reach an
ordered phase, %
we can also ask if such a
phase can be induced externally. Therefore, we 
consider %
an effective pinning control strategy in which %
a single %
pinned node
(agent) %
facilitates the alignment of the whole group. %
This is achieved by adding to 
the Vicsek model %
an additional
agent, labeled 0, which acts as the group's \emph{leader}. Agent 0
moves at the same constant speed $v_0$ as its $N$ \emph{followers} but with
a fixed direction %
$\theta_0$, representing %
the desired direction for the whole system. 
Each follower's neighbor set %
includes the leader whenever it 
is within the follower's %
circle of radius
$r$. Hence we have%
\be
 \theta_i(t+1)= \frac{1}{1+k_i(t)+b_i(t)} 
\left( \theta_i(t) + \sum_{j\in \mathcal{N}_i(t)} \theta_j(t) 
+ b_i(t) \theta_0
\right),
 \label{eq:Jadbabaie-leader}
\ee
where $b_i(t)=1$ whenever the leader is a neighbor of agent $i$ and 0
otherwise. 
It has been proved that if %
the $(N+1)$ agents are \emph{linked together} for each time interval, then %
alignment will be asymptotically
reached~\citep{Jadbabaie-IEEE-2003}. 
In other words, %
if the union of graphs of the $(N+1)$
  agents encountered along each time interval is \emph{connected},
  then eventually all the follower agents will align with the leader. 
 \subsubsection{Distributed flocking protocols}

Alignment, addressed by the Vicsek model, %
is only one component of %
flocking behavior. 
Indeed, there are three heuristic rules for flocking~\citep{Reynolds-87}: %
(i) \emph{Cohesion}: attempt to stay close to nearby flockmates;
(ii) \emph{Separation}: avoid collisions with nearby flockmates; and
(iii) \emph{Alignment}: attempt to match velocity with nearby
flockmates. 

We therefore need a general theoretical framework to %
design and analyze 
distributed flocking algorithms or protocols that embody these
three rules. 
The formal approach described next extracts the %
interaction rules that can ensure %
the emergence of flocking behavior~\citep{Olfati-Saber-IEEETAC-06}. 

\begin{figure}[t!]
  \begin{center}
   \includegraphics[width=0.5\textwidth]{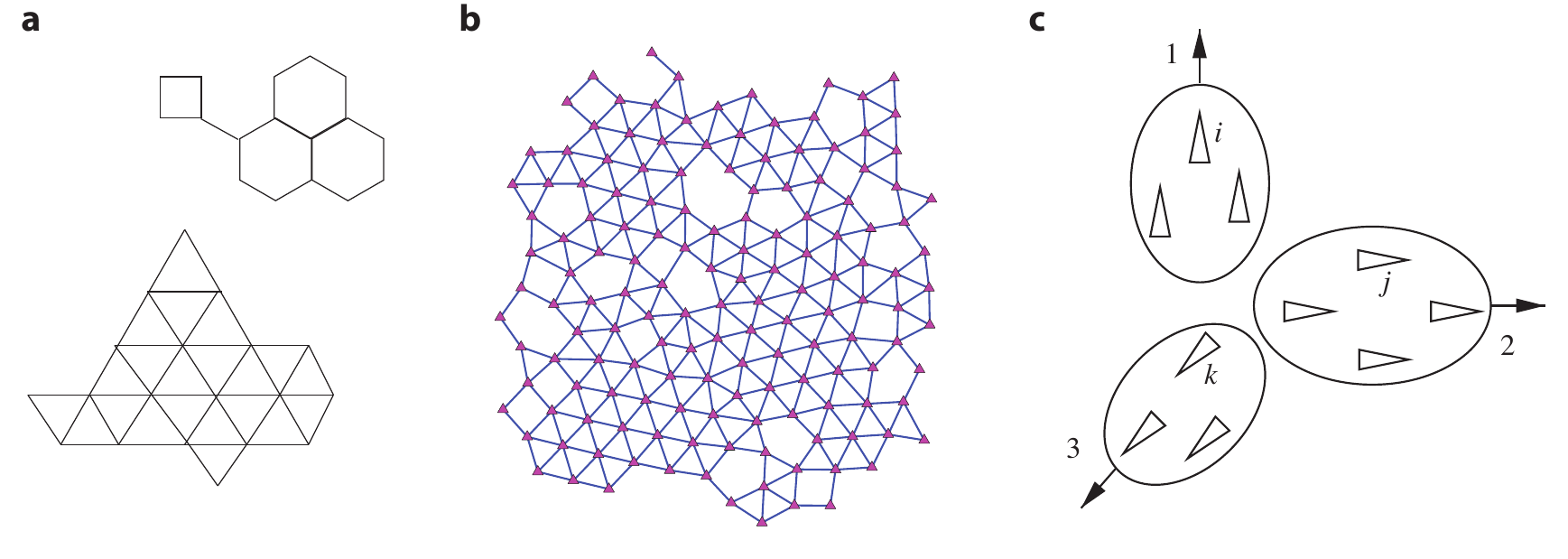}
  \end{center}
  \caption{Geometry of flocking and fragmentation.  
(a) Lattice-type flocking configuration in $D=2$. In this ideal case,
each agent is at the same distance from all of its neighbors on the
proximity graph. %
(b) A quasi-lattice for $D=2$ with $N=150$ nodes. %
(c) Fragmentation phenomenon, where agents merge form %
a few groups and
different groups are moving in %
different directions. This configuration will never
lead to flocking behavior. 
After \citep{Olfati-Saber-IEEETAC-06}.
  }    \label{fig:flocking_0}
\end{figure}

Consider a gradient-based flocking protocol equipped with a velocity
consensus mechanism, where each agent is steered by  
the control input %
\be
{\bf u}_i = {\bf f}_i^\mm{g} + {\bf f}_i^\mm{d}.
\label{eq:Reza-Flocking-1}
\ee
The first term %
 \be 
{\bf f}_i^\mm{g}\equiv - \nabla_{{\bf q}_i} V_i({\bf
   q})
\ee 
is gradient-based and %
regulates the distance between agent $i$ and its neighbors, %
avoiding the collision and cohesion of the agents. This term is derived from a
smooth collective potential function $V_i({\bf q})$, which has a
unique minimum when 
each agent is at the same distance %
from all of its neighbors on the proximity graph $G(\bf q)$,
representing %
the ideal case for flocking.   
The second term 
\be {\bf f}_i^\mm{d}=\sum_{j \in \mathcal{N}_i(t)}
a_{ij}(t) ({\bf p}_j - {\bf p}_i)
\ee regulates the velocity of agent $i$
to match the average velocity of its neighbors, being %
responsible for the velocity alignment. Here the weighted spatial adjacency matrix ${\bf
  A}(t)=[a_{ij}(t)]$ is calculated from the proximity network $G({\bf
  q})$. 
The flocking protocol (\ref{eq:Reza-Flocking-1}) embodies all three rules of
Reynolds. However, %
for a generic initial state and a 
large number of agents (e.g., $N>100$), the protocol (\ref{eq:Reza-Flocking-1})
leads to %
fragmentation, rather
than flocking%
~\citep{Olfati-Saber-IEEETAC-06}, meaning 
that the agents spontaneously form several groups, where %
different groups move in different directions (Fig.~\ref{fig:flocking_0}c). 
To resolve this fragmentation issue, we introduce a navigational feedback
term to the control input of each agent 
\be
{\bf u}_i = {\bf f}_i^\mm{g} + {\bf f}_i^\mm{d} + {\bf
  f}_i^\mm{\gamma}, 
\label{eq:Reza-Flocking-2}
\ee
where %
\be
{\bf f}_i^\mm{\gamma} =  -c_1 ({\bf q}_i - {\bf
  q}_\gamma) - c_2 ({\bf p}_i - {\bf
  p}_\gamma)
\label{eq:navigationalfeedback}
\ee drives agent $i$ to follow a group objective. The group objective can be considered
as a virtual leader with the following equation of motion: 
\be
\begin{cases}
\dot{\bf q}_\gamma &= {\bf p}_\gamma \\
\dot{\bf p}_\gamma &= {\bf f}_\gamma ({\bf q}_\gamma,{\bf p}_\gamma)
\end{cases},
\ee
where ${\bf q}_\gamma, {\bf p}_\gamma, {\bf f}_\gamma ({\bf
  q}_\gamma,{\bf p}_\gamma) \in \mathbb{R}^D$ are the position,
velocity, and acceleration (control input) of the virtual leader,
respectively. 
By taking into account the navigational
feedback, %
the protocol (\ref{eq:Reza-Flocking-2}) enables a
group of agents to track a virtual leader that moves %
at a constant
velocity, and hence leads to %
flocking behavior~\citep{Olfati-Saber-IEEETAC-06}. 

\begin{figure}[t!]
  \begin{center}
    \includegraphics[width=0.5\textwidth]{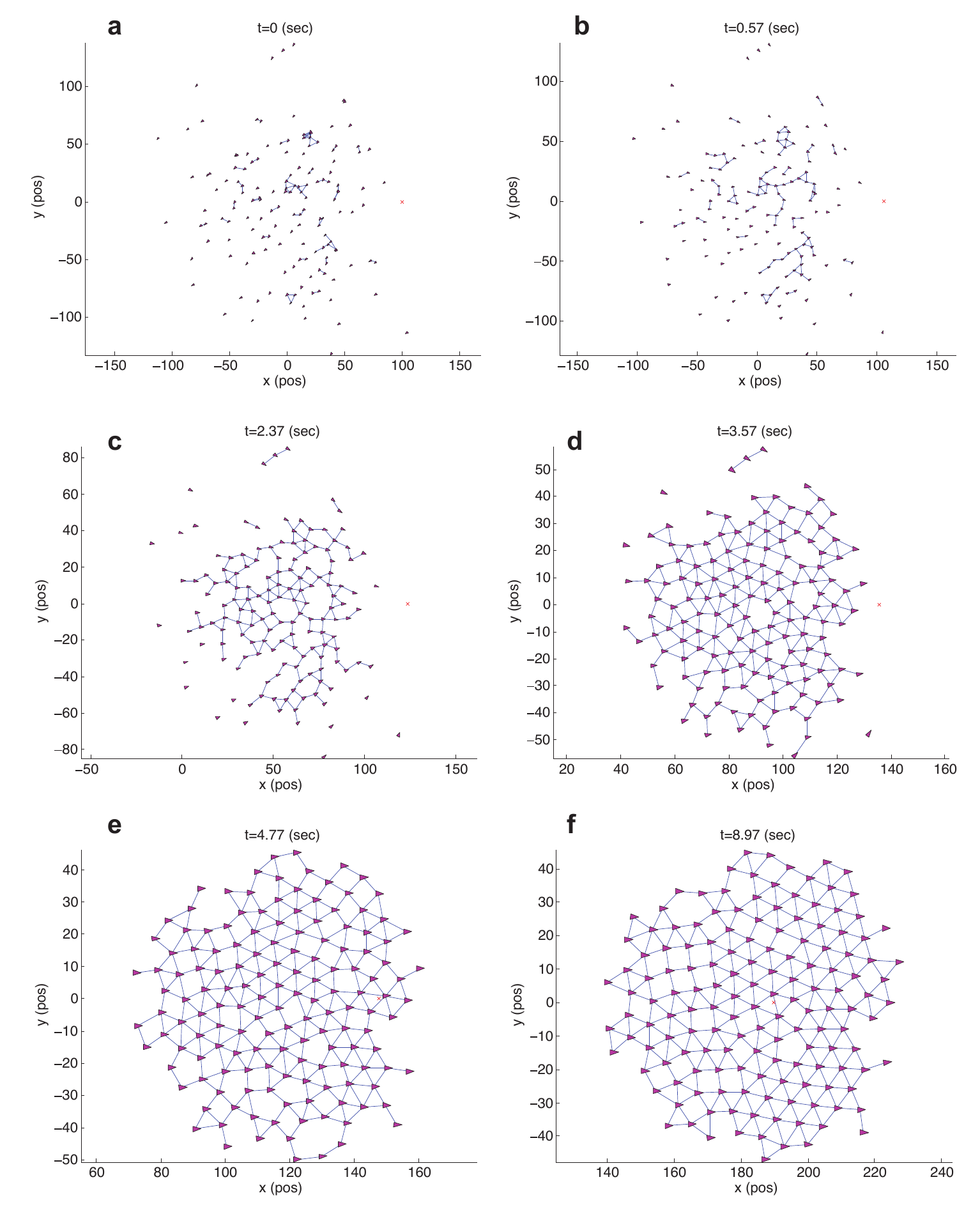}
  \end{center}
  \caption{(Color online) Flocking behavior in mutli-agent systems. 
After the application of 
the flocking algorithm
(\ref{eq:Reza-Flocking-2}) for a few seconds, the flocking of $N=100$
agents in 2$D$ is observed. After \citep{Olfati-Saber-IEEETAC-06}. 
  }    \label{fig:flocking}
\end{figure}

Note that protocol (\ref{eq:Reza-Flocking-2}) 
requires all agents to be \emph{informed}, i.e., to know the %
group objective, or equivalently, the current state $({\bf q}_\gamma,
{\bf p}_\gamma)$ of the virtual leader. 
It turns out that this is not necessary for flocking.   
Motivated by the idea of pinning control, %
it has been shown that, even when %
only a fraction of agents are %
informed (or pinned), the flocking protocol (\ref{eq:Reza-Flocking-2}) still
enables all the informed agents to move with the desired constant
velocity. An \emph{uninformed} agent will also move with the %
desired velocity if it can be influenced by the informed agents from
time to time~\citep{Su-IEEE-09}. 
Numerical simulations suggest that %
the larger the informed group is, the bigger
fraction of agents will move with the desired velocity~\citep{Su-IEEE-09}. 
If the virtual leader travels with a varying velocity ${\bf p}_\gamma(t)$, the
flocking protocol (\ref{eq:Reza-Flocking-2}) %
enables all agents to
eventually achieve a common velocity. Yet, this common velocity is not
guaranteed to match ${\bf p}_\gamma(t)$. 
To resolve this issue, we can incorporate the %
acceleration of the virtual leader into the navigational feedback
(\ref{eq:navigationalfeedback}) as follows 
 \be
 {\bf f}_i^\mm{\gamma} =  {\bf f}_\gamma({\bf q}_\gamma,{\bf p}_\gamma) -c_1 ({\bf q}_i - {\bf
   q}_\gamma) - c_2 ({\bf p}_i - {\bf
   p}_\gamma).
 \label{eq:navigationalfeedback2}
\ee
The resulting protocol enables the asymptotic
tracking of the virtual leader with a varying velocity, ensuring that
the %
position and velocity of the center of mass of all agents will
converge exponentially to those of the virtual leader~\citep{Su-IEEE-09}. 

In summary, the combination of control theoretical and
network science approaches can help us understand the
emergence of order in multi-agent systems. These tools are
indispensable if we wish to understand how to induce order externally, 
aiming %
to control the collective behavior of the system.

\section{Outlook}
Given the rapid advances in the %
control of complex
networks, we have chosen %
to focus on a group of results that will
likely stay with us for many years to come. The process of organizing
the material has %
also exposed obvious gaps in our knowledge.  
Therefore, %
next %
we highlight several 
research topics that %
must be addressed to realize the potential of the control of complex
systems. Some of these may be addressed shortly, others, however, may
continue to %
challenge the community %
for many years to come. %
\subsection{Stability of Complex %
Systems}%

Stability is a fundamental issue in the analysis and the design of a control
system, because an unstable system is %
extremely difficult and costly to control, and such a system can also be potentially
dangerous~\citep{Slotine-Book-91,Chen-inbook-2001}.  
Loosely speaking, a system is stable if its trajectories do not change
too much under small perturbations. 

The stability of a %
nonlinear dynamical systems $\dot{\mx} =
\mf (\mx, t)$ can be analyzed by the Lyapunov Stability
Theory (LST), 
without explicitly integrating the differential equation.  
LST includes two methods: (i) The indirect (or linearization) method, 
concerned with small perturbation %
around a system's %
equilibrium points $\mx^*$ 
and the stability
conclusion is inferred %
from a %
linear approximation of the nonlinear systems around this equilibrium
point. This justifies the use of linear control for the
design and analysis of weakly nonlinear systems.  
(ii) The direct method is based on the so-called Lyapunov function--- an
``energy-like'' scalar function whose time
variation can be viewed as ``energy dissipation''. It is not
restricted to small perturbations %
and in principle can be 
applied to any dynamical system. Yet, %
we lack a general theory to find a suitable Lyapunov function for
an arbitrary system. We have to rely on our experience and intuition
to formulate Lyapunov functions, like exploiting %
physical properties (such as energy conservation) and
physical insights 
~\citep{Slotine-Book-91}. 

For a wide range of complex %
systems certain diagonal-type
Lyapunov functions are %
useful for stability analysis~\citep{Kaszkurewicz-Book-00}. 
More importantly, in many cases the necessary and sufficient conditions
for the stability of nonlinear systems are also the necessary and
sufficient conditions for the diagonal stability of a certain matrix
associated to the nonlinear system. This matrix naturally captures the
underlying network structure of the nonlinear dynamical
system. 

Matrix diagonal stability is %
a well-known notion in stability
analysis since its introduction %
by Volterra around 1930 in the context of %
ecological systems~\citep{Volterra-1931}. Yet, its usefulness is %
limited by the difficulty of characterizing the class of large
diagonally stable matrices. %
Though there are %
efficient optimization-based algorithms to
numerically check if a given matrix is diagonally stable%
~\citep{Boyd-Book-1994}, 
there are no %
effective theoretical tools to characterize %
\emph{general} large
diagonally stable matrices. %
Recently, however, necessary and sufficient diagonal
stability conditions for matrices associated with \emph{special}
interconnection or network structures were 
studied~\citep{Arcak-Automatica-06, Arcak-MBE-08, Arcak-IEEETAC-11},  
improving our understanding of the
stability of gene regulatory and ecological networks. 
More research is required to understand stability, 
an important %
prerequisite for control. 

{\color{red}

The stability concepts we discussed above consider perturbations of
initial conditions for a \emph{fixed} dynamical system. There is
another important notion of stability, i.e.  
\emph{structural stability}, which concerns whether the
qualitative behavior of the system trajectories will be affected by
small perturbations of the system model
itself~\citep{Andronov-1937,Kuznetsov-Book-04}.  
To formally define structural stability, we introduce the concept of
topologically equivalence of dynamical systems. Two dynamical systems
are called \emph{topologically equivalent} if there is a homeomorphism
$h: \mathbb{R}^N \to \mathbb{R}^N$ mapping their phase portraits,
preserving the direction of time.  
Consider two smooth continuous-time dynamical systems  
(1) $\dot{\mx} = \mf (\mx)$; 
and
(2) $\dot{\mx} = \mg (\mx)$. 
Both (1) and (2) are defined in a closed region $D \in
\mathbb{R}^N$ (see Fig.~\ref{fig:structuralstability}). 
System (1) is called \emph{structurally stable} in a region $D_0
\subset D$ if for any system (2) that is sufficiently $C^1$-close to
system (1) there are regions $U,V \subset D$, and $D_0 \subset U$,
$D_0 \subset V$ such that system (1) is
topologically equivalent in $U$ to system (2) in
$V$ (see Fig.~\ref{fig:structuralstability}a).  
Here, the systems (1) and (2) are $C^1$-close if their ``distance'', defined as    
$
d_1 \equiv \sup_{x\in D} \left\{ 
    \| \mf(\mx) - \mg(\mx) \| 
+  \left\| \frac{\ud \mf(\mx)}{\ud \mx} - \frac{\ud \mg(\mx)}{\ud \mx} \right\|
\right\}
$ 
is small enough.

For a two-dimensional continuous-time dynamical system, the
Andronov-Pontryagin criterion offers sufficient and necessary
conditions for structural stablility~\citep{Andronov-1937}. 
A smooth dynamical system 
$\dot{\mx} = \mf (\mx), \mx \in \mathbb{R}^2$,  
is structurally stable in a region $D_0 \subset \mathbb{R}^2$ if and
only if (i) it has a finite number of equilibrium points and limit
cycles in $D_0$, and all of them are hyperbolic; 
(ii) there are no saddle separatrices returning to the same saddle
(see Fig.~\ref{fig:structuralstability} b,c) or
connecting two different saddles in $D_0$ (see
Fig.~\ref{fig:structuralstability}d).  
It has been proven that a typical or generic two-dimensional system
always satisfies the Andronov-Pontryagin criterion and hence is structurally
stable~\citep{Peixoto-1962}. In other words, structural stability is a
generic property for planar systems. Yet, this is
not true for high-dimensional systems.

For $N$-dimensional dynamical systems, Morse and Smale established the
sufficient conditions of structural stability
~\cite{Smale-1961,Smale-1967}. Such systems, often called Morse-Smale
systems, have only a finite number of equilibrium points and limit
cycles, all of which are hyperbolic and satisfy a transversaility
condition on their stable and unstable invariant manifolds.

\begin{figure}[t!]
\centering
\includegraphics[width=0.45\textwidth]{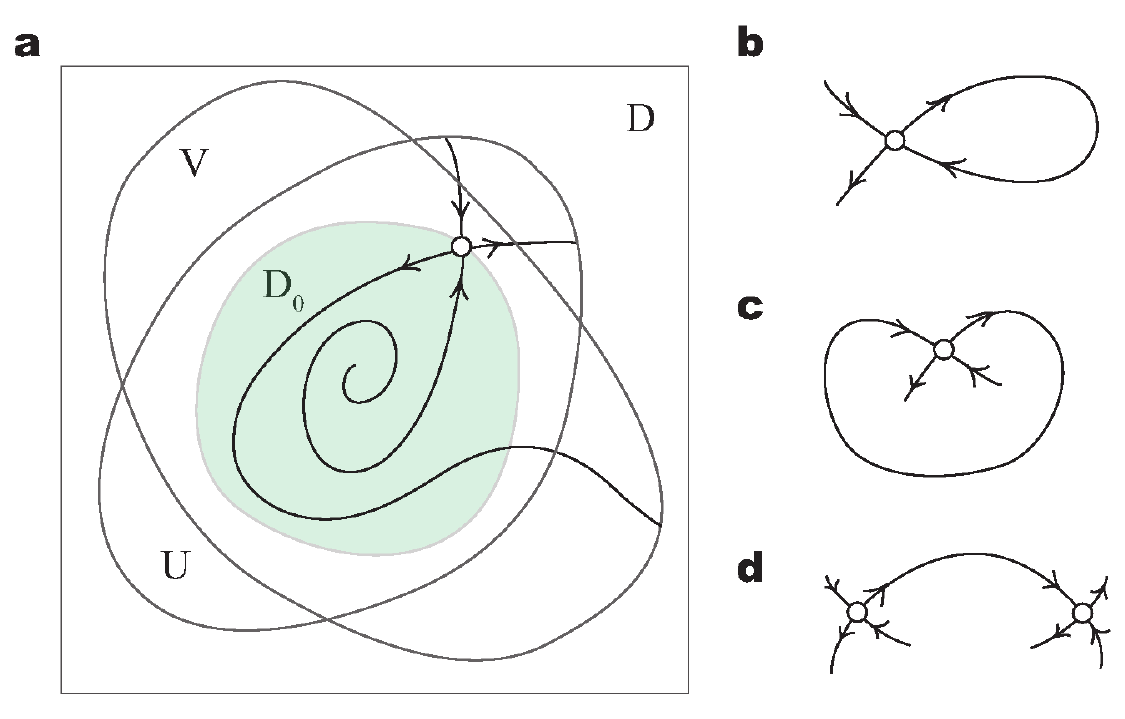}
\caption{(Color online)  {\bf Structural stability.} (a) Andronov's definition of
  structural stability. (b-d) Phase portraits of structurally unstable planar systems. 
This figure is redrawn from Figures 2.19 and 2.20 of
\citep{Kuznetsov-Book-04}. 
}\label{fig:structuralstability}
\end{figure}

The notion of structural stability has not been well explored in
complex networked systems. 

\iffalse
Recently, Rohr \emph{et al.} applied it to the
ecological system, where an interesting system
behavior is the stable coexistence of species --- the existence of an
equilibrium point that is both feasible (i.e. all species have
non-negative abdundance) and dynamically
stable~\citep{Rohr-Science-14}. In particular, they investigate the range of conditions
necessary for the stable coexistence of all species in mutualistic
systems. 
%
They recast the mathematical definition
of structural stability to that in which an ecological system is more structurally
stable, the greater the area of parameter values leading to both a
dynamically stable and feasible equilibrium. This implies that a
highly structurally stable ecological system is more likely to be 
stable and feasible by handling a wider range of conditions before the
first species becomes extinct. 
%
They show that the apparently
contradictory conclusions reached by previous studies  based on dynamical
stability and numerical simulations arise as a
consequence of overseeing either the necessary conditions for
persistence or its dependence on model parameterization. They further
show that the observed network architectures maximize the range of conditions for
species coexistence. The applicability of structural
stability to study other types of interspecific interactions, e.g., predator-prey is still
an open question.
\fi
}

\subsection{Controlling Adaptive Networks}
Adaptability, representing a system's ability to respond to changes in the external
conditions, is a key characteristic of complex systems. Indeed, the
structure of many real networks co-evolves with the dynamics that
takes place on them, naturally adapting to shifting environments~\citep{Gross-book-09}.

Adaptive networks, also known as state-dependent
dynamic networks in %
control theory~\citep{Mesbahi-IEEE-05,Mesbahi-Book-10}, are collections of units that interact
through a network, whose topology evolves as the state of the 
units changes with time. 
Adaptive networks are %
a special class of
\emph{temporal
  networks}, whose edges are not continuously
active~\citep{Karsai-PRE-11,Holme-PR-12,Pan-PLosONE-14,Marton-NJP-14}. 
If the temporal order of the network snapshots at different time
points depend on the states of the nodes, then the temporal network is
adaptive.   
A special case of %
adaptive networks are %
\emph{switched systems}, which 
consist of a family of subsystems
and a switching law that orchestrates the switching among
them~\citep{Xie-IEEETAC-02,Xie-SCL-03}. For switching systems, we can design the
switching signal among different subsystems and hence the switching law may
be independent from the states of the nodes. %

Mycelial fungi and acellular slime molds grow as
self-organized networks that explore new territory for food sources,
whilst maintaining an effective internal transport system to resist
continuous attacks or random damage~\citep{Fessel-PRL-12}.  
Honed by evolution, these biological networks are examples of adaptive
transportation networks, balancing 
real-world compromises between search 
strategy and transport efficiency%
~\citep{Tero-Science-10}.  

The genome is also an intriguing example of an adaptive %
network, where the chromosomal geometry directly relates to the genomic activity, which
in turn strongly correlates with geometry~\citep{Rajapakse-PNAS-11}. 
Similarly, neuronal connections (synapses) in our brains can
strengthen or weaken, 
and form %
in response to changes in brain activity, a
phenomenon called \emph{synaptic
  plasticity}~\citep{Bayati-PRE-12,Perin-PNAS-11}.  

A comprehensive analytical framework is needed %
to address the
control of %
adaptive, temporal and co-evolutionary networks. This framework must
recognize %
the network structure itself as a dynamical system, together with the
nodal or edge dynamics on the network, capturing 
the feedback mechanisms linking the structure and dynamics.   
Studying the controllability of such systems 
would be a natural starting point because seemingly mild limitations on either the network structure or
the dynamical rules may place severe constraints on the
controllability of the whole system~\citep{Rajapakse-PNAS-11}. 
Identifying these %
constraints is crucial if we want to refrain
from %
improving systems that already operate close to
their fundamental limits. 
\subsection{Controlling Networks of Networks}

Many natural and engineered systems are composed of a set of coupled
layers or a network of subsystems, %
characterized by %
different time scales and structural patterns. %
New notions, from \emph{multiplex
  networks}~\citep{Kivela-2014,Boccaletti-2014} to \emph{networks of
  networks}~\citep{DAgostino-Book-2014,Gao-NSR-14}, have been recently
proposed to explore the properties of these systems, %
focusing mainly on their structural integrity and robustness.  
Consider a multiplex network, i.e. a set of coupled layered
networks, whose different layers %
have %
different characteristics. %
We can model such a system as a 
layered network, whose interconnections between layers capture the interactions
between a node in one layer and its counterpart in another layer. %
Similarly, in a network of networks each node itself is a network or a
multi-input/multi-output (MIMO) subsystem. Different nodes/subsystems
could have totally different dimensions and %
dynamics. This is rather %
different from %
the control framework discussed in much of this paper, where
we typically assumed that all the nodes share the same type of dynamics or
even just scalar dynamics (with state variables $x_i \in \mathbb{R}$ for all
nodes).

Developing a framework to control %
networks of networks is a necessary
step if we wish to understand %
the control principles
of complex %
systems. 
Early attempts have focused %
on the issues of controllability or
observability with linear
dynamics~\citep{Yuan-NJP-14,Chapman-IEEETAC-14,Zhou-Automatica-2015,Wang-arXiv-15,Menichetti-15,Zhang-PRE-15}.  
For example, some controllability conditions on the overall network topology,
the node dynamics, the external control inputs and the inner
interactions have been derived for a %
networked MIMO
system~\citep{Wang-arXiv-15}. Interestingly, the controllability of
the networked MIMO system %
is an integrated result of multiple %
factors, which cannot be decoupled into the controllability 
of the individual subsystem or the properties solely determined by
the network topology. 
Despite these efforts, we %
lack a general framework to
systematically explore the control %
of networks of networks. 
Yet, the problem's importance will likely 
trigger more research in %
both network science and %
control theory. %

\subsection{Noise}

Complex systems, especially biological systems, are %
noisy. %
They are affected by %
two kinds of noise: the intrinsic randomness of
individual events and the extrinsic influence of changing environments~\citep{Lestas-Nature-10,Hilfinger-PNAS-11}. 
Consider, for example, regulatory processes in a cell. 
The intrinsic noise is rooted in the low copy number of biomolecules or diffusive cellular
dynamics. In particular, if $N$ is the number of 
molecules in the system, fluctuations in $N$ lead to %
statistical noise with intensity in the order of $N^{-1/2}$. For large $N$,
we can assume that a continuous deterministic dynamics effectively
describes the changes of the average concentrations. However, for small $N$ the
statistical noise cannot be ignored. For example, gene regulation may
be affected by large fluctuations due to the low copy number of
transcription factors. %
The extrinsic noise of a biological system is mainly due to the
changing environments experienced by the system. 
The environmental change may have microscopic origin (like cellular
age/cell cycle stage and organelle distributions) or can be %
related to the macroscopic physical or chemical environment (like illumination
conditions, temperature, pressure and pH level). %
To %
infer or reconstruct the states of a biological system, we also need
to deal with the measurement error, which is independent of the
biological system and can also be considered as extrinsic noise. 

Both internal and external noises are known to affect the
control of complex systems. 
At this time we %
lack a full understanding on the role of noise or stochastic
fluctuations on the control of complex systems.  

\subsection{Controlling Quantum Networks}
Quantum control theory aims to %
offer %
practical methods to control quantum
systems. 
Despite %
recent progress, %
quantum control theory is still
in its infancy~\citep{Dong-IET-10}, for %
several %
reasons. First, %
in classical control it is assumed that the measurement
does not affect the measured system. %
In contrast, in quantum control it is
difficult, if not impossible, to acquire information about quantum
states without destroying them. 
Second, some classes of quantum control tasks, like controlling
quantum entanglement and protecting quantum coherence, are unique for
quantum systems. In other words, there are no corresponding tasks
in classical control theory. %
The notion of quantum networks has been recently proposed by the quantum
information
community~\citep{Acin2007,Perseguers2008,Cuquet2009,Lapeyre2009,Czekaj2012,Perseguers2013,Perseguers-NP-2010,Perseguers2010,Cuquet2012,Hahn2012}, 
offering fresh perspectives in the field of
complex networks. 
In a quantum network, each node possesses exactly one qubit for each
of its neighbors. Since nodes can act on these qubits, they are often
called ``stations''. 
The edge between two %
nodes represents the \emph{entanglement} between two qubits. 
The degree %
of entanglement between
two nodes can be considered as the \emph{connection probability} ($p$)
in the
context of classical random graphs.

In a classical random graph %
if we let $p$ scale with the graph size %
as $p\sim N^z$, increasingly %
complex subgraphs appear as $z$ exceeds %
a series of
thresholds. For example,  for $z \le -2$ almost all graphs contain
only isolated nodes and edges. 
When $z$ passes through $-3/2$ (or $-4/3$), trees of order 3 (or 4)
suddenly appear. %
As $z$ approaches $-1$, trees and cycles of all orders appear~\citep{Albert-RMP-02}. 
Surprisingly, in quantum networks %
any %
subgraph can be
generated by local operations and classical communication, provided
that the entanglement between pairs of nodes scales with the graph size as $p~\sim
N^{-2}$~\citep{Perseguers-NP-2010}.    
In other words, thanks to the superposition principle and the ability
to coherently manipulate the qubits at the stations, even for the lowest non-trivial connection
probability that is just sufficient to get simple connections in a 
classical graph, we obtain 
quantum subgraphs of any complexity. 

This result illustrates that quantum %
networks have unique %
properties %
that are %
impossible in their classical counterparts.  
Hence, the control of %
quantum complex networks 
will %
require new methodologies. 

\subsection{Conclusion}

Revealing the control principles of complex networks %
remains %
a challenging problem that, given its %
depth and %
applications, will probably engage multiple %
research 
communities for the next decade.   
In this review we aimed %
to %
summarize %
in a coherent fashion the current body of knowledge on this
fascinating topic. This forced us to
explore 
key notions in control theory, like %
controllability
and observability, but also to explore %
how to steer a complex networked
system to a desired final state/trajectory or a desired collective
behavior. 
There are many outstanding open questions to be addressed, advances on
which will require %
interdisciplinary collaborations. %
We hope that this review will catalyze new interdisciplinary
approaches, %
moving our understanding of control forward and enhancing our ability
to control complex systems. %

 \begin{acknowledgments}
  We wish to thank Jianhua Xing, Haijun Zhou, Wenxu Wang, Tam\'{a}s
  Vicsek, Edward Ott, David Luenberger, Justin Ruths,
  Atsushi Mochizuki, Daniel J. Gauthier, Andrew Whalen, Adilson
  Motter, and Francesco Sorrentino for allowing us to reproduce their figures.
  We are grateful to Marco Tulio Angulo, Guanrong Chen, Gang Yan, 
  Aming Li, Zolt\'{a}n Toroczkai, Frank Schweitzer, S\'{e}rgio Pequito, Francesco
  Sorrentino, Atsushi Mochizuki, Bernold Fiedler, Jean-Jacques
  Slotine, Hiroki Sayama, Travis E. Gibson, and Chuliang Song for
  reading our manuscript and providing helpful suggestions.  
This research was supported by the John Templeton Foundation (award
\#51977); Army Research Laboratories (ARL) Network Science (NS)
Collaborative Technology Alliance (CTA) grant: ARL NS-CTA
W911NF-09-2-0053; European Union grant no. FP7 317532 (MULTIPLEX). 
\end{acknowledgments}

\bibliographystyle{apsrmp4-1}

\end{document}

